\documentclass[twocolumn, usenatbib]{mnras}
\usepackage{savesym}
\usepackage{graphicx}
\usepackage{longtable}
\usepackage{changepage}

\expandafter\let\csname equation*\endcsname\relax
  \expandafter\let\csname endequation*\endcsname\relax 
\usepackage{subfig}
\usepackage{amsmath}
\usepackage{amssymb}
\usepackage{verbatim}
\usepackage[yyyymmdd,hhmmss]{datetime}
\usepackage{array}
\usepackage{times}

\newcommand{\beq}{\begin{equation}}
\newcommand{\eeq}{\end{equation}}
\newcommand\tv {{t_{\rm visc}}}
\newcommand\M {{M_{\bullet}}}
\newcommand\tvc {{t_{{\rm visc}, c}}}
\newcommand\W {{W^r_{\ \phi}}}

\newcommand\Ntot{{63}} 
\newcommand\Nplat{{49}} 
\newcommand\Nnot{{14}} 
\newcommand\Nswift{{60}} 
\newcommand\Nztf{{46}} 

\title[TDE scaling relationships ]{ Fundamental scaling relationships revealed in the optical light curves of  tidal disruption events   }
\author [Andrew Mummery, et al. ]{Andrew Mummery$^1$\thanks{E-mail:
andrew.mummery@physics.ox.ac.uk}, Sjoert van Velzen$^2$, Edward Nathan$^3$, Adam Ingram$^4$, \newauthor
Erica Hammerstein$^5$, Ludovic Fraser-Taliente$^1$, Steven Balbus$^1$ 
\\
$^1$ Oxford Theoretical Physics, Beecroft Building,  Clarendon Laboratory, Parks Road, Oxford, OX1 3PU, United Kingdom \\
$^2$ Leiden Observatory, Leiden University, Postbus 9513, 2300 RA Leiden, The Netherlands. \\
$^3$ Cahill Center for Astronomy and Astrophysics, California Institute of Technology, Pasadena, CA 91125, USA. \\
$^4$ School of Mathematics, Statistics and Physics, Newcastle University, Herschel Building, Newcastle upon Tyne, NE1 7RU, UK \\
$^5$ Department of Astronomy, University of Maryland, College Park, MD 20742, USA. }
\begin{document}

\date{}

\pagerange{\pageref{firstpage}--\pageref{lastpage}} \pubyear{2023}

\maketitle

\label{firstpage}

\begin{abstract} 
We present fundamental  scaling relationships between properties of the optical/UV light curves of tidal disruption events (TDEs) and the mass of the black hole that disrupted the star.
We have uncovered these relations from the late-time emission of TDEs. Using a sample of \Ntot\ optically-selected TDEs, the latest catalog to date, we observed flattening of the early-time emission into a near-constant late-time plateau for at least two-thirds of our sources. Compared to other properties of the TDE lightcurves (e.g., peak luminosity or decay rate) the plateau luminosity shows the tightest correlation with the total mass of host galaxy ($p$-value of $2 \times 10^{-6}$, with a residual scatter of 0.3 dex).
Physically this plateau stems from the presence of an accretion flow. We demonstrate theoretically and numerically that the amplitude  of this plateau emission is strongly correlated with black hole mass. By simulating a large population ($N=10^6$) of TDEs, we determine a plateau luminosity-black hole mass scaling relationship well described by  $ \log_{10} \left({\M/M_\odot}\right) = 1.50 \log_{10} \left({ L_{\rm plat}}/10^{43} \, {\rm erg\,s^{-1}}\right) + 9.0 $ (here $ L_{\rm plat}$ is measured at $6\times 10^{14}$~Hz in the rest-frame).  
The observed plateau luminosities of TDEs and black hole masses in our large sample are in excellent agreement with this simulation. Using the black hole mass predicted from the observed TDE plateau luminosity, we reproduce the well-known scaling relations between black hole mass and galaxy velocity dispersion. The large black hole masses of 10 of the TDEs in our sample allow us to provide constraints on their black hole spins,  favouring rapidly rotating black holes.   
Finally, we also discover two significant correlations between early time properties of optical TDE light curves (the $g$-band peak luminosity and radiated energy) and the TDEs black hole mass.  
This relation allows black hole mass measurements to be made of all optical TDEs, including sources without a late-time plateau detection. 
\end{abstract}

\noindent

\section{Introduction} 
The tidal disruption, and subsequent accretion, of unfortunate stars by supermassive black holes at the centre of galaxies offers near unparalleled opportunities to study the properties and demographics of massive black holes in the local universe. If the multi-wavelength emission observed from these ``TDEs'' can be reliably used to constrain the parameters of the supermassive black holes at their heart, then the demographics of the local massive black hole population can be probed \citep[e.g.,][]{Frank_Rees76}, a technique that will be particularly powerful at the uncertain low mass end of the black hole mass function \citep{Stone16}. 

The low mass end of the supermassive black hole mass function is of particular interest, as this regime has the potential to provide unique insight into the evolution of black holes from the stellar mass to supermassive scales  \citep[e.g.,][]{Kormendy13, Shankar16}, along
with other questions of broad astrophysical interest including the possible importance of feedback for dwarf galaxies \citep{Silk2017, Bradford18}, and the dynamical evolution of dense stellar systems \citep[e.g.,][]{MillerHamilton02, Gurkan14}.
The low-mass end of the black hole mass function contains black holes which will be prime sources of gravitational radiation for upcoming gravitational wave detectors in space (Laser Interferometer Space Antenna, e.g., \cite{Amaro-Seoane15_LISA, Amaro-SeoaneLISA17}). To determine predicted detection rates of gravitational wave signals independent measurements of the black hole number densities are required. A knowledge of the background massive black hole population will also aide in interpreting these future gravitational wave signals. 
TDEs naturally probe this low mass end of the black hole mass function, owing to the inverted dependence of TDE rate on black hole mass \cite{Rees88}.

While detailed modelling of X-ray bright TDEs, both spectrally (e.g., \cite{Wen20}, \cite{Mummery_Wevers_23}), and in the time domain \citep{MumBalb20a}, results in constraints on the central black hole parameters which are typically consistent with those inferred from galactic scaling relationships, efforts to utilise the observed optical-UV emission from the TDE population have proven less successful. {While original analyses (e.g. \cite{Mockler19} and \cite{Ryu20}) argued that different models of the early time emission of TDEs produced black hole mass estimates ``consistent'' with galactic scaling relationships, these results where based on small sample sizes (9 and 12 TDEs respectively), and no statistical significance was quoted in these works. Later population studies with larger samples find low, or no, statistical significance.}
\citet{Hammerstein23} applied {these} two different  models to the light curves of 30 TDEs discovered by the Zwicky Transient Facility (ZTF; \citealt{Graham19}), and found no correlation between the black hole masses predicted by these models and the masses of their host galaxies.  This is despite a strong correlation being known to exist between central black hole mass and galactic mass \citep{Magorrian98,Ferrarese00}; see \citet{Greene20} for recent compilation. {Similarly, \cite{Ramsden22} found only a very weak correlation between 29 TDE black hole masses estimated from the early time optical emission and their host galaxy bulge masses. The gradient of the TDE-only scaling relationship was at a $5\sigma$ tension with the pre-existing galactic scaling relationship \citep[e.g.,][]{Kormendy13} while being only $2\sigma$ away from an anti-correlation.    }

Forthcoming optical surveys such as Rubin/LSST are expected to discover potentially tens of thousands of TDEs \citet{vanVelzen19_ZTF, Bricman20}. 
Motivated by this, in this paper we highlight how the {\it late time} optical/UV emission of a TDE can be used as a powerful probe of the central black hole's mass {\it and} spin. 

While the physical origin of the early time optical/UV emission from TDEs is still uncertain, with no consensus having been reached in the community \citep[for a review see][]{Roth20_ISSI}, the late time emission of TDEs is dominated by direct emission from an optically thick accretion flow \citep{vanVelzen18_FUV, MumBalb20a}.  This late time optical emission from TDEs is observed to undergo a plateau \citep{Brown17a, vanVelzen18_FUV}, becoming near time-independent, a result of the competing effects of the accretion flow cooling while also spreading to larger radii to conserve angular momentum \citep{MumBalb20a, Lodato11, Cannizzo90}.  

In this paper we demonstrate that the amplitude of the plateau emission (the {principal} observational degree of freedom in this time-independent phase) is strongly correlated with the central black hole's mass.  We demonstrate this fact analytically (section \ref{sec:2}, appendix \ref{disctheory}), and numerically (section \ref{sec:4}). This opens up the possibility of utilising this observed phase of emission to measure the masses of TDE black holes.  By simulating a large sample ($N = 10^6$) of TDE systems we demonstrate that an observation of the late time plateau luminosity of a TDE provides an estimate of the central black hole mass with typically $\sim 0.5$ dex of scatter. 

There is by now a large population ($\sim 10^2$  sources) of optically bright TDEs, many of which have sufficient coverage to extract the properties of this late time plateau phase. By measuring the late-time optical/UV light curves of all of these TDEs, we are able to extract the amplitude of the plateau of \Nplat\ sources (out of a total of \Ntot\ systems). This plateau luminosity correlates strongly with the mass of the TDE's host galaxy.  Utilising our numerical simulations of the late time plateau flux of TDE disc systems, we are able to measure the black hole masses of all of these \Nplat\ systems.   The black hole masses inferred using this technique correlate strongly with both the host galactic mass and galactic velocity dispersion, as
is expected for TDE host black hole masses. However,
these masses are typically at the  low mass end ($10^5 \lesssim M_\bullet/M_\odot \lesssim 10^7$) of the total black hole population, where there are very few measurements available by other techniques. 

Using the black hole masses of our TDE with plateau detections, we determine two empirical scaling relationships between early time TDE  optical/UV light curve properties (the peak luminosity and radiated energy) and the central black hole mass. This allows us to provide black hole masses for the remaining TDE sources without a clear plateau detection. We again find that these masses correlate  with both the host galactic mass, and galactic velocity dispersion.  

The layout of this paper is as follows. In section \ref{sec:2} we present the results of theoretical disc calculations of the correlation between the amplitude of the late time TDE plateau and central black hole properties (derived in appendix \ref{disctheory}).  In section \ref{sec:3} we summarise our simulation procedure, the results of which we present in section \ref{sec:4}.  In section \ref{sec:5} we analyse the properties of the \Ntot\ TDEs observed at optical/UV sources, extracting the plateau luminosities of \Nplat\ TDEs.  In section \ref{sec:6} we compare the predictions of disc theory and the observed plateau luminosities of the TDE population. In section \ref{sec:7} we present updated scaling relationships between observed TDE properties and central black hole parameters, and between galaxy properties and black hole parameters. We conclude in section \ref{sec:8}. Some technical results are presented in appendices. 

\section{ Disc Theory }\label{sec:2}
It is by now clear that TDEs settle down at late times into an evolving accretion flow which dominates the late time emission at both optical and ultra-violet frequencies. There is by now observational evidence for late-time disc emission from a growing sample of TDEs \citep{vanVelzen18_FUV}, these observations are well described by evolving relativistic discs \citep{MumBalb20a}.   First principles simulations of the stellar disruption itself \citep{Steinberg22} suggest that the circularisation of debris streams into a disc at times post the peak of optical/UV emission is a run away process, despite the complexity of the early time debris evolution. 

While detailed modelling of the disc-dominated X-ray spectra of TDEs has been used throughout the literature to infer the properties of  TDE  black holes (\citet{MumBalb20a, Wen20, Wen21, Mummery_Wevers_23}), only one TDE, namely ASASSN-15lh (\citet{Dong16, Leloudas16}), has  published black hole mass and spin constraints derived from disc modelling exclusively at optical/UV frequencies (\citet{MumBalb20b}).  

Naive steady state disc theory predicts an optical/UV disc luminosity with amplitude which scales as  $(\M \dot M)^{2/3}$, and is therefore highly degenerate between the (unknown) mass accretion rate $\dot M$, and the black hole mass $\M$ \citep{Frank02}. However, we shall demonstrate in this paper that the late time disc temperature profile in a TDE disc is much more highly constrained than this naive ``free-$\dot M$'' model, as the total initial mass, radial and temporal scales of the disc are known apriori for a given stellar disruption. This initial mass content must then propagate radially according to the standard constraints of mass and angular momentum conservation, and the late-time optical/UV luminosity of a {\it TDE  disc} is as a result strongly constrained. 

In Appendix \ref{disctheory} we derive the properties of the fully time-dependent thin disc optical/UV luminosity scaling relationships in both the mid-frequency and Rayleigh-Jeans spectral regions. The required spectral integrals can be solved analytically by taking the Newtonian limit, and simplifying the physics of the radiative transfer of photons through the disc atmosphere. For the remainder of this paper we will solve the disc and photon equations numerically in full general relativity, but these Newtonian calculations act as a useful benchmark which captures the important physics. In Appendix \ref{disctheory} we further demonstrate mathematically {\it why} disc-dominated TDE light curves are close to time-independent.  

Our theoretical results can be summarised as follows. In the mid-frequency part of the disc spectrum $\big($i.e., discs which are observed at frequencies that satisfy $kT_{\rm in}(t) \gg h\nu \gg k T_{\rm out}(t)$, where $T_{\rm in/out}$ are the inner (hottest) and outer (coldest) disc temperatures respectively$\big)$, we find 
\beq
\nu L_\nu \sim \M^{2/3}  \left({M_\star \over R_\star} \right) \beta \cos(i) .
\eeq
In this expression $M_\star$ and $R_\star$ are the mass and radius of the disrupted star respectively, $\beta$ is the orbital penetration factor (defined as the ratio of the disrupted star's tidal and pericentre distances  $\beta \equiv R_T/R_p \geq 1$; $\beta$ is also sometimes called the impact parameter), and $i$ is the disc-observer inclination angle.  Whereas in the Rayleigh-Jeans tail  $(h\nu \ll kT_{\rm out}(t))$, we find
\beq
\nu L_\nu \sim \M^{2/3}  M_\star^{-1/24}  R_\star^{7/8} \beta^{-7/8} \cos(i) .
\eeq
The key result here is that there are two factors which will determine the properties of an observed population of late time optical/UV TDEs: the general trend across the population will be dominated by the range of  black hole masses $\M$, while the inclination $\cos(i)$ and stellar properties $(M_\star, R_\star)$ will introduce scatter at fixed $\M$.  We stress that the stellar properties do not dominate the properties of a population of late time optical/UV TDEs at leading order. To see this, note that main sequence stars satisfy a mass-radius relationship of $R_\star \propto M_\star^{0.56}$ \citep{Kippenhahn90}, and therefore $\nu L_\nu \propto M_\star^x$, with $x = 0.44$ in the mid-frequency range, and $x \simeq 0.45$ in the Rayleigh-Jeans tail. As the stellar masses of those stars involved in TDEs are dominated by low-mass stars (e.g., \citet{Stone16}), which do not vary in mass by more than a factor of a few, the dominant trends across a population will be driven by $\M$ (most TDEs are expected to have $\beta \approx 1$). 

We therefore have a clear prediction from time-dependent disc theory: a population of TDEs observed at late times in the optical/UV will have plateau luminosities with dominant scaling given by the central black hole masses of the different events, and with scatter at fixed black hole mass dominated by the random inclinations of TDE systems, with an additional contribution from the different stellar properties involved in these events. 

This analytical Newtonian calculation motivates a full numerical calculation of the optical/UV plateau luminosity as a function of TDE system parameters. We describe this numerical calculation in the following section. 

\section{Simulation procedure  }\label{sec:3}
In this section we numerically simulate the late time optical/UV luminosity of a large population of TDEs. We introduce our procedure in detail below, but in brief, for each TDE we sample a set of stellar (mass and radius), orbital (penetration factor $\beta$) and black hole (mass and spin) parameters from pre-determined distributions, which determine the initial condition (mass and radial location) of the disc. We then solve the relativistic disc equations out to $t = 1000$ days, before observing the disc at a random inclination, at which time the UV luminosity is  recorded. 

In the following sub-sections we first introduce the fundamental equations to be solved, then our stellar, black hole and orbital parameter sampling procedures, before presenting our results. 

\subsection{The disc evolution equation}
The underlying disc model describes the evolution of the azimuthally-averaged, height-integrated disc surface density $\Sigma (r, t)$.    Standard cylindrical Kerr geometry Boyer-Lindquist coordinates are used: $r$ (radius), $\phi$ (azimuth), $z$ (height), $t$ (time), and $\text{d}\tau$ (invariant line element).  The contravariant four velocity of the disc fluid is denoted $U^\mu$ (related to coordinate $x^\mu$ by $U^\mu=\text{d}x^\mu/\text{d}\tau$); its covariant counterpart is $U_\mu$.  The specific angular momentum corresponds to $U_\phi$, a covariant quantity.     We assume that there is an anomalous stress tensor present, $\W$, due to low-level disk turbulence.   The stress is a measure of the correlation between the fluctuations in $U^r$ and $U_\phi$ (Balbus 2017), and could also include correlated magnetic fields.  As its notation suggests, $W^\mu_\nu$ is a mixed tensor of rank two.    

It is convenient to introduce the quantity $\zeta$,
\beq\label{z}
\zeta \equiv \sqrt{g}\Sigma \W / U^0=  r \Sigma \W/U^0,
\eeq
where $g>0$ is the absolute value of the determinant of the (mid-plane) Kerr metric tensor $g_{\mu\nu}$.  The Kerr metric  describes the spacetime external to a black hole of mass $\M$ and angular momentum $J$.  For our choice of (midplane) coordinates, $\sqrt{g}=r$.   The ISCO radius, inside of which the disc is rotationally unstable, is denoted as $r_I$.  Other notation is standard: the gravitation radius is  $r_g = G\M/c^2$, and the black hole spin parameter (with dimensions of length) is $a = J/\M c$. We denote the dimensionless black hole spin $a_\bullet \equiv a/r_g$. The Kerr metric describes black holes whenever $|a_\bullet| \leq 1$. 

Under these assumptions, the governing equation for the evolution of the disc may generally be written \citep{EardleyLightman75, Balbus17}:
\beq\label{fund}
{\partial  \zeta\over \partial  t} =  \mathcal{W} {\partial \ \over \partial  r}\left({U^0\over U'_\phi}    {\partial  \zeta \over \partial  r} \right), 
\eeq
where the primed notation denotes a radial gradient, and we have defined the stress-like quantity
\beq
\mathcal{W} \equiv  {1 \over (U^0)^2} \left(\W + \Sigma {\partial \W\over \partial  \Sigma} \right)  .
\eeq
The functional forms of $U^0, U_\phi'$, etc. relevant for the Kerr metric are listed in Appendix \ref{eq_sol_app}. This equation is the fundamental evolution equation for the disc surface density. See Appendix \ref{eq_sol_app} for further details on the approaches used to solve this equation. The observed properties of a TDE disc are determined  by the evolving temperature profile of the flow, which may be related to the disc surface density through the constraints of energy conservation. 

The dominant $r-\phi$ component of the turbulent stress tensor $\W$ serves to transport angular momentum outward as well as to extract the free energy of the disc shear, which is then thermalised and radiated from the disc surface.   In standard $\alpha$-disc modelling, which we follow here, both the the extraction and the dissipation are assumed to be local processes.
With these assumptions, the profile of the disc surface temperature $T$ is given by \citep{Balbus17}
\beq\label{temperature}
\sigma T^4 = \frac{3\sqrt{G\M}}{4\,r^{5/2}} \W \Sigma(r, t) \frac{1 + a \sqrt{r_g / r^3}}{\left(1 - {3r_g}/{r} + 2a\sqrt{{r_g}/{r^3}}\right)^{3/2}} .
\eeq
where $\sigma$ is the  Stefan-Boltzmann constant. 
Once equation (\ref{fund}) is solved for a given initial condition, the disc temperature profile is specified at all  radii at all future times (eq. \ref{temperature}). 

\subsection{The optical/UV luminosity integral}\label{lum_deriv}
\subsubsection{Photon orbits and ray-tracing }
\begin{figure}
  \includegraphics[width=.5\textwidth]{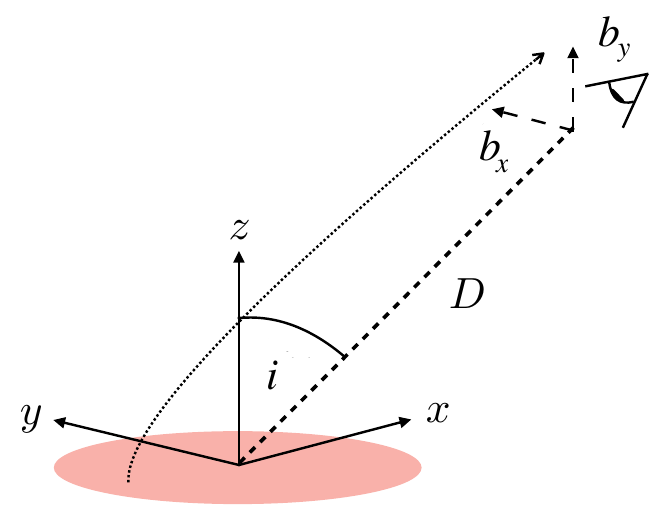} 
 \caption{Ray tracing geometry.  The coordinates $b_x$ and $b_y$ lie in the observer plane; $x$ and $y$  in the disc plane.   A schematic photon trajectory from the inner disc is shown.  The observer-disc inclination angle is denoted $i$.} 
 \label{RT}
\end{figure}

The specific flux density $F_\nu$ of the disc radiation, as observed by a distant observer at rest (subscript ${\rm obs}$), is given by
\beq
F_{\nu}(\nu_{\rm obs}) = \int I_\nu (\nu_{\rm obs} ) \, \text{d}\Theta_{{{\rm obs} }} .
\eeq 
Here, $\nu_{\rm obs} $ is the photon frequency and $I_\nu(\nu_{\rm obs} )$ the specific intensity,  both measured at the location of the distant observer.   The differential element of solid angle subtended on the observer's sky by the disk element is $\text{d}\Theta_{{{\rm obs} }}$. 
Since $I_\nu / \nu ^3$ is a relativistic invariant \citep[e.g.,][]{MTW}, we may write
\beq
F_{\nu}(\nu_{\rm obs} ) = \int f_\gamma^3 I_\nu (\nu_{\rm emit}) \, \text{d}\Theta_{{{\rm obs} }},
\eeq 
where we define the frequency ratio factor $f_\gamma$ as the ratio of $\nu_{\rm obs} $ to the emitted local rest frame frequency $\nu_{\rm emit}$:
\begin{equation}\label{redshift}
f_\gamma(r,\phi) \equiv \frac{\nu_{{\rm obs} }}{\nu_{\rm emit}} = {p_\mu U^\mu\ ({\rm Ob})\over p_\lambda U^\lambda\ ({\rm Em})}=\frac{1}{U^{0}} \left[ 1+ \frac{p_{\phi}}{p_0} \Omega \right]^{-1} ,
\end{equation}
where (Ob) and (Em) refer to observer and emitter, respectively.   The covariant quantities $p_\phi$ and $-p_0$ (on the far right) correspond to the angular momentum and energy of the {\em emitted photon} in the local rest frame.   These may be conveniently regarded as constants of the motion for a photon propagating through the Kerr metric.    Except for special viewing geometries, these quantities must in general be found by numerical ray tracing calculations (see  Appendix \ref{raytrace}). Our ray tracing geometry is summarised in Fig. \ref{RT}. 

\subsubsection{Radiative transfer in the disc atmosphere} 
The surface temperature of the disc $T(r,t)$ is given by the constraints of energy conservation (equation \ref{temperature}), and corresponds physically to the temperature of the disc surface at a height above the midplane where the optical depth of the disc equals 1.  It is important to note, however, that the disc's central temperature is given  by \citep[e.g.,][]{Frank02} 
\beq
T_c^4 = {3\over 8} \kappa \Sigma T^4 ,
\eeq
where for standard astrophysical parameters $\kappa \Sigma \gg 1$ \citep[e.g.,][]{Shakura73}. 
This result highlights that the energy of the disc photons produced in the disc midplane is higher than the surface temperature, taking a value roughly $E_\gamma \sim kT_c$. Only if the liberated disc energy can be fully thermalised in the disc atmosphere do the photons emerge with temperature $T$ (eq. \ref{temperature}). On their path through the disc atmosphere, photons can either be absorbed and re-emitted (thus totally thermalising their energy), or they can undergo elastic scattering. Elastic scattering however, by definition, does not change the energy of the photon, and so if this process dominates in the disc atmosphere, photons will be observed to have the “hotter” temperatures associated with the altitudes closer to the disc midplane, not the disc’s $\tau = 1$ surface. This modifies the emergent disc spectrum, a result which is typically modelled with a so-called colour-correction factor $f_{\rm col}$, which  can be thought of as quantifying the relative dominance of these two different opacities in the disc atmosphere.

More precisely, the specific intensity of the locally emitted radiation is assumed to be given by a modified Planck function $B_\nu$, of the form 
\begin{align}\label{planck}
I_\nu(\nu_{\rm emit}) &= f_{\rm col}^{-4} B_\nu(\nu_{\rm emit},f_{\rm col} T) \\ &= \frac{2h\nu_{\rm emit}^3}{c^2f_{\rm col}^4 } \left[ \exp\left( \frac{h\nu_{\rm emit}}{k f_{\rm col} T} \right) - 1\right]^{-1} .
\end{align}
Note that the normalising factor $1/f_{\rm col}^{4}$ here ensures that, despite the temperature of  the emission being increased, the total (integrated over all frequencies)  emitted luminosity remains $\sigma T^4$ (and therefore energy is conserved).  The value of the colour-correction factor $f_{\rm col}$ depend on the local properties of the emitting region. In this work we use the \citet{Done12} model,  which we summarise below. 

For the lowest disc temperatures, below a critical temperature $T = 3\times10^4 {\rm K}$,  Hydrogen is neutral and the Hydrogen absorption opacity is extremely  large. This results in the full thermalisation of the liberated disc energy, meaning that the emitted disc spectrum is well described by a pure blackbody function with temperature $T$, i.e., 
\beq\label{col3}
f_{\rm col}(T) = 1, \quad T(r, t) < 3\times 10^4\,  {\rm K} . 
\eeq
As the temperature increases above $3 \times 10^4$ K, the colour correction factor begins to increase. This results from the growing  ionisation fraction's  of both Hydrogen and Helium, which acts to reduce the total disc absorption opacity. As a result the electron scattering opacity begins to dominate, more photons are scattered out of the disc atmosphere,  and the typical temperature of observed photons increases.  \citet{Done12} model the colour correction factor in this regime as 
\beq\label{col2}
f_{\rm col}(T) =  \left(\frac{T}{3\times10^4 {\rm K}} \right)^{0.82}, \, \quad 3\times10^4 \, {\rm K} < T(r, t) < 1\times10^5 \, {\rm K} .
\eeq
It should be noted that the choice of temperature index and normalisation in this expression were set so that the colour-correction factor was equal to $1$ at $T = 3 \times 10^4$ K, and was continuous in joining onto the Compton-scattering regime discussed below, and was not determined by fundamental atomic physics. This parameterisation did however accurately reproduce the results of full radiative transfer simulations \citep{Done12}. 
For the highest disc temperatures $T > 1\times10^5 {\rm K}$, electron scattering completely dominates the absorption opacity and the colour correction factor begins to saturate, a result of Compton down-scattering in the disc atmosphere, to
\beq\label{col1}
f_{\rm col}(T) =  \left(\frac{72\, {\rm keV}}{k_B T}\right)^{1/9}, \quad  T(r, t) > 1\times10^5 \, {\rm K}. 
\eeq
This saturation leads to a maximum value of $f_{\rm col} \approx 2.7$.

\subsubsection{The optical/UV spectral integral}
For an observer at a large distance $D$ from the source, the differential solid angle into which the radiation is emitted is
\beq
 \text{d}\Theta_{{{\rm obs} }} = \frac{\text{d}b_x \, \text{d} b_y}{D^2} ,
\eeq 
where $b_x$ and $b_y$ are the impact parameters at infinity \citep{Li05}.  See {Fig.\  (\ref{RT})}  for further details.   
The observed flux from the disc surface ${\cal S}$ is therefore formally given by
\beq\label{flux}
F_\nu(\nu_{\rm obs} ,t) = {1\over D^2} \iint_{\cal S} {f_\gamma^3 f_{\rm col}^{-4} B_\nu (\nu_{\rm obs} /f_\gamma , f_{\rm col} T)}\,  {\text{d}b_x \text{d} b_y} .
\eeq
Note that $f_\gamma$ will generally depend upon $b_x$ and $b_y$.   With $T$ given by equations (\ref{fund}) and (\ref{temperature}),  ray tracing calculations determining  $f_\gamma(b_x,b_y)$, and the colour-correction given by eqs. (\ref{col3}--\ref{col1}) the observed spectrum may be obtained with all relativistic effects (kinematic and gravitational Doppler shifts, and gravitational  lensing) included.  The late-time optical/UV luminosity is then given by $\nu L_\nu \equiv 4\pi D^2 \nu F_\nu$, or explicitly 
\beq\label{lum}
\nu L_\nu = 4\pi \nu _{\rm obs}  \iint_{\cal S} {f_\gamma^3 f_{\rm col}^{-4} B_\nu (\nu_{\rm obs} /f_\gamma , f_{\rm col} T)}\,  {\text{d}b_x  \text{d} b_y} .
\eeq

\subsection{Parameter distributions }
Once a set of stellar, orbital and black hole parameters are specified, the late time optical/UV luminosity is determined as described above. In this sub-section we discuss how we  sample different black hole, stellar and orbital parameters.   

\subsubsection{Stellar parameters }
Stars of different  masses are both formed at intrinsically different rates, but are also tidally disrupted at intrinsically different rates, due to their differing structures.  The total rate at which different stars will appear in a population of TDEs is then given by the product of the stellar mass function and intrinsic TDE rate function.  

We use the Kroupa initial mass function \citep{Kroupa01}  to determine the intrinsic rate at which stars of different masses are formed. We assume that this equals the probability of a given star existing in the galactic centre. As the plateau luminosity is only a weak function of stellar mass/radius, any differences between the Kroupa IMF and the present day mass function of galactic centres should have minimal effect on our results.  The Kroupa IMF takes the form of a multiply broken power-law, with each power-law section taking the form
\beq
p_{\rm IMF}(M_\star) \propto M_\star^{k_i}.
\eeq
The values of $k_i$ are the following: $k_1 = -1.8$ for $M_\star < 0.5 M_\odot$; $k_2 = -2.7$ for $0.5 M_\odot < M_\star < M_\odot$; and $k_3 = -2.3$ for $M_\star > M_\odot$. We do not include stars with masses $M_\star < 0.08 M_\odot$ in our sample. The intrinsic rate at which TDEs occur for different stellar parameters, for a given black hole mass and spin,  is a quantity which may be calculated theoretically. We use the rate calculation of \citet[][see also \citet{Magorrian99, Rees88}]{Wang04}  whereby the intrinsic rate of tidal disruptions scales as 
\beq
p_{\rm rate}(M_\star, R_\star) \propto M_\star^{-1/3} R_\star^{1/4} .
\eeq
In other words this calculation encapsulates  the intuitive result that more massive stars are harder to disrupt, but stars with larger radii are easier to disrupt. The masses and radii of stars on the main sequence are related. We use the mass-radius relationship of \citet{Kippenhahn90} 
\beq\label{star_mass}
R_\star \propto 
\begin{cases}
R_\odot \left({M_\star / M_\odot}\right)^{0.56}, \quad M_\star \leq M_\odot, \\
\\
R_\odot \left({M_\star / M_\odot}\right)^{0.79}, \quad M_\star > M_\odot. 
\end{cases} 
\eeq
Note that we do not include giant (evolved) stars in our analysis, which are significantly more rare than main sequence stars. 
This then allows us to determine the rate at which stars of different masses enter our TDE distribution, namely:
\beq\label{pstar}
p_\star(M_\star) \propto p_{\rm IMF}(M_\star) \times p_{\rm rate}(M_\star, R_\star(M_\star)) . 
\eeq
We sample stellar masses for our simulation from $p_\star$, and then use Eq. (\ref{star_mass}) to compute stellar radii. 

\subsubsection{Black hole parameters }\label{BH_DIST}
A Kerr black hole is entirely described by just two parameters, the black hole's mass and spin. For this simulation, we are aiming to understand the properties of the optical/UV plateau luminosity {\it as a function of black hole mass. } As such, we take a completely agnostic distribution of black hole masses, assuming that they are uniformly distributed between $10^4$ and $10^9$ solar masses:
\beq\label{pbhm}
p_{\M} \propto 1, \quad 10^4 < \M/M_\odot < 10^9.
\eeq
This is of course not a realistic description of the population of black holes expected to be involved in TDEs (we stress that this is not the purpose of the chosen distribution). 

We choose an equally agnostic black hole spin distribution, again assuming a flat distribution covering the entire range of possible values:
\beq \label{pbha}
p_{a_\bullet} \propto 1 , \quad -1 < a_\bullet < 1.
\eeq
We will demonstrate in later sections that the black hole spin is not an important parameter when it comes to the direct production of the late-time disc luminosity (i.e., the optical/UV luminosity of the disc is an extremely weak function of black hole spin), however the black hole spin is a fundamentally important parameter for determining which black holes may tidally disrupt stars of a given mass, as we discuss below. 

\subsubsection{Orbital parameters }\label{beta}
Any orbit of a star about a central black hole which takes the star within its tidal radius will result in a tidal disruption.   However, there are broadly two orbits on which a  star can enter this tidal region. The first is through the slow diffusive evolution of it's orbital energy and angular momentum by the many-body gravitational interactions of the galactic centre, where the orbital pericentre slowly reaches the tidal radius over many orbits. The second are more extreme so-called ``pinhole'' events, when the star is scattered onto an orbit with pericentre potentially much smaller than the tidal radius, having previously resided on a ``safe'' orbit.  

The probability of a given TDE occurring from each type of stellar orbit depends on the properties (density profile, velocity dispersion, etc.) of a given galaxy.  \citet{Stone16} analysed a population of 144 galaxies, computing the ``pinhole fraction'' for each galaxy. This quantity encapsulates the probability that a given TDE in that galaxy would occur via the pinhole route, as opposed to the diffusive route. By using a $\M-\sigma$ relationship \citet{Stone16} related this fraction to the black hole masses at the centre of the galaxy, finding an empirical relationship given by: 
\beq
f_{\rm pinhole} = \min\left[ 1, 0.22 \left(\M /10^8M_\odot\right)^{-0.307} \right] , 
\eeq
Note that in this model all TDEs around black holes with masses $\M \lesssim 7 \times 10^{5} M_\odot$ are ``pinhole'' TDEs. In our simulation, once a black hole mass $\M$ has been sampled, we pick orbital penetration parameters, defined as 
\beq
\beta \equiv {r_T \over r_{\rm peri}} \geq 1,
\eeq
as follows. With probability $f_{\rm pinhole}$ we sample a pinhole TDE, with penetration factor probability distribution given by 
\beq
p_{\rm pinhole}(\beta) \propto 1/\beta^2 , \quad \beta > 1, 
\eeq
\citep[see][for a discussion of why this is the relevant probability distribution for $\beta$]{Stone16}. With probability $1 - f_{\rm pinhole}$ we sample a diffusive TDE, which always results in $\beta = 1$, i.e.,  
\beq
p_{\rm diffusive }(\beta) \propto \delta(\beta - 1). 
\eeq
With $\beta$, the black hole and stellar parameters all sampled, the pericentre distance of the star's orbit can be calculated. For a tidal disruption event to produce observable electromagnetic emission, this radius must be exterior to the black hole's event horizon, as we now discuss.

\subsubsection{The tidal radius and Hills mass}\label{hill_mech}
In an approximate  Newtonian framework the tidal radius represents the black hole mass at which the differential tidal force of the black hole on the object $(F_T)$ becomes equal to the self-gravity of the object $(F_g)$
\beq
F_T \simeq {G \M M_\star R_\star \over r^3} ,\quad F_g \simeq {G M^2_\star \over R^2_\star } ,
\eeq
in other words
\beq
r_T \simeq R_\star \left({\M \over M_\star}\right)^{1/3} .
\eeq
Within this radius tidal disruption events occur.  For a given value of $\beta$ (see above), the pericentre of a tidally disrupted star's orbit is $r_T/\beta$, which must be outside of the black hole's event horizon to be observed. It is important to note however that the tidal radius only grows as $r_T \propto \M^{1/3}$, while the event horizon of the black hole grows linearly with it's mass
\beq
r_E = {G \M \over c^2} \left(1 + \sqrt{1 - a_\bullet^2}\right).
\eeq
There is therefore a maximum black hole mass, known as the Hills mass \citep{Hills75}, where even a $\beta = 1$ disruption will occur within the black hole's event horizon and be unobservable. 

The Hills mass represents the black hole mass at which the differential tidal force of the black hole on the object $(F_T)$ becomes equal to the self-gravity of the object $(F_g)$, precisely at the black hole's event horizon ($r_E$). Solving the above set of equations we find the Newtonian estimate for the Hills mass
\beq
{\M_H } =  \left({ c^6 R^3_\star \over 8 G^3 M_\star  } \right)^{1/2} \sim 10^8 M_\odot,
\eeq
where we have used $M_\star \simeq M_\odot, R_\star \simeq  R_\odot$ in forming this final numerical value. 

In a fully relativistic framework, \citet{Kesden12} derived the Kerr black hole spin dependent  tidal radius, incorporating  the effects of the increasing tidal force of a rotating black hole's spacetime. For our simulations we use the \citet{Kesden12} value of the tidal radius, which is a function of black hole mass and spin, and stellar mass and radius. {We restrict our attention to incoming stellar orbits in the black hole's equatorial plane, which drastically simplifies the relativistic calculations while maintaining the key dependence on the black hole's spin \citep{Marck83}}. We then compute the pericentre distance of the star's orbit, using the sampled parameter $\beta$. For a TDE to produce observable emission, the pericentre of the star's orbit must be exterior to the IBCO (innermost bound circular orbit) radius of the black hole.  The IBCO radius is given by 
\beq\label{ibco}
r_{\rm ibco} = {G\M\over c^2} \left(1 + \sqrt{1-a_\bullet}\right)^2 ,
\eeq 
and represents the limiting radius which separates parabolic test particle orbits which escape to infinity as $t \to + \infty$, and those which terminate at the singularity of the black hole \citep[e.g.,][]{Chandrasekhar83}. If the pericentre orbit of the incoming star is smaller than the IBCO radius, then even if the disruption occurs outside of the black hole's event horizon, all of the stellar debris will quickly cross the black hole's event horizon, producing minimal emission. We therefore do not simulate TDE systems where 
\beq
{1\over \beta }r_T \leq r_{\rm ibco} . 
\eeq

As the IBCO grows linearly with black hole mass, a limiting Hills mass can be determined for the general Kerr metric. 
Although not presented explicitly in \citet{Kesden12}, the exact limiting Hills mass can be written in closed form, and is given by  the remarkably simple formula 
\begin{equation}\label{fullhills}
\M_H =\left({5 c^6 R^3_\star \over G^3 M_\star} \right)^{1/2}  {1 \over \left(1 + \sqrt{1-a_\bullet}\right)^3} .
\end{equation}
The simplification of the \citet{Kesden12} analysis used here is noting that the limiting tidally disrupt-able orbit of a star is that of the parabolic orbit with angular momentum equal to that of the innermost bound circular orbit, evolving in the equatorial plane.   Note that the ratio of the Hills masses of a maximally rotating and Schwarzschild black hole are related exactly by a factor 8
\beq
{\M_H(a_\bullet = 1) \Big/ \M_H(a_\bullet = 0) } = 8. 
\eeq  
We therefore expect high black hole mass {\it observed} TDEs to be dominated by rapidly rotating black holes, even for our assumption of a uniform distribution  of the background population of supermassive  black hole spins. 

{Finally, note that in this paper we do not include partial TDEs in our sample. Partial TDEs are events characterised by stellar orbits with pericentre greater than the tidal radius ($\beta < 1$). If the pericentre radius is only slightly larger than the tidal radius, some stellar material (presumably from the outer less-bound layers of the star) may still be stripped off and potentially observational emission produced. We stress that these events are not included for modelling convenience and not because there is some reason that partial TDEs will not produce an accretion flow and a late time plateau.  It is unclear in a partial TDE how the fraction of stellar debris that forms into a disc $f_d$ will depend on the orbital penetration factor $\beta$. In Appendix \ref{disctheory} we demonstrate that the Rayleigh-Jeans flux from a TDE disc  scales as  }
\begin{equation}
    \nu L_\nu \sim f_d^{1/4} \beta^{-7/8} , 
\end{equation}
{and as $f_d$ will decrease for $\beta < 1$ it appears unlikely that partial TDEs will have drastically lower plateau fluxes than full TDEs. While we do not expect the inclusion of partial TDEs to modify the results of the simulations substantially this possibility certainly warrants future study.  }

\begin{figure}
\includegraphics[width=0.5\textwidth]{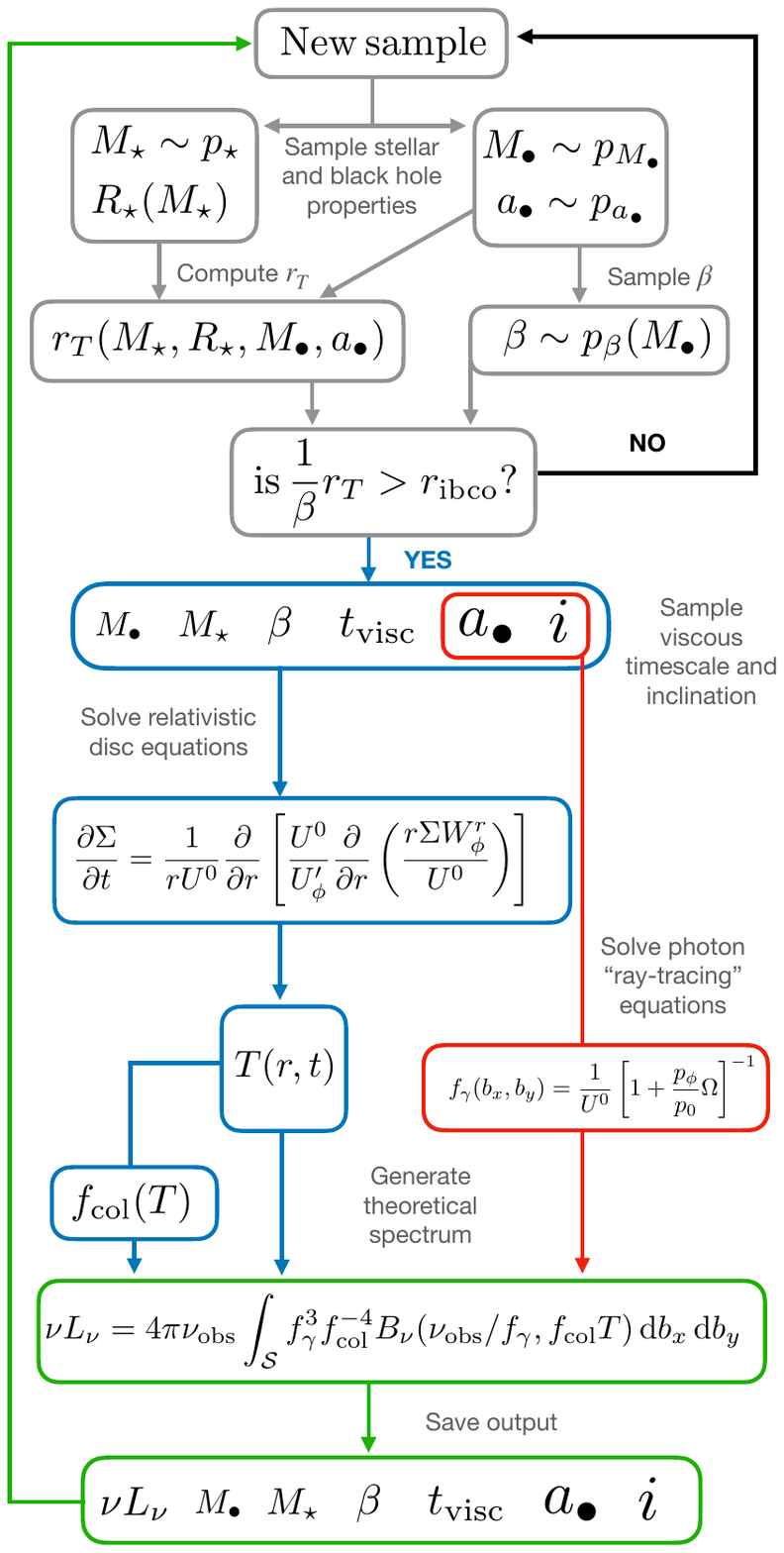}
\caption{A schematic of each step of the population simulation procedure. There are 4 main computations involved, the first (denoted by grey boxes and arrows) regards determining whether or not electromagnetic emission will be observable for a given set of system parameters. The second (blue boxes and arrows) involves solving the relativistic disc equations, and determining the late-time disc temperature profile. Solving the photon geodesics is the third required computation (red boxes and arrows). Finally, we compute the late time luminosity, which we save along with the system parameters (green boxes and arrows).      }
\label{SimProc}
\end{figure}

\begin{figure*}
\includegraphics[width=0.45\textwidth]{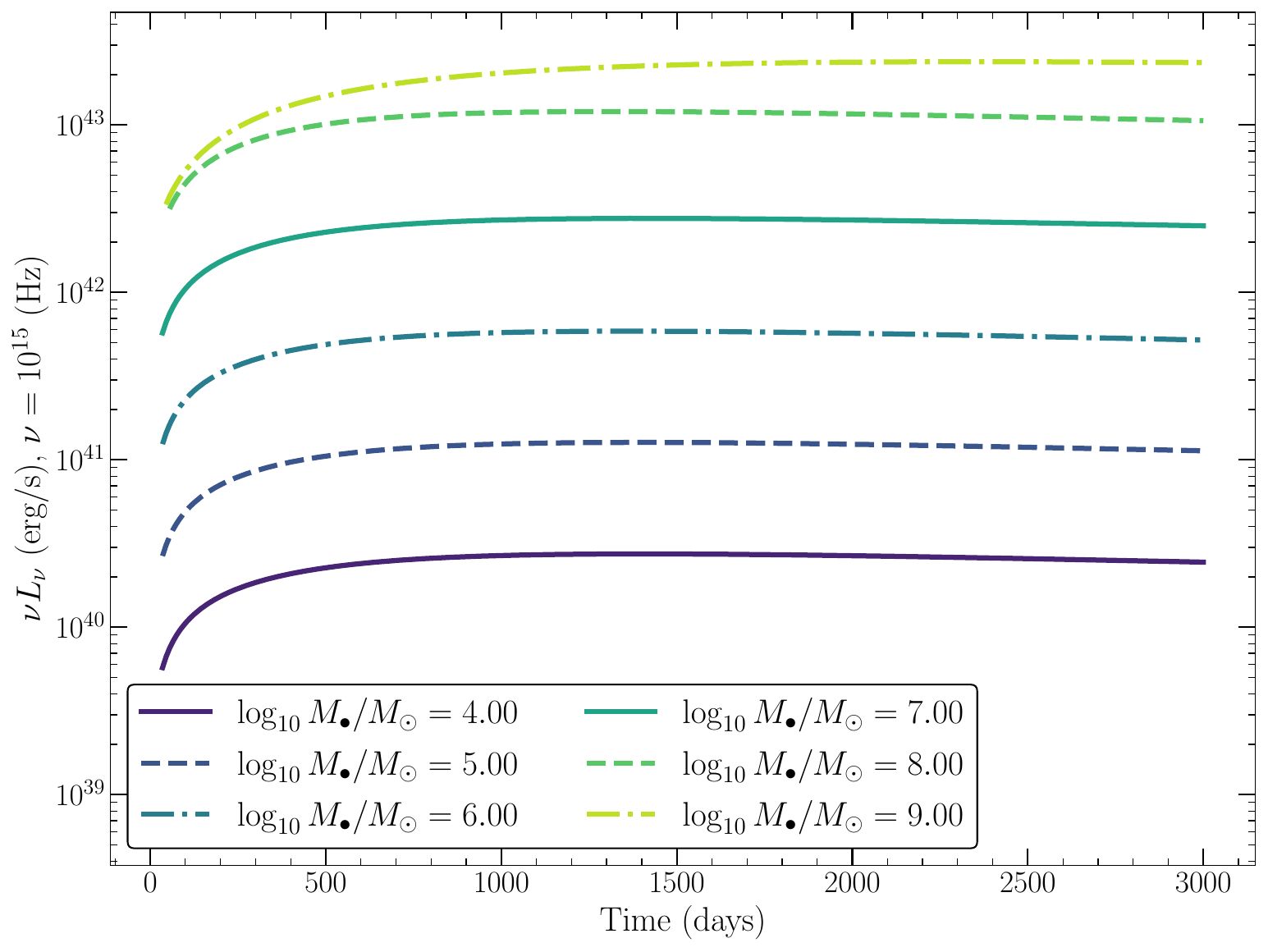}
\includegraphics[width=0.45\textwidth]{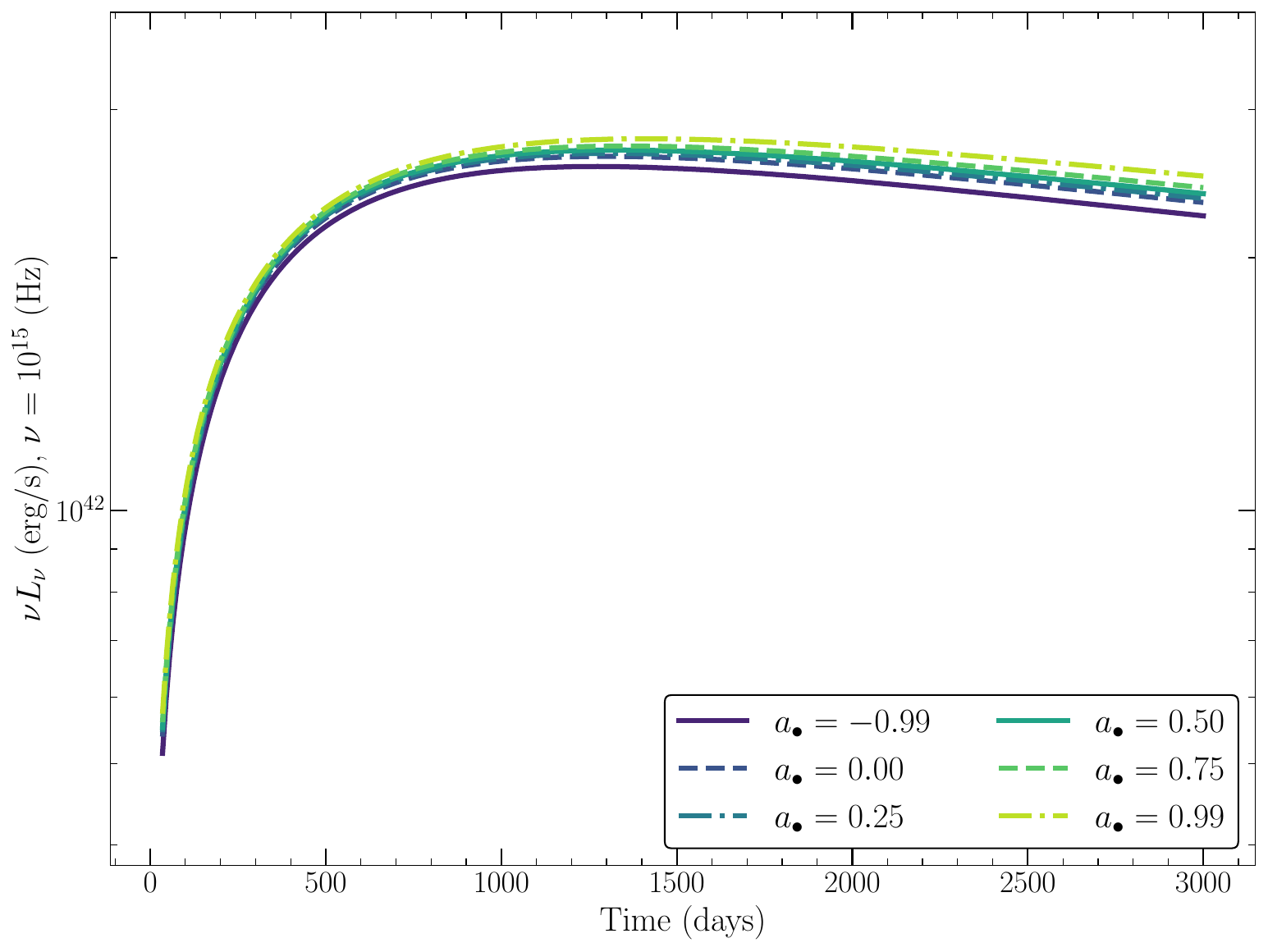}
\includegraphics[width=0.45\textwidth]{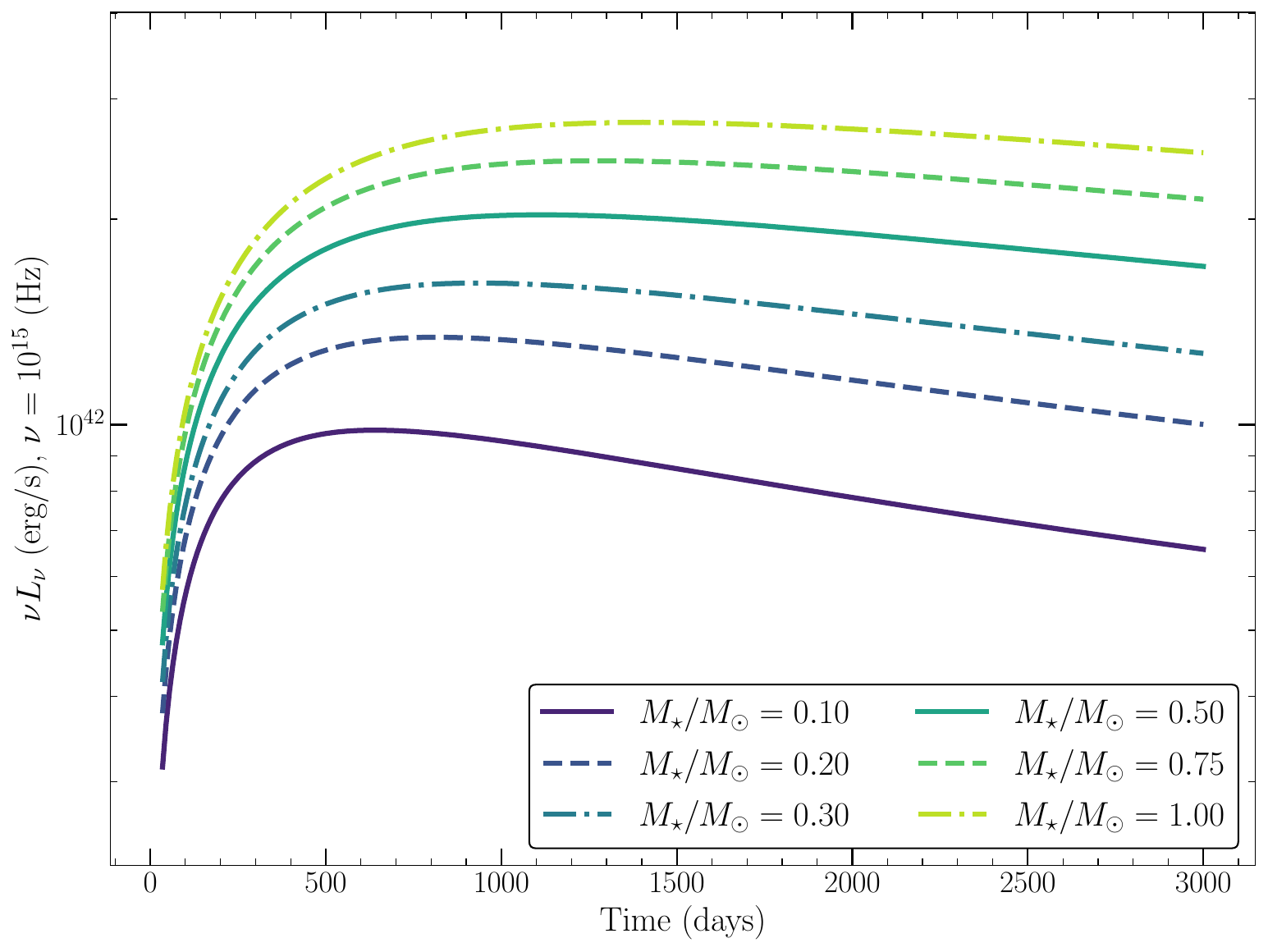}
\includegraphics[width=0.45\textwidth]{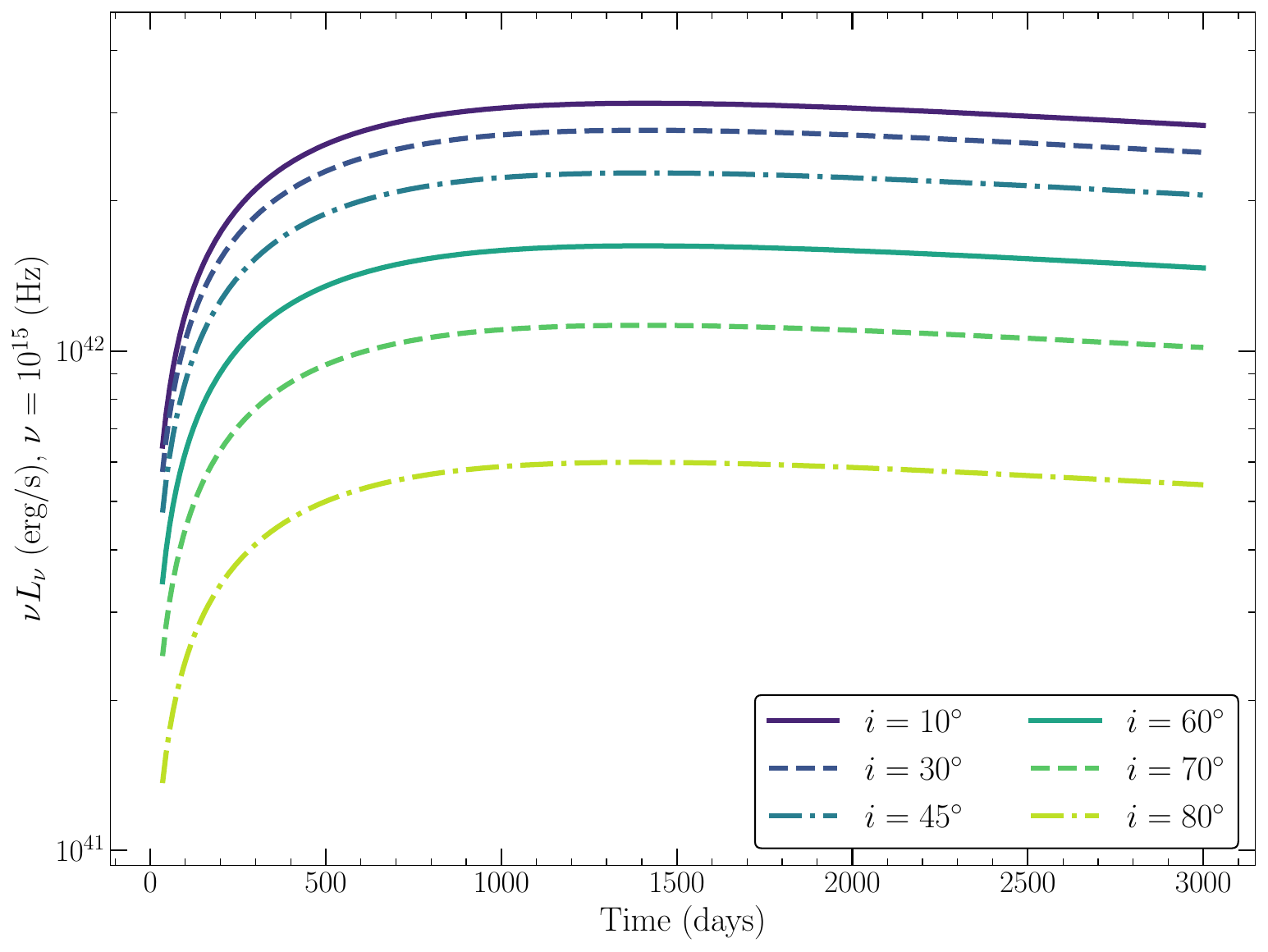}
\caption{Example UV light curves $(\nu L_\nu$ vs time) observed at $\nu = 10^{15}$ Hz for TDE disc systems of differing parameters. Except when specified in the figure legend, the parameters used are $M_\star = M_\odot$, $\M = 10^7 M_\odot$, $i = 45^\circ$, ${\cal V} = 1500$, $a_\bullet = 0.99$, $\beta = 1$.   }
\label{elc}
\end{figure*}

\subsubsection{Viscosity parameterisation } 
If the pericentre radius of the incoming star's orbit is exterior to the IBCO, then the returning stellar debris will eventually circularise into an accretion disc. The so-called circularisation radius $r_c$ is the radial scale at which the stellar debris are expected to return, and is given by conservation of angular momentum to be 
\beq\label{rc} 
r_c = {2 \over \beta } r_T . 
\eeq
The extra factor of 2 results from angular momentum conservation as a parabolic orbit is turned into a circular orbit. 

The so-called ``viscous'' timescale of an evolving accretion flow is given by the following simple expression:
\beq
\tv = \alpha^{-1} \left({H\over R}\right)^{-2} t_{\rm orbital} = \alpha^{-1} \left({H\over R}\right)^{-2} \sqrt{r^3 \over G\M} .
\eeq
where $r$ is the radius at which the flow begins, $\M$ is the black hole's mass, $H/R$ is the disc aspect ratio, $\alpha$ the \citet{Shakura73} alpha parameter, and $t_{\rm orbital}$ is the time it takes for the disc material to complete one orbit of the black hole at radius $r$.  A simple substitution of the circularisation radius into the expression for the viscous timescale demonstrates that 
\beq
\tvc = \alpha^{-1}\left({H\over R}\right)^{-2} \sqrt{8R_\star^3 \over \beta^3GM_\star} ,
\eeq
and we see that any explicit dependence on the properties of the black hole has dropped out of this expression.  The ratio of the viscous and orbital timescales is a dimensionless number, expected to be large 
\beq
{\tvc \over t_{\rm orbital}} = \alpha^{-1} \left({H\over R}\right)^{-2}  \equiv {\cal V} .
\eeq
The value of ${\cal V}$ will change over a population of TDEs, but we do not expect this change to be systematically linked to the properties of the TDE's black hole. To see this, note that the classic \citet{Shakura73} result for the disc aspect ratio for an accretion flow with fixed Eddington ratio is 
\beq
{H \over R} \propto \M^b, 
\eeq
where $b = 0$ if radiation pressure dominates within the flow, and $b = -1/10$ if gas pressure dominates. It is unlikely that $\alpha$ will systematically vary with black hole parameters. 

As such, we anchor our values of ${\cal V}$ in the range of values observed in the TDE population. Unlike the optical/UV light curves of typical TDEs, the X-ray light curves of thermal TDEs vary rapidly, and disc modelling of their evolving X-ray luminosity allows the viscous timescale to be determined for a given TDE.  The fastest evolving TDE which has been modelled in the X-ray is AT2019dsg \citep{Mummery21}, with value ${\cal V} \sim 1000$, while the slowest evolving modelled TDE is ASASSN-15oi \citep{Mummery21}, with ${\cal V} \sim 12500$. The TDE ASASSN-15lh has rapidly evolving optical-UV light curves, a result of its much cooler disc \citep{MumBalb20b}, with viscosity parameter ${\cal V} \sim 100$. Clearly TDE systems satisfy a broad range of ${\cal V}$ values, and we therefore allow the values of ${\cal V}$ to vary between these observed values. We uniformly sample values of ${\cal V}$ in the range 
\beq\label{pv}
p_{\cal V} \propto 1, \quad 10^2 < {\cal V} < 10^4. 
\eeq
With ${\cal V}$ specified, the amplitude of the turbulent stress $\W$ is determined. We solve the disc equations with simple radius-dependent profiles
\beq
\W = w \left({ r \over r_0}\right)^\mu ,
\eeq
where $w$ is uniquely determined by the choice of ${\cal V}$ and $r_0$. 

We find numerically that the choice of stress index $\mu$ has minimal effect on the resulting optical/UV light curves.  {The effects of the value of ${\cal V}$ on the disc light curves is discussed further in section \ref{sec:elc}.  }

\subsubsection{The disc-observer inclination }
We assume that TDEs occur in the equatorial plane of the central black hole's spin axis, and further assume that this axis will be completely randomly orientated with respect to the observer. This means that the inclination angle of TDE discs will be randomly orientated on the observing sphere, and therefore $\cos(i)$ will be uniformly distributed 
\beq\label{pi}
p_i(\cos(i)) \propto 1 .
\eeq

\subsection{Simulation procedure }

\begin{figure*}
\includegraphics[width=0.49\textwidth]{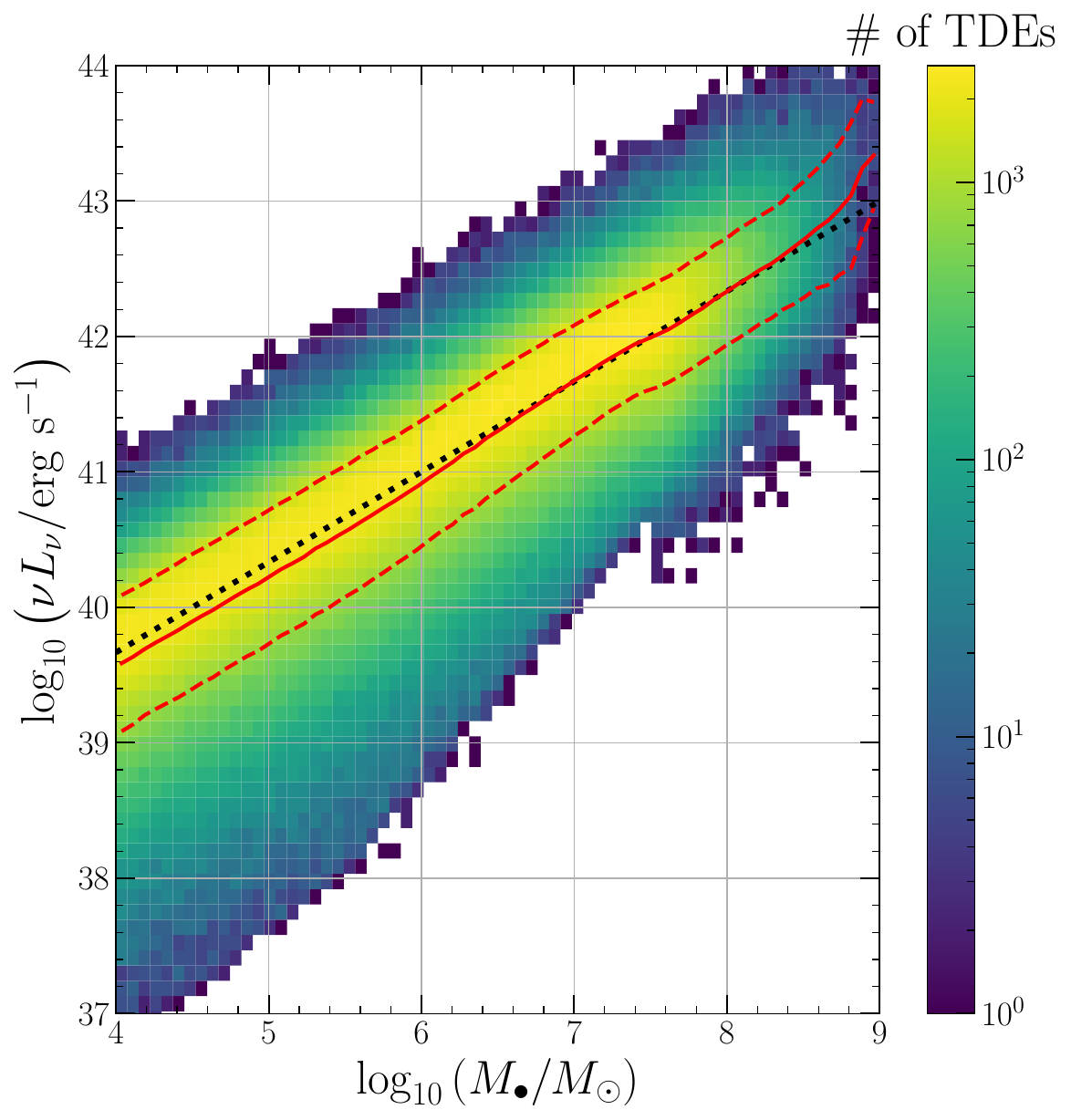}
\includegraphics[width=0.49\textwidth]{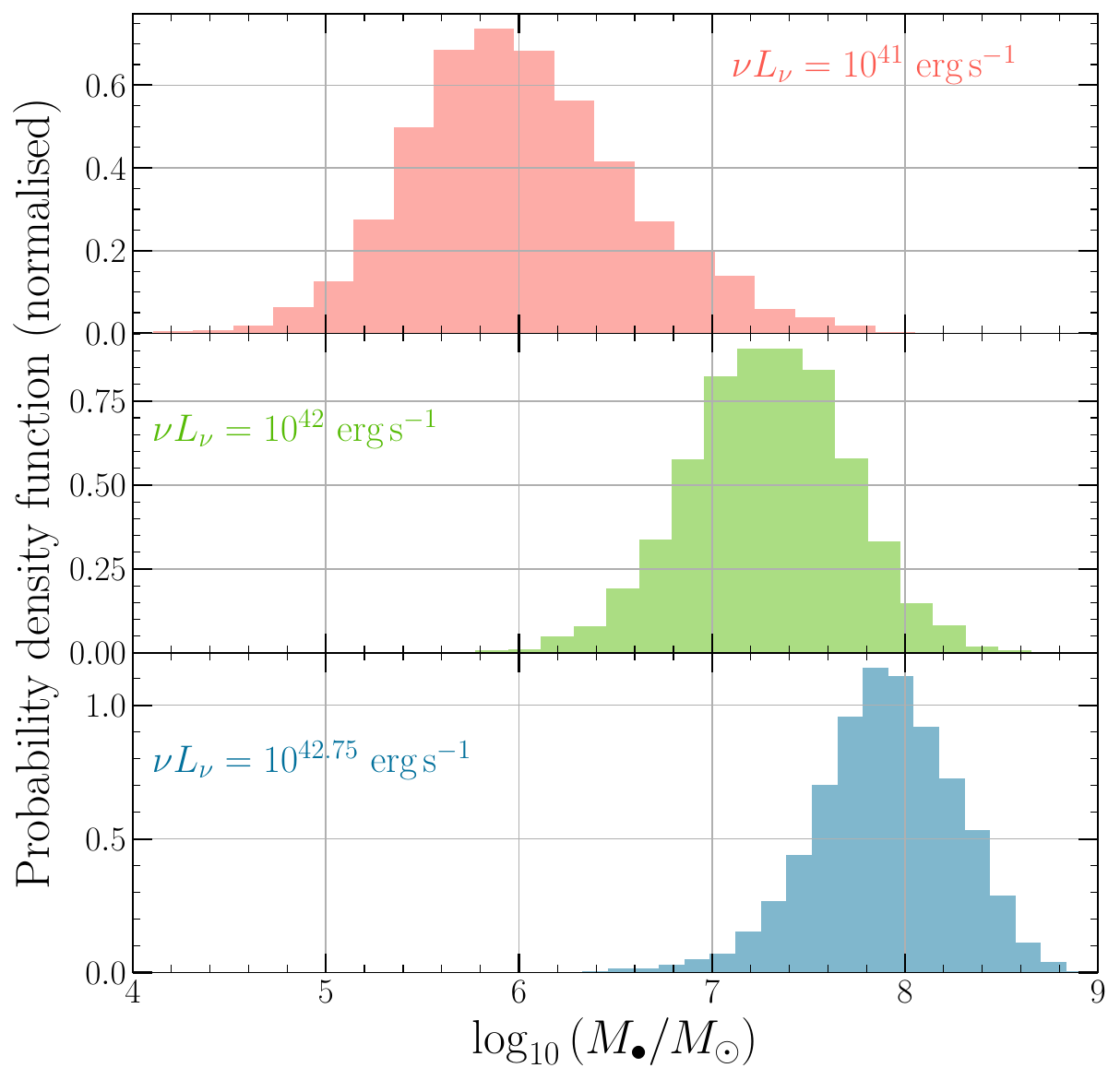}
\caption{The results of our simulated TDE population, presented as plateau luminosity $\nu L_\nu$ (measured in the rest-frame $g$-band) as a function of black hole mass $\M$. We see, as predicted by disc theory, a clear correlation between black hole mass and TDE plateau luminosity.   In the left panel we display a density plot of the total number of TDE systems in each mass and luminosity bin. This panel  demonstrates that while the total  spread in values of $L_{\rm plat}$ for a given $\M$  is formally large (as demonstrated by the outermost contour), the majority of points lie in a more compact central region of the plot (the red dashed curves show the $1\sigma$ region around the median, which is denoted by a red solid curve). The black dotted curve is $\nu L_\nu \propto \M^{2/3}$, exactly as predicted by disc theory (appendix \ref{disctheory}). On the right we show example theoretical probability density functions of the central black hole masses which produce late time plateaus at the levels displayed on each panel. Note that the width of the black hole mass distributions increase as the plateau luminosity decreases, a result of the contribution of highly inclined high black hole mass systems which are not present at higher luminosities due to the Hills mass effect. The observed plateau luminosity can be used to measure TDE black hole masses. }
\label{sim_plat}
\end{figure*}

\begin{figure*}
\includegraphics[width=0.8\textwidth]{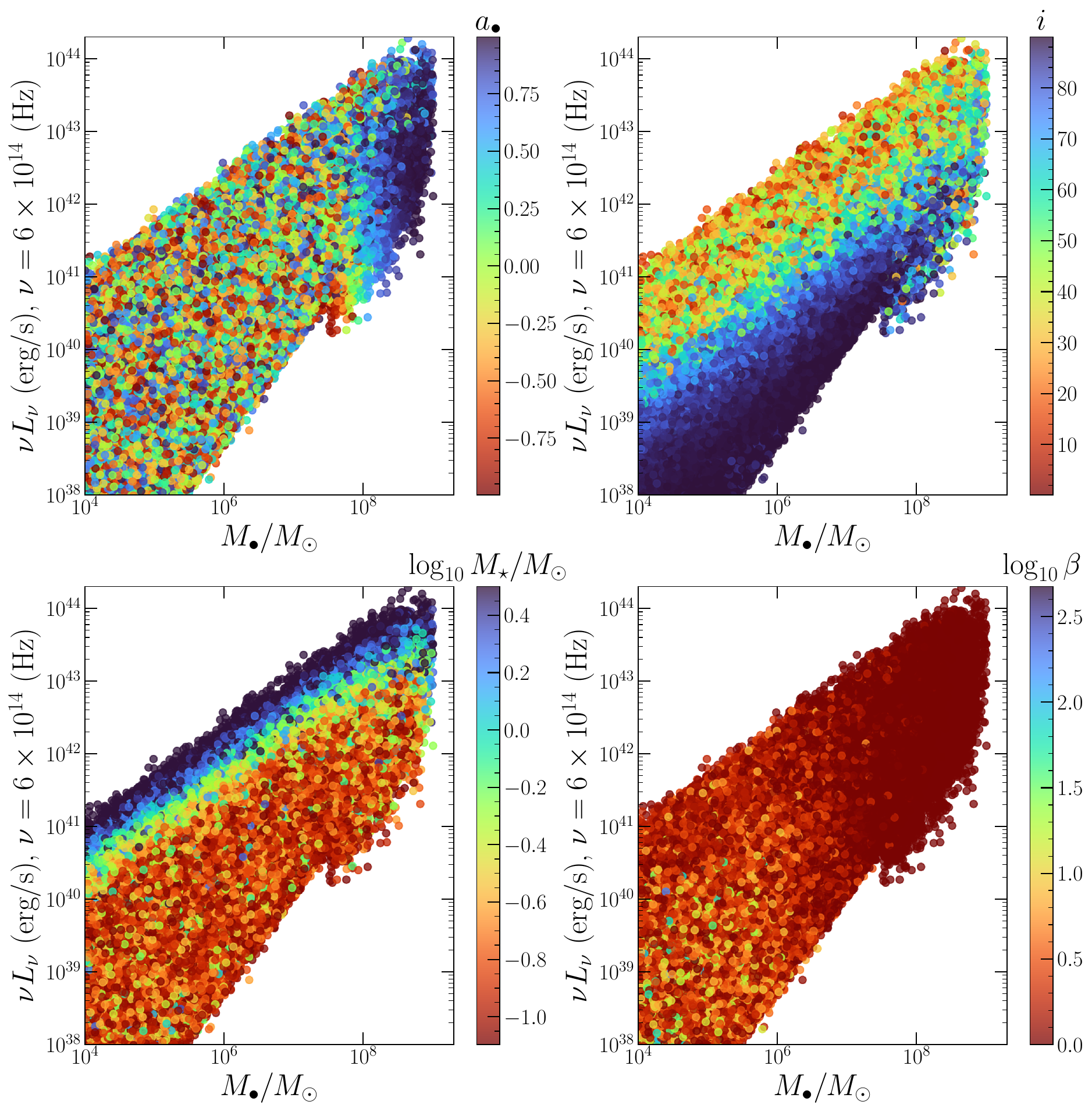}
\caption{The different trends of the late time luminosity with system parameters. Each panel is coloured by the system parameter of interest. Upper left: black hole spin, upper right: inclination, lower left: stellar mass, lower right: orbital penetration parameter. The trend with each parameter is readily understandable from standard disc theory, as we discuss in the main body of the text. 
}
\label{scatter}
\end{figure*}

We simulate large populations of TDEs in the following manner, shown schematically in Fig. \ref{SimProc}. 

The first important question for each simulated system is whether or not observable electromagnetic emission could be produced.  For each simulated TDE, we sample a stellar mass from $p_\star$ (eq. \ref{pstar}), and black hole mass and spin  (eqs. \ref{pbhm}, \ref{pbha}). With the black hole mass specified, we sample a value of $\beta$ as described in section \ref{beta}. The pericentre radius of the star's orbit is then computed following the \citet{Kesden12} relativistic formalism. If the pericentre radius is greater than the black hole's IBCO radius (eq. \ref{ibco}), then we proceed to solving the disc equations. If not, we restart the sampling procedure. 

For those systems which would produce observable emission, we sample a viscosity parameter from eq. \ref{pv}, and a disc-observer inclination angle from \ref{pi}. We assume that half of the initial stellar mass forms into a circularised disc, with initial radius equal to the circularisation radius (eq. \ref{rc}). With this initial condition specified, we solve the relativistic disc equation (eq. \ref{fund}), and propagate the disc temperature profile (eq. \ref{temperature}) out to 1000 days. 

We solve the photon geodesic trajectories, and observe the disc at an inclination $i$, computing the 1000 day luminosity $\nu L_\nu$ (eq. \ref{lum}) at rest-frame frequencies $\nu = 5 \times 10^{14}$ Hz, $6\times 10^{14}$ Hz and $1 \times 10^{15}$ Hz. These frequencies correspond broadly to the $r$-band, $g$-band and $u$/NUV-band respectively.  We then save the luminosity, and each of the system parameters $(\M, a_\bullet, M_\star, \beta,  \tv, i)$ for this sample. The process is then repeated for a new set of parameters. Before we present the results of these simulations, we plot some example light curves below.

\subsection{Example light curves }\label{sec:elc}
In Fig. \ref{elc} we plot example UV light curves $(\nu L_\nu$ vs time) observed at $\nu = 10^{15}$ Hz for TDE disc systems of differing parameters. Except when specified in the figure legend, the parameters used are $M_\star = M_\odot$, $\M = 10^7 M_\odot$, $i = 45^\circ$, ${\cal V} = 1500$, $a_\bullet = 0.99$, $\beta = 1$. 

In the upper left panel we display TDE disc light curves of systems evolving around black holes of differing masses.  The differing luminosities of these systems represent the key theoretical result of this paper. As predicted by simple disc theory (Appendix \ref{disctheory}) there is a strong correlation between central black hole mass and the amplitude of the late time UV plateau luminosity. 

In the upper right plot we display light curves for differing black hole spins. As can be seen the black hole spin does not effect the amplitude of the disc UV luminosity, as the vast majority of the emission comes from much further out than the near-ISCO region. As we discussed earlier (eq. \ref{fullhills}), the black hole spin is crucially important for determining which black holes are able to tidally disrupt stars of a given mass. We will return to this point later. 

In the lower left panel we display disc light curves formed from the disruption of stars of differing masses.  The resulting light curves display a positive, but relatively weak, correlation between stellar mass and late time UV luminosity, as expected from disc theory (Appendix \ref{disctheory}). On a more subtle level, there are stellar mass dependent properties of the light curve morphology themselves, with low mass ($M_\star = 0.1 M_\odot$) TDEs declining after $\sim 400$ days, while higher mass $(M_\star = M_\odot)$ systems continue to rise out to 1000 days. 

Finally, in the lower right panel we display the effects of inclination on the TDE UV light curves. More inclined disc systems are fainter, due to the smaller projected area of the disc in the observing plane. 

{While not displayed here, the optical/UV light curves of TDE discs do show some slight dependence on viscosity parameter ${\cal V}$. Firstly, the rise to plateau of the disc light curves (Fig. \ref{elc}) is shortened by decreasing ${\cal V}$. This phase of TDE light curve evolution is typically unobservable however (being much fainter than the early time component). On very long timescales the time at which the plateau begins to decline is also shortened by decreasing ${\cal V}$, this might be of some interest to the handful of sources which show plateau evolution (section \ref{sec:non_dec}).   Of more relevance, the amplitude of the Rayleigh-Jeans disc flux is weakly dependent on ${\cal V}$, namely $\nu L_\nu \sim {\cal V}^{-1/4}$, and so the more rapidly rising disc light curves are also brighter. This 1/4 exponent however means that the factor 100 range in ${\cal V}$ we consider translates to a factor $\sim 3$ in plateau luminosity variance.    } 

\section{Simulation results } \label{sec:4}
In Fig. \ref{sim_plat} we display the results of our simulation of $N = 10^6$ TDE systems, generated using the procedure discussed above. We plot the late time $g$-band ($\nu = 6 \times 10^{14}$ Hz; see Fig. \ref{sim+real} for NUV band) luminosity from each system as a function of the system's black hole mass in the left hand panel of Fig. \ref{sim_plat}. 

In the left panel of Fig. \ref{sim_plat} we display a density plot of the total number of TDE systems in each mass and luminosity bin. This panel  demonstrates that while the total possible spread in values of $L_{\rm plat}$ for a given $\M$  is formally large (as demonstrated by the outermost contour), the majority of points lie in a more compact central region of the plot (the red dashed curves show the $1\sigma$ region around the median, which is denoted by a red solid curve). For a given observed value of $L_{\rm plat}$ the typical $1\sigma$ scatter in the black hole mass which produced this plateau luminosity is of order $0.5$ dex, comparable to the $\M-\sigma$ relationship. 

 A power-law fit to the median of luminosities which are (logarithmically) binned by black hole mass is well described by the following expression 
\beq
\log_{10} \left({L_{\rm plat} \over {\rm erg/s}}\right) = 0.67 \log_{10}\left( {\M \over M_\odot}\right) + 37.0 ,
\eeq
where $L_{\rm plat} \equiv \nu L_\nu$  (black dashed curve, Fig. \ref{sim_plat}). This is precisely the theoretical prediction derived in Appendix \ref{disctheory}
\beq
L_{\rm plat} \propto \M^{2/3} .
\eeq

As an example of the power of the late time plateau in constraining TDE black hole masses, in the right hand panel we plot three normalised black hole mass probability density functions for observed plateau luminosities at $10^{41}, 10^{42}$ and $10^{42.75}$ erg/s. These distributions are generated by recording all systems which produce a plateau luminosity within $10\%$ of the quoted luminosity (this ten percent range is typically much larger than the observational uncertainties on measured plateaus).  Clearly a single observation of the late time plateau luminosity of a TDE can be used to place strong constraints on its central black hole mass.  Note that the width of the black hole mass distributions increase as the plateau luminosity decreases, a result of the contribution of highly inclined high black hole mass systems which are not present at higher luminosities  due to the Hills mass effect.

\subsection{Sources of scatter}
There are clearly a number of sources of scatter in the $\nu L_\nu - \M$ relationship plotted in Fig. \ref{sim_plat}.  To understand the sources of this scatter in more detail we plot the trends of the late time UV luminosity with different system parameters across the entire TDE population in Fig. \ref{scatter}. To aid in interpreting these trends we remind the reader that from classical disc theory one can derive the Rayleigh-Jeans flux scaling (Appendix \ref{disctheory}) 
\beq
\nu L_\nu \sim \M^{2/3}  M_\star^{-1/24}  R_\star^{7/8} \beta^{-7/8} \cos(i) ,
\eeq
and by further assuming a main sequence mass-radius relationship $R_\star \propto M_\star^{0.56}$ \citet{Kippenhahn90}, 
\beq
\nu L_\nu \sim \M^{2/3} M_\star^{11/25} \beta^{-7/8} \cos(i) .
\eeq

In the upper left panel of Fig. \ref{scatter} we plot the $L_{\rm plat} - \M$ population, coloured by central black hole spin. While we have already argued (Fig. \ref{elc}, upper right panel) that varying the black hole spin has only a very weak effect on the luminosity of the disc, the black hole spin is critically important for determining which TDE systems will produce observable electromagnetic emission (eq. \ref{fullhills}).  This Hills effect can be seen clearly in the upper left panel of Fig. \ref{scatter}, where despite the input black hole spin distribution being uniform (as can be clearly seen for low mass $\M \sim 10^6 M_\odot$ TDEs), the observed TDE population of high black hole mass TDEs  (i.e., $\M \gtrsim 5 \times 10^7 M_\odot$) is dominated by rapidly rotating black holes (upper right corner of plot). 

The dominant source of scatter in the observed luminosity  can be identified  in the upper right panel of Fig. \ref{scatter}. Namely, the chief source of scatter results from a simple $\cos(i)$ projected disc area effect.  Those discs which are observed edge-on have a much smaller observed emitting area which contributes to the late time disc luminosity, and are correspondingly dimmer. TDE discs which are observed face-on are the brightest. 

The effects of stellar mass are displayed in the lower left panel of  Fig. \ref{scatter}. While the entire population is dominated by low-mass stars (as is expected from a the steep fall off in the stellar population at high masses ${\rm d}N_\star /{\rm d}M_\star \propto M_\star^{-2.3}$), for a given black hole mass the brightest TDE discs in the UV are dominated by those formed from the most massive stars.  This effect can be clearly seen in the uppermost boundary of the $L_{\rm plat} - \M$ population. 

Finally, in the lower right panel of Fig. \ref{scatter} we colour the $L_{\rm plat}-\M$ population by impact parameter $\beta$. The (diagonal) trend across the population as a whole simply represents the increased probability of low mass galaxies producing so-called pinhole TDEs \citep[see section \ref{beta};][]{Stone16}, and the larger range of 
$\beta$s available to low mass TDEs which do not result in a TDE occurring within the black hole's event horizon. For a given black hole mass, higher $\beta$ TDEs typically produce lower UV luminosities, a result of their resultant discs having smaller radial extents. This has a very minor effect on the population level.  

\section{The light curves of real TDEs } \label{sec:5}
In this section we collate and analyse the light curves of all published optically bright TDEs, with the aim of extracting the late-time plateau luminosity from as large a sample as is possible.  The properties of this plateau luminosity population will then be compared to TDE disc theory in section \ref{sec:6}. 

\subsection{Source sample}
We collect all optically-selected TDEs from the literature by combining catalogs from three different papers: \citet{vanVelzen20_ISSI}, who list all optical TDEs up mid-2019; \citet{Hammerstein23} which presents the TDEs that are detected in the first half of the ZTF survey; and \citet{Yao23} who presents the latest TDEs from ZTF. After selecting sources with  detections at least 1 year post peak (in the source rest-frame), we obtain \Ntot\ TDEs. 

\subsection{Photometry}
Below we describe the details of the ZTF and UVOT data reduction. The ZTF and UVOT/UV lightcurves that we obtain will be available for download at the journal website. 

\subsubsection{Swift/UVOT}
Of the \Ntot\ TDEs in our sample, \Nswift\ have been observed and detected at UV wavelengths by the Neil Gehrels {\it Swift} Observatory. We  download the latest {\it Swift}/UVOT data for all sources and reprocess all photometry (using the 20201215 UVOT calibration files). Following \citet{vanVelzen18_FUV,vanVelzen19_ZTF}, we estimate the host galaxy flux in {\it Swift}/UVOT filters by fitting a stellar-population synthesis model to the pre-TDE photometry. This baseline is subtracted from the aperture photometry obtained from the {\it Swift} images and the uncertainty on the baseline flux is propagated into the resulting difference flux. {The default aperture radius is 5~arcsec. If needed, this  radius is manually adjusted to capture the flux of larger host galaxies. For each source, the aperture radius and the host galaxy baseline magnitudes are available in the file that contains the difference  photometry.}

\subsubsection{ZTF forced photometry}
Of the \Ntot\ TDE in our sample, \Nztf\ are detected in ZTF. For these sources we obtained forced photometry \citep{Masci19} light curves using ZTF DR18 (which was released in July 2023 and gives access to ZTF forced photometry light curves up to May 2023). We reduce the lightcurves following the steps outlined in \citet{Hammerstein23}. 

An important step in the forced photometry reduction is baseline subtraction, which removes any residual flux of the host galaxy from the difference images.  The time window for the baseline is defined from the observations obtained after the last image used to build the reference frame and before the onset of the TDE (this onset is assumed to be at most 100 days before the peak; for each source we confirm this by visual inspection and adjust if needed). For each ZTF field and filter combination, the median difference flux inside the baseline window is subtracted from the entire difference flux light curve.  If fewer than 10 observations are available for the baseline window we reject this field and filter combination. This requirement removes all forced photometry data for two sources (AT2018zr and AT2018hyz), for these we only use the near-peak ZTF photometry as published in \citet{vanVelzen20}. 

\subsubsection{Other photometry}
{A handful TDEs are not observed with  ZTF or UVOT. For these we include data from other sources (e.g., SDSS Stripe 82, PTF, Pan-STARRS), see \citet{vanVelzen20_ISSI} and references therein. Of particular importance for the plateau detections are the late-time {Hubble Space Telescope} (HST) near-UV \citep{Gezari15} far-UV \citep{vanVelzen18_FUV} detections. For the HST photometry, no host baseline subtraction is needed since the nuclear UV emission on sub-arcsecond scales should be dominated by the TDE \citep{vanVelzen18_FUV}. }

\subsubsection{Flux addition in ZTF forced photometry}\label{sec:compgal}
At a redshift of $z=0.1$, the typical distance of ZTF TDEs, an optical luminosity of $10^{41}\,{\rm erg}\,{\rm s}^{-1}$corresponds to a magnitude of 24. This predicted plateau luminosity (Fig.~\ref{sim_plat}) is much fainter compared the ZTF single-epoch ($5\sigma$) flux limit of $m\approx 21$. However, by {measuring} the mean forced photometry flux of $\sim 10^2$ observations {(e.g., by fitting a straight line to the data)}, we should be able to detect a source with $m=24$ at a signal-to-noise ratio (SNR) close to three.

To test the quality of our forced photometry lightcurves, we introduce a collection of ``comparison galaxies". These are selected to have a similar quality of the lightcurve compared to each TDE. To achieve this, we pick the galaxy with nearest $r$-band flux within 10~arcmin of each TDE host. To mimic the detection of a late-time plateau, we compute  the inverse-variance weighted mean forced photometry flux using only the last year of observations (this window typically contains 35 observations in the $g$-band). We shall denote this measurement with $f_{\rm late}$.    

The comparison galaxies should have no net late-time flux. Hence collection of measurements of $f_{\rm late}$ for these sources provides direct insight into the accuracy and precision of the forced photometry. The median $f_{\rm late}$ all comparison galaxies is $-0.28\mu$Jy. We find three significant outliers relative to the sample variance. All three of these outliers have a negative flux around $-3 \mu$Jy. For two of these comparison galaxies with negative late-time flux, the late-time TDE light curve also shows a significant negative flux. This suggests the native flux is caused by a problem with the calibration of some epochs of the ZTF field that contains the TDE and its comparison galaxy. 

Not all ZTF fields have this problem. After removing the three outliers, the standard deviation of $f_{\rm late}$ of the comparison galaxies is 0.61$\mu$Jy, which corresponds to a 2-$\sigma$ detection threshold of 23.4. This is consistent with the expected improvement over the single-epoch detection limit by a factor $\sqrt{N}$, with $N=35$ the typical number of observations at are used to compute $f_{\rm late}$. 

\begin{figure}
\includegraphics[width=0.5\textwidth, trim={0 30 0 50}, clip]{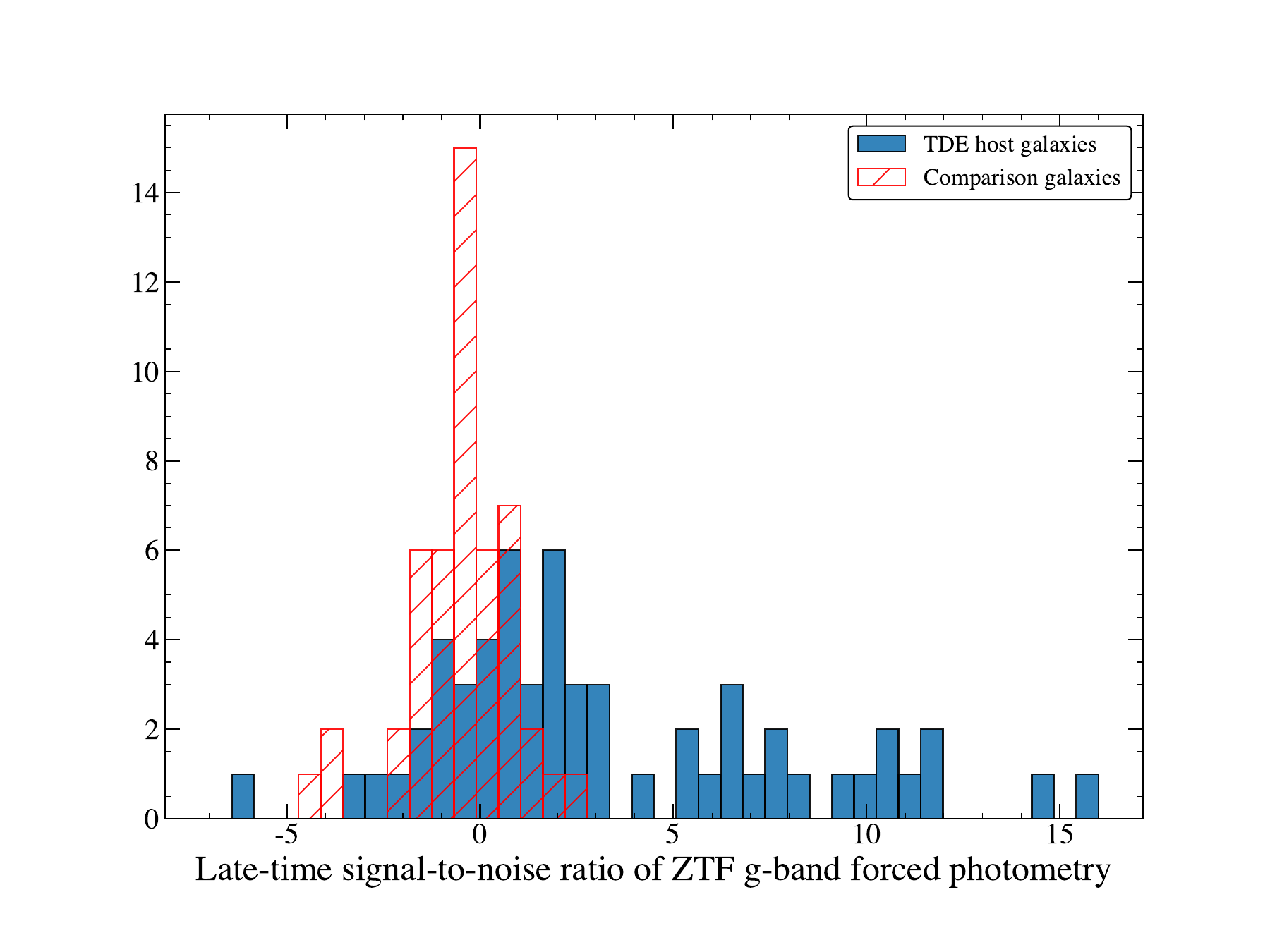}
\caption{The signal-to-noise ratio (SNR) of the mean $g$-band difference image flux measured for the last 365 days of the light curve. The comparison galaxies are selected to have a similar total flux and sky position of the TDE host galaxies. We see a large number of significant late-time detection for the TDE light curves. A tail of negative SNR, due to systematic errors, is seen for both the TDEs and comparison galaxies. }
\label{fig:snr_check}
\end{figure}

Using the standard deviation of $f_{\rm late}$ for the comparison galaxies we estimate the significance of  $f_{\rm late}$ for both the TDEs and their comparison galaxies. The result is shown in Fig.~\ref{fig:snr_check}. Only one of the comparison galaxies has a positive $f_{\rm late}$ with a significance greater than 2$\sigma$, while 29 of the ZTF-detected TDEs pass this threshold. 

To conclude, the experiment with comparison galaxies yields no evidence that positive residuals in the difference images could lead to spurious late-time TDE plateau detections. However, we do find a modest negative offset in the typical late-time difference flux of $-0.3\mu$Jy, with occasional outliers up to $-3\mu$Jy. 

Because the origin of the negative late-time flux is currently unknown (and under investigation), we have not attempted to correct the TDE lightcurves for this systematic effect. While this decreases our sensitivity to accretion disk signatures in optical TDE light curves, we are still able to obtain a large number of plateau detections, as discussed in the next section. {Adding an offset of $+0.3\mu$Jy would affect the inferred flux of the plateau by less than 0.02~dex for 50\% of TDEs in our sample and by more than 0.2~dex for five sources.}

\begin{figure}
\includegraphics[width=0.5\textwidth, trim={0 30 0 50}, clip]{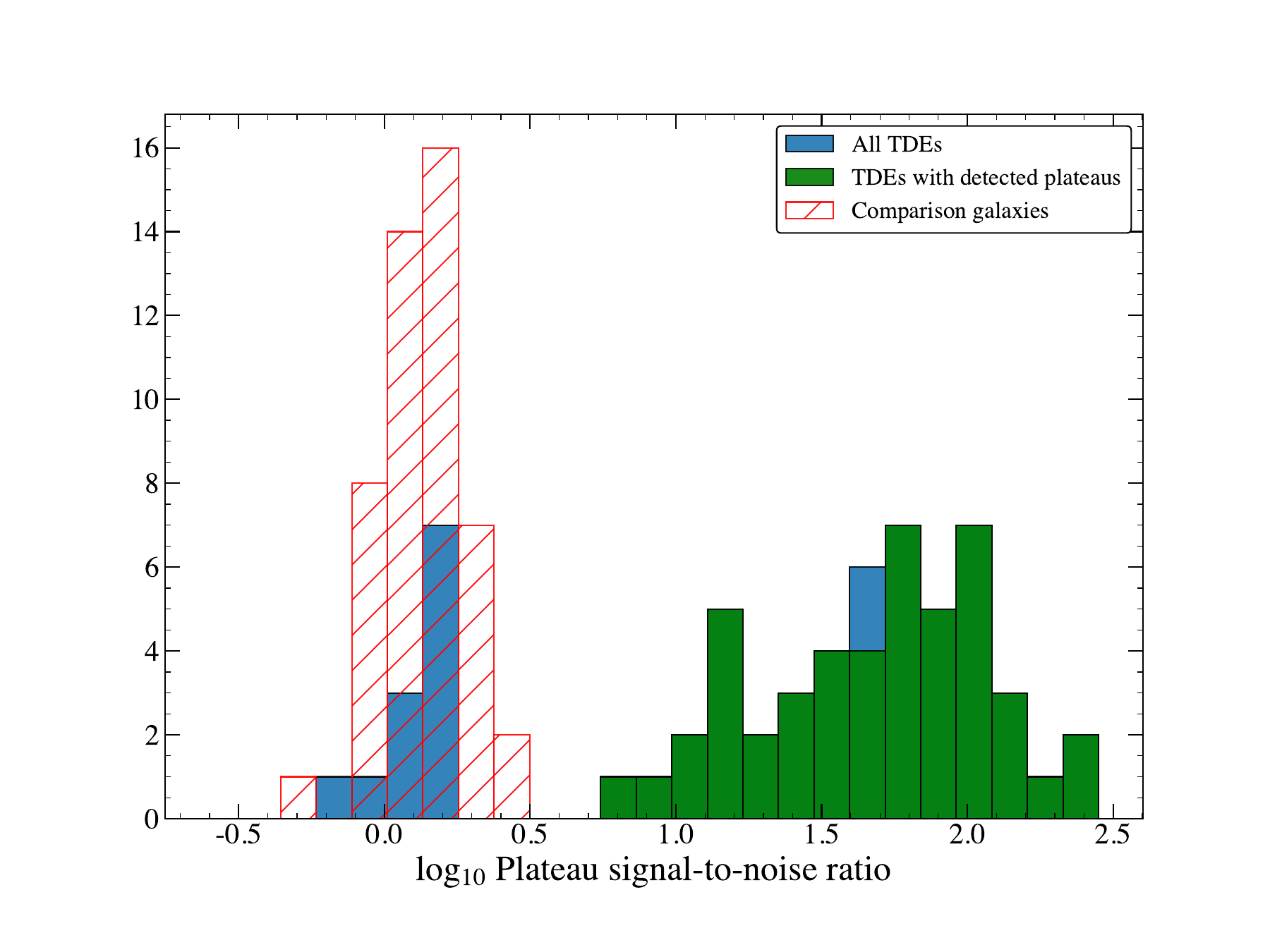}
\caption{The signal-to-noise ratio (SNR) of the plateau luminosity ($L_{\rm plat}$ in Eq.~\ref{eq:lclate}) using optical and UV observations.  }
\label{fig:p_snr_check}
\end{figure}

\subsection{Light curve modelling }

\begin{table*}
    \centering
    \begin{tabular}{l c c c c c}
    Lightcurve property & N  &Kendall's $\tau$ & Significance & Power-law index & Scatter (dex) \\
    \hline\hline
    {Plateau $g$-band luminosity}        & 49&  0.47&$1.8\times 10^{-6}$&  1.14&  0.30\\
Plateau NUV luminosity        & 49&  0.46&$2.5\times 10^{-6}$&  1.09&  0.30\\
Fallback timescale            & 63&  0.36&$3.8\times 10^{-5}$&  0.39&  0.38\\
Exponential decay timescale   & 63&  0.34&$7.6\times 10^{-5}$&  0.23&  0.39\\
Peak blackbody luminosity     & 61&  0.31&$3.4\times 10^{-4}$&  0.78&  0.36\\
Radiated energy ($g$-band)    & 63&  0.31&$2.9\times 10^{-4}$&  0.88&  0.34\\
$g$-band peak luminosity      & 63&  0.24&$5.6\times 10^{-3}$&  0.65&  0.37\\
Peak blackbody radius         & 63&  0.22&$1.1\times 10^{-2}$&  0.31&  0.40\\
Gaussian rise timescale       & 51&  0.19&$5.0\times 10^{-2}$&  0.13&  0.40\\
Peak blackbody temperature    & 58&  0.08&$0.38$    &  0.04&  0.45\\
   
    \end{tabular}
\caption{TDE lightcurve properties compared to the host galaxy stellar mass. The rows are sorted by the significance of the Kendall's $\tau$ test between the TDE property and the host mass. We require that each property is measured with an uncertainty of at most 0.3 dex. The second column gives the number of TDEs that pass this requirement. In the fifth column we list the best-fit power-law index for the relation $\log_{10}(X)\propto p \log_{10}(M_{\rm gal})$, with $X$ each of the properties considered. The final column lists the root-mean-square scatter of the residuals for this fit, measured along the mass axis.}
\label{tab:masscorr}
\end{table*}

We first correct the observed lightcurves for Galactic extinction. For each TDE, we find the $E(B-V)$ from the maps of \citet*{Schlegel98} and compute the extinction in each filter using a blackbody spectrum with $T=3\times 10^4$ (this fixed temperature is justified because over the range of observed optical/UV TDE temperatures, the extinction changes by only a few percent). The luminosity is computed using a flat cosmology with $\Omega_\Lambda=0.7$ and $H_0=70~{\rm km}\,{\rm s}^{-1}{\rm Mpc}^{-1}$.

Following \citet{vanVelzen18_FUV,vanVelzen19_ZTF} we model the light curves with a Gaussian-rise, exponential decay model. The spectrum is described by a blackbody, and the lightcurve model if fit to all optical/UV photometry simultaneously. To measure the late-time plateau luminosity we simply add one more component: a second blackbody with a constant flux that starts at the time of maximum light. Putting this together we obtain:

\begin{align}
\label{eq:lcsum}
 L(t) &= L_{\rm early}(t) + L_{\rm late}(t) , \\[6pt]
\nonumber
    L_{\rm early}(t) &= L_{\rm peak}(\nu_0)~\frac{B(\nu, T_{\rm early})}{B(\nu_0, T_{\rm early})} , \\
    &\times \begin{cases}  e^{-(t-t_{\rm peak})^2/2\sigma_{\rm rise}^2},  & \quad t\leq t_{\rm peak},  \\ 
     e^{-(t-t_{\rm peak})/\tau_{\rm decay}},  & \quad t>t_{\rm peak} , 
     \label{eq:lcearly}
      \end{cases}\\[6pt]
\nonumber
    L_{\rm late}(t)  &= L_{\rm plat}(\nu_0)~\frac{B(\nu, T_{\rm plat})}{B(\nu_0, T_{\rm plat})} , \\
    &\times \begin{cases}  0, & \qquad\qquad\qquad\qquad t\leq t_{\rm peak} , \\[6pt]
     1, & \qquad\qquad\qquad\qquad t>t_{\rm peak} .
     \end{cases}
     \label{eq:lclate}
\end{align}
Our model thus has 7 free parameters: 
\begin{itemize}
    \item rise time ($\sigma_{\rm rise}$);
    \item time of peak ($t_{\rm peak}$);
    \item luminosity at peak ($L_{\rm peak}$, at a reference frequency, $\nu_{0, \rm peak}$);
    \item post-peak exponential decay rate ($\tau_{\rm decay}$);
    \item temperature near peak ($T_{\rm early}$);
    \item plateau luminosity ($L_{\rm plat}$,  at a reference frequency, $\nu_{0, \rm plat}$); 
    \item plateau blackbody temperature ($T_{\rm plat}$).
\end{itemize}
To find the posterior distribution of the model parameters we use MCMC with a Gaussian likelihood function that allows for additional variance (see e.g., \citealt{vanVelzen18_FUV}). 

{We apply our model to the linear flux at the full time resolution, i.e., no binning is applied to the observations}.  We optimize for the log$_{10}$ of the model parameters, as such the resulting 68\% credible intervals are dimensionless and measured in ``dex". 

Before applying this model to the entire light curve, we first consider  only the first 180 days of post-peak observations and use only the exponential decay model (i.e., $L_{\rm early}$ in Eq.~\ref{eq:lcsum}). For this step, flat priors are used for all parameters. The priors are uninformative (i.e., well outside the range of the final posterior distributions). We make an exception for TDEs only detected in the post-peak phase. For these sources, the time of peak is fixed to the date of the first observation. 

The posterior distribution of the model parameters obtained from the first 180 days of post-peak observations are used to inform the full model that includes the plateau. For the decay and rise parameters we use Gaussian priors whose width is given by posterior distribution obtained of the first 180~days. These prior encode our believe that the early-time optical/UV emission is not dominated by the accretion disk.  A Gaussian prior with  $\sigma=0.1$~dex, centered on the early-time temperature is used for the plateau temperature. In addition, we require that the plateau temperature is greater than $10^4$~K. These priors on the temperature encode the theoretically expected disk temperature and also take into account that both the light at peak and the light from the disk will be affected by the same amount of reddening due to dust in the host galaxy (host galaxy extinction is not a parameter in our  lightcurve model). 

Besides the exponential decay model, we also consider a power-law decay. That is, the post-peak term of Eq.~\ref{eq:lcearly} is replaced with $[(t-t_{\rm peak}+t_0)/t_0]^{p}$. We fix the index to $p=-5/3$ and apply this model to the first year of post-peak observations without including a plateau. The inferred value of $t_0$ from this analysis will be referred to as the `fallback timescale'.

\subsection{Measured TDE Plateaus}
After we apply our exponential decay plus plateau model (Eq.~\ref{eq:lcsum}) to the optical and UV data of the \Ntot\ TDEs in our sample, we find 51 sources with a $L_{\rm plat}$ measured with SNR $>5$ {(i.e., less than $\ln(10)/5=0.46$~dex uncertainty on $\log_{10}L_{\rm plat}$)}, see Fig.~\ref{fig:p_snr_check}. To define a secure plateau detection, we further require that the plateau luminosity at the time of the last observation of the TDE exceeds the prediction for a $-5/3$ power-law decay model. Applying this requirement leaves \Nplat\ sources, about 80\% of the original population.  In Appendix~\ref{sec:lcfits} we show the lightcurve models and the data.

{Of these \Nplat\ TDEs with detected plateaus, 31 have observations in UV bands at late times ($t > 1$yr), and 31 have observations in optical bands at late times.  Sources lacking late time optical observations are typically those which were detected prior to the start of the ZTF survey.    }

We also applied our plateau model to the population of comparison galaxies (selected to be spatially close to the TDE and with a similar total flux, see Sec.\ref{sec:compgal}). For each comparison galaxy, we attempted to detect a plateau using all ZTF data ($g$-band and $r$-band) obtained 100~days after the peak of its corresponding TDE. As expected, we find no significant plateau detections for the comparison galaxies (Fig.~\ref{fig:p_snr_check}). 

\begin{figure}
    \includegraphics[width=0.5\textwidth, trim={0 70 0 50}, clip]{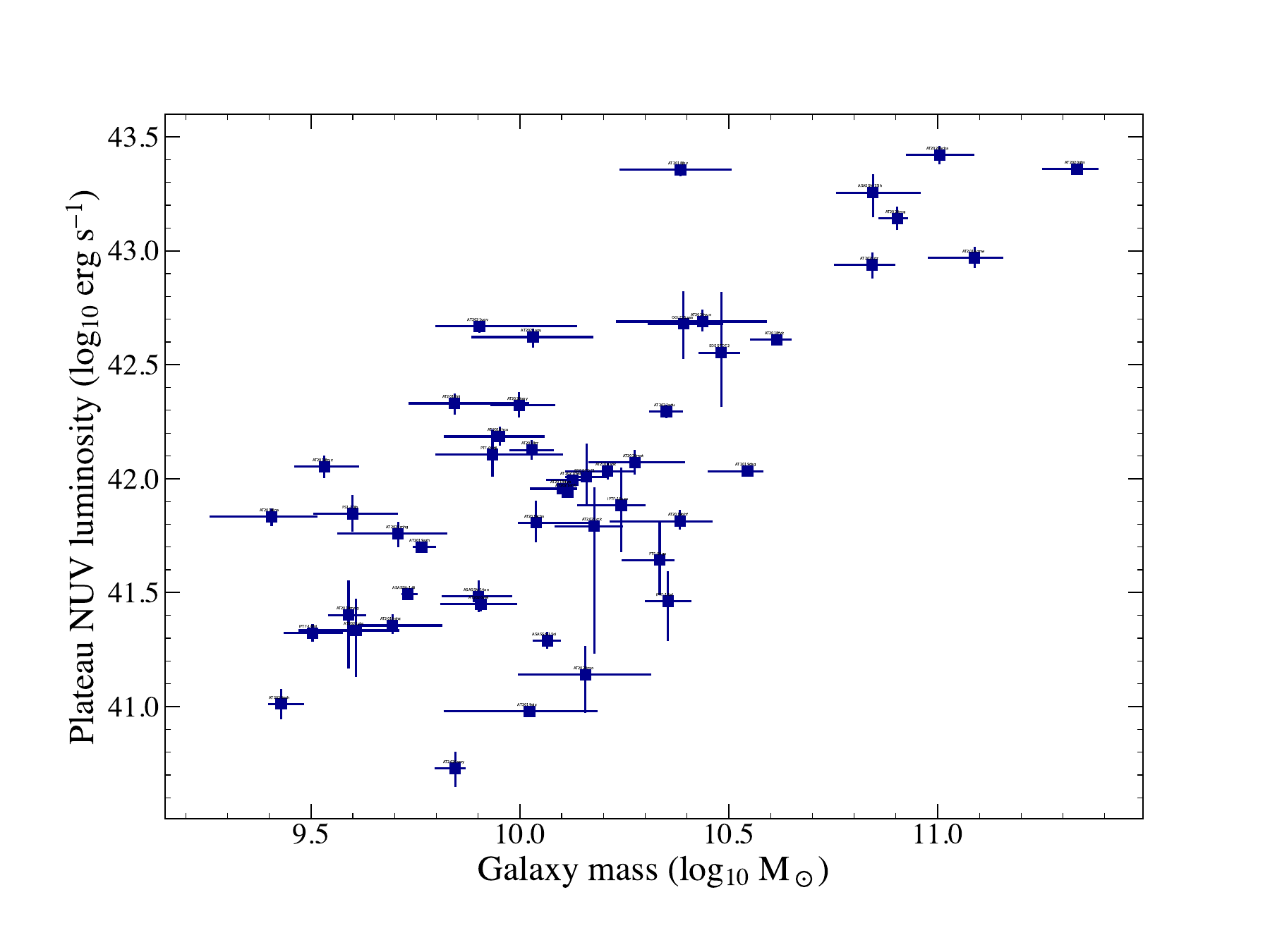}
    \includegraphics[width=0.5 \textwidth, trim={0 70 0 50}, clip]{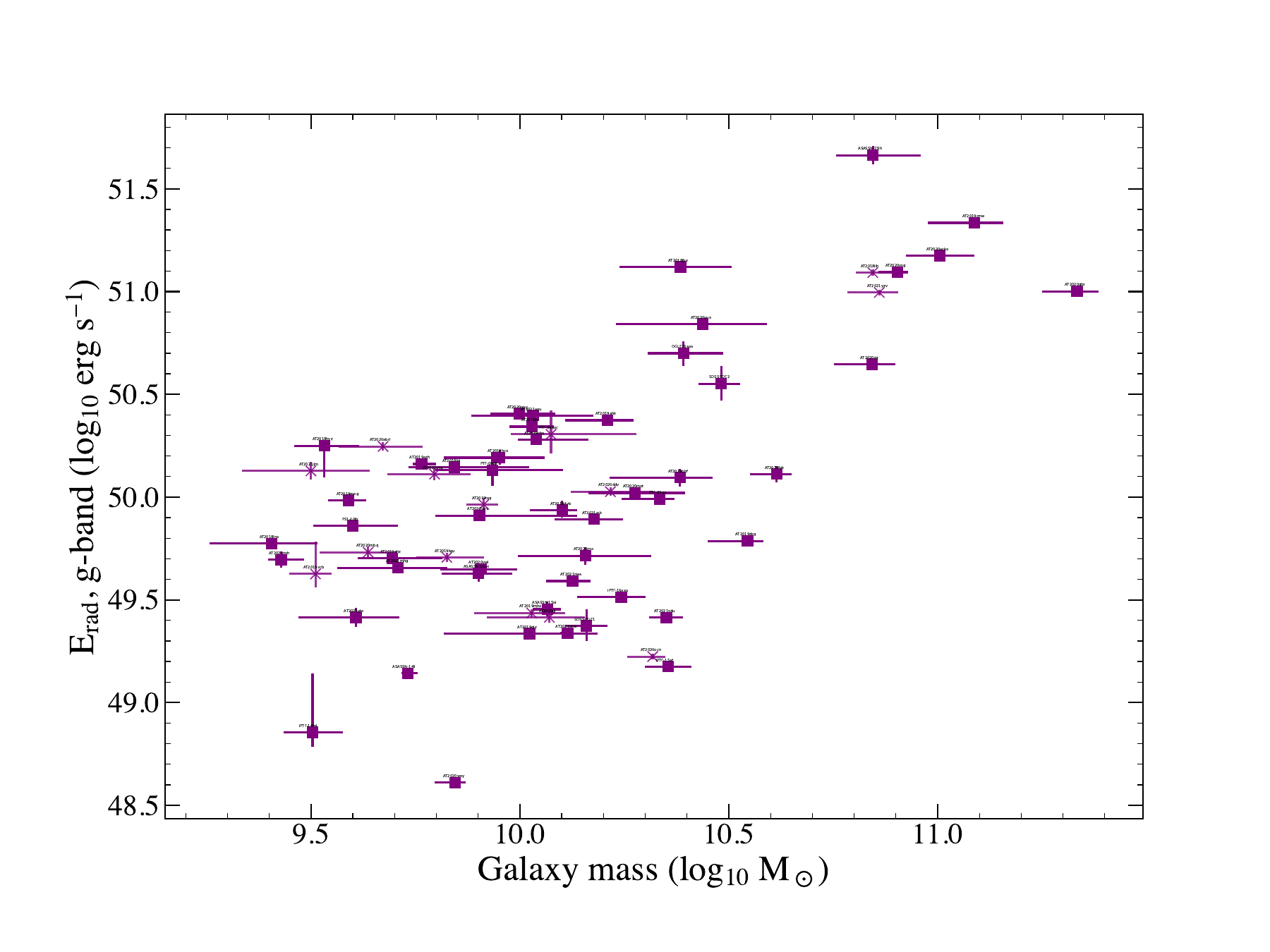}
    \includegraphics[width=0.5 \textwidth, trim={0 30pt 0 50}, clip]{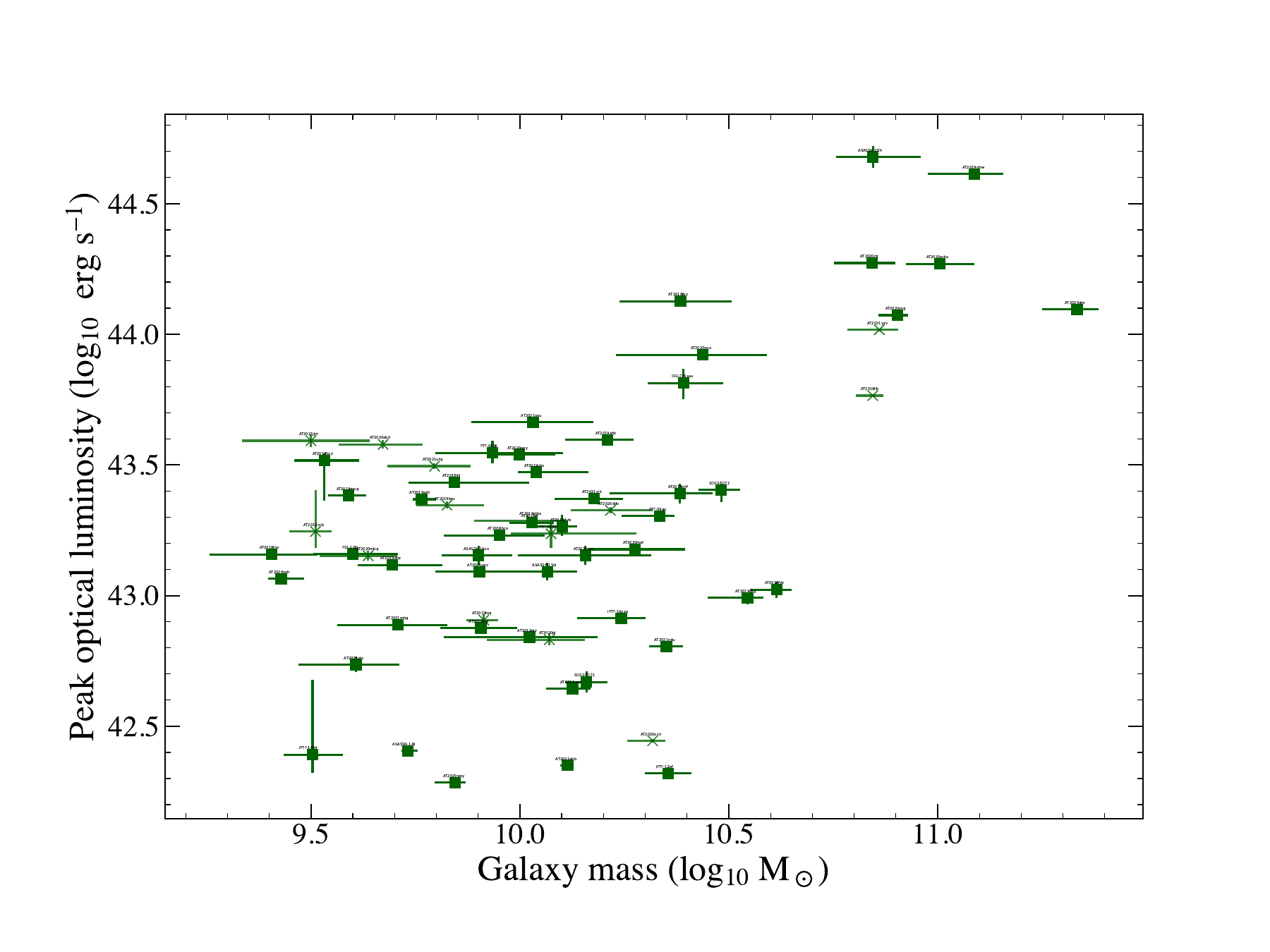}
\caption{TDE light curve properties versus host galaxy stellar mass.  In the middle and lower panel, the sources without a detected plateau are indicated with {a cross} symbol. }\label{fig:masscorr}    
\end{figure}

\subsubsection{Comparison to host galaxy mass}
Since the stellar mass of the host galaxy correlates with black hole mass, it will be instructive to compare the TDE lightcurve properties we extracted from the data to the mass of their host galaxies. The results are summarized in Table~\ref{tab:masscorr}. We find that the strongest correlation (i.e., lowest scatter and highest significance) is found between host galaxy mass and plateau luminosity. The significance of the correlation exceeds $4\sigma$, as measured with the non-parametric Kendall's $\tau$ test. 

We also recover correlations that have been reported in the literature, such as host galaxy mass and peak luminosity \citep{Hammerstein23}, fallback rate \citep{vanVelzen20_ISSI}, or e-folding time \citep{Blagorodnova17}.

In Fig.~\ref{fig:masscorr} we show the correlation between host galaxy mass and plateau luminosity. We also show two other light curve properties that are more easily extracted from the data: the peak-optical luminosity and the early-time energy radiated in the $g$-band (measured from the peak luminosity and the $e$-folding time). 

\subsection{Plateau non-detections}\label{sec:non_dec}
For \Nnot\ of the \Ntot\ TDE we obtain no clear plateau detections. This could be due to limited sensitivity (in part due to the negative flux residuals that sometimes plague ZTF data, see Sec.\ref{sec:compgal}) or due to a lack of intrinsic plateau emission in a subset of TDEs.  Comparing the detections to the non-detections in an Anderson-Darling test, we find no statistically significant difference in the redshift, host galaxy mass or peak luminosity of the two populations. There is evidence ($p=0.02$) that the non-detections occur in TDEs with a lower peak flux. This should be expected because the plateau flux can be two orders of magnitude lower than the peak flux. Hence for TDEs with a relatively low peak flux, it will be harder to detect a plateau. We note that all TDEs with an optical peak flux brighter than $m=18.4$ (21 sources) have a detected plateau. 

It is worth noting that two TDEs show evidence for a disappearing plateau. First of all, the source SDSS-TDE1 is detected in the UV 600 days post peak, yet much more sensitive late-time HST UV observations obtained 2600 days later yield no detection \citep{vanVelzen18_FUV}. Second, the source AT2018dyb \citep[ASASSN-18pg;][]{Holoien20} shows evidence for a plateau at 400~days post-peak , but has faded by at least a factor of 10 about 1000 days later. And finally, the source AT2018fyk \citep{Wevers19a} shows a dramatic decrease of the late-time UV emission followed by a rebrightening in the most recent observations \citep{Wevers2021}.  

While these 3 sources make up a small fraction of the total population of TDEs (\Ntot), this behaviour is of intrinsic  interest.  While it is beyond the scope of this work to examine the light curves of individual TDEs, we note that this behaviour may be indicative of instabilities  in the accretion flow (of a possible viscous \citet{LightmanEardley74}, or thermal \citet{Shakura73}, origin), or a state transition in the disc as the Eddington ratio of these sources falls with time. The presence (or lack thereof) of instabilities in accretion flows (which have long been predicted by simple 1D disc theory) is a long and controversial topic, as they are often recovered in MHD simulations \citep[e.g.,][]{Jiang13, Fragile18}, yet observations of accreting X-ray binaries are typically well described by stable thermal discs \citep[e.g.,][]{Done07}. Future detailed modelling of these 3 sources is of interest, and may offer some insight into fundamental questions of accretion physics.

\begin{figure}
\includegraphics[width=0.5\textwidth]{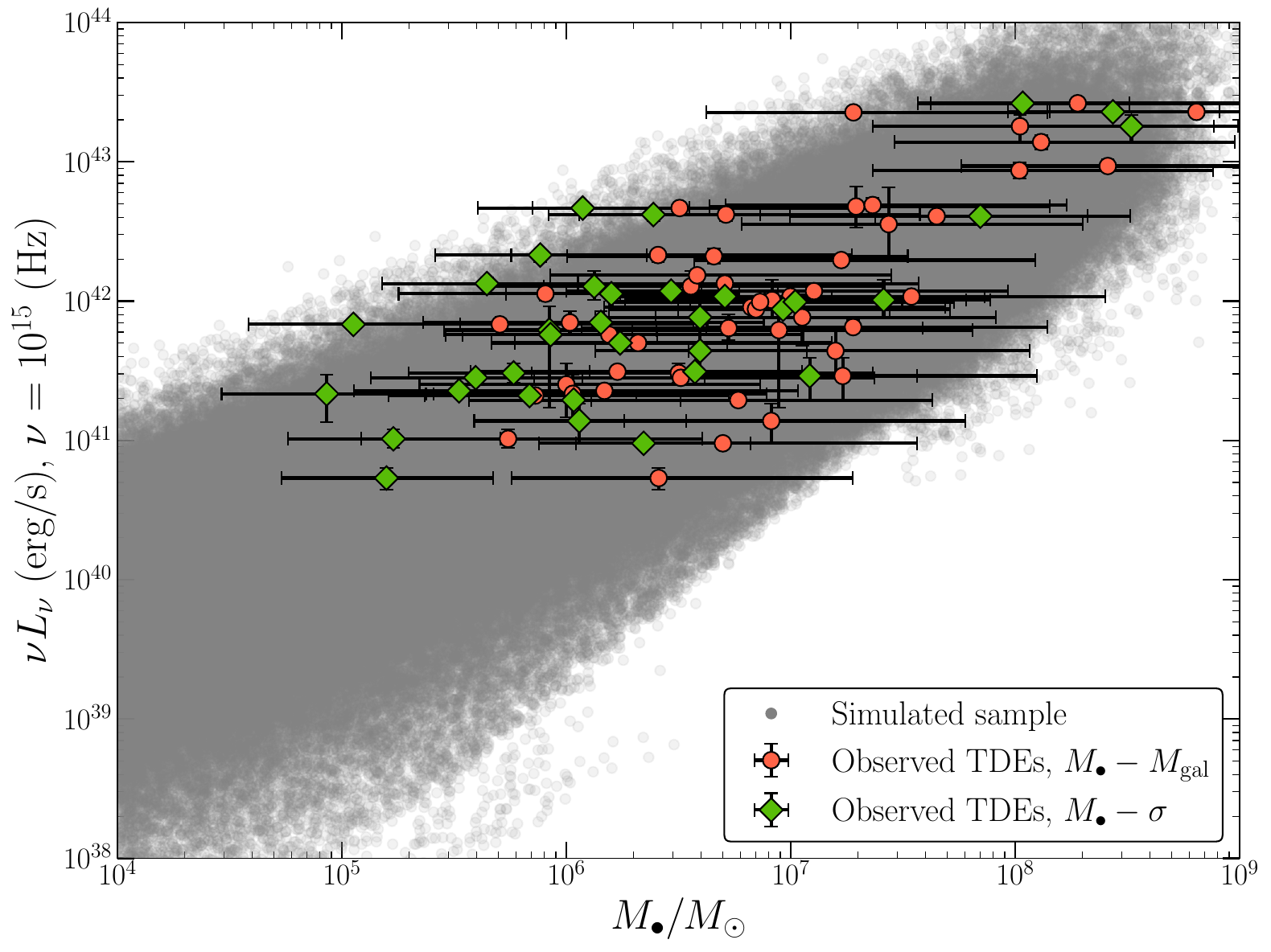}
\includegraphics[width=0.5\textwidth]{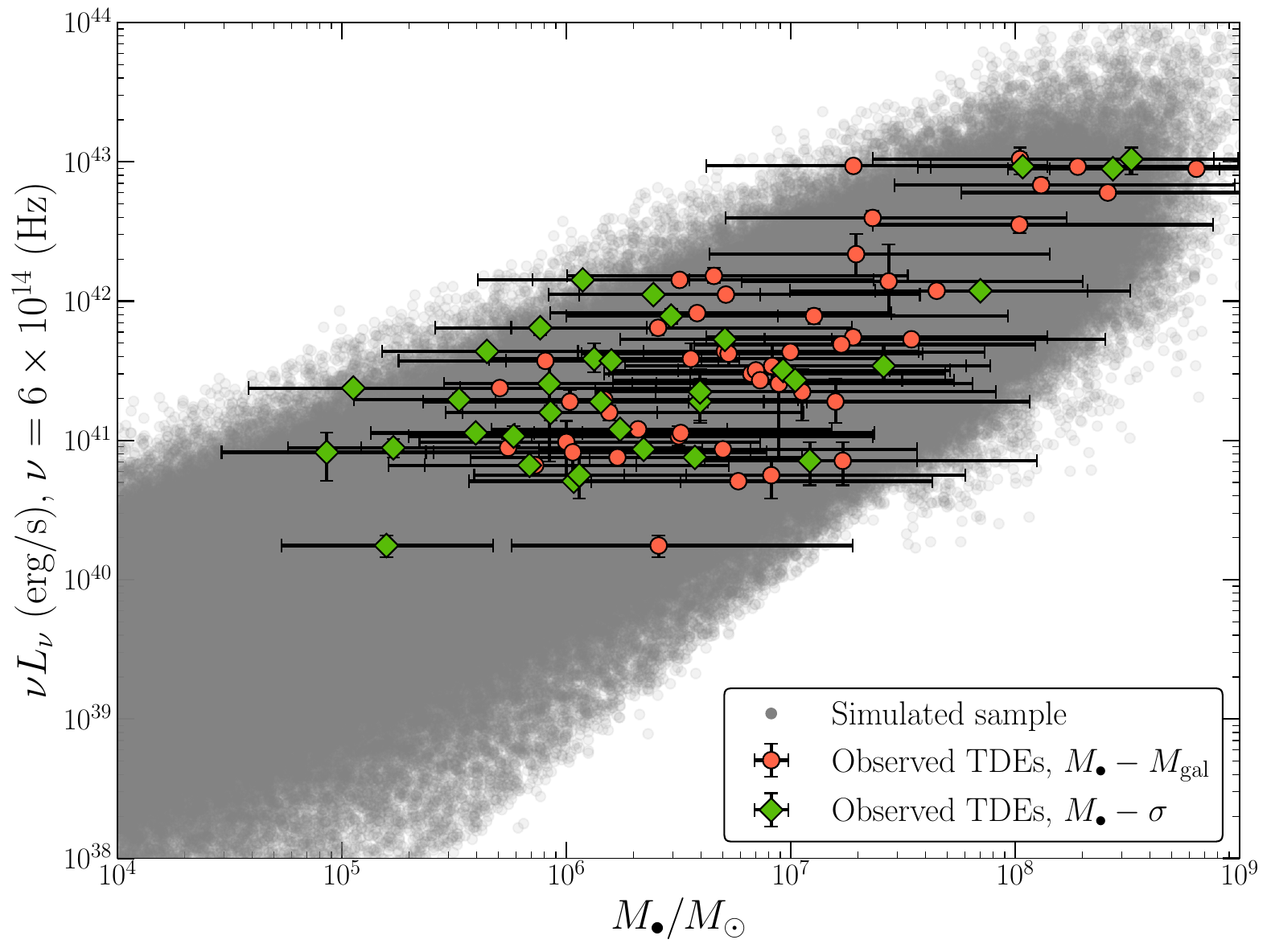}
\caption{A comparison of the theoretical and observational  $\nu L_\nu-\M $ TDE populations, at 2 different observing frequencies (top = NUV wavelength, bottom = optical $g$-band wavelength, all calculated in the source rest frame). In grey points we display the simulated population. We denote by green diamonds the observed TDEs with a  $\M-\sigma$ estimate of their black hole mass, and by orange circles those with $\M- M_{\rm gal}$ mass estimates. Some TDEs appear twice if they reside in galaxies with both a $\sigma$ and $ M_{\rm gal}$ measurement. The error bars on the plateau luminosities may be smaller than the marker sizes.   It is clear to see that the observed population of TDEs fit exactly with the theoretical distribution.   }
\label{sim+real}
\end{figure}

\section{Comparing theory and observations }\label{sec:6}
In this section we compare the properties of the observed population of TDEs with our simulated distribution. 

\subsection{Verification of relationship }
The first important test of our analysis is whether those TDEs with both a measured plateau luminosity and a black hole mass estimate from a galactic scaling relationship lie on the theoretical $\nu L_\nu - \M$ distribution.  

To test this, we plot in Fig. \ref{sim+real} the simulated $\nu L_\nu - \M$ distribution, and over-plot the observed TDE distribution. In Fig. \ref{sim+real} we display by grey points the simulated population. We plot the entire population at different observing frequencies.  
For the observed distribution of TDEs, the black hole masses are computed from either the galactic mass scaling relationship \citep{Greene20} 
\beq\label{sig_scale}
\log_{10} \left[\M/M_\odot \right]= 7.43 + 1.61 \log_{10} \left[ M_{\rm gal}\big/(3\times 10^{10} M_\odot) \right],
\eeq
or the $\M-\sigma$ relationship \citep{Greene20}
\beq\label{galmass_scale}
\log_{10} \left[\M/M_\odot \right] = 7.87 + 4.38 \log_{10} \left[\sigma\big/(160\, {\rm km\,s^{-1}}) \right].
\eeq
The intrinsic scatter in the $\M- M_{\rm gal}$ relationship is 0.8 dex, while the intrinsic scatter in the $\M-\sigma$ relationship is 0.5 dex.


In Fig. \ref{sim+real} we denote by green diamonds those TDEs with a  $\M-\sigma$ estimate of their black hole mass, and by orange circles those with $\M- M_{\rm gal}$ mass estimates. Some TDEs appear twice if they reside in galaxies with both a $\sigma$ and $ M_{\rm gal}$ measurement. The error bars on the plateau luminosities may be smaller than the marker sizes.   It is clear to see that the observed population of TDEs fit exactly with the theoretical distribution.  The observed amplitude of emission is consistent at multiple observing frequencies (the two panels), and therefore the late time spectral shape of the optical/UV emission is also consistent between theory and observation, although it is important to bear in mind that the two observing frequencies are only separated by a factor of 5/3.  

 As we argued previously (Fig. \ref{sim_plat}), a measurement of the late-time UV plateau luminosity can now be used as a method of measuring the masses of the supermassive black holes at the heart of a TDE. 
In Table \ref{mass_plat_table} we record the black hole masses inferred from each of the \Nplat\ TDEs with a measured late time plateau luminosity.

\begin{figure*}
\includegraphics[width=0.49\textwidth]{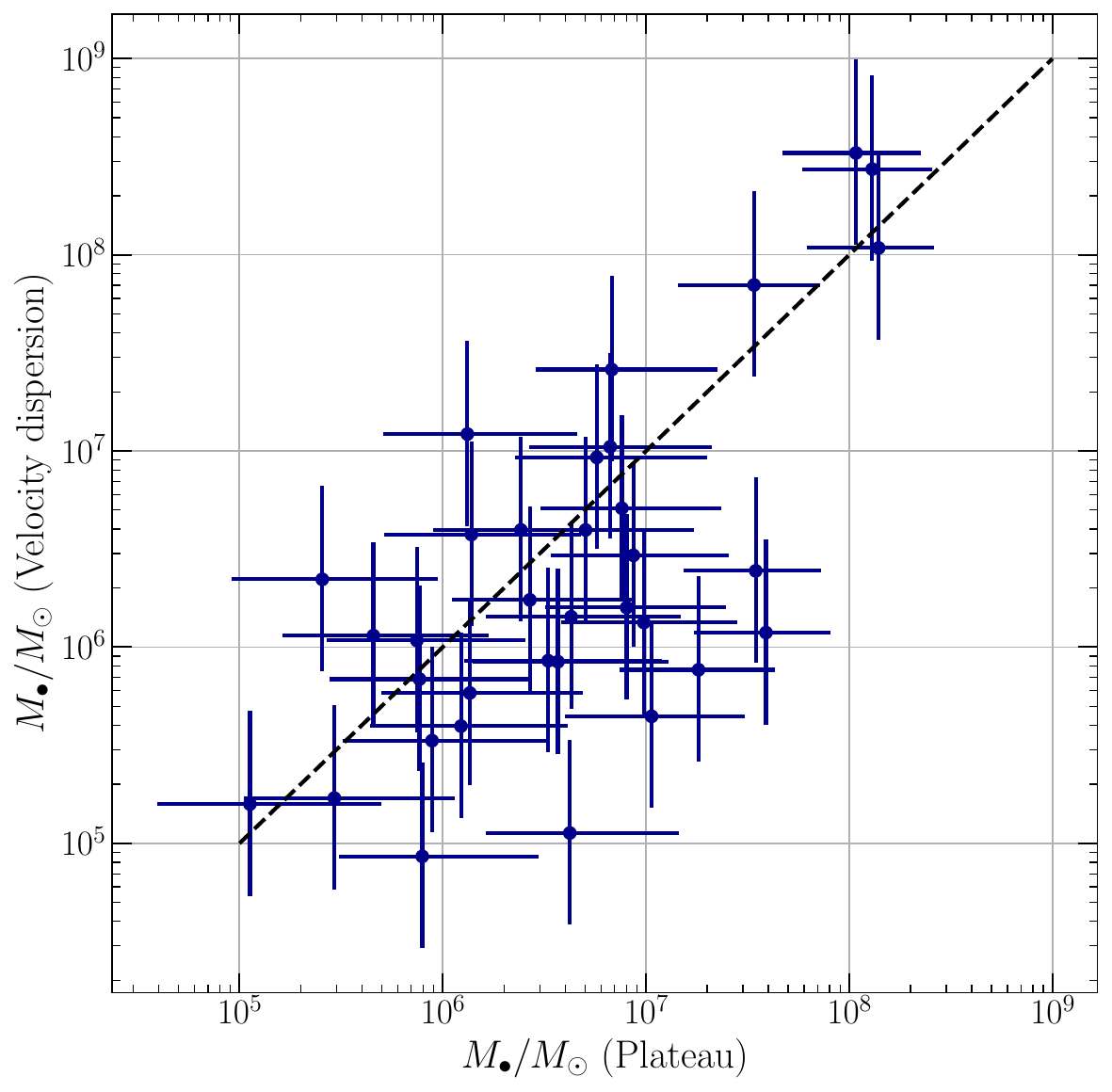}
\includegraphics[width=0.49\textwidth]{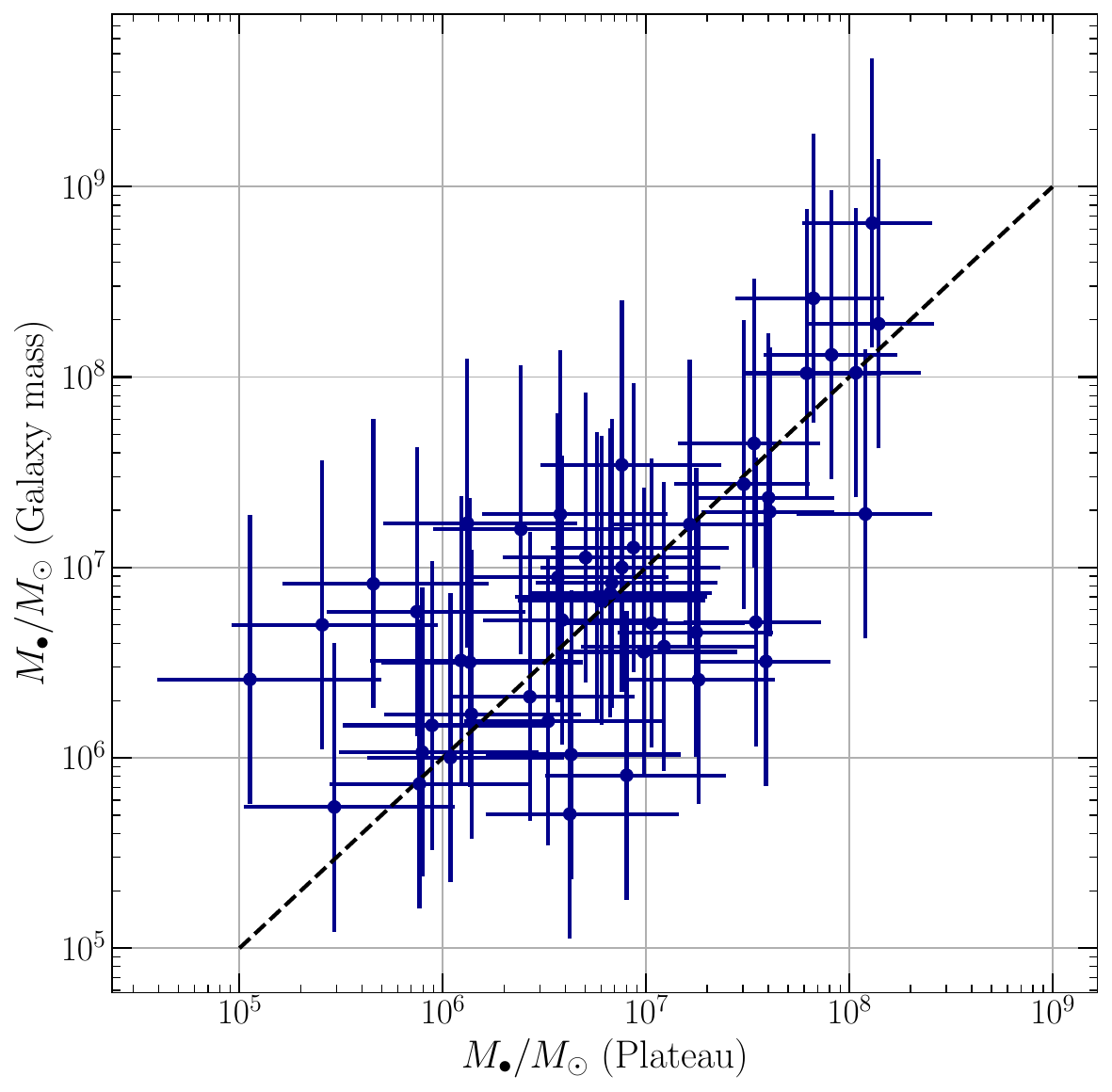}
\includegraphics[width=0.49\textwidth]{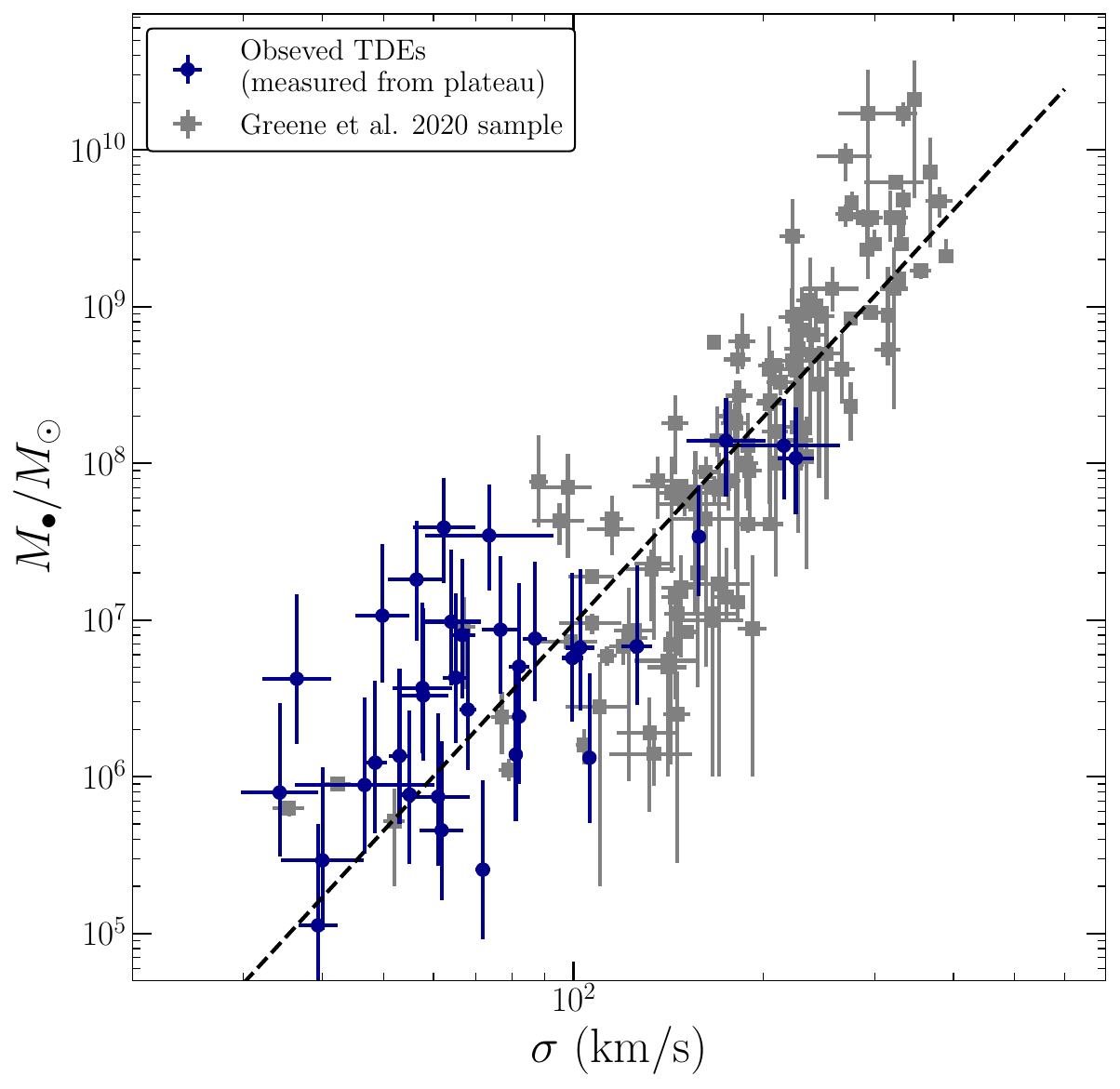}
\includegraphics[width=0.49\textwidth]{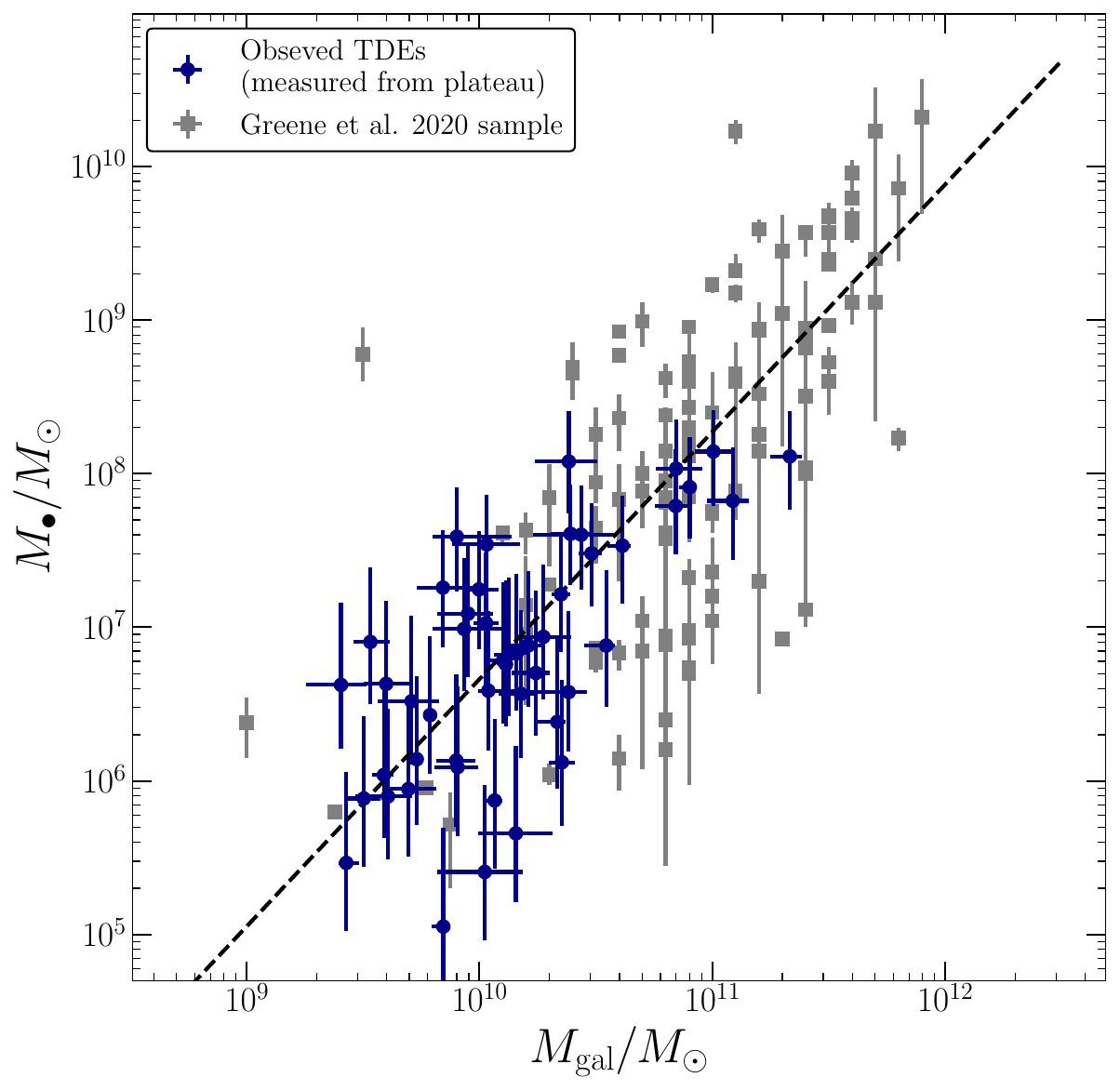}
\caption{ Upper: TDE black hole mass-mass plots, where on the horizontal axis we plot the mass as inferred from the TDE plateau, and on the vertical axis we plot the mass as inferred from a galactic scaling relationship (on the left we use the velocity dispersion $\sigma$, and on the right the host galaxy mass $M_{\rm gal}$). The black dashed line shows $\M = \M$, i.e., perfect agreement between the independent approaches. Lower: the combined populations of black hole masses and galactic properties (again on the left we  display velocity dispersion $\sigma$, and on the right the host galaxy mass $M_{\rm gal}$). The points in grey are taken from the paper \citet{Greene20}, while the points in blue are the TDEs we are able to add in this analysis. The black dashed lines in these two plots are the scaling relationships presented in \citet[our equations \ref{sig_scale} and \ref{galmass_scale}]{Greene20}. The black hole masses inferred from the plateau luminosity  correlate strongly with both the black hole masses inferred from the host galaxy mass (Kendall's $\tau = 0.46, p = 2.5 \times 10^{-6})$ and host velocity dispersion $(\tau = 0.39, p = 1.3 \times 10^{-3})$.   It is clear that the black hole masses inferred from the TDE plateaus fit as is expected with the pre-existing galactic populations.  }
\label{MvsM}
\end{figure*}

\subsection{Galactic scaling relationships  } 
\begin{table}
    \centering
    \begin{tabular}{l p{140pt}}
    Reference & TDEs   \\
    \hline\hline

\citet{Yao23} & AT2018iih, \mbox{AT2019azh}, \mbox{AT2019dsg}, \mbox{AT2020acka}, \mbox{AT2020mot}, \mbox{AT2020vwl}, \mbox{AT2020wey}, \mbox{AT2021axu}, \mbox{AT2021crk}, \mbox{AT2021ehb}, \mbox{AT2021mhg}, \mbox{AT2021nwa}, \mbox{AT2021uqv}, \mbox{AT2021yte} \\
\citet{Wevers17} & ASASSN-14ae, \mbox{ASASSN-14li}, \mbox{PS1-10jh}, \mbox{PTF-09axc}, \mbox{PTF-09djl}, \mbox{PTF-09ge}, \mbox{SDSS-TDE1}, \mbox{iPTF-15af}, \mbox{iPTF-16axa}, \mbox{iPTF-16fnl} \\
\citet{Hammerstein23b} & AT2018hyz, \mbox{AT2018lna}, \mbox{AT2018lni}, \mbox{AT2018zr}, \mbox{AT2019ehz}, \mbox{AT2019qiz}, \mbox{AT2020ddv}, \mbox{AT2020ocn}, \mbox{AT2020qhs}, \mbox{AT2020zso} \\
\citet{Wevers20} & ASASSN-15oi, \mbox{AT2018fyk} \\
\citet{Kruhler18} & ASASSN-15lh \\
   
    \end{tabular}
\caption{Origin for velocity dispersion measurements.}
\label{tab:sigmarefs}
\end{table}

Using measurements of the late time TDE plateau luminosity, we now have \Nplat\ TDE systems with black hole masses estimated directly from their observed light curves. This allows us to pose and answer a number of interesting questions. In Fig. \ref{MvsM} we display four tests of our analysis. In the upper two panels we display mass-mass plots, where on the horizontal axis we plot the mass as inferred from the TDE plateau, and on the vertical axis we plot the mass as inferred from a galactic scaling relationship (on the left we use the velocity dispersion $\sigma$, and on the right the host galaxy mass $M_{\rm gal}$; see equations \ref{sig_scale} and \ref{galmass_scale} for the explicit galactic scaling relationships). The TDE host galaxy velocity dispersion measurements are obtained from the literature (see Table~\ref{tab:sigmarefs}). The black dashed line shows $\M = \M$, i.e., perfect agreement between the independent approaches.  The black hole masses inferred from TDE plateaus are correlated with the black hole masses inferred from galactic properties. Quantitatively, a Kendall $\tau$ test finds the black hole masses inferred from the plateau luminosity  correlate strongly with both the black hole masses inferred from the host galaxy mass $(\tau = 0.46, p = 2.5 \times 10^{-6})$ and host velocity dispersion $(\tau = 0.39, p = 1.3 \times 10^{-3})$. Note that the lower significance of the velocity dispersion correlation stems principally from the fewer sources (34) with velocity dispersion measurements than galaxy mass measurements (\Nplat).

\begin{figure*}
\includegraphics[width=0.49\textwidth]{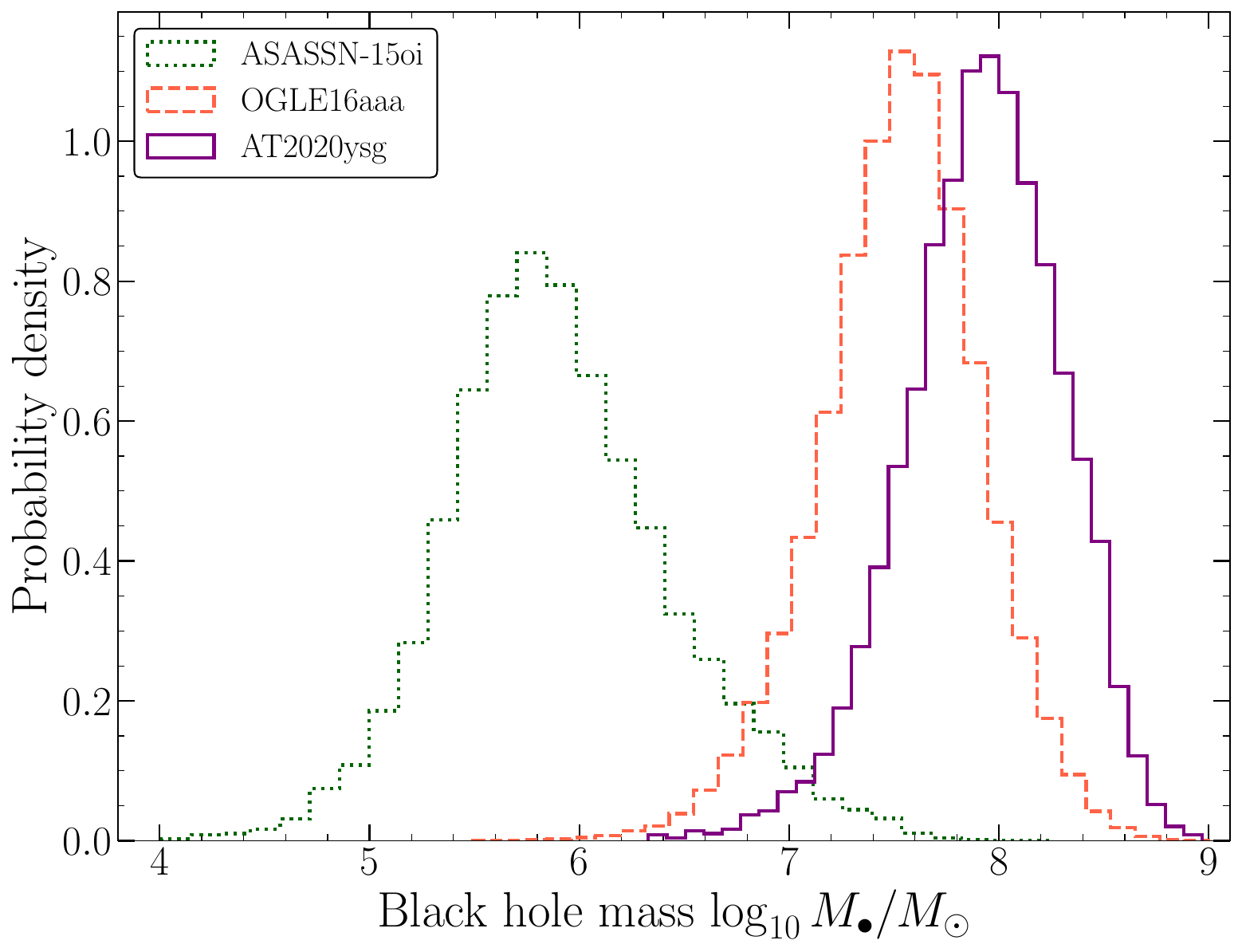}
\includegraphics[width=0.49\textwidth]{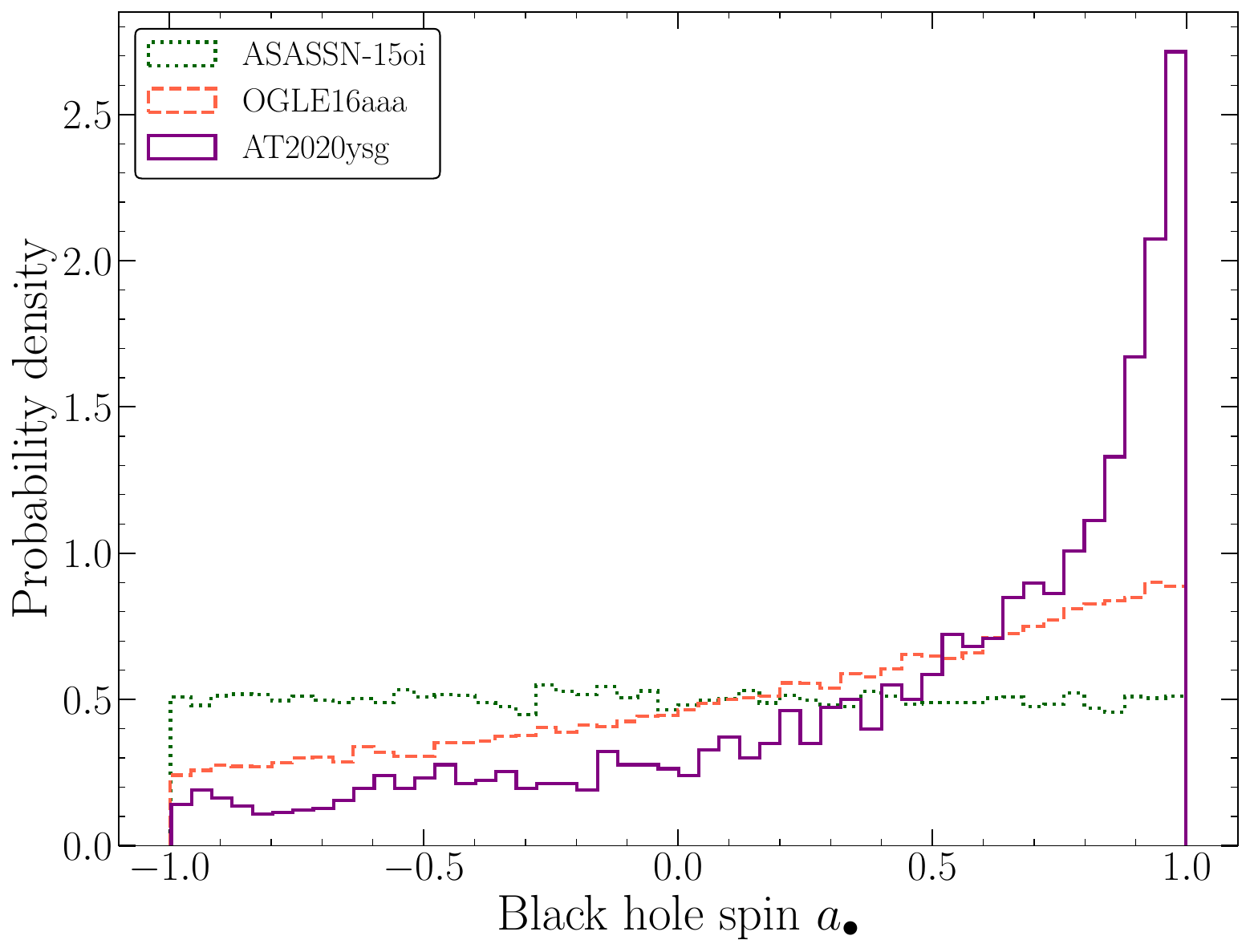}

    \caption{Spin and mass constraints of 3 TDEs with varying levels of plateau luminosities. The brightest plateaus (like AT2020ysg; purple solid curves) in our sample correspond to large mass and rapidly rotating black holes, an effect driven by the Hills {mass} (see text). While slightly less bright plateaus (such as OGLE16aaa; orange dashed curves) require only slightly less massive black holes, this has a strong effect on the spin constraints. Those TDEs around low mass black holes with dim plateaus (such as ASASSN-15oi; green dotted curves) have zero spin-constraining power. We stress that these constraints are somewhat conservative, as we have assumed a uniform background spin distribution of supermassive black holes.  }
    \label{SpinDistsTDE}
\end{figure*}

\begin{figure}
    \centering
    \includegraphics[width=\linewidth]{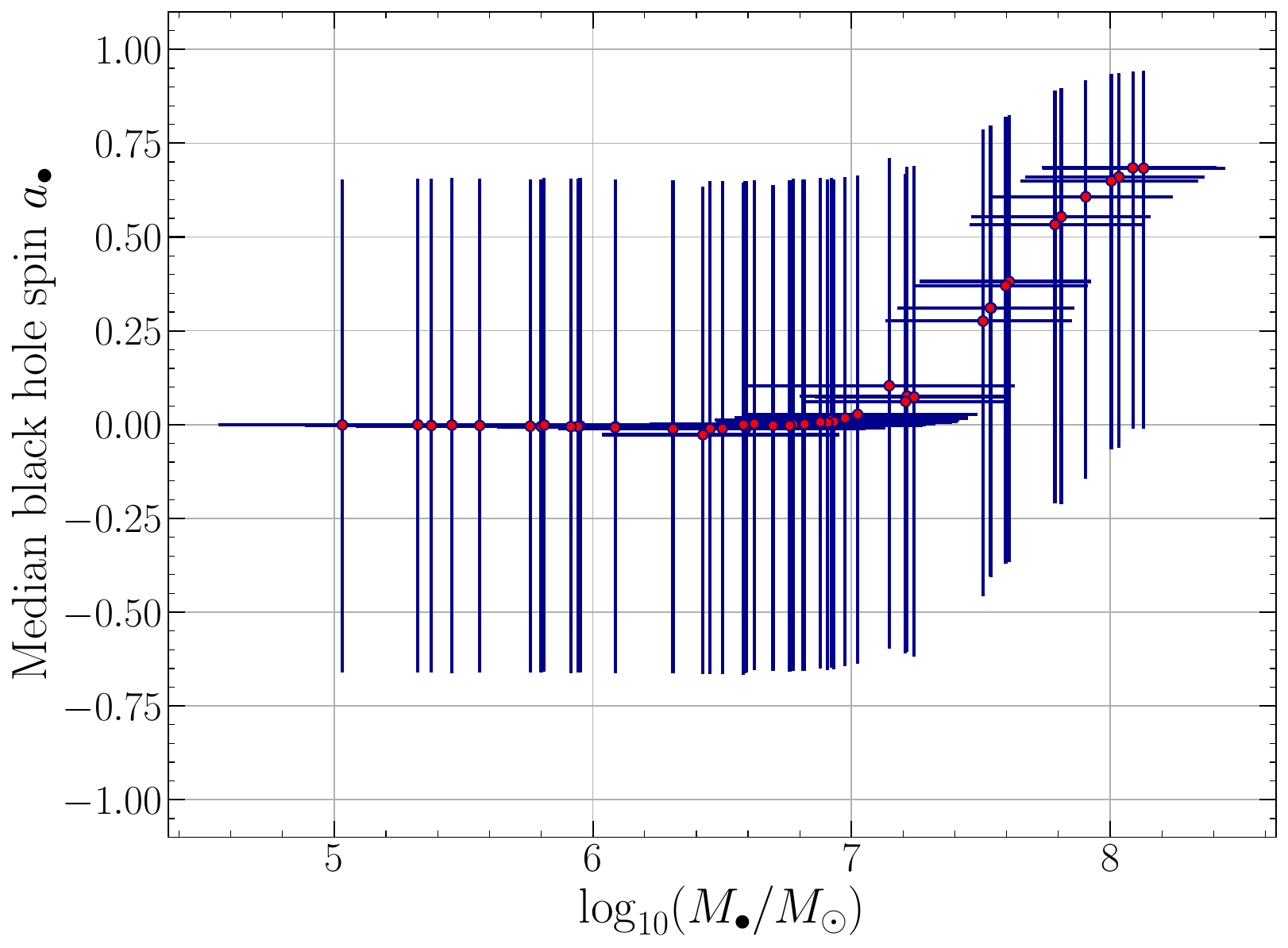}
    \caption{Inferred black hole spin (median and 1$\sigma$ uncertainty) plotted as a function of black hole mass for the \Nplat\ TDEs in our sample with plateau luminosity measurements. For sources with inferred masses below $10^7 M_\odot$ we have no black hole spin constraining power from the plateau, and the distribution returns the input flat distribution (with median $a_\bullet = 0$ and one-sigma range $\pm 0.67$). At high inferred masses the TDEs must be rapidly rotating.   }
    \label{mass_v_spin}
\end{figure}

In the lower two panels we show the combined populations of black hole masses and galactic properties (again on the left we  display velocity dispersion $\sigma$, and on the right the host galaxy mass $M_{\rm gal}$). The points in grey are taken from the paper \citet[their tables 2, 6, 7, 8]{Greene20}, while the points in blue are the TDEs we are able to add in this analysis. The black dashed lines in these two plots are the scaling relationships presented in \citet[our equations \ref{sig_scale} and \ref{galmass_scale}]{Greene20}.   It is clear that the black hole masses inferred from the TDE plateaus fit as is expected with the pre-existing galactic populations.

\subsection{Spin constraints on TDE black holes }

In addition to providing constraints on the mass of the central black hole in a TDE the late time plateau luminosity can, for particularly bright plateau luminosities, also place constraints on the spin of the central black hole. The reason for this is the {dependence of the} Hills {mass on black hole spin} \citep{Hills75}, discussed in section \ref{hill_mech}. In effect, owing to their smaller tidal forces and larger event horizons, more slowly rotating black holes can disrupt a given star only if they are less massive than a more rapidly rotating black hole. As we demonstrated earlier, a maximally rotating Kerr black hole can disrupt a given star at masses 8 times higher than a Schwarzschild black hole. Therefore at the highest black hole masses there will be an over representation of rapidly rotating black holes in the observed TDE population.

This is shown most concretely in the upper left panel of Fig. \ref{scatter}, where we colour each of the plateau luminosity measurements of our sample by black hole spin. At high black hole masses the observed distribution of TDE plateaus is completely dominated by rapidly rotating black holes, {\it despite the input distribution of black hole spins being assumed to be uniform (section \ref{BH_DIST})}. This uniform input distribution can be seen at lower black hole masses (Fig. \ref{scatter}).  

In the observed TDE plateau luminosity population, there are a number of sources at high luminosities ($L_{\rm plat} > 10^{43}$ erg/s in the NUV-band; Fig. \ref{sim+real}). These sources also have large masses as inferred from their galactic scaling relationships. These sources therefore are highly likely to contain rapidly rotating and massive black holes. 

In Fig. \ref{SpinDistsTDE} we display three examples of joint spin and mass constraints for TDEs with differing levels of plateau luminosities. These distributions were computed by recording all TDE systems in our simulated sample within 1$\sigma$ of the observed plateau luminosity of each TDE. The 10 brightest sources in our sample all have spin distributions with median $a_\bullet > 0.3$, indicative of rapid rotation.  Example joint black hole mass-spin constraints are highlighted by Fig. \ref{SpinDistsTDE}. 

We stress that these constraints are somewhat conservative, as we have assumed a uniform background spin distribution of supermassive black holes. As the Hills mass is a stronger function of black hole spin than stellar properties (eq. \ref{fullhills}), to explain these high mass sources with Schwarzschild black holes would require a large population of very high mass stars, and an unreasonably top-heavy IMF. 
The spin distributions of the remaining 39 TDEs are effectively indistinguishable from the flat input distribution (see Table \ref{spin_plat_table}). 

The inferred (median and 1$\sigma$ uncertainty) black hole spins and masses of our sample are displayed in Figure. \ref{mass_v_spin}.  For sources with inferred masses below $10^7 M_\odot$ we have no black hole spin constraining power from the plateau, and the distribution returns the input flat distribution (with median $a_\bullet = 0$ and one-sigma range $\pm 0.67$). At high inferred masses the TDEs must be rapidly rotating.

\begin{figure}
    \centering
    \includegraphics[width=\linewidth]{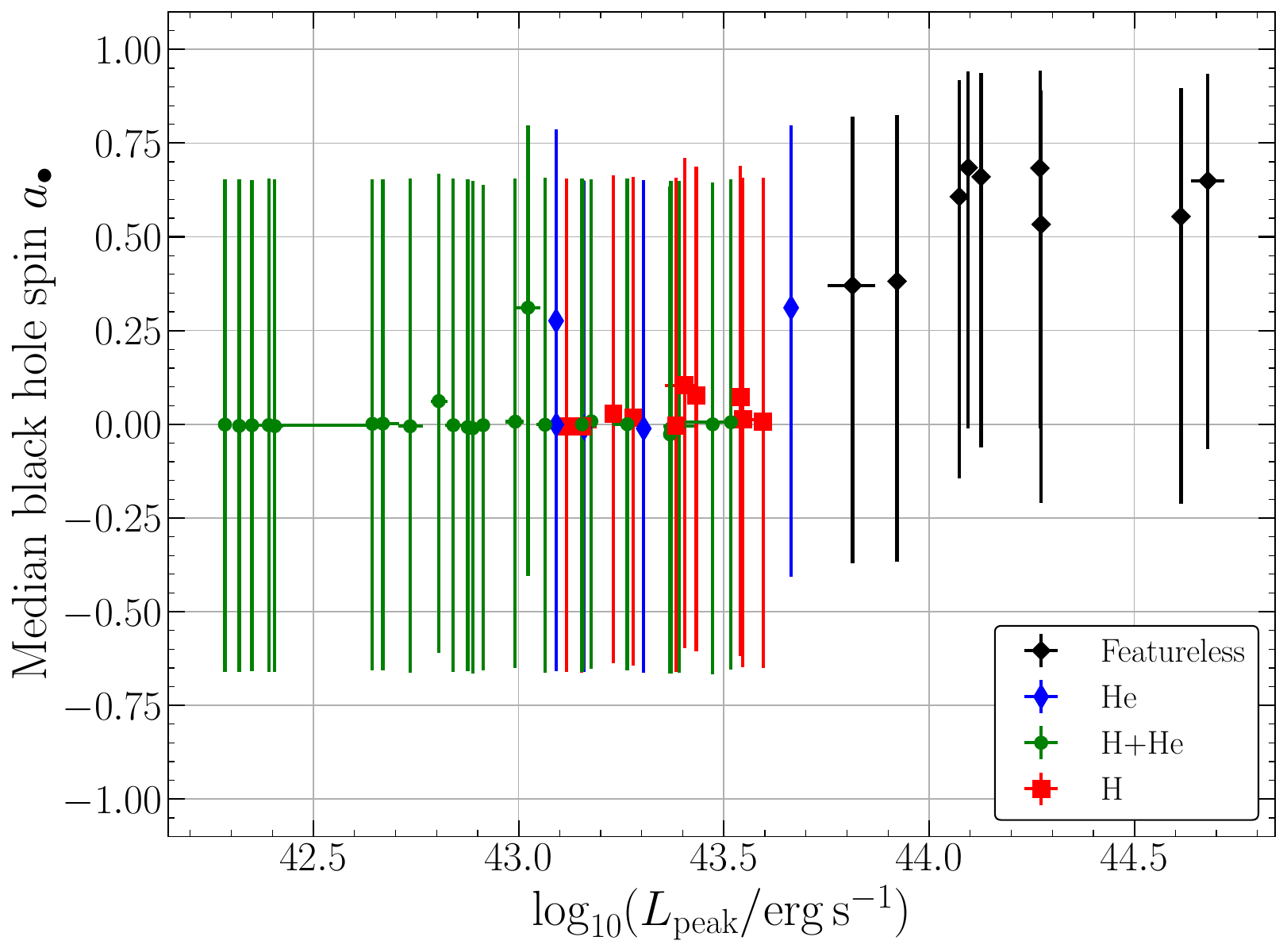}
    \caption{Inferred black hole spin (median and 1$\sigma$ uncertainty) plotted as a function of peak $g$-band luminosity, and coloured by TDE spectral type,  for the \Nplat\ TDEs in our sample with plateau luminosity measurements. For sources with inferred masses below $10^7 M_\odot$ we have no black hole spin constraining power from the plateau, and the distribution returns the input flat distribution (with median $a_\bullet = 0$ and one-sigma range $\pm 0.67$). These sources typically have peak $g$-band luminosities lower than $10^{43.5}$ erg/s. At high inferred masses the TDEs must be rapidly rotating, corresponding to the brightest $g$-band sources in our sample. All but one of our brightest and most rapidly rotating black holes produce TDEs with featureless spectra.   }
    \label{peak_v_spin}
\end{figure}

In Fig. \ref{peak_v_spin} we plot the inferred black hole spin (median and 1$\sigma$ uncertainty) plotted as a function of peak $g$-band luminosity, and coloured by TDE spectral type,  for the \Nplat\ TDEs in our sample with plateau luminosity measurements. {The spectral type describes which broad lines are detected in the optical spectra of the source, just Hydrogen (H), Hydrogen and Helium (H + He), just Helium (He), or no lines (Featureless) \citep{vanVelzen21}.  } For sources with inferred masses below $10^7 M_\odot$ we have no black hole spin constraining power from the plateau, and the distribution returns the input flat distribution (with median $a_\bullet = 0$ and one-sigma range $\pm 0.67$). These sources typically have peak $g$-band luminosities lower than $10^{43.5}$ erg/s. At high inferred masses the TDEs must be rapidly rotating, corresponding to the brightest $g$-band sources in our sample. All but one of our brightest and most rapidly rotating black holes produce TDEs with featureless spectra (a new class of TDEs first reported in \citealt{Hammerstein23}).  

\section{TDE scaling relationships }\label{sec:7}
\begin{figure*}
    \includegraphics[width=0.48\textwidth]{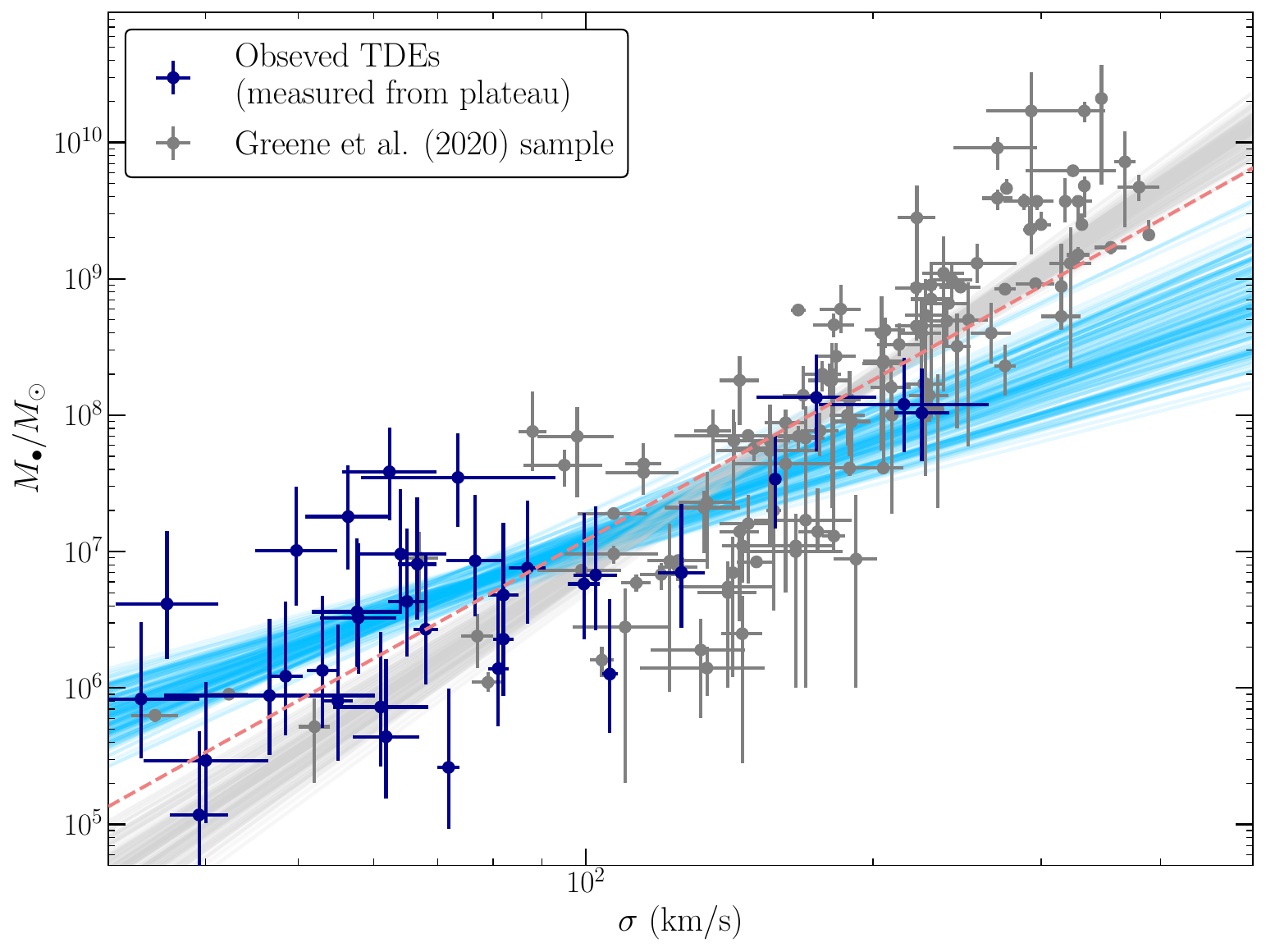}
    \includegraphics[width=0.48\textwidth]{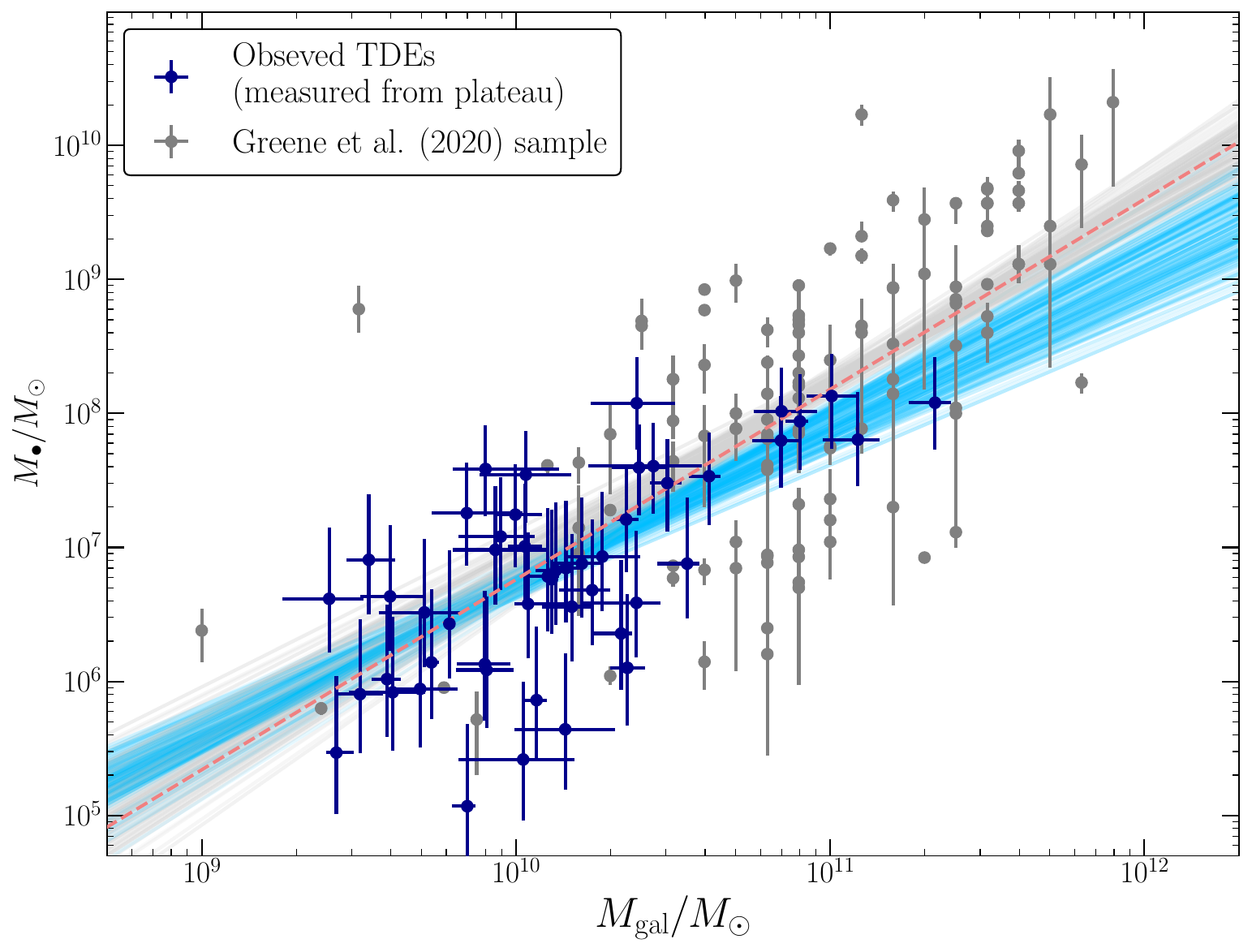}
    \includegraphics[width=0.48\textwidth]{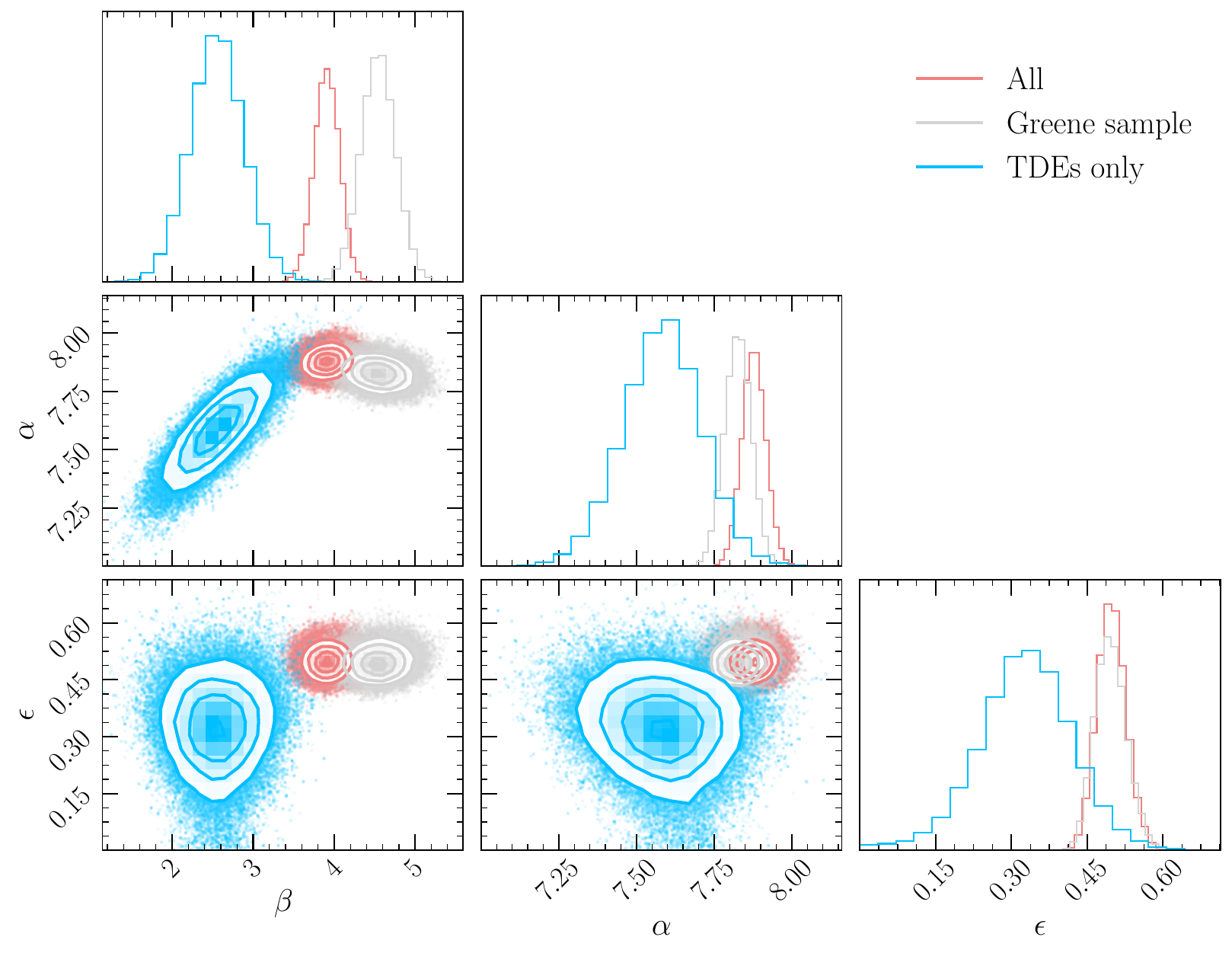}
    \includegraphics[width=0.48\textwidth]{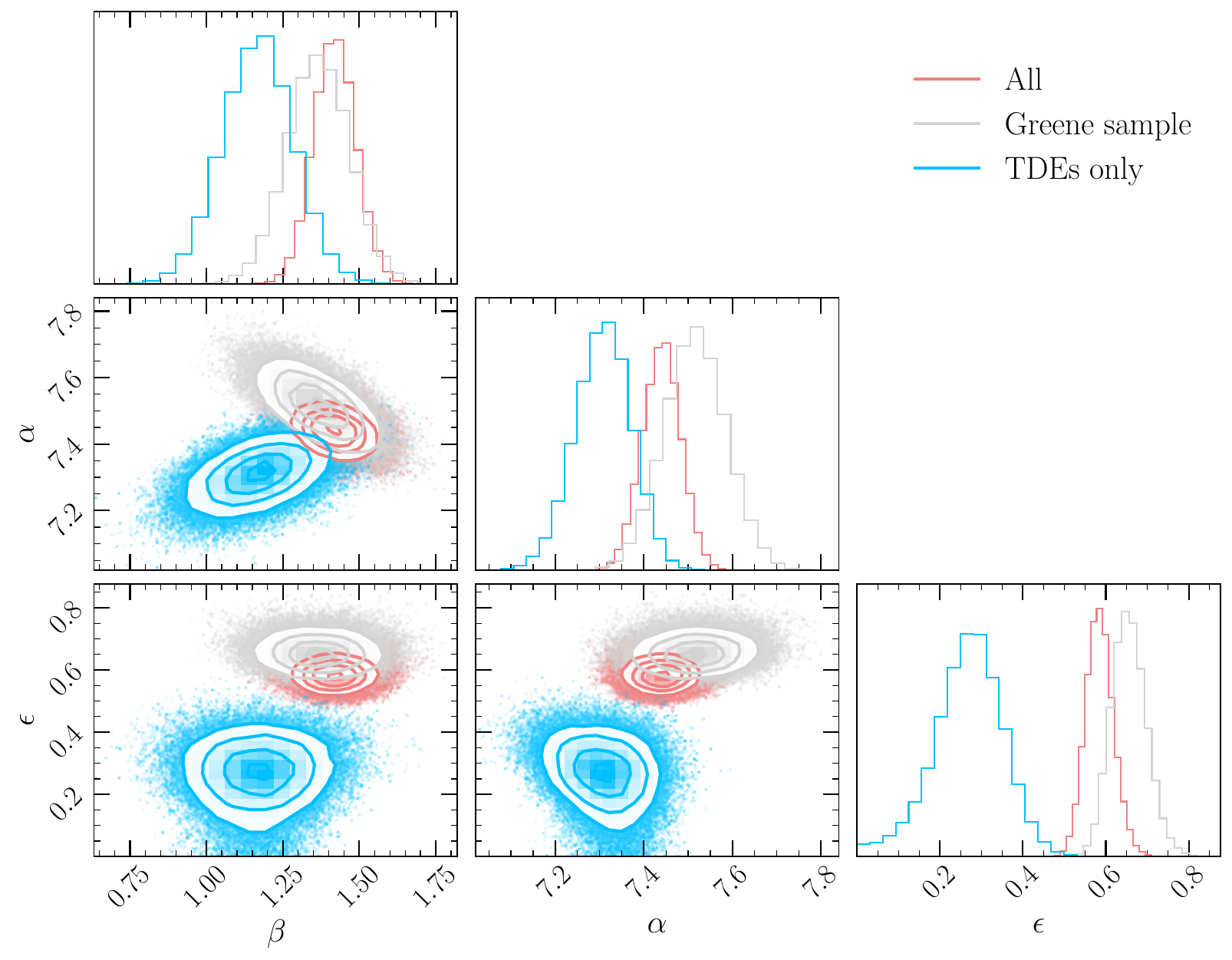}
    \caption{MCMC fits to galactic scaling relationships, including the \citet{Greene20} sample (grey points) and the new additions of TDEs (blue points; on the left we display the velocity dispersion $\sigma$, and on the right the host galaxy mass $M_{\rm gal})$. Contours shown in blue display fits to only the population of TDEs, while contours shown in grey are fits to only the \citet{Greene20} sample. Red contours (lower panels) display a joint fit, the median of which is displayed by a red dashed line in the upper panels. Fits to the TDE-only population produce generally shallower scaling relationships than fits to the entire population. This may be a result of a break in the  scaling relationships themselves, or a systematic effect whereby more massive TDE black holes (at fixed galactic properties) are easier to observe at late times.    }
    \label{GalPropFits}
\end{figure*}

In this section we present various scaling relationships between the observed peak and plateau luminosities, and the total energy radiated at early times, and the black hole mass at the heart of the TDE. We also present updated galactic scaling relationships between velocity dispersion, galaxy mass and black hole mass by combining dynamical black hole mass estimates with our TDE-based mass estimates. 

\subsection{Theoretical plateau scaling }
As we have demonstrated in this paper, the late time plateau luminosity correlates strongly with the TDE's central black hole mass. By taking the simulated population (Fig. \ref{sim_plat}), binning in logarithmically spaced mass bins, and fitting a power law profile to the binned luminosities we find a best fitting relationship of 
\beq
 \log_{10} \left({\M \over M_\odot}\right) = 1.50 \, \log_{10} \left({ L_{\rm plat} \over 10^{43} \, {\rm erg}\,{\rm s}^{-1}}\right) + 9.0  , 
\eeq
where $L_{\rm plat} \equiv \nu L_{\nu, {\rm plat}}$ in the rest-frame $g$-band $(\nu = 6 \times10^{14} \, {\rm Hz})$. 

In the rest-frame NUV band at $\nu = 10^{15}$ Hz we find 
\beq
 \log_{10} \left({\M \over M_\odot}\right) = 1.50 \, \log_{10} \left({ L_{\rm plat} \over 10^{43} \, {\rm erg}\,{\rm s}^{-1}}\right) + 8.3  .  
\eeq
\subsection{Updated galactic scaling relationships}
With \Nplat\ measurements of black hole masses estimated from late time TDE plateaus, we are able to extend and update galactic scaling relationships between velocity dispersion, galactic mass and black hole mass.

In this and following sections we will fit power-law profiles of the following general form 
\beq\label{power_law_fit}
\log_{10}\left(Y\right) = \alpha + \beta \log_{10}\left(X\right) ,
\eeq
where 
\beq
Y \equiv {\M \over M_\odot} ,
\eeq 
and $X$ will be some normalised scaling variable. To understand the intrinsic scatter in these scaling relationships, we incorporate an intrinsic scatter $\epsilon$ into the uncertainty of the black hole mass measurements 
\beq
\left(\delta \log_{10} Y\right)^2 \to \left( \delta \log_{10} Y\right)^2 + \epsilon^2 , 
\eeq
where $\delta \log_{10} Y \equiv \delta \log_{10} \M/M_\odot$, the uncertainty in the logarithm of each black hole mass measurement. We then minimise the likelihood 
\begin{multline}
 {\cal L} = \sum_{i}  { \left( \log_{10}(Y_i) - \alpha - \beta \log_{10}\left(X_i\right) \right)^2 \over \left(\delta \log_{10}\left(Y_i\right) \right)^2 + \epsilon^2  } \\ + \ln \Big[ \left(\delta \log_{10}\left(Y_i \right)  \right)^2 + \epsilon^2 \Big] ,
\end{multline}
where the summation is over all pairs $(X_i, Y_i)$ of normalised scaling variables and black hole masses. 

We begin with known galactic scaling relationships, combining our TDE sample with the black hole population of \citet{Greene20}.  In common with Greene et al., for the velocity dispersion scaling relationship we define 
\beq
X \equiv {\sigma \over 160 \, {\rm km\, s^{-1}}}, 
\eeq 
while for galactic mass we define 
\beq
X \equiv {M_{\rm gal} \over 3 \times 10^{10} \, M_\odot } .
\eeq

In Fig. \ref{GalPropFits} we display MCMC (performed using \verb|emcee|, \citet{EMCEE}) fits between galactic properties and black hole mass.  On the left we display the velocity dispersion $\sigma$, and on the right the host galaxy mass $M_{\rm gal}$.  In the upper panels we show posterior samples of the power law fits to both the TDE-only population, the Greene-only population, and a combined population.  Posteriors shown in blue display fits to only the population of TDEs, while posteriors shown in grey are fits to only the \citet{Greene20} sample. We display the median of the joint fit by a red dashed line in the upper panels.

In the lower panels we show corner plots of the fits to the different populations.  Contours shown in blue display fits to only the population of TDEs, while contours shown in grey are fits to only the \citet{Greene20} sample. Red contours (lower panels) display a joint fit. Fits to the TDE-only population produce generally shallower scaling relationships than fits to the entire population. This may be a result of a break in the fundamental scaling relationships themselves, or a systematic effect whereby more massive TDE black holes (at fixed galactic properties) are easier to observe at late times. 

For the fit to the combined (\citet{Greene20} and TDE plateau) population we find the following black hole scaling law parameters  
\begin{align}
\alpha &=  7.88 \pm 0.04, \\
\beta &= 3.91 \pm 0.16, \\
\epsilon &= 0.50 \pm 0.03.,
\end{align}
for the velocity dispersion relationship, and 
\begin{align}
\alpha &= 7.44 \pm 0.04, \\
\beta &= 1.42 \pm 0.07, \\
\epsilon &= 0.58 \pm 0.03 ,
\end{align}
for the galaxy mass relationship. 

An interesting result to note is that the intrinsic scatter $\epsilon$ inferred from fits to the TDE only population is generally much smaller than fits to the total, or \citet{Greene20} populations. We do not believe this is reflecting a change in the intrinsic scatter of the systematically lower mass black holes probed by TDEs, but shows that that the statistical uncertainties we assign to TDE-based black hole  mass estimates $(\sim 0.5$ dex) appear to explain the observed variance, while the dynamical mass estimates show more outliers outliers relative to the reported uncertainties.  

\subsection{Empirical radiated energy scaling}
With the plateau luminosity now providing a robust measurement of \Nplat\ TDE black hole masses (e.g., Table \ref{mass_plat_table}), we can now calibrate  other, empirical, correlations between observed TDE light curve parameters and central black hole mass.  We expect these correlations to be present as we learnt earlier Table \ref{tab:masscorr} that the radiated $g$-band energy and peak $g$-band luminosities both correlated with galaxy mass (albeit to a lesser extent than plateau luminosity), which acts as a tracer of black hole mass. 

\begin{figure}
\includegraphics[width=0.5\textwidth]{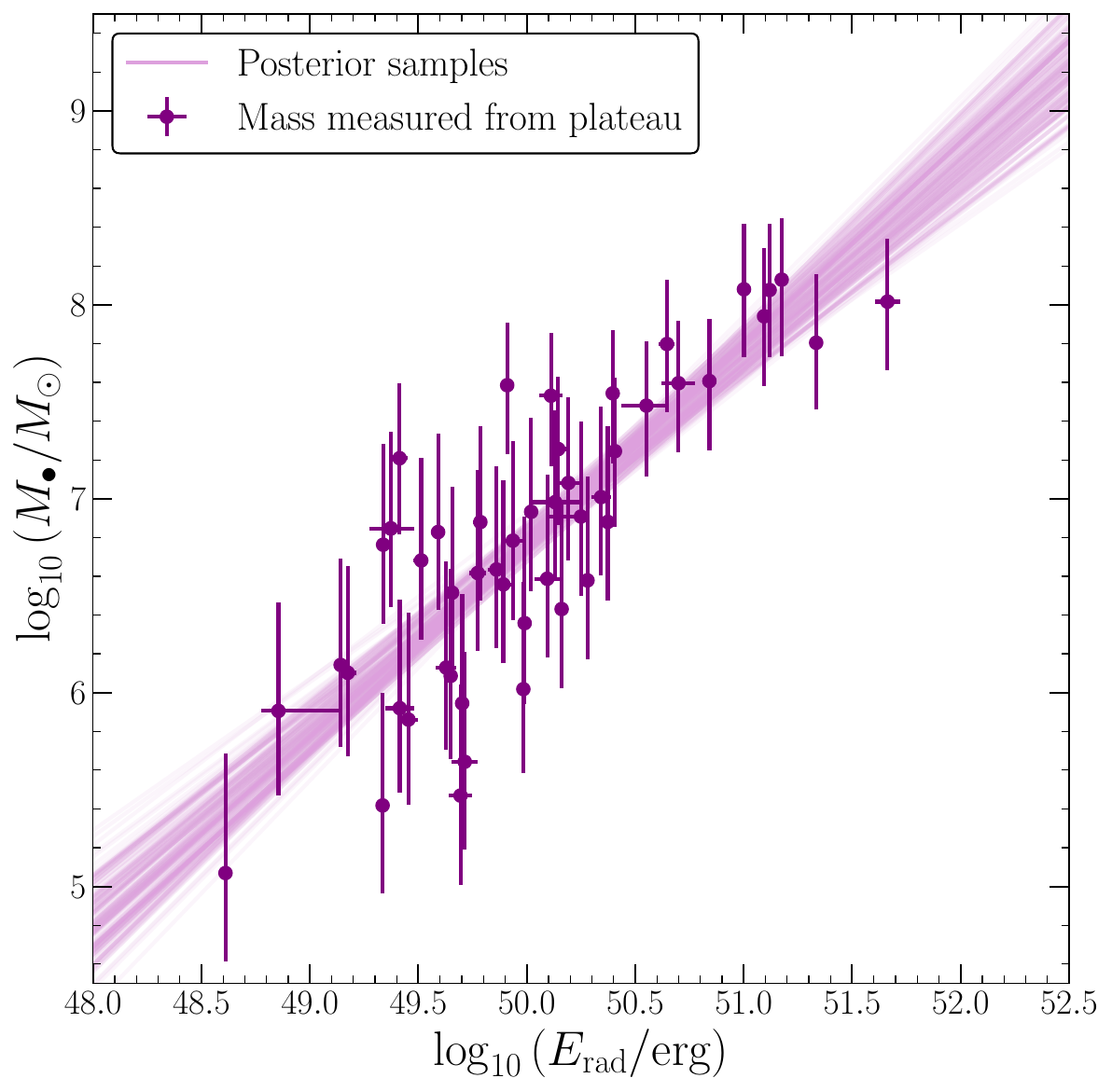}
\includegraphics[width=0.5\textwidth]{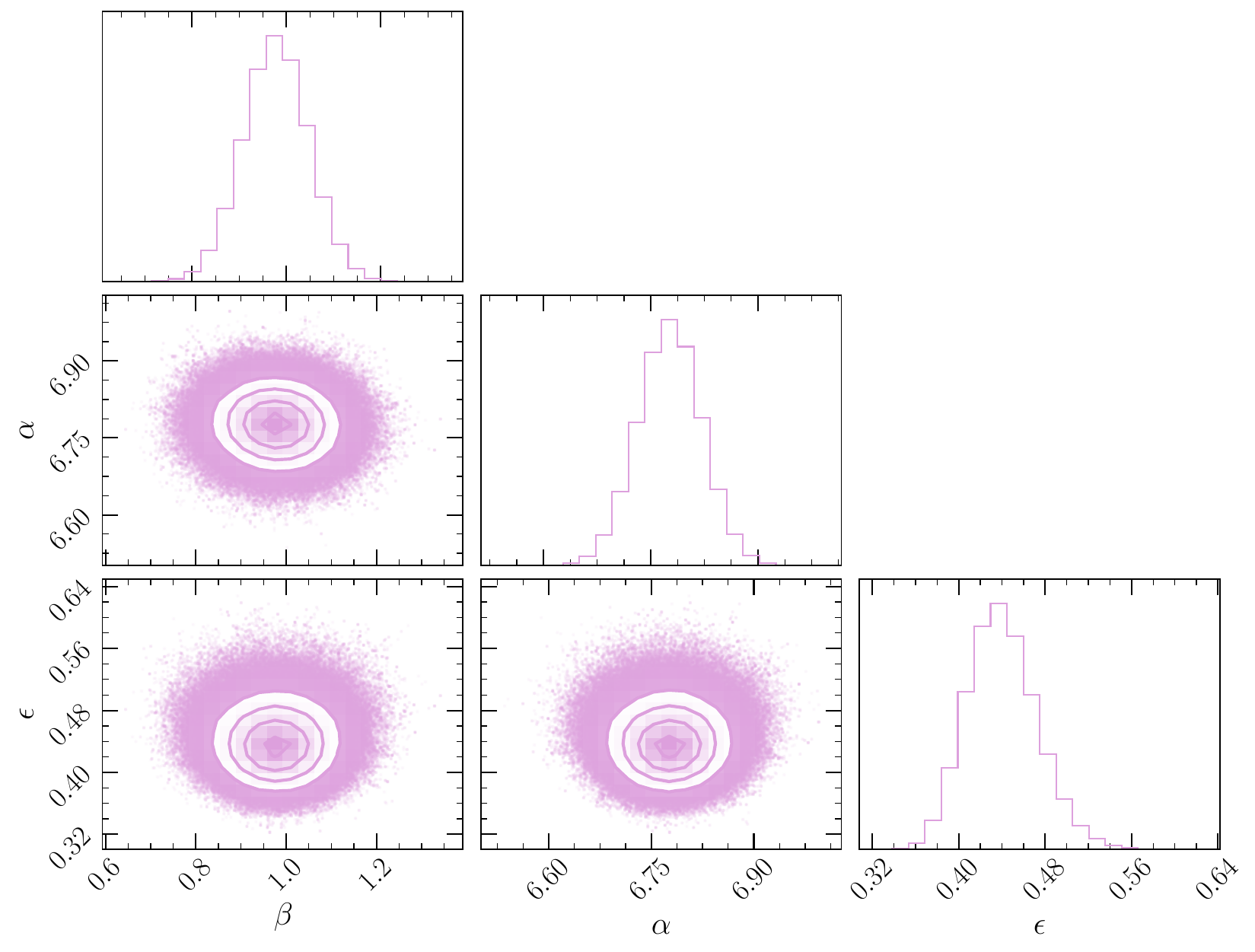}
\caption{Upper: the black hole masses (inferred from their plateau luminosities) of our TDE sample, plotted against their observed  $g$-band radiated energy. The radiated energy is defined as $E_{\rm rad} \equiv L_{\rm peak} \times \tau_{\rm decay}$. The functional form fit is $\log(\M/M_\odot) = \alpha + \beta \log(E_{\rm rad}/10^{50}\,{\rm erg})$. We note a strong positive relationship between  black hole mass and $g$-band radiated energy, which is well described by a single power-law relationship. Lower: MCMC posterior contours of a power law fit between black hole mass and $g$-band radiated energy.  A Kendall $\tau$ test finds a strong correlation between radiated energy and black hole mass $(\tau = 0.61, p <  10^{-7})$.  This highlights how $g$-band radiated energy can be used as an empirical scaling relationship with $\sim 0.3$ dex intrinsic scatter.   }
\label{EradScale}
\end{figure}
\begin{figure*}
\includegraphics[width=0.49\textwidth]{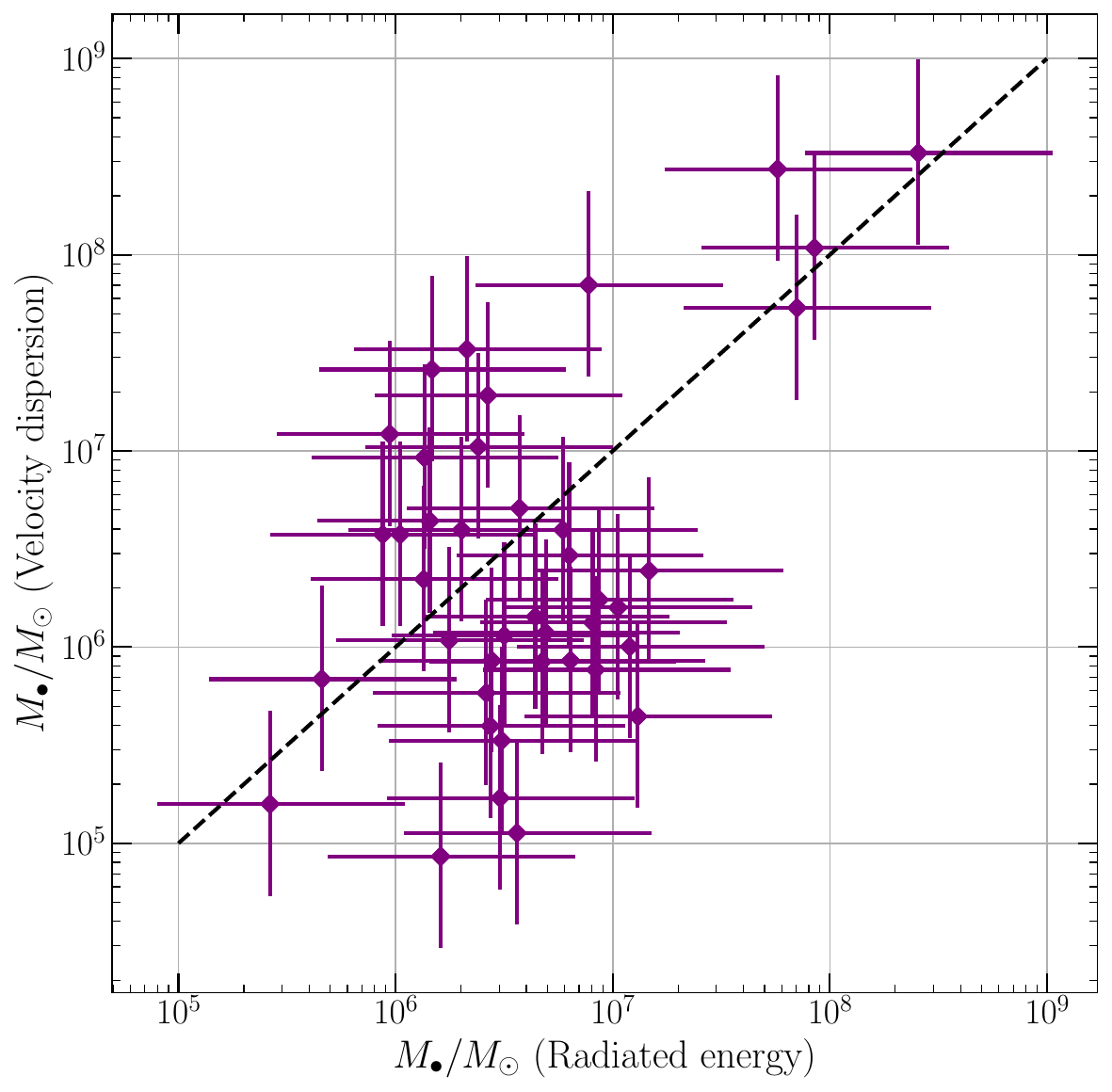}
\includegraphics[width=0.49\textwidth]{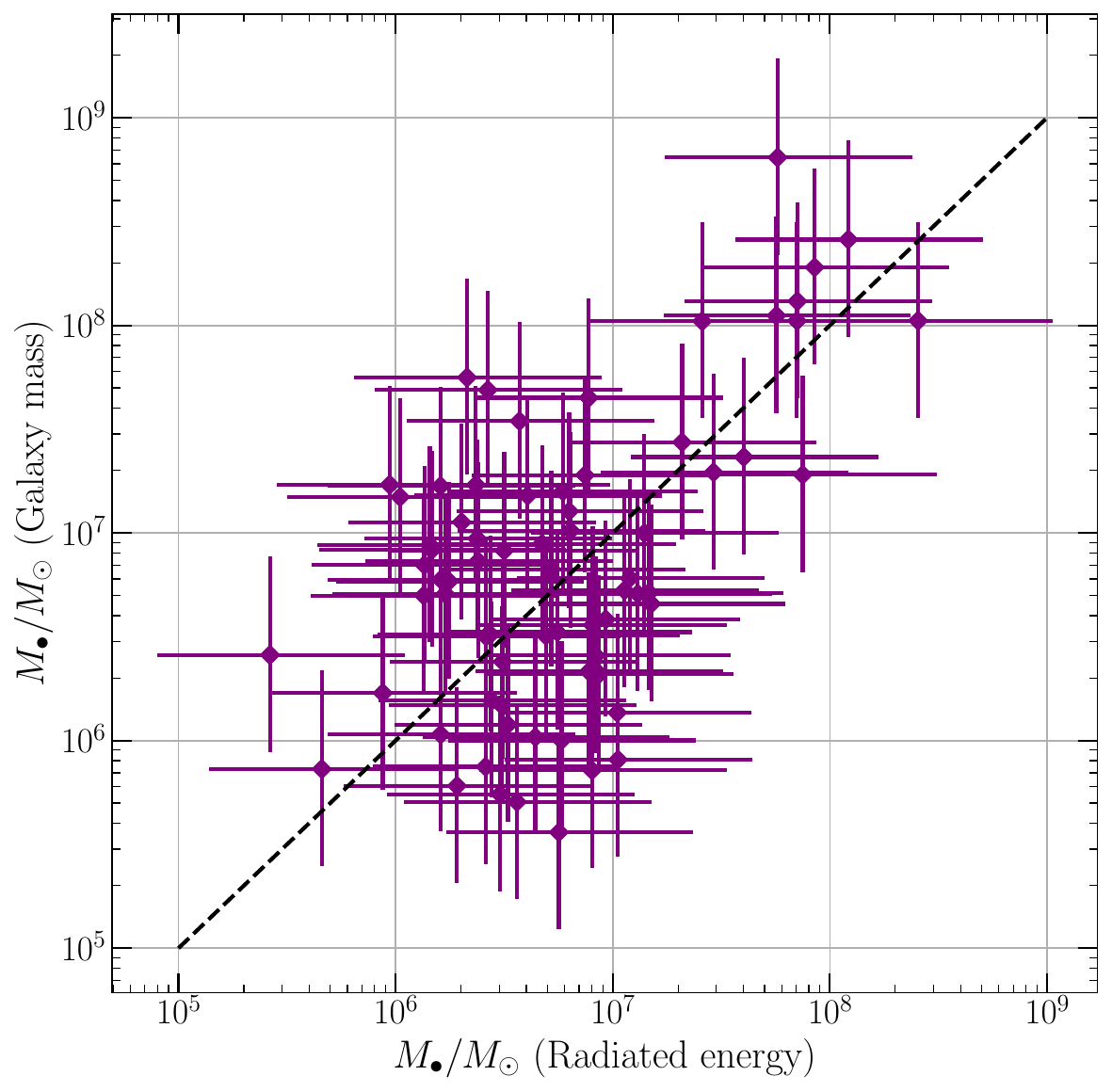}
\includegraphics[width=0.49\textwidth]{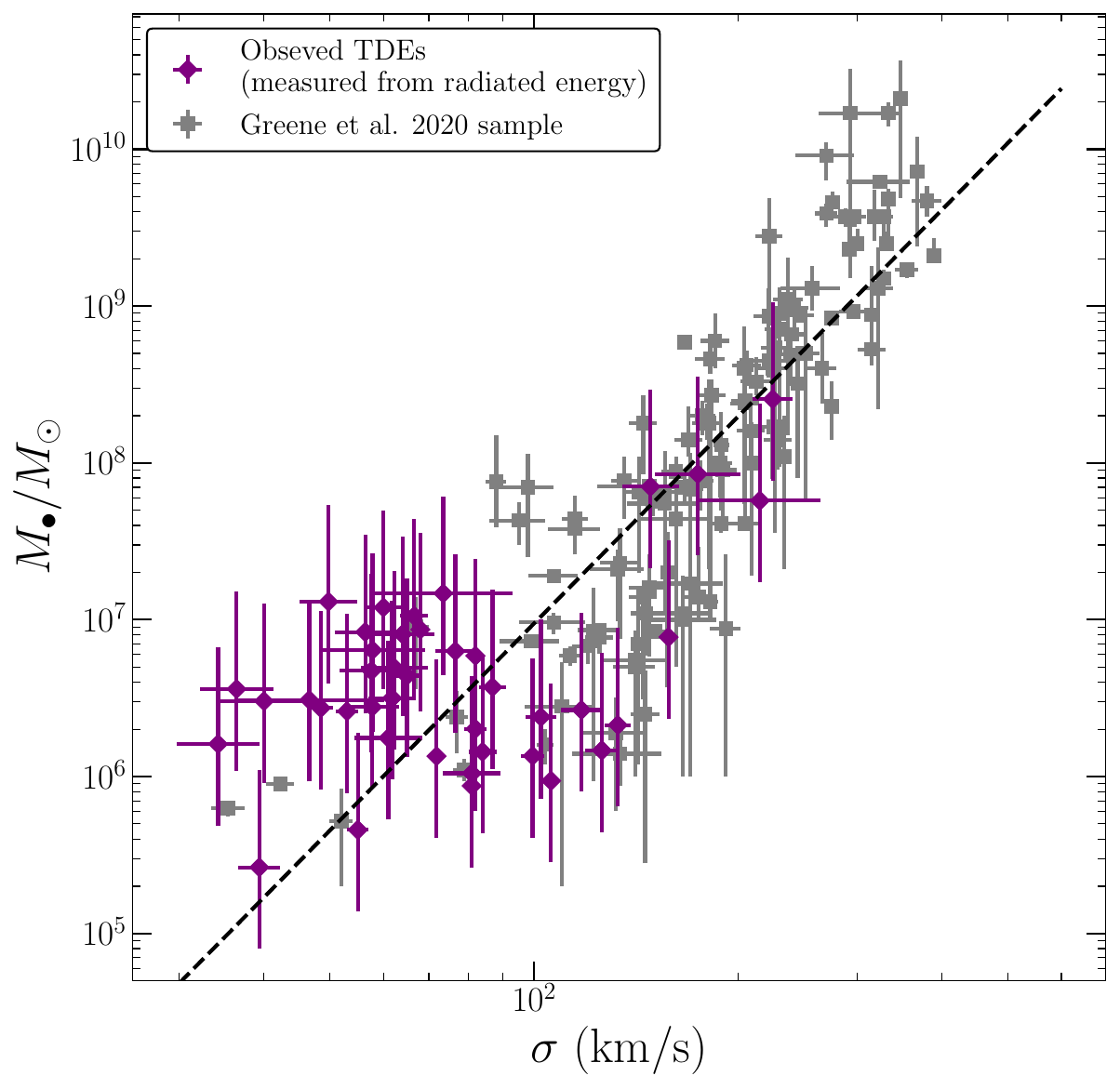}
\includegraphics[width=0.49\textwidth]{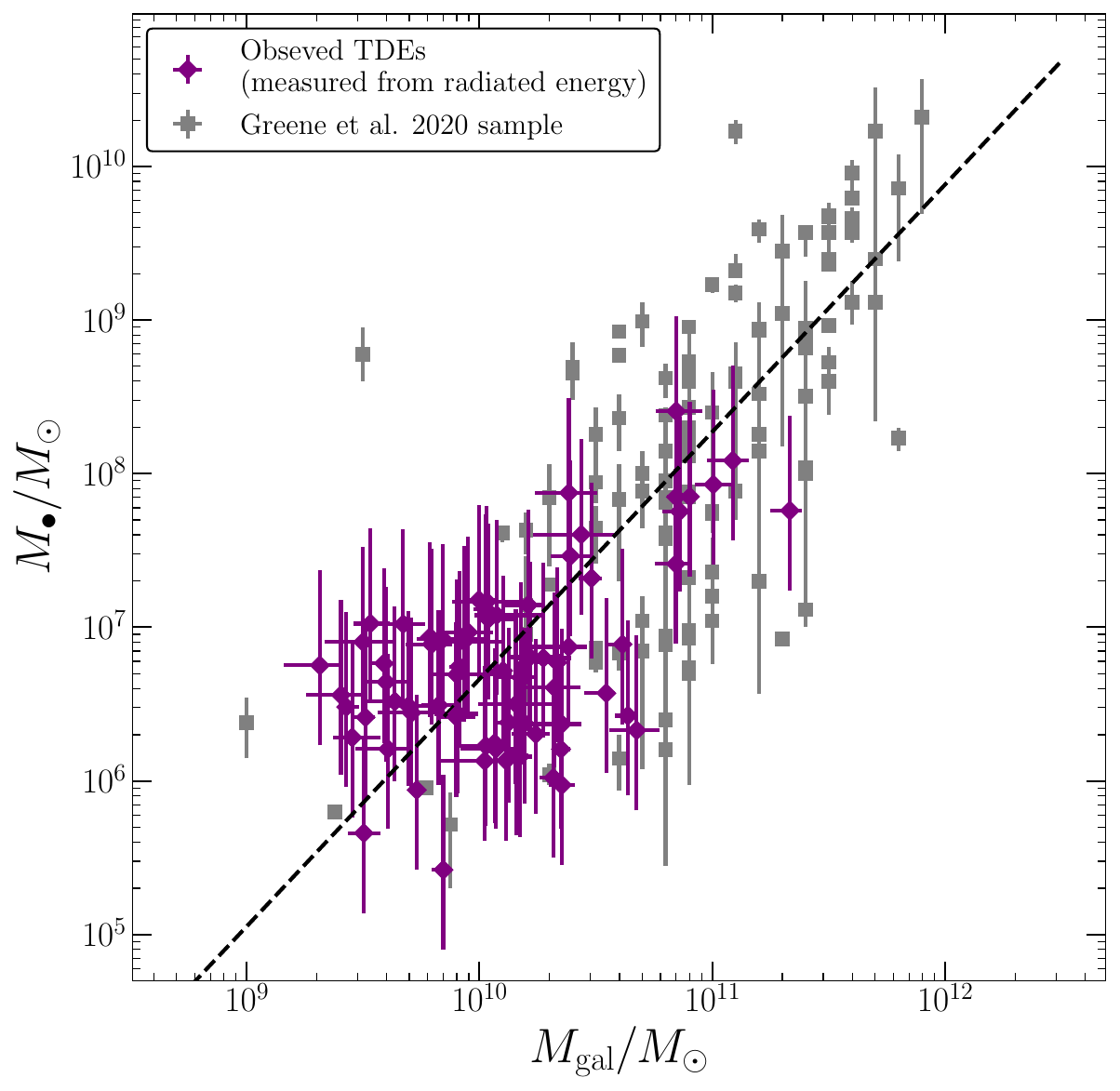}
\caption{ Upper: TDE black hole mass-mass plots, where on the horizontal axis we plot the mass as inferred from the TDEs $g$-band radiated energy, and on the vertical axis we plot the mass as inferred from a galactic scaling relationship (on the left we use the velocity dispersion $\sigma$, and on the right the host galaxy mass $M_{\rm gal}$). The black dashed line shows $\M = \M$, i.e., perfect agreement between the independent approaches. Lower: the combined populations of black hole masses and galactic properties (again on the left we  display velocity dispersion $\sigma$, and on the right the host galaxy mass $M_{\rm gal}$). The points in grey are taken from the paper \citet{Greene20}, while the points in purple are the TDEs we are able to add using the  empirical energy radiated scaling relationship. The black dashed lines in these two plots are the scaling relationships presented in \citet[our equations \ref{sig_scale} and \ref{galmass_scale}]{Greene20}.   The black hole masses inferred from this empirical scaling relationship fit as is expected with the pre-existing galactic populations. }
\label{MvsM_e}
\end{figure*}

The radiated energy is defined by 
\beq
E_{\rm rad} \equiv \nu L_{\nu, {\rm peak}} \times \tau_{\rm decay} ,
\eeq
with $L_{\nu, {\rm peak}}$ measured in the rest-frame $g$-band.
We define 
\beq
X \equiv {E_{\rm rad} \over 10^{50} \, {\rm erg}}, 
\eeq 
a suitably dimensionless $g$-band radiated energy. 
In Figure \ref{EradScale} we plot the black hole mass (measured from the TDE plateau luminosity) as a function of early time radiated energy. A clear correlation is visible. MCMC fits to a power-law profile between black hole mass (estimated from the plateau luminosity) and early time radiated energy (equation \ref{power_law_fit}) return
\begin{align}
\alpha &= 6.78 \pm 0.04 , \\
\beta &= 0.98 \pm 0.07 , \\
\epsilon &= 0.44 \pm 0.03 . 
\end{align}
In the lower panel we display the posterior distributions of the parameters in these fits.  A Kendall $\tau$ test finds a strong correlation between radiated energy and black hole mass $(\tau = 0.61, p <  10^{-7})$.

Interestingly, the early time radiated energy in the $g$-band scales approximately linearly with the central black hole mass, and can therefore be used as a proxy for black hole mass in those TDEs in which we were unable to measure a plateau luminosity (see Table \ref{mass_plat_table} for explicit values). 

In Fig. \ref{MvsM_e} we display  four illustrations of this new black hole mass scaling relation for this early time relationship, in an identical fashion to Fig. \ref{MvsM}. In the upper two panels we display mass-mass plots, where on the horizontal axis we plot the mass as inferred from the TDE early time $g$-band energy, and on the vertical axis we plot the mass as inferred from a galactic scaling relationship (on the left we use the velocity dispersion $\sigma$, and on the right the host galaxy mass $M_{\rm gal}$; see equations \ref{sig_scale} and \ref{galmass_scale} for the explicit scaling relationships). The black dashed line shows $\M = \M$, i.e., perfect agreement between the independent approaches.  The black hole masses inferred from early time TDE $g$-band energies are correlated with the black hole masses inferred from galactic properties. 

In the lower two panels we show the combined populations of black hole masses and galaxy properties (again on the left we  display velocity dispersion $\sigma$, and on the right the host galaxy mass $M_{\rm gal}$). The points in grey are taken from the paper \citet{Greene20}, while the points in purple are the TDEs we are able to add using the early time radiated energy. The black dashed lines in these two plots are the scaling relationships presented in \citet[our equations \ref{sig_scale} and \ref{galmass_scale}]{Greene20}.

\subsection{Empirical peak scaling }

\begin{figure}
\includegraphics[width=0.5\textwidth]{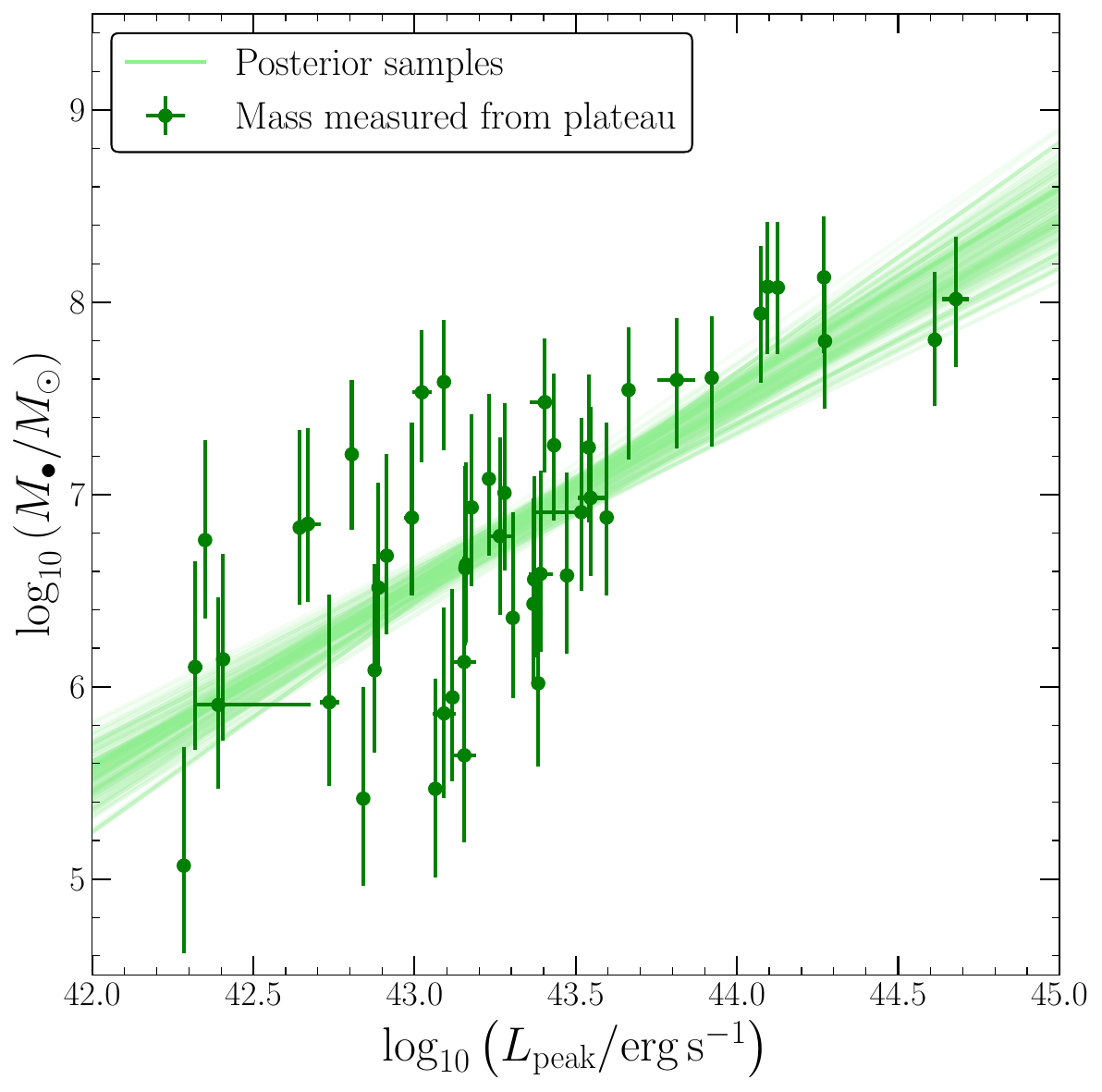}
\includegraphics[width=0.5\textwidth]{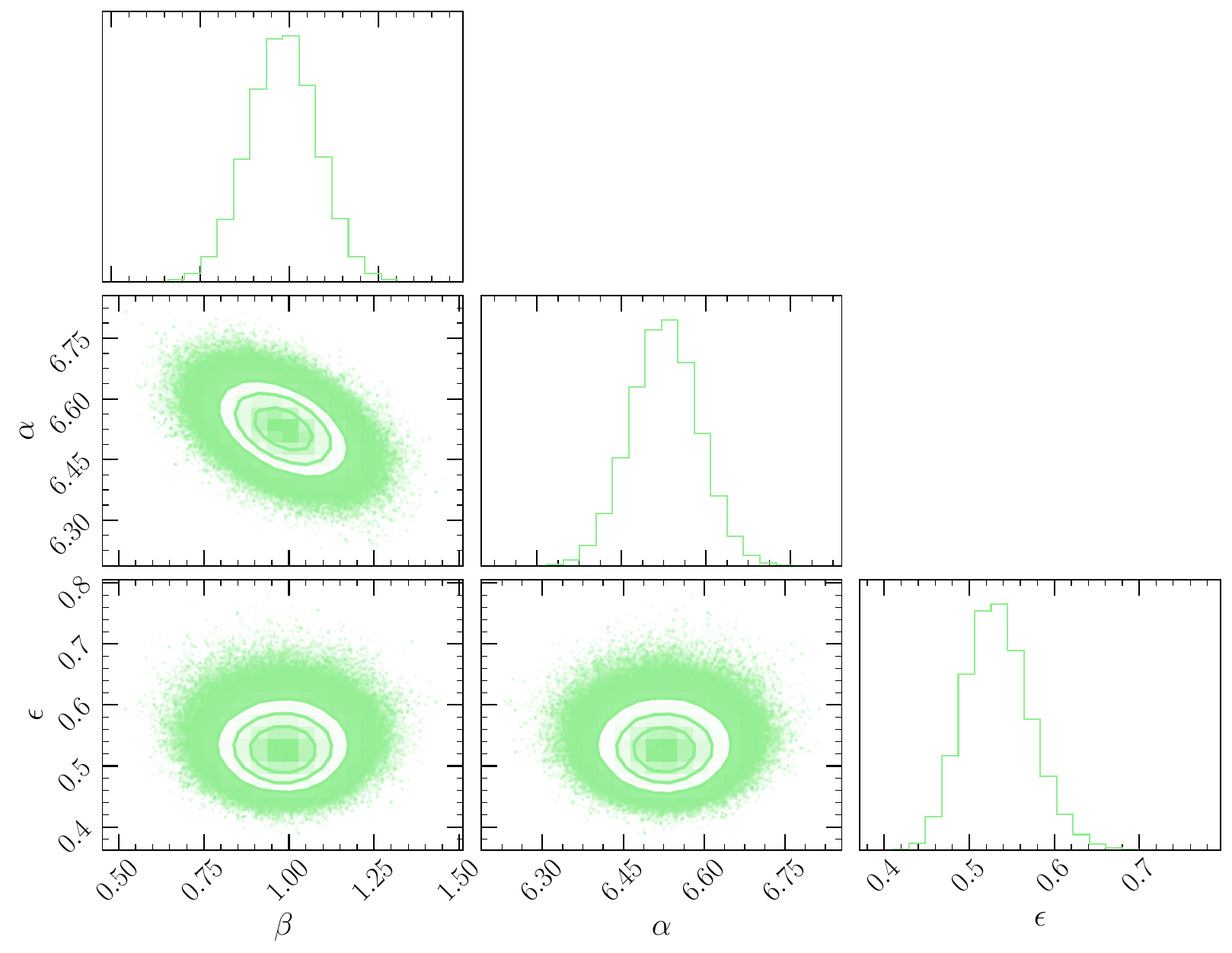}
\caption{Upper: the black hole masses (inferred from their plateau luminosities) of our TDE sample, plotted against their observed peak ($g$-band) luminosity. We note a strong positive relationship between  black hole mass and peak luminosity, which is well described by a single power-law relationship. Lower: MCMC posterior contours of a power law fit between black hole mass and peak $g$-band luminosity. The functional form fit is $\log(\M/M_\odot) = \alpha + \beta \log(L_{\rm peak}/10^{43}\,{\rm erg\, s^{-1}})$. A Kendall $\tau$ test finds a strong correlation between peak luminosity and black hole mass $(\tau = 0.53, p <  10^{-7})$.   This highlights how peak $g$-band luminosity can be used as an empirical scaling relationship with $\sim 0.4$ dex intrinsic scatter.   }
\label{PeakScale}
\end{figure}

Finally, we present an empirical scaling relationship between the peak $g$-band luminosity $(\nu L_\nu)$ observed from the \Nplat\ TDEs with measured plateaus, against the black hole masses of these TDEs (inferred from their plateau luminosity, Table \ref{mass_plat_table}). There is a clear  positive relationship between peak luminosity and black hole mass, which is well described by a single power-law profile. 

For our dimensionless luminosity variable we define 
\beq
X \equiv {L_{\rm peak} \over 10^{43} \, {\rm erg\, s^{-1}}}, 
\eeq 
where $L_{\rm peak} \equiv \nu L_{\nu, {\rm peak}}$. 
MCMC fits to a power-law profile between black hole mass and peak $g$-band luminosity (equation \ref{power_law_fit}) return
\begin{align}
\alpha &= 6.52 \pm 0.06 , \\
\beta &= 0.98 \pm 0.10 , \\
\epsilon &= 0.53 \pm 0.04 . 
\end{align}
In the lower panel we display the posterior distributions of the parameters in these fits.   A Kendall $\tau$ test finds a significant correlation between peak luminosity and black hole mass estimated from the TDE plateau luminosity $(\tau = 0.53, p <  10^{-7})$.

\begin{figure*}
\includegraphics[width=0.49\textwidth]{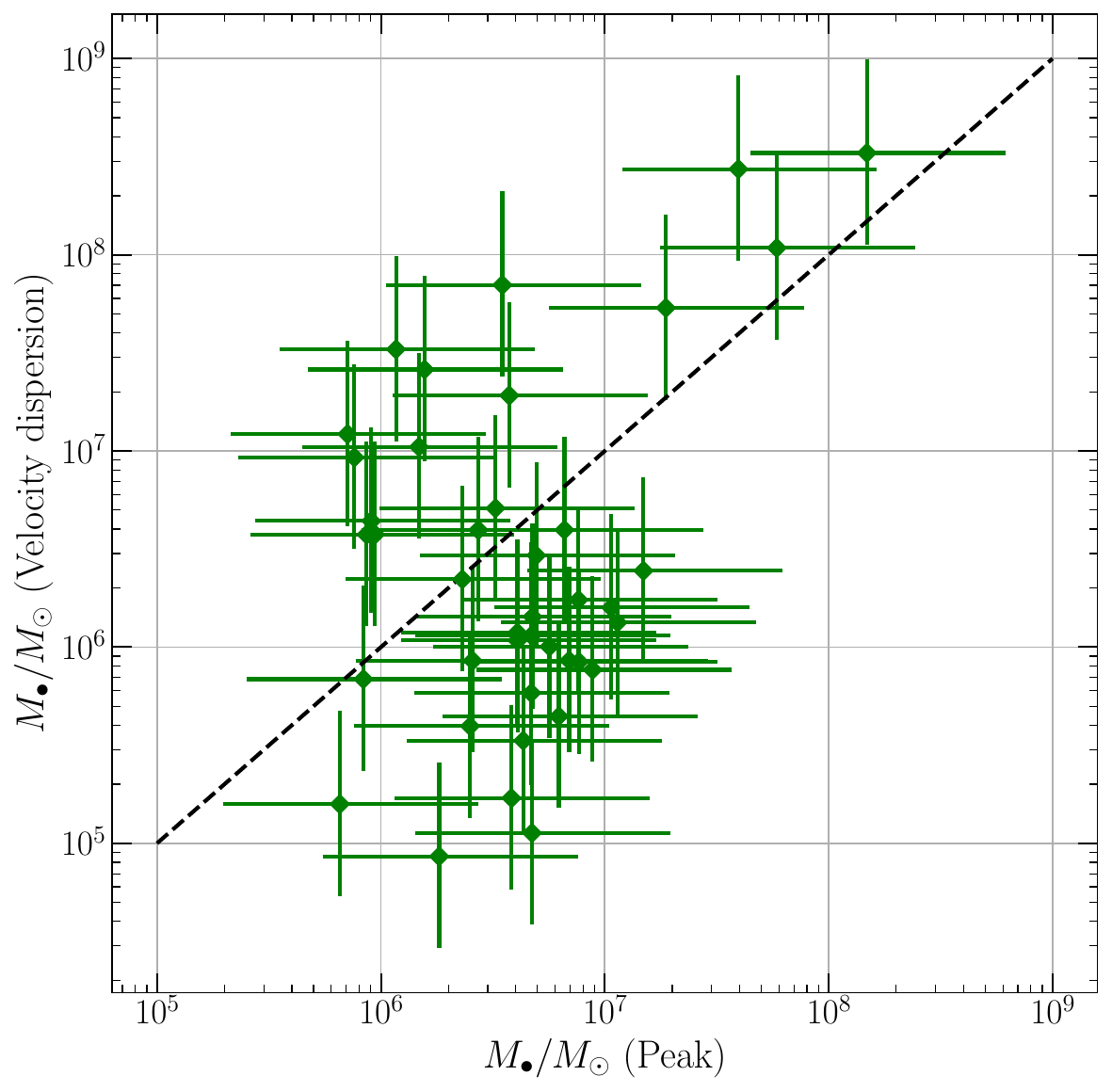}
\includegraphics[width=0.49\textwidth]{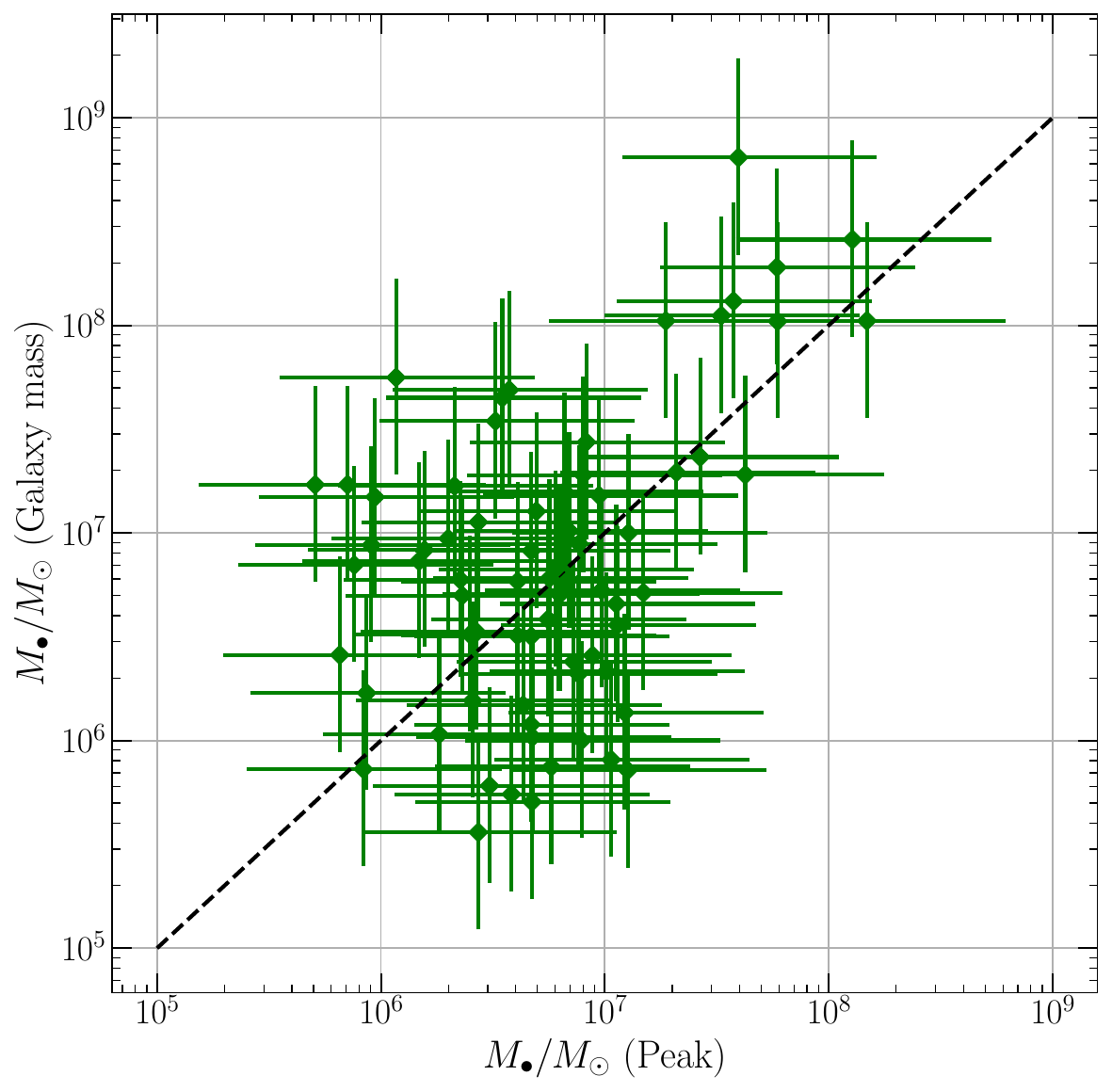}
\includegraphics[width=0.49\textwidth]{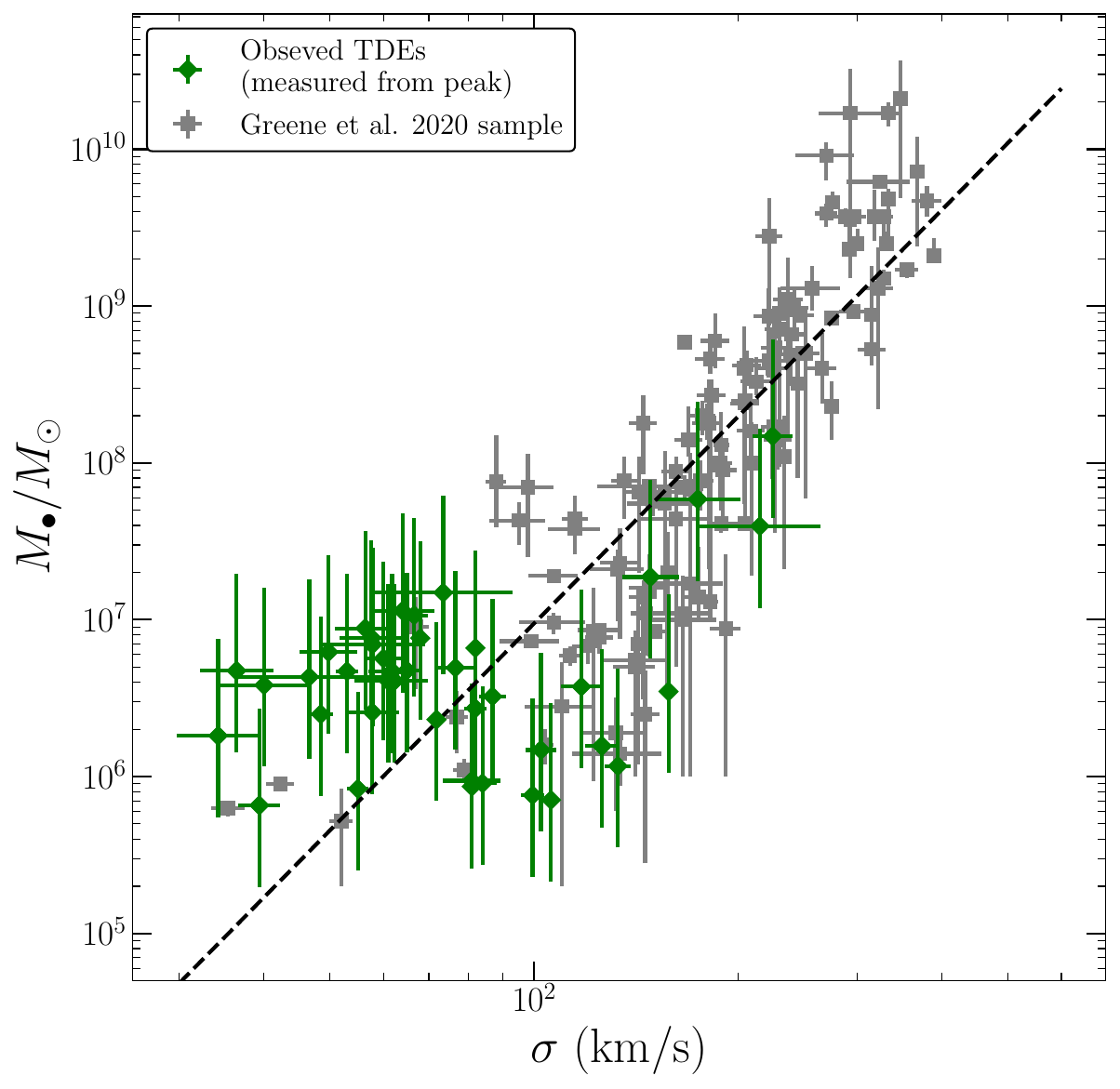}
\includegraphics[width=0.49\textwidth]{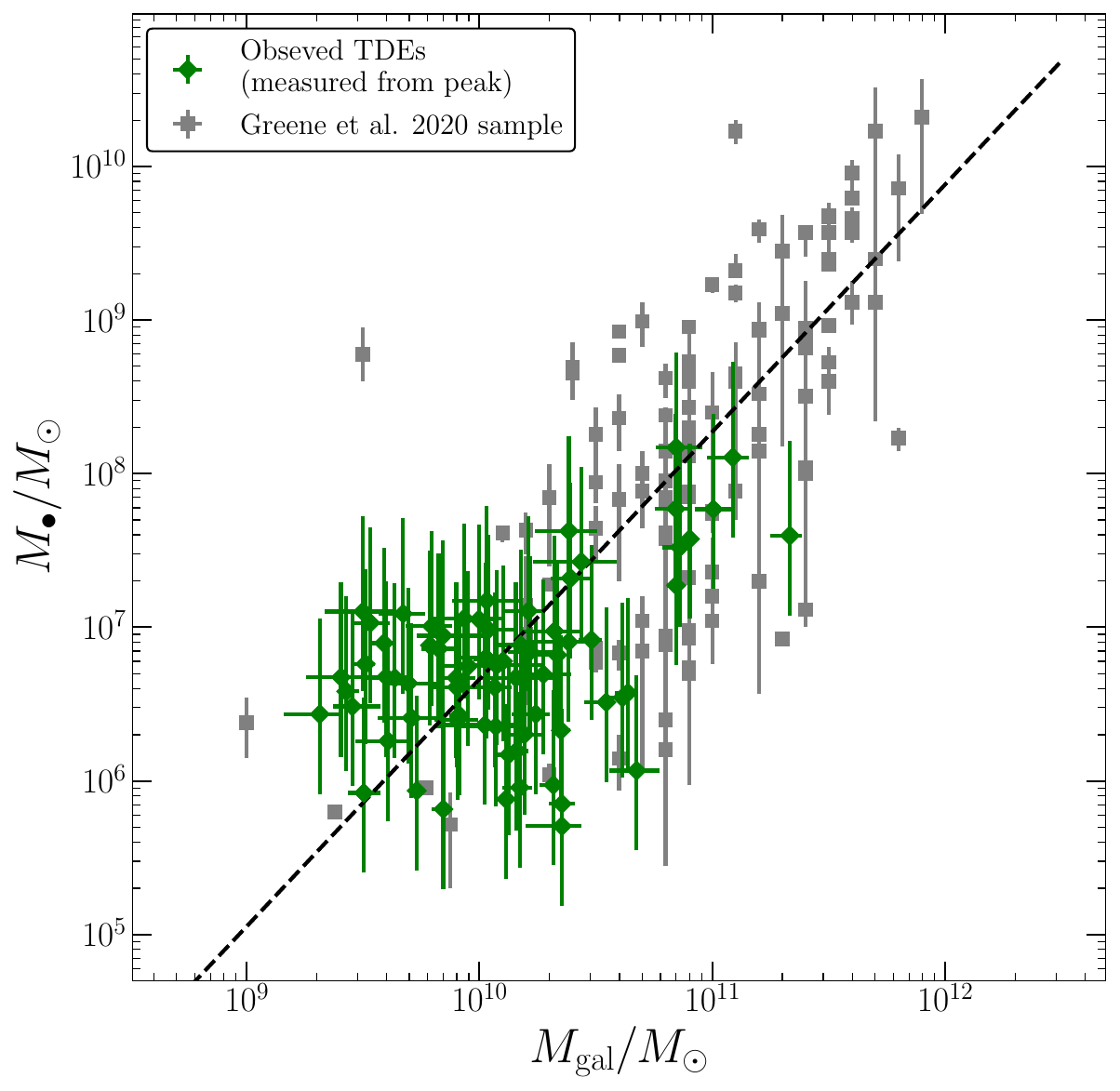}
\caption{Upper: TDE black hole mass-mass plots, where on the horizontal axis we plot the mass as inferred from the TDEs peak $g$-band luminosity, and on the vertical axis we plot the mass as inferred from a galactic scaling relationship (on the left we use the velocity dispersion $\sigma$, and on the right the host galaxy mass $M_{\rm gal}$). The black dashed line shows $\M = \M$, i.e., perfect agreement between the independent approaches. Lower: the combined populations of black hole masses and galactic properties (again on the left we  display velocity dispersion $\sigma$, and on the right the host galaxy mass $M_{\rm gal}$). The points in grey are taken from \citet{Greene20}, while the points in green are the TDEs we are able to add using the  empirical peak $g$-band luminosity scaling relationship. The black dashed lines in these two plots are the scaling relationships presented in \citet[our equations \ref{sig_scale} and \ref{galmass_scale}]{Greene20}.   The black hole masses inferred from this empirical scaling relationship fit as is expected with the pre-existing galactic populations. }
\label{MvsM_p}
\end{figure*}

This is an import result, as it means that even those TDEs with poorly sampled late time light curves may be used to measure TDE black hole masses, from a handfull of early time observations. In Fig. \ref{MvsM_p} we repeat the analysis of the proceeding section (and Fig. \ref{MvsM}), but now for black hole masses estimated entirely from the peak luminosity scaling relationship (see Table \ref{mass_plat_table} for the full list).

An interesting result is that while the early time peak luminosity and radiated energy do appear to be powerful probes of the central black hole mass, the mass estimates which utilise the plateau luminosity produce the least scatter in the galactic scaling mass-mass plots (Figs. \ref{MvsM}, \ref{MvsM_e} and \ref{MvsM_p}). Of the early time probes, the radiated energy has a tighter correlation than the peak luminosity; it appears that the harder one has to work to make an observation, the better an estimate of the TDEs black hole mass one is rewarded with. 

\subsection{Black hole mass scaling relationships from different TDE features}
We now have three different methods for estimating black hole masses at the heart of a tidal disruption event. In this sub-section we test whether the choice of TDE scaling relationship affects the resulting galactic scaling relationships one will infer from a joint fit of our population and the \citet{Greene20} compilation. 

\begin{figure*}
\includegraphics[width=0.67\textwidth]{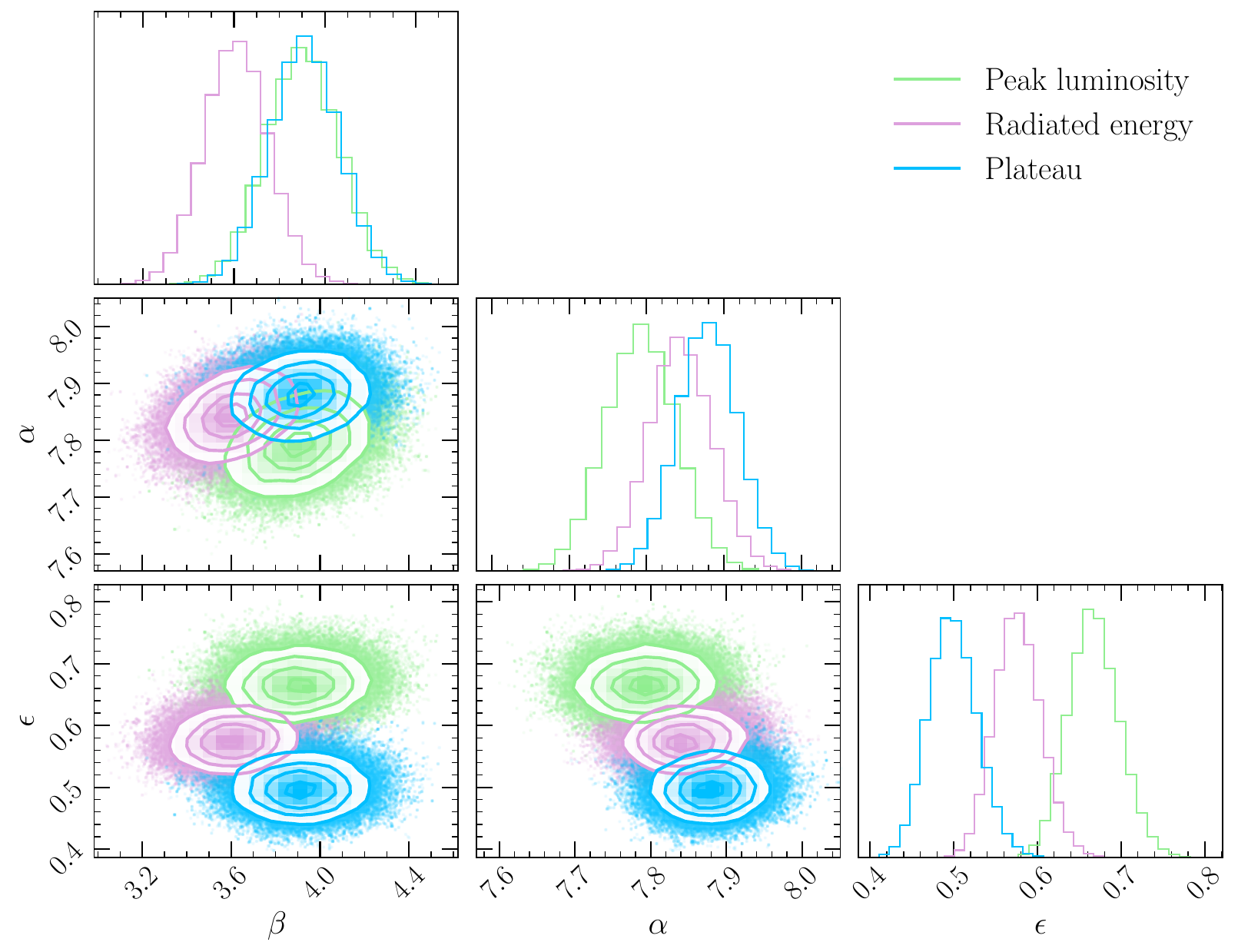}
\caption{ MCMC fits to the velocity dispersion galactic scaling relationship (defined as $\log(\M/M_\odot) = \alpha + \beta \log(\sigma/160\,{\rm km\,s^{-1}})$), including the \citet{Greene20} sample and the new additions of TDEs. Contours shown in blue display fits to a joint population of TDEs and the \citet{Greene20} sample, where the black hole masses of the TDE population are derived using the late time plateau. In green we display contours where the black hole masses of the TDE population are derived using the peak $g$-band luminosity, and in purple by the $g$-band radiated energy. All three techniques of measuring TDE black holes masses produce consistent scaling relationships, but with differing levels of intrinsic scatter. The technique of measuring masses from TDE plateaus produces the lowest intrinsic scatter, and is likely therefore most accurate, as well as being physically most understood.  }
\label{SigmaALL}
\end{figure*}

\begin{figure*}
\includegraphics[width=0.67\textwidth]{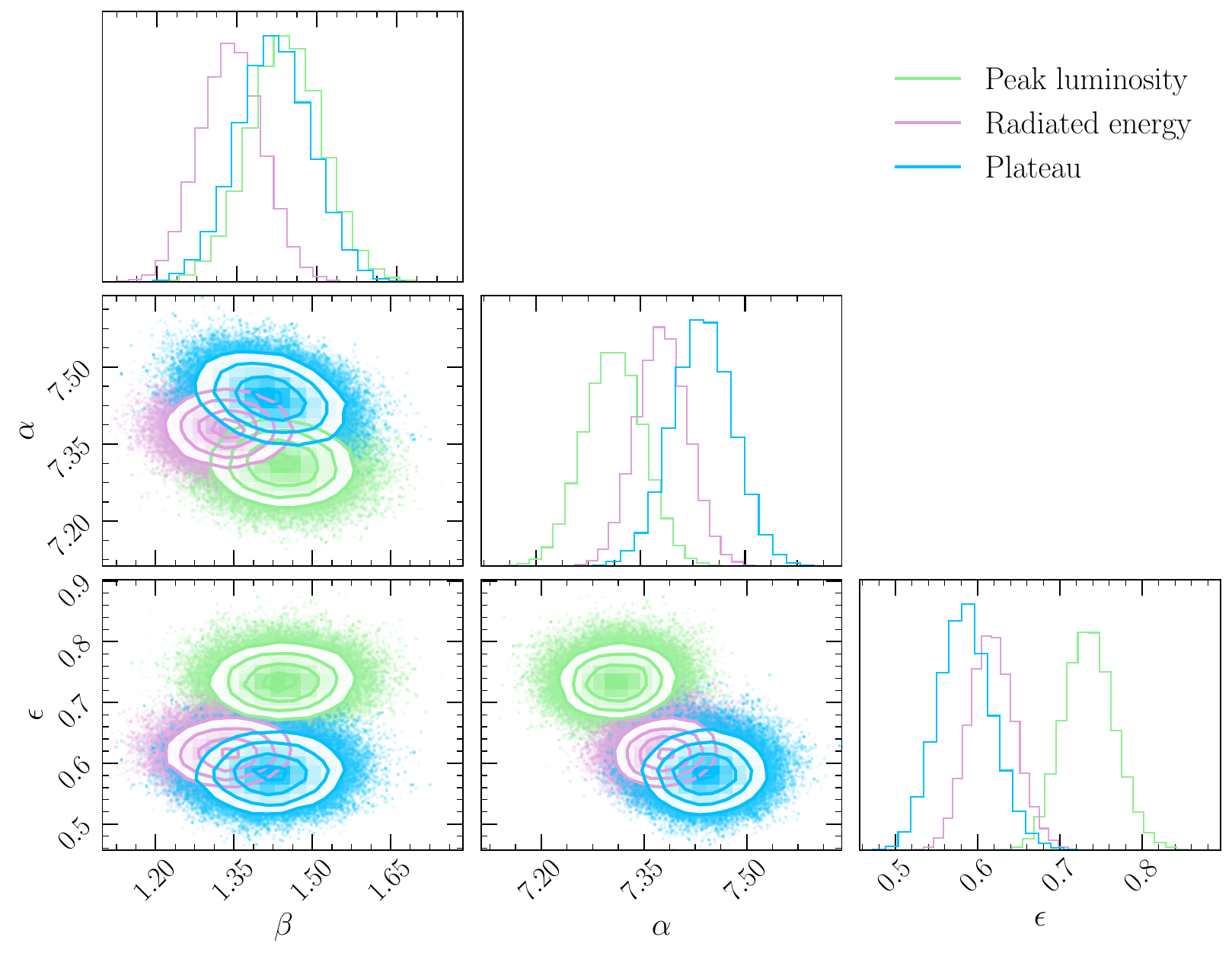}
\caption{ MCMC fits to the galaxy mass galactic scaling relationship (defined as $\log(\M/M_\odot) = \alpha + \beta \log(M_{\rm gal}/3 \times 10^{10}M_\odot)$), including the \citet{Greene20} sample and the new additions of TDEs. Contours shown in blue display fits to a joint population of TDEs and the \citet{Greene20} sample, where the black hole masses of the TDE population are derived using the late time plateau. In green we display contours where the black hole masses of the TDE population are derived using the peak $g$-band luminosity, and in purple by the $g$-band radiated energy. All three techniques of measuring TDE black holes masses produce consistent scaling relationships, but with differing levels of intrinsic scatter.   }
\label{MgalALL}
\end{figure*}

In Figs. \ref{SigmaALL} and \ref{MgalALL} we present the posterior distributions of MCMC fits to the velocity dispersion galactic scaling relationship (defined as $\log(\M/M_\odot) = \alpha + \beta \log(\sigma/160\,{\rm km\,s^{-1}})$), and galactic mass scaling relationship (defined as $\log(\M/M_\odot) = \alpha + \beta \log(M_{\rm gal}/3 \times 10^{10}M_\odot)$), including the \citet{Greene20} sample and the new additions of TDEs. Contours shown in blue display fits to a joint population of TDEs and the \citet{Greene20} sample, where the black hole masses of the TDE population are derived using the late time plateau. In green we display contours where the black hole masses of the TDE population are derived using the peak $g$-band luminosity, and in purple by the $g$-band radiated energy. All three techniques of measuring TDE black holes masses produce consistent scaling relationships, but with differing levels of intrinsic scatter. The technique of measuring masses from TDE plateaus produces the lowest intrinsic scatter, and is likely therefore most accurate, as well as being physically most understood. 


\subsection{Physics of the early time scaling relationships}\label{early_phys}

It is interesting to note that a positive scaling between the black hole mass and peak luminosity of a tidal disruption event is not expected from classical models of TDEs.  A simple mass fallback calculation, where the luminosity is sourced from the rate of returning debris $(\dot M_{\rm fb})$ with some efficiency $\eta$
\beq
L_{\rm peak} \sim \eta \dot M_{\rm fb} c^2 , 
\eeq
would have {incorrect} scaling \citep[e.g.,][]{Rees88}
\beq
L_{\rm peak} \sim \eta M_\star^2 R_\star^{-3/2} \M^{-1/2}   ,
\eeq
{as would the fall-back radiated energy 
\begin{equation}
    E_{\rm fb} \sim \eta \dot M_{\rm fb} t_{\rm fb} c^2 \sim \eta M_\star c^2, 
\end{equation}
with no black hole mass dependence. This is of course under the assumption that the efficiency $\eta$ has no black hole mass dependence. A roughly linear $\M$ dependence of $\eta$ could reproduce the empirical scaling relationships found here, but this would require some theoretical support so as to avoid over-fitting a fall-back driven model.  } 

We note however that there is an energy scale in the problem which scales with black hole properties in a manner similar to that found empirically for the peak luminosity and $g$-band radiated energy.  The tidally disrupted debris streams have a spread in specific energies given approximately by the work done by the tidal force over the tidal radius $r_T$ \citep[e.g.,][]{Rees88}. In other words the most tightly bound debris has 
\beq
\epsilon_T \sim - {G \M R_\star \over r_T^2} .
\eeq
To circularise and form into an accretion flow, the debris streams must reduce their specific energies to that of a circular orbit at $r_c$, explicitly 
\beq
\epsilon_c \sim - {\beta G \M \over 2 r_T} .
\eeq
The change in specific energy of a debris stream which is cricularised is therefore
\beq
\Delta \epsilon = \epsilon_T - \epsilon_c \sim {\beta G\M \over 2 r_T} \left( 1 - {2R_\star \over \beta r_T}\right),
\eeq
or equivalently 
\beq
\Delta \epsilon \sim {\beta G\M^{2/3} M_\star^{1/3} \over 2 R_\star} \left( 1 - {2 \over \beta} \left({M_\star \over \M}\right)^{1/3} \right) .
\eeq
Typically the second term here is negligible $M_\star \ll \M$, and so if $f_d M_\star$ is eventually circularised into a disc, then the total energy change of the debris is 
\begin{multline}
\Delta E \sim {\beta f_d G\M^{2/3} M_\star^{4/3} \over 2 R_\star} \\ \sim \left( {f_d \over 0.5} \right) \left({\beta \over 1}\right) \left({\M \over 10^6 M_\odot} \right)^{2/3}\left({M_\star \over M_\odot}\right)^{4/3} \left({R_\star \over R_\odot} \right)^{-1}\, 10^{52}\,  {\rm erg}  .
\end{multline}
It is interesting to note that the energy radiated in the early time $g$-band  light curves correlates with black hole mass with a linear $\M$ scaling (Fig. \ref{EradScale}), which is not dissimilar to this debris energy change scaling.  {However, the typical radiated energy in the $g$-band is a factor $\sim 100$ smaller than this value, assuming solar stellar parameters, at $\M = 10^6 M_\odot$.  }

If the peak luminosity inherits the same black hole mass scaling as the energy change of the debris streams, then the empirical scaling of Fig. \ref{PeakScale} could be understood. However, this would require that the timescale over which energy is liberated does not scale with black hole mass, something which is not expected from the classical fall-back timescale calculation 
\beq
t_{\rm fb} = 41 \left({\M \over 10^6 M_\odot}\right)^{1/2} \, \left({R_\star \over R_\odot}\right)^{3/2} \, \left({M_\star \over M_\odot}\right)^{-1} \,\, {\rm days}. 
\eeq

We note  that there is one timescale in the TDE system which is independent of black hole parameters: the orbital timescale of the debris once circularised, which is given by 
\beq
t_{\rm orb} = \sqrt{R_c^3 \over G\M} = \sqrt{8R_\star^3 \over \beta^3GM_\star} .
\eeq
{Note, however, that this is an extremely short timescale
\beq
t_{\rm orb} = 0.05 \,\left({\beta \over 1}\right)^{-3/2}  \left({R_\star \over R_\odot}\right)^{3/2} \, \left({M_\star \over M_\odot} \right)^{-1/2} \,\,  {\rm days} ,
\eeq
and a naive division $L \sim \eta  \Delta E / t_{\rm orb }$ would require small radiative efficiencies $\eta \sim 10^{-4}$.  }

Finally, a recent ``cooling-envelope'' model proposed by \citet{Metzger22} also (in a qualitative sense) recovers some of the early time luminosity scalings discovered in this work. The \citet{Metzger22} model predicts a peak optical luminosity which scales as $L_{\rm peak} \propto \M^{4/5}$, and a disc formation timescale $t_{\rm disc} \sim \M^{-1/3}$. It is possible that such a ``cooling envelope'' describes the early time evolution of the TDEs in our sample, but further work is required to test this hypothesis.  

To conclude, the early time luminosity and radiated energy of the TDEs in our sample grow strongly with central black hole mass, something which is not predicted by simple ``fallback driven'' models \citep{Rees88}.   Positive scalings between black hole mass and early time optical/UV luminosity could be understood if the source of this emission is either the change in orbital kinetic energy required for disc circularisation, or the cooling envelope model of \citet{Metzger22}.   It is too early at this stage to make quantitative statements in favour of either scenario, and further work is required to elucidate the nature of this early time emission more fully.   

\section{Conclusions}\label{sec:8}
In this work we have derived a new and powerful technique for determining the properties of the black holes at the centre of tidal disruption events.  A single observation of the late time plateau luminosity observed in a TDE's optical/UV light curve provides an estimate of the central black hole's mass with intrinsic scatter of only $\sim 0.5$ dex. We use this new technique to produce estimates of the black hole masses of \Nplat\ TDE systems, which are found to correlate strongly with both the host galactic mass, and galactic velocity dispersion. By adding \Nplat\ (34) sources to the black hole mass -- galactic host mass (velocity dispersion) scaling relationships, we are able to provide updated and extended (into the low black hole mass regime) galaxy scaling relationships.

Sources which display large plateau luminosities (with $\nu L_\nu \gtrsim 10^{43}$ erg/s at $\nu = 10^{15}$ Hz) can only be produced by TDE black holes with large masses ($M_\bullet \gtrsim 10^{8} M_\odot$), which must be rapidly rotating to be able to tidally disrupt a main sequence star. We provide spin constraints of the black holes of the 10 brightest TDE sources in our sample, finding rapid rotation is favoured to explain these observations.   

Complimenting this work are new, empirical (although with some physical basis, section \ref{early_phys}), scaling relationship discovered between the peak optical luminosity, and radiated energy, of TDE sources and their central black hole masses. Both of these results promise to be powerful tools for the analysis of the large data sets discovered by future optical survey instruments, such as Rubin/LSST.   

The results of this work will provide powerful probes of the demographics of the local supermassive black hole population, particularly in the previously uncertain low-mass end of the black hole mass function.

\section*{Acknowledgments} 
  
  The authors would like to thank G. Leloudas, M. Nicholl, E. Rossi, and B. Shappe for illuminating discussions. 
 
 This work was supported by a Leverhulme Trust International Professorship grant [number LIP-202-014]. For the purpose of Open Access, AM has applied a CC BY public copyright licence to any Author Accepted Manuscript version arising from this submission. This work is partially supported by the Hintze Family Charitable Trust and STFC grant ST/S000488/1. EN acknowledges support from NASA theory grant 80NSSC20K0540. AI acknowledges support from the Royal Society. EH acknowledges support by NASA under award number 80GSFC21M0002.

Parts of this work are based on observations obtained with the 48-inch Samuel Oschin Telescope and the 60-inch Telescope at the Palomar Observatory as part of the Zwicky Transient Facility project. ZTF is supported by the National Science Foundation under Grants No. AST-1440341 and AST-2034437 and a collaboration including current partners Caltech, IPAC, the Weizmann Institute for Science, the Oskar Klein Center at Stockholm University, the University of Maryland, Deutsches Elektronen-Synchrotron and Humboldt University, the TANGO Consortium of Taiwan, the University of Wisconsin at Milwaukee, Trinity College Dublin, Lawrence Livermore National Laboratories, IN2P3, University of Warwick, Ruhr University Bochum, Northwestern University and former partners the University of Washington, Los Alamos National Laboratories, and Lawrence Berkeley National Laboratories. Operations are conducted by COO, IPAC, and UW. 

\section*{Data accessibility statement}
All TDE light curves are available in the following repository  \href{https://github.com/sjoertvv/manyTDE}{https://github.com/sjoertvv/manyTDE}, including the inferred model parameters in Tables \ref{mass_plat_table}. 

\bibliographystyle{mnras}
\bibliography{andy, general_desk}

\appendix{}
\section{Thin disc theory predicts a $\nu L_{\nu} \propto \M^{2/3}$ scaling } \label{disctheory}
In this appendix we prove that thin disc theory, when specialised to the TDE context, predicts a $\nu L_\nu \propto \M^{2/3}$ scaling relationship. 

We start with the constraints of energy conservation.  The constraint of energy conservation can be formulated in the following manner. The product of the peak bolometric luminosity of the evolving TDE disc and the viscous timescale of the disc must be related by 
\beq
L_{\rm bol, peak} \, t_{\rm visc} \simeq f_d \epsilon_{\rm acc} M_\star c^2 .
\eeq
In this expression $f_d$ represents the disc formation efficiency (what fraction of the stellar material eventually forms into a disc $f_d = M_d/M_\star$), and $\epsilon_{\rm acc}$ represents the mass to light accretion efficiency. 

 The so-called ``viscous'' timescale of an evolving accretion flow is given by the following simple expression:
\beq
\tv = \alpha^{-1} \left({H\over R}\right)^{-2} \sqrt{r^3 \over G\M} ,
\eeq
where $r$ is the radius at which the flow begins, and $\M$ is the black hole's mass. A tidal disruption event occurs when a star passes within the tidal radius of a supermassive black hole, which is the point at which the tidal force $(F_T)$ becomes comparable to the stars own self-gravity $(F_g)$
\beq
F_T \simeq {G \M M_\star R_\star \over r^3}, \quad F_{g} \simeq {G M_\star^2 \over R_\star^2} , 
\eeq 
or explicitly 
\beq
r_T \simeq R_\star \left({\M \over M_\star}\right)^{1/3} .
\eeq
A star on an incoming orbit with pericenter radius $r_p$ which satisfies 
\beq
r_p = {1\over \beta} r_T ,\quad \beta \geq 1, 
\eeq
will be fully tidally disrupted, and the debris from such an event will circularise at a radius $r_c$
\beq
r_c = {2 \over \beta} r_T = {2R_\star \over \beta} \left({\M \over M_\star}\right)^{1/3} ,
\eeq
where the extra factor of 2 results from angular momentum conservation as a parabolic orbit is turned into a circularised orbit. A simple substitution of the circularisation radius into the expression for the viscous timescale demonstrates that 
\beq
\tvc = \alpha^{-1}\left({H\over R}\right)^{-2} \sqrt{8R_\star^3 \over \beta^3GM_\star} ,
\eeq
and we see that any explicit dependence on the properties of the black hole has dropped out of this expression. Thus, by combining with energy conservation, we find that 
\beq
L_{\rm bol, peak}  \simeq f_d \epsilon_{\rm acc} {M_\star c^2 \over \tvc} \simeq  \alpha \left({H\over R}\right)^{2} f_d \epsilon_{\rm acc}  \sqrt{ \beta^3GM_\star^3 \over 8R_\star^3 } c^2 .
\eeq
The peak bolometric luminosity of a TDE disc will be dominated by the emission from regions close to the black hole's ISCO, and therefore 
\beq
L_{\rm bol, peak}  \simeq \pi R_I^2 \sigma T_p^4 ,
\eeq
meaning 
\beq
T_p \simeq \left( {\alpha  f_d \epsilon_{\rm acc} \over \sigma \pi R_I^2} \left({H\over R}\right)^{2} \sqrt{ \beta^3GM_\star^3 \over 8R_\star^3 } c^2 \right)^{1/4} .
\eeq
It is well known that after a viscous timescale the thin disc temperature profiles approach a self-similar profile of the following general form \citep[e.g.,][]{Pringle91, MumBalb19a}
\begin{multline}
T_{\rm disc}(R, t) \simeq T_p \left({R\over R_I}\right)^{-3/4} \left({t \over \tvc}\right)^{-n/4} \\ \times \Theta(R_{\rm out}(t) - R) \Theta(R - R_I) .
\end{multline}
Here $n$ is the rate at which the bolometric luminosity decays \citep[$n$ is expected to take a value $n = 19/16$ for a canonical TDE disc,][]{Cannizzo90}, and the outer edge of the disc grows as a power law in time $R_{\rm out}(t) \sim t^{\gamma}$, with index which depends weakly on the choice of disc stress parameterisation \citep{MumBalb19a}. For $n = 19/16$ we have $\gamma = 3/8$. The two Heaviside theta functions (defined as $\Theta(z>0)=1, \Theta(z<0)=0$) simply enforce the inner and outer boundary conditions on the disc. With a given disc temperature profile the observed luminosity at an observed frequency $\nu$ is defined as 
\beq
\nu L_\nu \equiv 4\pi D^2 \nu F_\nu  
\eeq
where (in the Newtonian limit) 
\beq
F_\nu(\nu) =  \frac{2\pi \cos(i)}{D^2}\int  {R } B_\nu (\nu, T) \, {\rm d}R .
\eeq
Explicitly
\beq
\nu L_\nu =  {8\pi^2 \cos(i)}\, \nu \int {R } B_\nu (\nu, T) \, {\rm d}R, 
\eeq
and after substituting for the Planck function 
\beq
B_\nu = {2\pi h \nu^3 \over c^2}  {1 \over \exp\left({h\nu / kT}\right) - 1 },
\eeq
 we find 
\beq
\nu L_\nu =  \frac{16\pi^2 h \nu^4 \cos(i)}{c^2}\int {R\, {\rm d}R \over \exp\left( {h\nu}/{k  T(R, t)} \right) - 1}  .
\eeq
{Note that we have neglected the effects of the disc colour-correction in deriving this expression. The colour-correction factor's temperature (and therefore radius) dependence complicates the integral analysis without changing the parameter scalings, and so is left out for convenience.  } Using the self-similar temperature profile defined above, we have 
\begin{multline}
\nu L_\nu =  \frac{16\pi^2 h \nu^4 \cos(i)}{c^2} R_I^2 \\ \int_{1}^{R_{\rm out}(t)/R_I} {X\, {\rm d}X \over \exp\left( {h\nu  X^{3/4} \tau^{n/4} /  k  T_p}\ \right) - 1}  ,
\end{multline}
where we have defined $X \equiv R/R_I, \tau \equiv t/\tvc$. Defining now 
\beq
Y \equiv  {h\nu  X^{3/4} \tau^{n/4} \over   k  T_p} ,
\eeq
we are left with 
\begin{multline}
\nu L_\nu =  \frac{16\pi^2 h \nu^4 \cos(i)}{c^2} R_I^2 \left(  {k  T_p   \over  h\nu } \right)^{8/3}   \\ \underbrace{\vphantom{ \int_{\Lambda(t)}^{\Lambda(t) X_{\rm out}^{3/4}(t)} {Y^{5/3}\, {\rm d}Y \over \exp\left(Y \right) - 1}  } \tau^{-2n/3}}_{\rm disc \, cooling}\,\, \underbrace{ \int_{\Lambda(t)}^{\Lambda(t) X_{\rm out}^{3/4}(t)} {Y^{5/3}\, {\rm d}Y \over \exp\left(Y \right) - 1}  }_{\rm disc \, spreading },
\end{multline}
where we have defined the dimensionless scales 
\beq
\Lambda(t) \equiv {h \nu \over kT_p} \tau^{n/4} , \quad X_{\rm out}(t) \equiv  \left({R_{\rm out}(t)\over R_I}\right) \simeq \left({R_{\rm circ} \over R_I}\right) \tau^{\gamma}  .
\eeq
In the above expression we have highlighted the terms associated with the cooling and spreading of the disc. Disc cooling is trivial to understand: as the disc cools the overall normalisation of the bolometric luminosity decreases. Disc spreading is simple to understand qualitatively: increasing the emitting area of a blackbody surface increases the emitted flux. Mathematically we have a strictly positive integrand integrated over a {\it growing} interval. Provided we restrict our observations to frequencies below the peak of the disc spectrum $\Lambda(t) < 1$, then this integral increases as a function of time, and in fact offset's the cooling of the disc. This delicate balancing of disc cooling and spreading is what leads to the prolonged plateau in TDE UV light curves. 

If the inner temperature of the disc is hot $\Lambda(t) \ll 1$, and the outer radius of the disc is large $\Lambda(t) X^{3/4}_{\rm out}(t) \gg 1$, then this integral is approximately 
\begin{align}
I_Y &= \int_{\Lambda(t)}^{\Lambda(t) X_{\rm out}^{3/4}(t)} {Y^{5/3}\, {\rm d}Y \over \exp\left(Y \right) - 1} \\
&\simeq  \int_0^{\infty} {Y^{5/3}\, {\rm d}Y \over \exp\left(Y \right) - 1} = \Gamma\left({8 \over 3} \right) \zeta_{\cal R} \left({8 \over 3} \right) \simeq 1.93 ,
\end{align}
which is a simple constant independent of disc and black hole properties. The disc spectrum in this regime is then well approximated by 
\beq
\nu L_\nu =  \frac{16\pi^2 h \nu^4 \cos(i)}{c^2} R_I^2 \left(  {k  T_p   \over  h\nu } \right)^{8/3} \tau^{-2n/3} \Gamma\left({8 \over 3} \right) \zeta_{\cal R} \left({8 \over 3} \right) .
\eeq
The amplitude of the above relationship then tells us the predicted scaling of the UV luminosity with system parameters. Taking $R_I \propto \M$, and substituting for the earlier scaling of $T_p$, we find 
\beq
\nu L_\nu \sim \M^{2/3}  \left({M_\star \over R_\star} \right) \beta \cos(i) .
\eeq
Finally, by assuming a main sequence mass-radius stellar relationship $R_\star \propto M_\star^{0.56}$, we have 
\beq
\nu L_\nu \sim \M^{2/3}  {M_\star}^{0.44}  \beta \cos(i) .
\eeq
Only one of these parameters will scale by many orders of magnitude across a population of tidal disruption events: the black hole mass $\M$. This will therefore be the principal parameter driving the observed distribution of TDE $\nu L_\nu$'s.

If however the outer disc edge is not so large and all disc temperatures are larger than the observing frequency $\Lambda(t) X^{3/4}_{\rm out}(t) \ll 1$, then 
\begin{align}
I_Y &= \int_{\Lambda(t)}^{\Lambda(t) X_{\rm out}^{3/4}(t)} {Y^{5/3}\, {\rm d}Y \over \exp\left(Y \right) - 1} \\
&\simeq  \int_0^{\Lambda(t) X^{3/4}_{\rm out}(t) } {Y^{2/3}\, {\rm d}Y } ={3\over 5} \left(\Lambda(t) X^{3/4}_{\rm out}(t)\right)^{5/3},
\end{align}
we then recover the classic Rayleigh-Jeans tail result 
\begin{align}
\nu L_\nu &=  \frac{48\pi^2 h \nu^4 \cos(i)}{5 c^2} R_I^2 \left(  {k  T_p   \over  h\nu } \right) \tau^{-n/4} X^{5/4}_{\rm out}(t) \\
&\simeq \frac{48\pi^2 h \nu^4 \cos(i)}{5 c^2} R_I^2 \left(  {k  T_p   \over  h\nu } \right)  \left({R_{\rm circ} \over R_I}\right)^{5/4} \tau^{5\gamma/4-n/4}  .
\end{align}
Note that for the canonical model, $5\gamma/4 - n/4 = 11/64$, and the Rayleigh-Jeans flux formally grows with time.

Once again we may substitute the scaling relationships $R_I \propto \M$, $R_{\rm circ} \propto R_\star/\beta (\M/M_\star)^{1/3}$, and for the temperature scaling to find 
\beq
\nu L_\nu \sim \M^{2/3}  M_\star^{-1/24}  R_\star^{7/8} \beta^{-7/8} \cos(i) .
\eeq
Finally, by assuming a main sequence mass-radius stellar relationship $R_\star \propto M_\star^{0.56}$, we have 
\beq
\nu L_\nu \sim \M^{2/3}  {M_\star}^{0.45}  \beta^{-7/8} \cos(i) .
\eeq
We see that independent of the regime (Rayleigh-Jeans or mid-frequency), the simple thin-disc prediction for the $L_{\rm UV} - \M$ scaling is $L_{\rm UV} \propto \M^{2/3}$. 

We have, of course, used many simplifications in deriving this argument, and this in no way should replace the full numerical simulation  performed in the main body of the paper. It is however useful to see that these results may be understood with reference to classical theory.  

\section{Solutions of the relativistic disc equations }\label{eq_sol_app}
\subsection{Functional forms of orbital parameters }
Both the numerical and analytical solutions of the relativistic disc equations require the functional forms of the orbiting fluids 4-velocity and 4-momenta. These expressions correspond to a fluid element undergoing circular motion in the equatorial plane of a Kerr black hole, and are given explicitly by 
 \begin{align}
U_0 &= -\frac{1-2r_g/r +a\sqrt{r_g/r^3}}{\left( 1- 3r_g/r + 2a\sqrt{r_g/r^3}\right)^{1/2} }, \\
  U^0 &= \frac{1+a\sqrt{{r_g}/{r^3}}}{\left({1 - {3r_g}/{r} + 2a\sqrt{{r_g}/{r^3} } }\right)^{1/2}},  \\
 U_\phi & = \sqrt{G\M r}\ \frac{1 + {a^2}/{r^2} - 2a\sqrt{{r_g}/{r^3}}}{\left({1 - {3r_g}/{r} + 2a\sqrt{{r_g}/{r^3} } }\right)^{1/2}},  \\
 U^\phi &= \frac{\sqrt{{G\M}/{r^3}}}    {\left( 1 - {3r_g}/{r} + 2a\sqrt{{r_g}/{r^3} } \right)^{1/2}},  \\
 \Omega &= \frac{U^\phi}{U^0} = \frac{\sqrt{{G\M }/{r^3}}}{1 + a\sqrt{{r_g}/{r^3}} } .
 \end{align}

\subsection{Numerical}

Numerical solutions of the relativistic thin disc equations were found using the finite element method, as implemented in \verb|DOLFIN|, part of the finite element software package \verb|FEniCS| (\citet{LoggWells2010, LoggEtal_10_2012, LoggEtal2012}). 

\subsubsection{Finite element method overview}

The finite element method enables efficient, stable, numerical solution of nonlinear partial differential equations over complex geometries, handling arbitrary boundary conditions effectively. In particular, we have a flexible mesh, which is useful for the disc problem.

For the simulation of TDE discs, we reformulate the relativistic disc evolution equation \eqref{fund} as an integral over the spatial volume of the disc system at each timestep - this is called the weak form. Boundary conditions are then directly inserted at the boundaries of the integration region; in this way, arbitrary disc edge conditions and initial profiles of infalling matter can be solved. With the geometry discretised into a mesh, which can have an arbitrary density profile, the mesh can be concentrated efficiently where high derivatives of the matter profile solution are expected - crucial near the  inner edge of the disc. Applying the calculus of variations, a linear combination of basis functions (the 'finite elements') that minimises the error of this integral is identified: this gives the time evolution of the TDE disc.

\subsubsection{Weak formulation}
The \verb|FEniCS| software package solves non-linear differential equations in their so-called ``weak'' form. Some manipulation of our governing equation is required so as to derive a particularly stable numerical weak form of the disc equations, which we discuss below. 


The $\zeta= \sqrt{g} \Sigma W_{\phi}^{r}/U^0$ evolution equation, to be solved on the domain ${\cal D} = (r_I, r_{\rm EXT}]$, is
\beq
\frac{\partial \zeta}{\partial t}=\left(W_{\phi}^{r}+\Sigma \frac{\partial W_{\phi}^{r}}{\partial \Sigma}\right) \frac{1}{\left(U^{0}\right)^{2}} \frac{\partial}{\partial r}\left(\frac{U^{0}}{U_{\phi}^{\prime}} \frac{\partial \zeta}{\partial r}\right),
\eeq
with outer boundary condition taken to be $ {\partial \zeta}/{\partial r} = 0$. Note that we ensure that the outer disc radius $r_{\rm EXT}$ is sufficiently far from the initial disc radius that the outer disc boundary condition does not effect the disc evolution.  

Following the standard notation of the \verb|FEniCS| package, we define $u = \zeta$, $x = r-r_I$, and use prime $'$ and dot $^{.}$ to denote differentiation with respect to $x$ and $t$ respectively. Then, if we define the following:
\begin{align}
A(r, u) &\equiv \left(W_{\phi}^{r}+\Sigma \frac{\partial W_{\phi}^{r}}{\partial \Sigma}\right) \frac{1}{\left(U^{0}\right)^{2}} , \\
B(r) &\equiv \frac{U^{0}}{U_{\phi}^{\prime}}, \\
\dot{u} &= A(B u')' = AB'u' + ABu''. 
\end{align}
Additionally, we multiply through by $x^2$ to suppress the ISCO singularity in $B(r)$; this is done in order to avoid a division by zero. Using a backwards difference in time, and discretising time in steps of $\Delta t$, we define $u = u(r, t=\Delta t^{n+1})$, which is to be solved for, and $u_n = u(r, t=\Delta t^n)$, which is known. We can approximate the time derivative as follows:
\beq
x^2 \frac{\partial \zeta}{\partial t} \simeq x^2 (u - u_n)/\Delta t = x^2AB'u' + x^2 ABu'' .
\eeq
This yields a series of pseudo-stationary problems, one at each time step, for $u$. To obtain the weak form, which is what is solved by \verb|DOLFIN|, we multiply by an arbitrary `test function' $v$ and integrate:
\begin{multline}
    F(u; v) = \int_{\cal D} \Big(x^2 uv - x^2 u_n v -  x^2 AB'u'v \Delta t \\ -  x^2 ABu''v \Delta t \Big) \, \mathrm{d}x .
\end{multline}
The \verb|DOLFIN| package numerically finds solutions to the equation 
\beq
F(u; v) = 0, \quad\quad \forall v . 
\eeq 
It is standard practice with the finite element method to minimise the order of the derivatives, so we integrate the $\partial_x^2 u $ term by parts.
\beq
\int_{\cal D} x^2 ABu''v \,\mathrm{d}x = \left[x^2 AB v u'\right]_{r_I}^{r_{\rm EXT}} - \int_{\cal D} (x^2 AB v )' u' \,\mathrm{d}x,
\eeq
The boundary conditions applied to the disc lead to the term in square brackets vanishing (note that $x^2 AB$ vanishes at $r_I$, as $B \propto 1/x$). We thus obtain the weak form:
\begin{multline}
    F(u; v) = \int_{\cal D} \Big(x^2 uv - x^2 u_n v -  x^2AB'u'v \Delta t \\ + (x^2 AB v)'u' \Delta t 
 \Big)\, \mathrm{d}x. 
\end{multline}

This can be solved using \verb|DOLFIN| directly at each timestep. Given some mesh, \verb|DOLFIN| automatically discretises and solves this equation by the calculus of variations, giving a numerical solution $u(r,t=\Delta t^{n+1})$. Iterating this procedure, we obtain a $u$ distribution for all times required. 
 
\subsection{Analytical}
Analytical solutions of the relativistic thin disc equations were presented in \citet{Mummery23a}. They take the functional form 
\begin{multline}\label{green_s}
G_\Sigma(x, x_0, \tau) = \\ {M_d \over 2\pi r_I^2 c_0}  \sqrt{x^{-\alpha} f_\alpha(x) \exp\left(-{1 \over x} \right) \left(1 - {2\over x}\right)^{5/2 - 3/4\alpha} }\\ 
 {x^{-3/4 - \mu} \over \tau} \exp\left({-f_\alpha(x)^2 - f_\alpha(x_0)^2 \over  4\tau} \right) I_{1\over 4\alpha} \left({ f_\alpha(x) f_\alpha(x_0) \over  2\tau}\right),
\end{multline}
where 
\beq
c_0 = x_0^{(1 + 14\mu)/8} \left(1 - {2\over x_0}\right)^{3/4 - 3/8\alpha} \exp\left(-{1\over 2x_0}\right) \sqrt{1\over f_\alpha(x_0)} ,
\eeq
and $M_d$ is the total mass of the disc at $t = 0$. These solutions are approximate, and represent asymptotic leading order solutions. \citet{Mummery23a} verified that these solutions reproduce full numerical calculations of the relativistic disc equations with an accuracy  at the $\mathcal{O}(1\%)$ level. In this expression, $I_\nu$ is the modified Bessel function of order $\nu$, and $\alpha$ is related to the stress index $\mu$ via $\alpha = (3-2\mu)/4$. The function $f_\alpha(x)$ is given by 
\begin{multline}
f_\alpha(x) = {x^\alpha \over 2 \alpha} \sqrt{1 - {2\over x}}\left[1  - {x^{ - 1} \over { (\alpha - 1)}} {}_2F_1\left(1, {3\over 2}-\alpha; 2-\alpha; {2\over x}\right) \right] \\ + {2^{\alpha - 2} \over \alpha (\alpha - 1)}\sqrt{\pi} {\Gamma(2-\alpha)  \over  \Gamma({3/ 2} - \alpha)} ,
\end{multline}
where $_2F_1(a, b, c, z)$ is the hypergeometric function, $\Gamma(z)$ is the gamma function, and 
\beq
x \equiv 2 r / r_I. 
\eeq
The variable $x_0 = 2r_0/r_I$ is the (normalised) initial location of the disc material. The time variable $\tau$ is given by 
\beq
\tau \equiv \sqrt{2 \over GMr_I} {w t \over r_I} \left(1 - {r_I \over r_0}\right),
\eeq 
where $t$ is measured in physical units. It is important to note that $\tau$ as defined here is {\it not} equal to the time in units of the viscous timescale at the initial radius $\tau \neq t/t_{\rm visc}(r_0)$.

\section{The ray-tracing algorithm }\label{raytrace}
The luminosity of the disc emission is given by (see section \ref{lum_deriv} for a derivation)
\beq\label{lum_app}
\nu L_\nu = 4\pi \nu _{\rm obs}  \iint_{\cal S} {f_\gamma^3 f_{\rm col}^{-4} B_\nu (\nu_{\rm obs} /f_\gamma , f_{\rm col} T)}\,  {\text{d}b_x  \text{d} b_y} ,
\eeq
where $f_{\rm col}$ and $T$ are functions of disc radius and time, and depend only on the solutions of the relativistic disc equations.  The photon energy shift factor $f_\gamma$ is given by 
\beq
f_\gamma = \frac{1}{U^{0}} \left[ 1+ \frac{p_{\phi}}{p_0} \Omega \right]^{-1} .
\eeq
Here $U^0$ and $\Omega = U^\phi / U^0$ are 4-velocity components of the rotating  disc fluid, and $p_\phi$ and $p_0$ are photon 4-momentum components.  The ratio $p_\phi / p_0$ is a constant of motion for a photon propagating through the Kerr metric. As a conserved quantity, $p_\phi / p_0$  can  be calculated from the photon initial conditions.  

We assume a distant observer orientated at an inclination angle $i$ from the disc plane at a distance $D$. We set up an image plane perpendicular to the line of sight centred at $\phi = 0$ (Fig. \ref{RT}), with image plane cartesian coordinates $(b_x, b_y)$. A photon at an image plane coordinate  $(b_x, b_y)$ has a corresponding spherical-polar coordinate $(r_i, \theta_i, \phi_i)$, given by \citep{Psaltis12}
\begin{align}
r_i &= \left(D^2 + b_x^2 + b_y^2\right)^{1/2}, \label{r0} \\
\cos\theta_i &=  r_i^{-1} \left( D \cos i + b_y\sin i\right) ,\\
\tan\phi_i &=   b_x \left(D\sin i - b_y\cos i \right)^{-1}  .
\end{align}
The only photons which will contribute to the image have 3-momentum which is perpendicular to the image plane. This orthogonality condition uniquely specifies the initial photon 4-velocity \citep{Psaltis12}
\begin{align}
u^r_i &\equiv \left(\frac{\text{d} r}{\text{d}\tau'}\right)_{\text{obs}} = \frac{D}{r_i} , \\
u^\theta_i &\equiv \left(\frac{\text{d} \theta}{\text{d}\tau'}\right)_{\text{obs}}  = \frac{ D\left(D\cos i+ b_y\sin i \right) - r_i^2\cos i  }{r_i^2 \left( r_i^2 - \left(D \cos i   + b_y\sin i  \right)^2\right)^{1/2}} , \\
u^\phi_i &\equiv \left(\frac{\text{d} \phi}{\text{d}\tau'}\right)_{\text{obs}} = \frac{- b_x \sin i}{\left( D \sin  i - b_y\cos  i\right)^2 + b_x^2} . \label{up0}
\end{align}
We note that the normalisation of these 4-velocity components can all be scaled by an arbitrary factor without effecting the trajectories.

We trace the rays back from the observer to the disc by solving the null-geodesics of the Kerr metric using the code \verb|YNOGK|, which is based on \verb|GEOKERR| \citep{YangWang13, DexterAgol09}. 

Starting from a finely spaced grid of points $(b_x, b_y)$ in the image plane, we trace the geodesics of each photon back to the disc plane, recording the location at which the photon intercepts the disc plane $(r_f)$, and the ratio $p_\phi/p_0$ for each photon.  The parameter $r_f$ allows the disc temperature $T$ to be calculated at a given time $t$ (equation \ref{temperature}). The parameters $r_f$ and $p_\phi/p_0$ together uniquely define the energy-shift factor $f_\gamma$. The integrand  (of eq. \ref{lum_app}) can therefore be calculated at every grid point in the image plane, and the integral (\ref{lum_app}) is then calculated numerically. 

\section{TDE black hole mass and spin measurements }
See Tables~\ref{mass_plat_table} and \ref{spin_plat_table}.

\begin{table}
    \centering
    \begin{tabular}{p{90pt} p{140pt}}
    Reference & TDEs   \\
    \hline\hline

\citet{vanVelzen10} & SDSS-TDE1, \mbox{SDSS-TDE2} \\
\citet{Gezari12} & PS1-10jh \\
\citet{Arcavi14} & PTF-09axc, \mbox{PTF-09djl}, \mbox{PTF-09ge} \\
\citet{Holoien14} & ASASSN-14ae \\
\citet{Miller_14li,Holoien16} & ASASSN-14li \\
\citet{Holoien16b} & ASASSN-15oi \\
\citet{Dong16,Leloudas16} & ASASSN-15lh \\
\citet{Hung17} & iPTF-16axa \\
\citet{Blagorodnova17} & iPTF-16fnl \\
\citet{Wyrzykowski17} & OGLE16aaa \\
\citet{Blagorodnova19} & iPTF-15af \\
\citet{vanVelzen18_NedStark,Holoien18a} & AT2018zr \\
\citet{Holoien19} & AT2019ahk \\
\citet{Leloudas19} & AT2018dyb \\
\citet{Wevers19a} & AT2018fyk \\
\citet{Short20,vanVelzen20} & AT2018hyz \\
\citet{Nicholl20,vanVelzen20} & AT2019qiz \\
\citet{vanVelzen20} & AT2018hco, \mbox{AT2018iih}, \mbox{AT2018lna}, \mbox{AT2018lni}, \mbox{AT2019bhf}, \mbox{AT2019cho}, \mbox{AT2019dsg}, \mbox{AT2019ehz}, \mbox{AT2019lwu}, \mbox{AT2019meg}, \mbox{AT2019mha} \\
\citet{vanVelzen20,Hinkle21,Liu22} & AT2019azh \\
\citet{Wevers22,Hammerstein23} & AT2020zso \\
\citet{Yao22} & AT2021ehb \\
\citet{Angus22} & AT2020neh \\
\citet{Hammerstein23} & AT2018jbv, \mbox{AT2019teq}, \mbox{AT2019vcb}, \mbox{AT2020ddv}, \mbox{AT2020mbq}, \mbox{AT2020mot}, \mbox{AT2020ocn}, \mbox{AT2020opy}, \mbox{AT2020pj}, \mbox{AT2020qhs}, \mbox{AT2020riz}, \mbox{AT2020wey}, \mbox{AT2020ysg} \\
\citet{Goodwin23,Yao23} & AT2020vwl \\
\citet{Yao23} & AT2019cmw, \mbox{AT2020abri}, \mbox{AT2020acka}, \mbox{AT2020yue}, \mbox{AT2021axu}, \mbox{AT2021crk}, \mbox{AT2021jjm}, \mbox{AT2021mhg}, \mbox{AT2021nwa}, \mbox{AT2021sdu}, \mbox{AT2021uqv}, \mbox{AT2021utq}, \mbox{AT2021yte}, \mbox{AT2021yzv} \\
   
    \end{tabular}
\caption{Literature references (``discovery papers") for the TDEs used in work. }
\label{tab:tderefs}
\end{table}

\section{Lightcurve data and models}\label{sec:lcfits}
\begin{figure*}
\includegraphics[width=.32\textwidth, clip=30 10 30 10, clip]{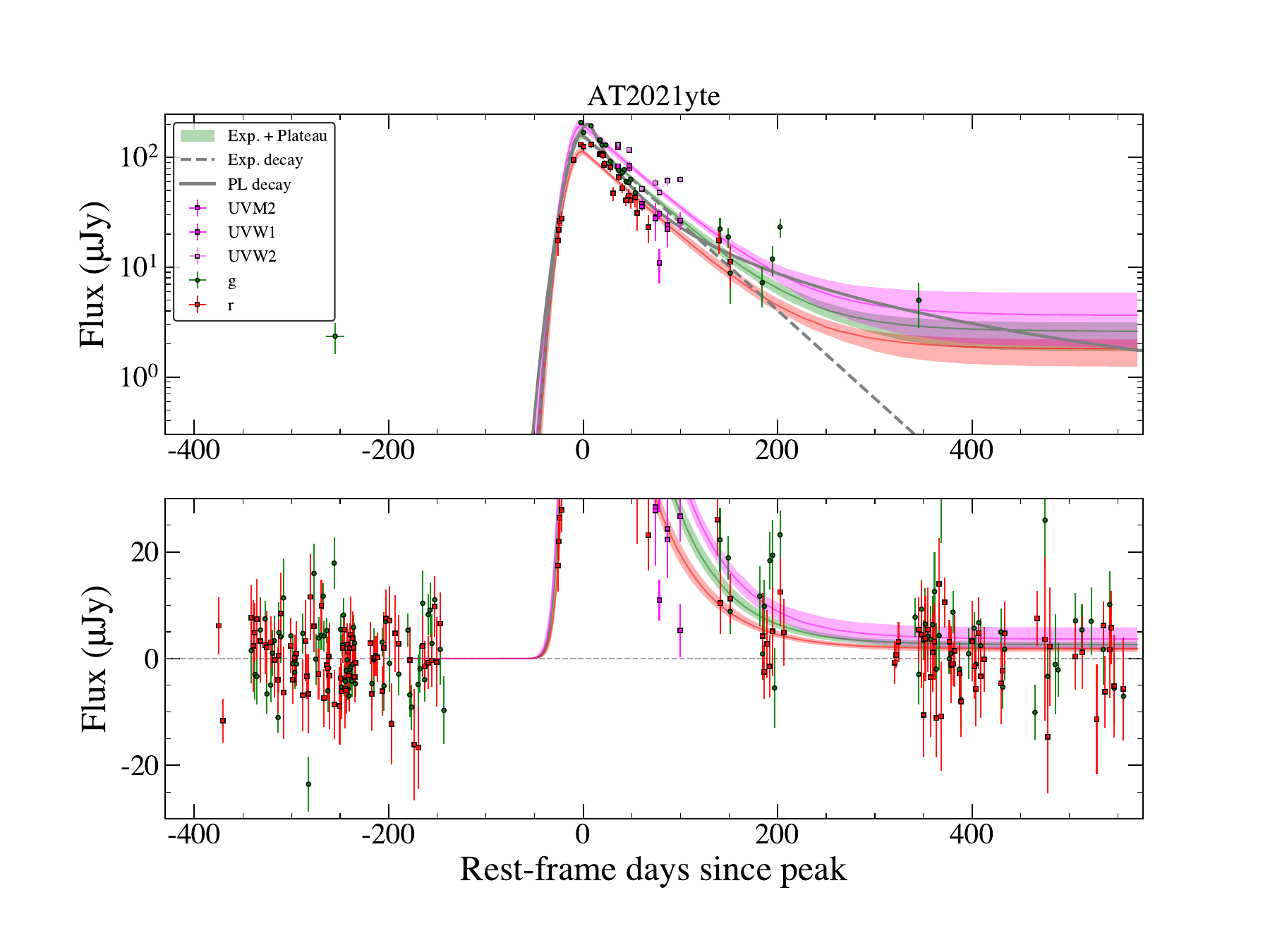}\quad
\includegraphics[width=.32\textwidth, clip=30 10 30 10, clip]{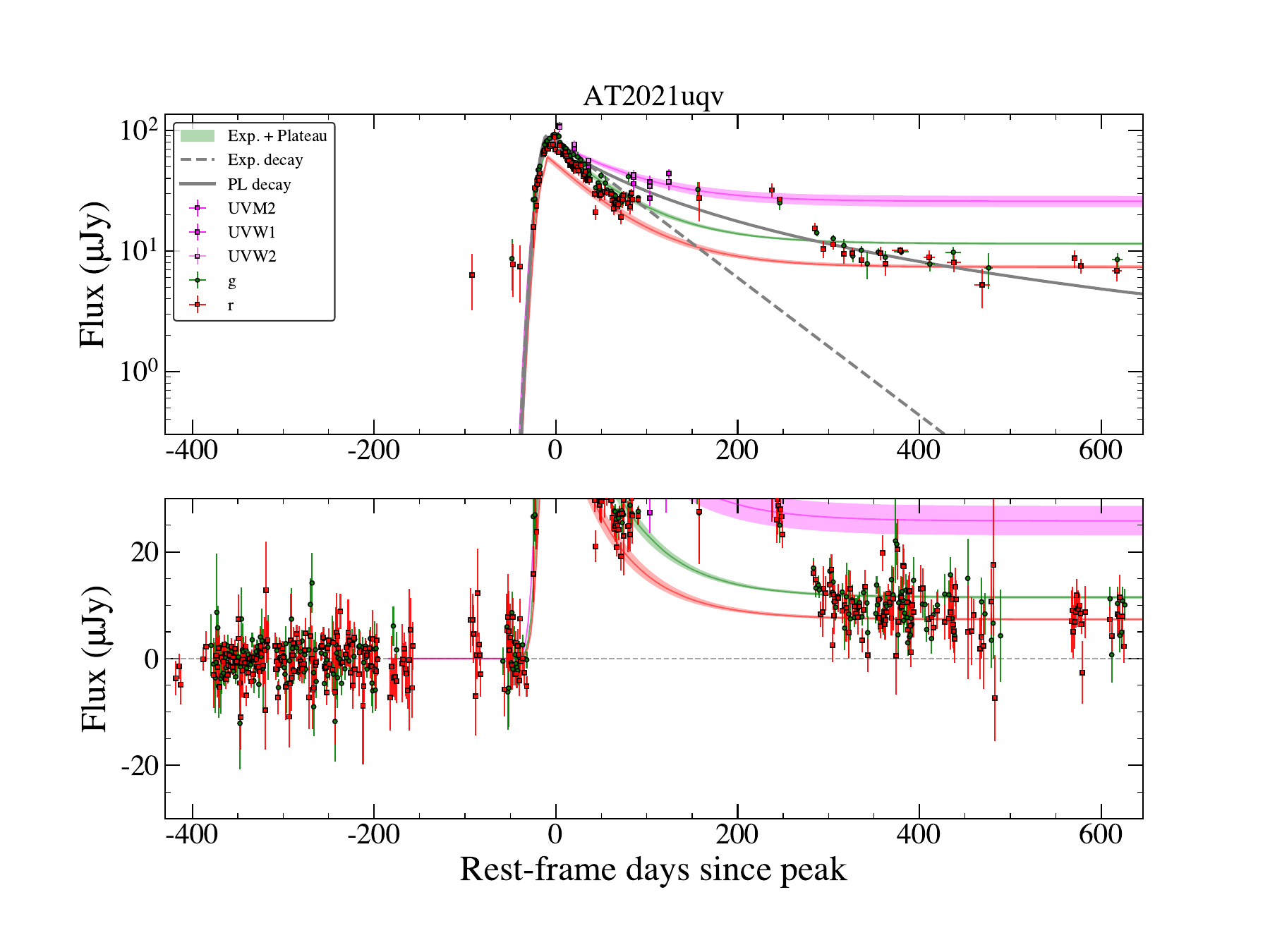}\quad
\includegraphics[width=.32\textwidth, clip=30 10 30 10, clip]{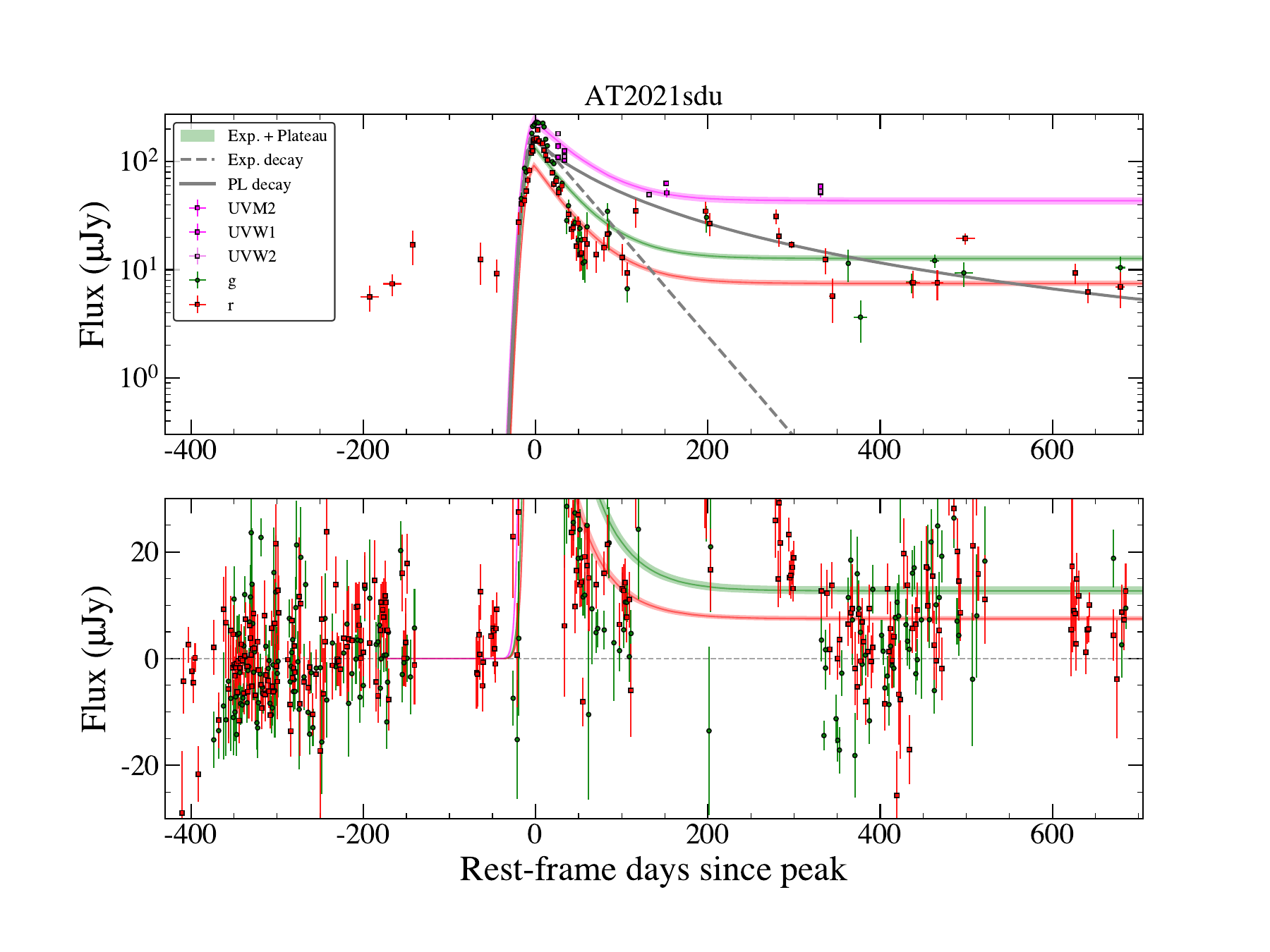}\\
\includegraphics[width=.32\textwidth, clip=30 10 30 10, clip]{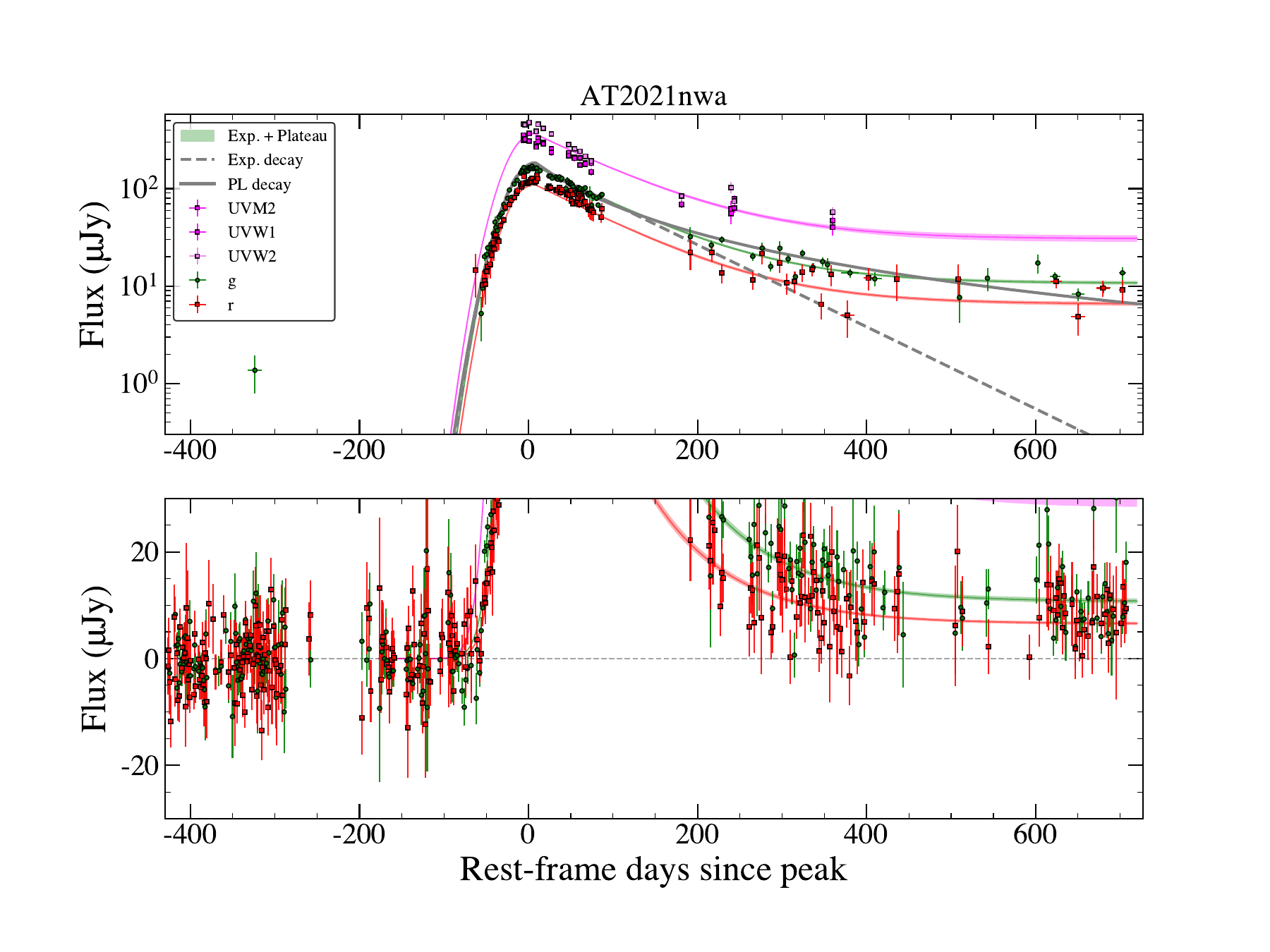}\quad
\includegraphics[width=.32\textwidth, clip=30 10 30 10, clip]{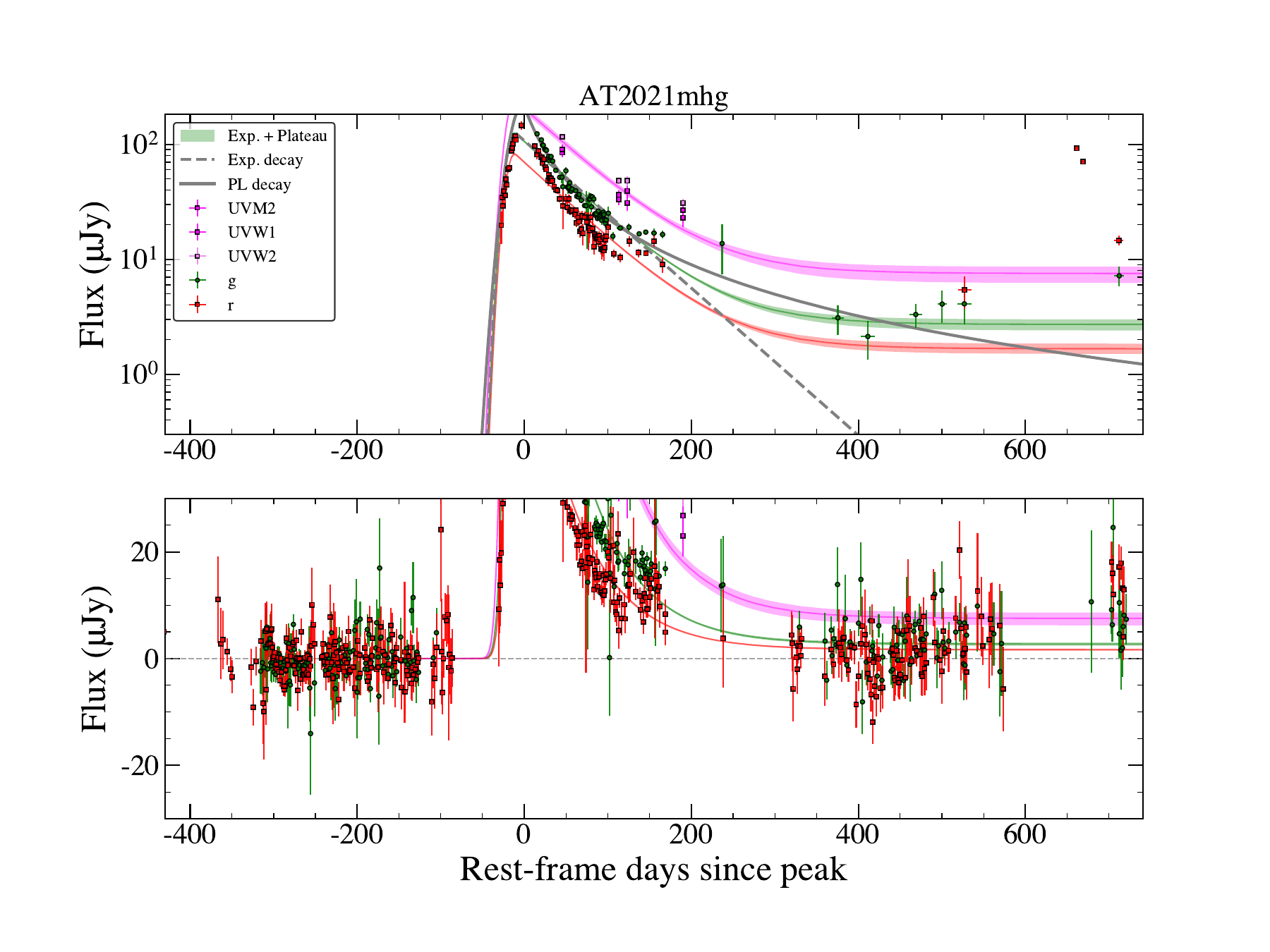}\quad
\includegraphics[width=.32\textwidth, clip=30 10 30 10, clip]{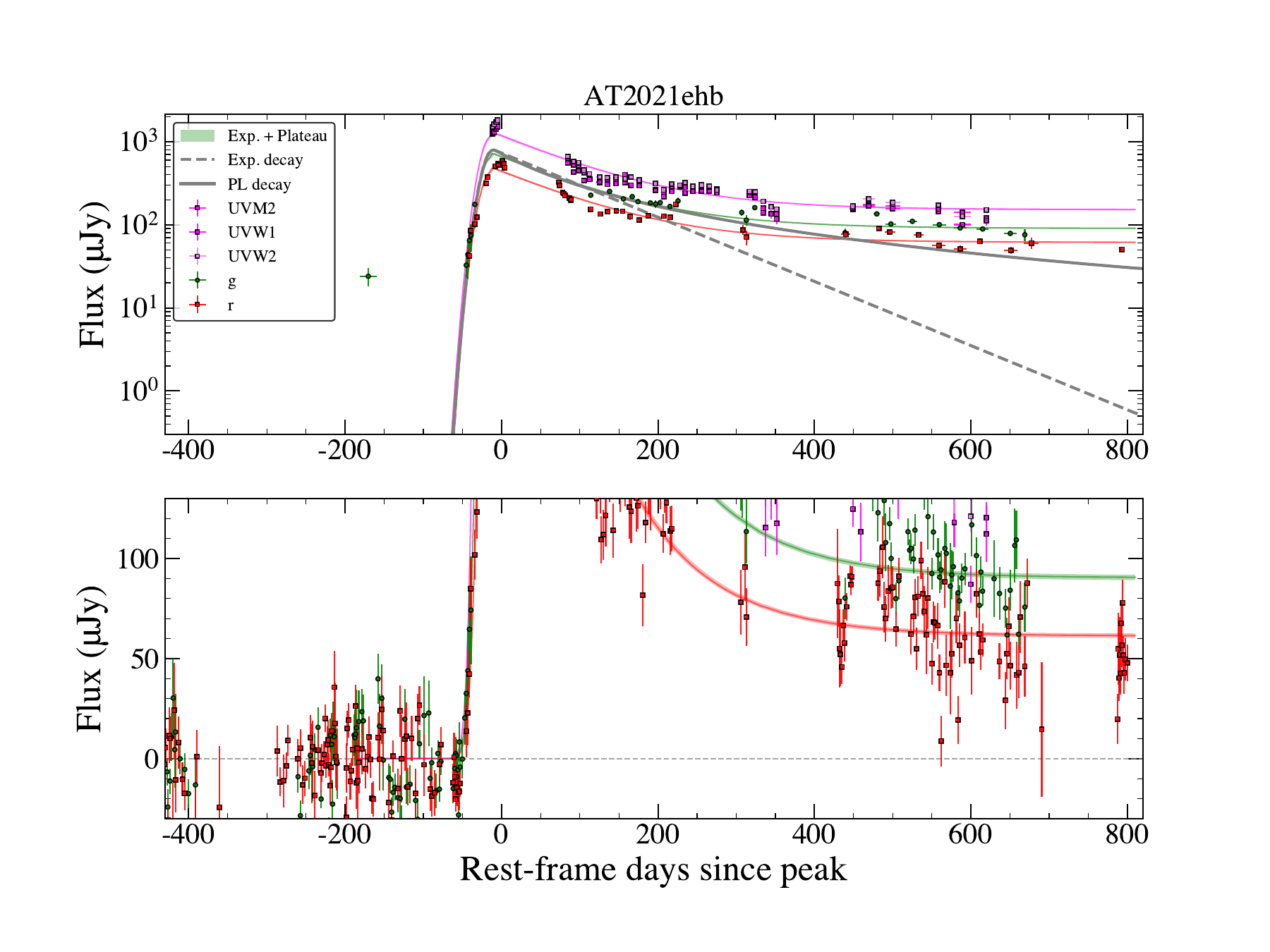}\\
\includegraphics[width=.32\textwidth, clip=30 10 30 10, clip]{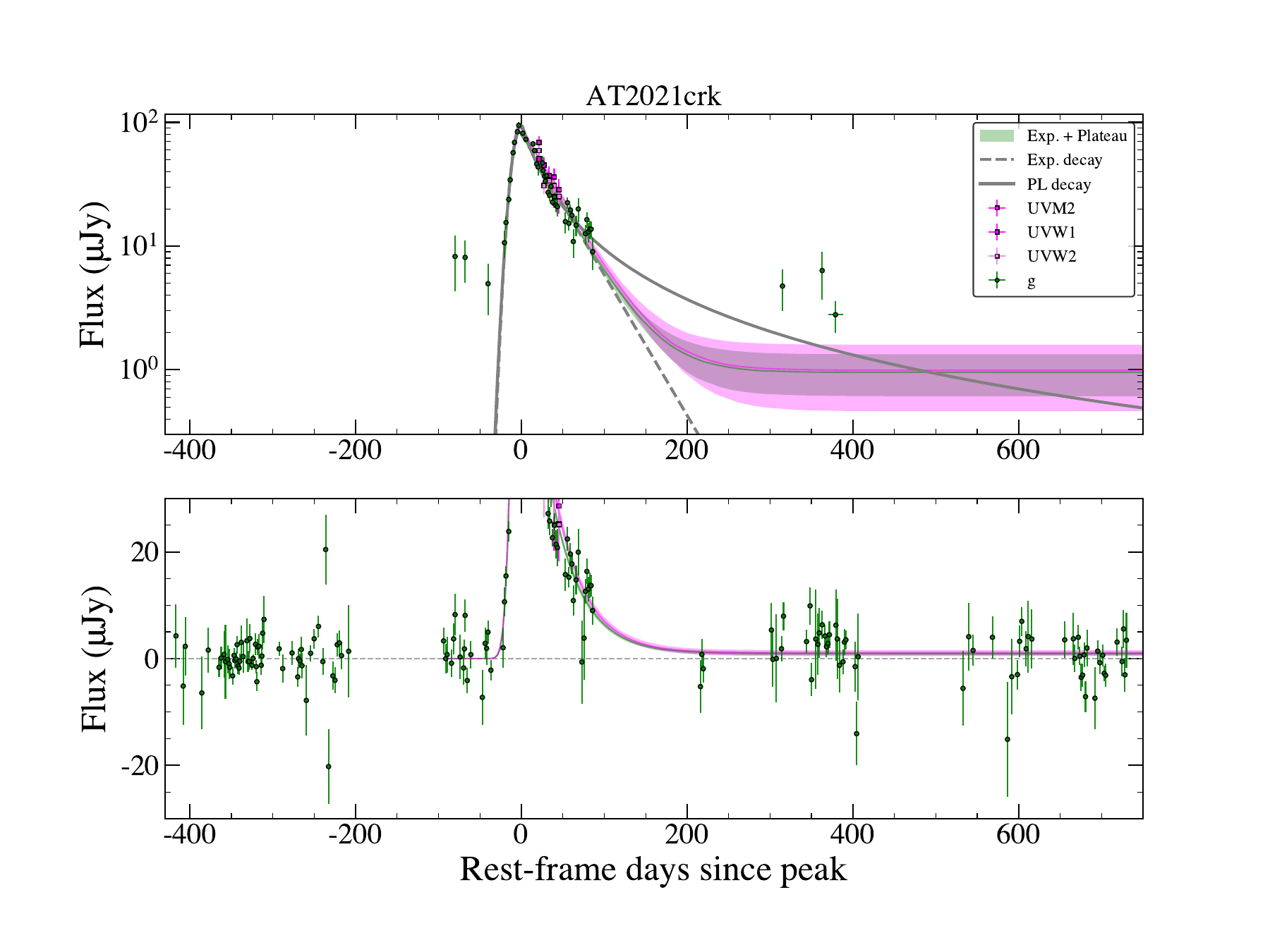}\quad
\includegraphics[width=.32\textwidth, clip=30 10 30 10, clip]{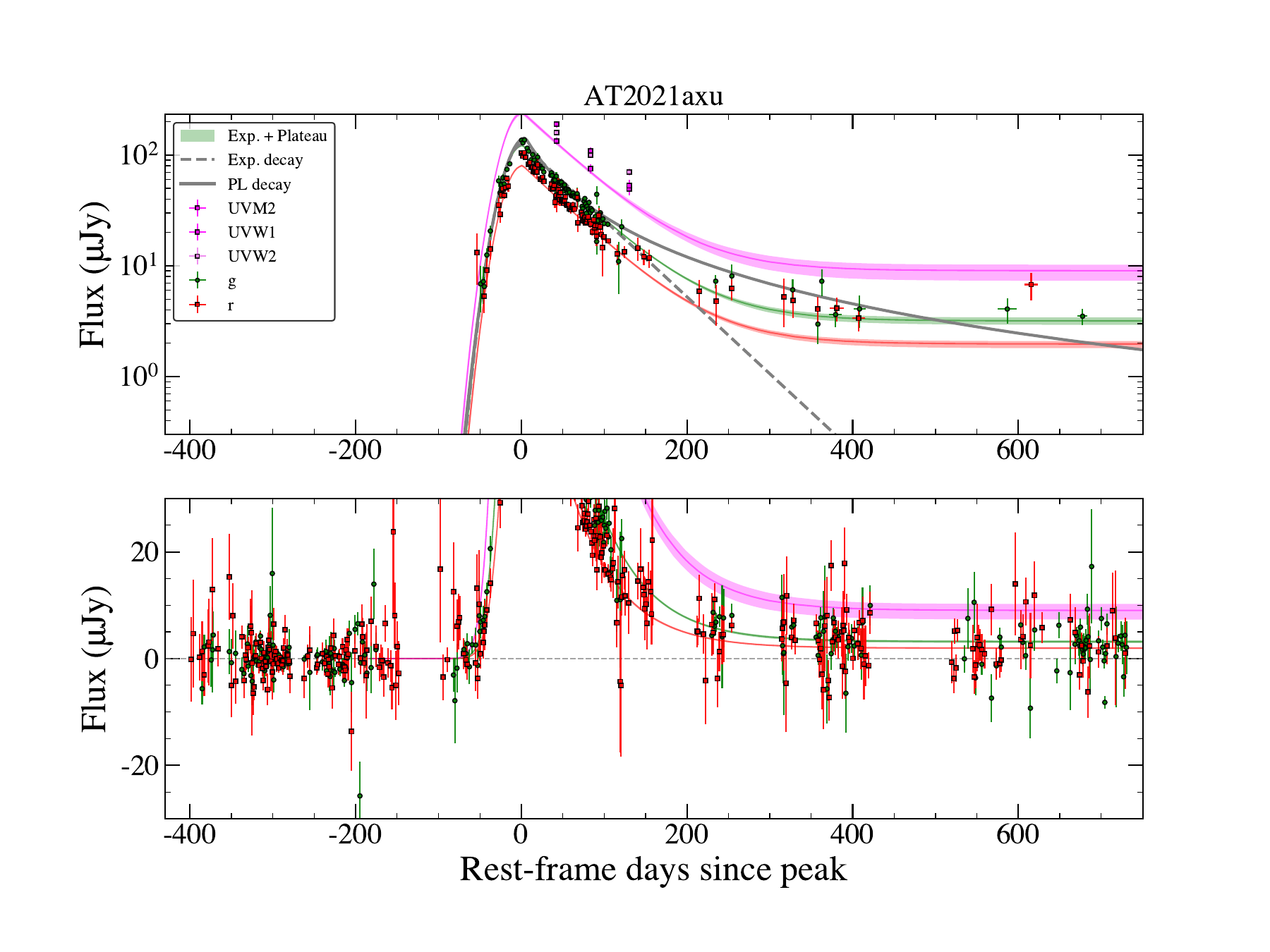}\quad
\includegraphics[width=.32\textwidth, clip=30 10 30 10, clip]{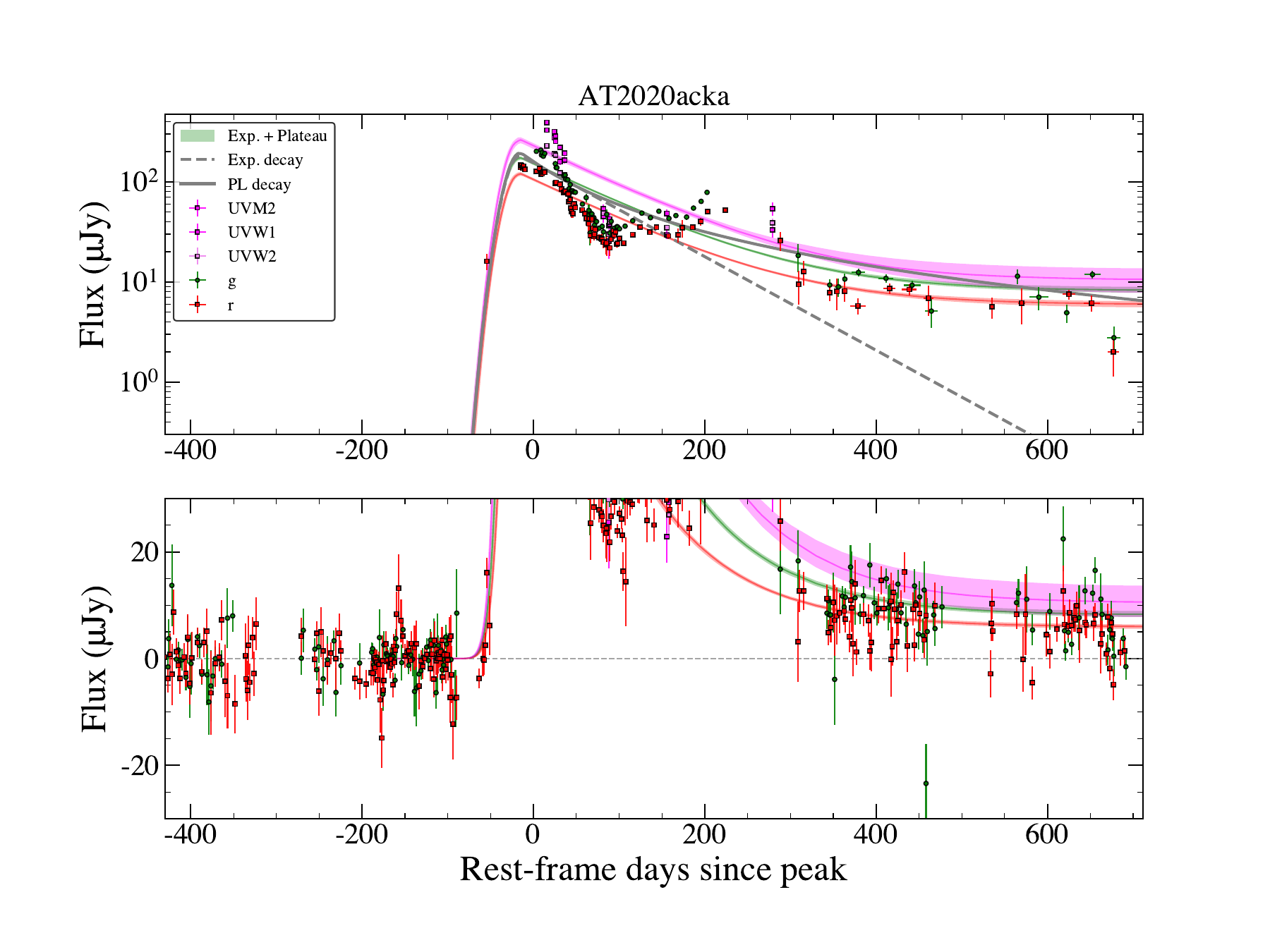}\\
\includegraphics[width=.32\textwidth, clip=30 10 30 10, clip]{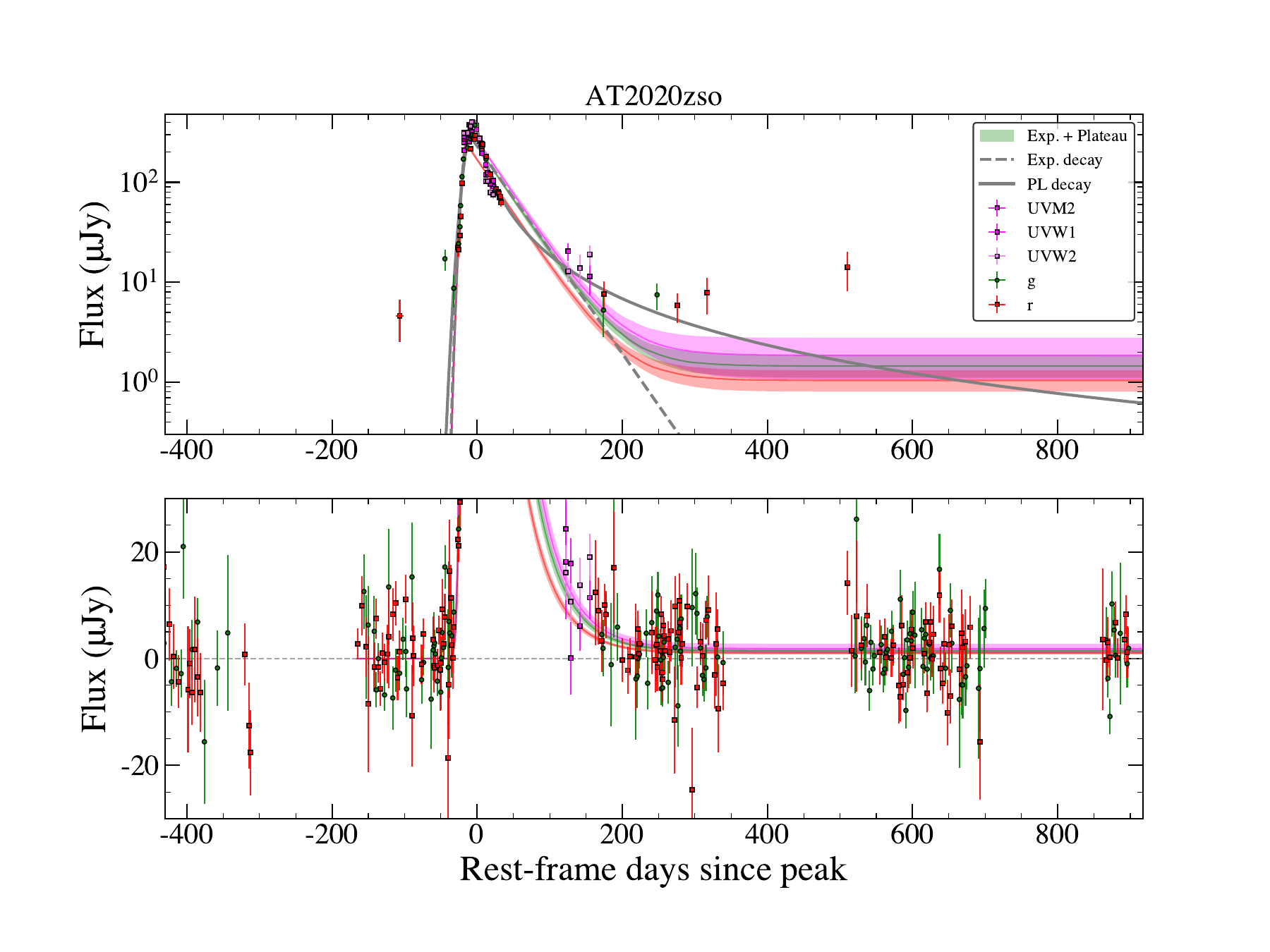}\quad
\includegraphics[width=.32\textwidth, clip=30 10 30 10, clip]{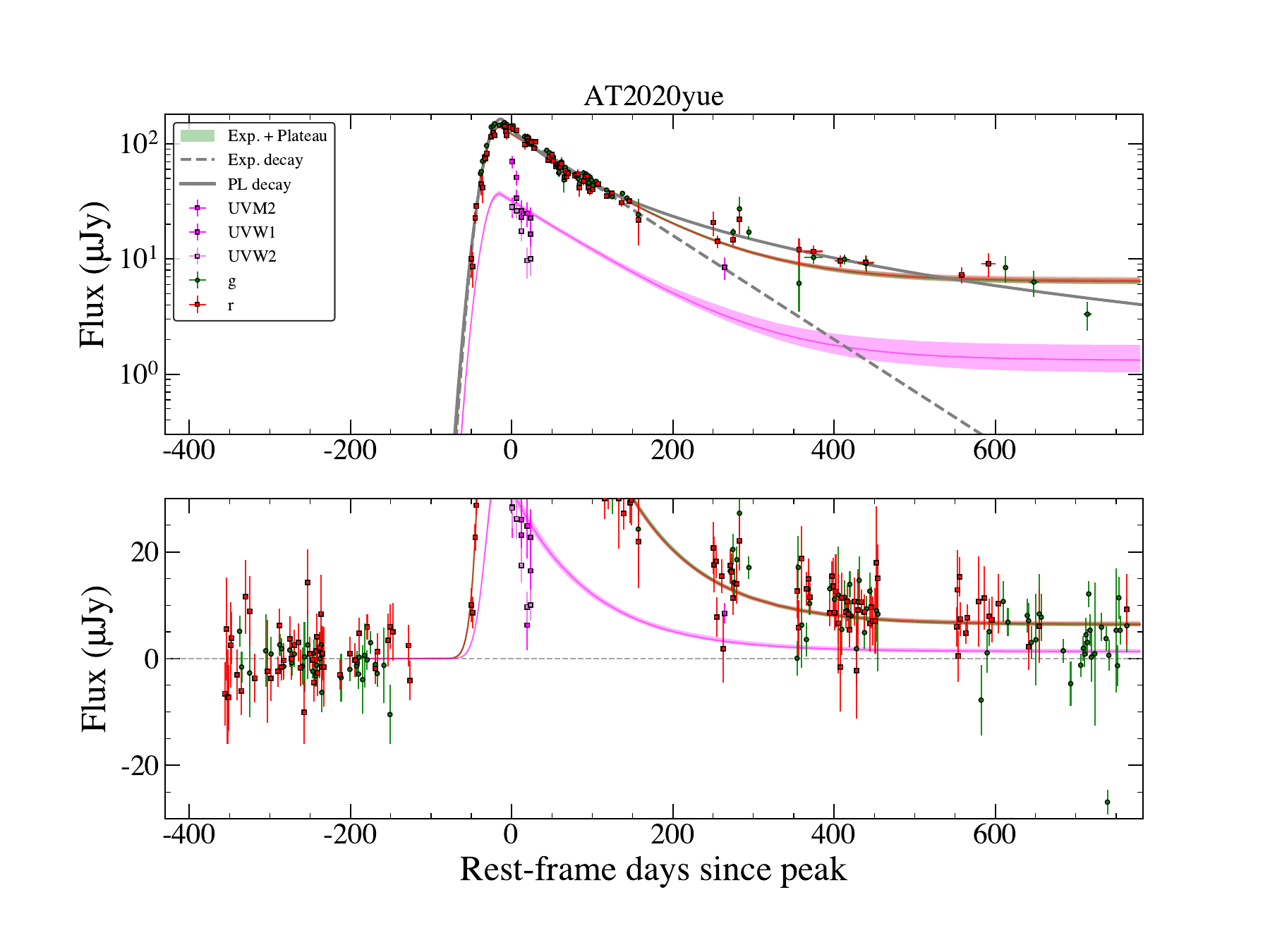}\quad
\includegraphics[width=.32\textwidth, clip=30 10 30 10, clip]{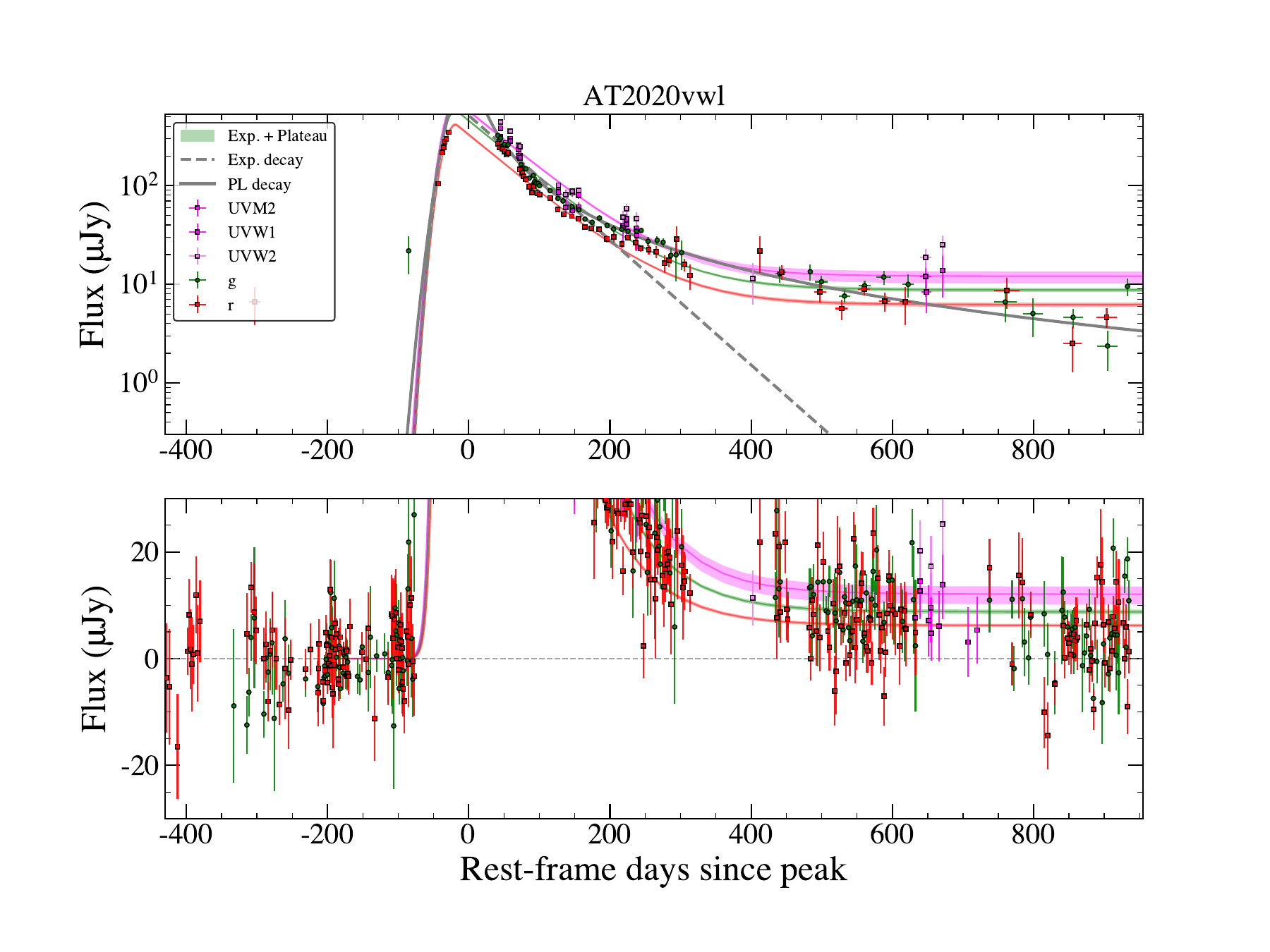}\\
\includegraphics[width=.32\textwidth, clip=30 10 30 10, clip]{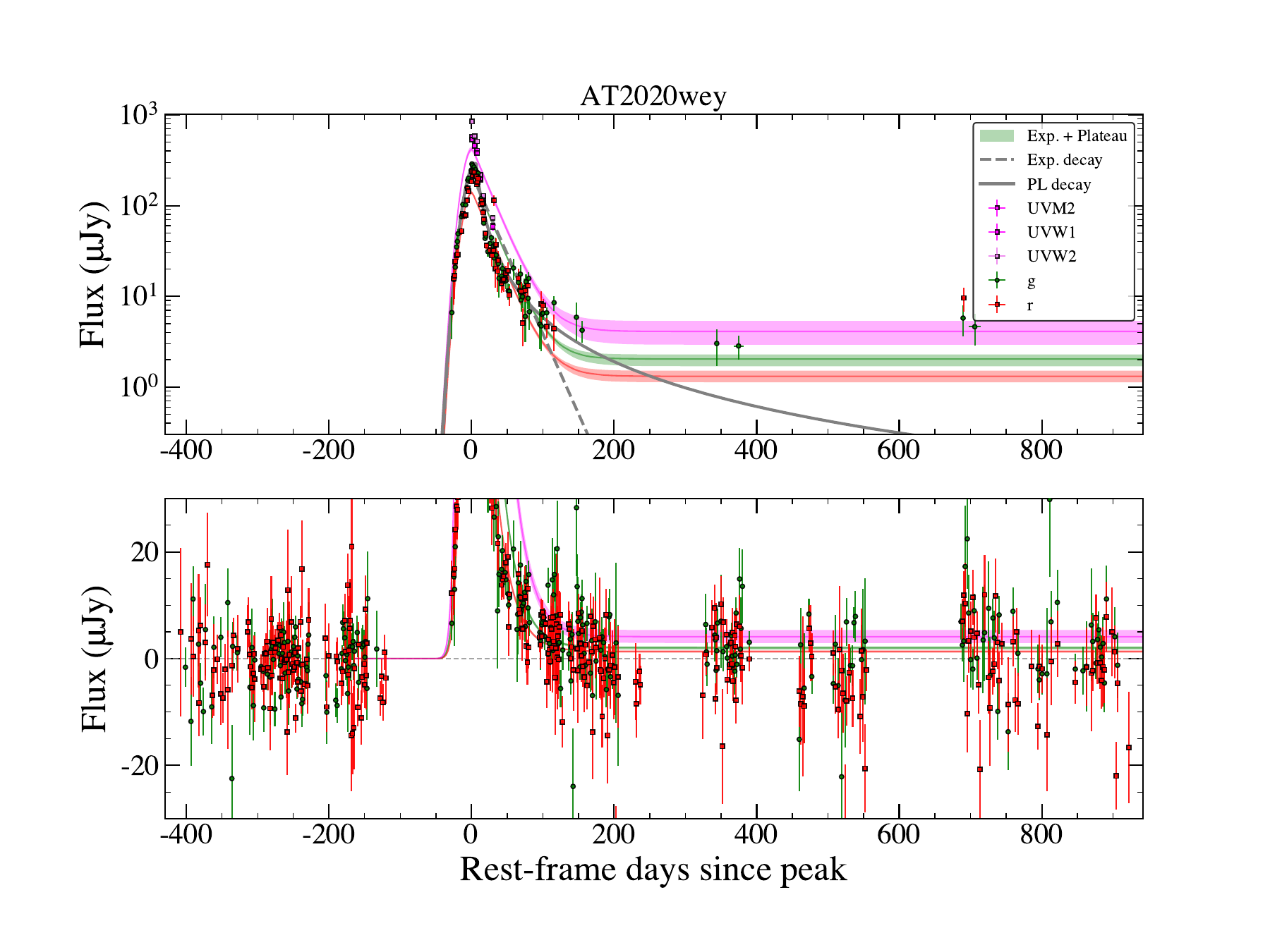}\quad
\includegraphics[width=.32\textwidth, clip=30 10 30 10, clip]{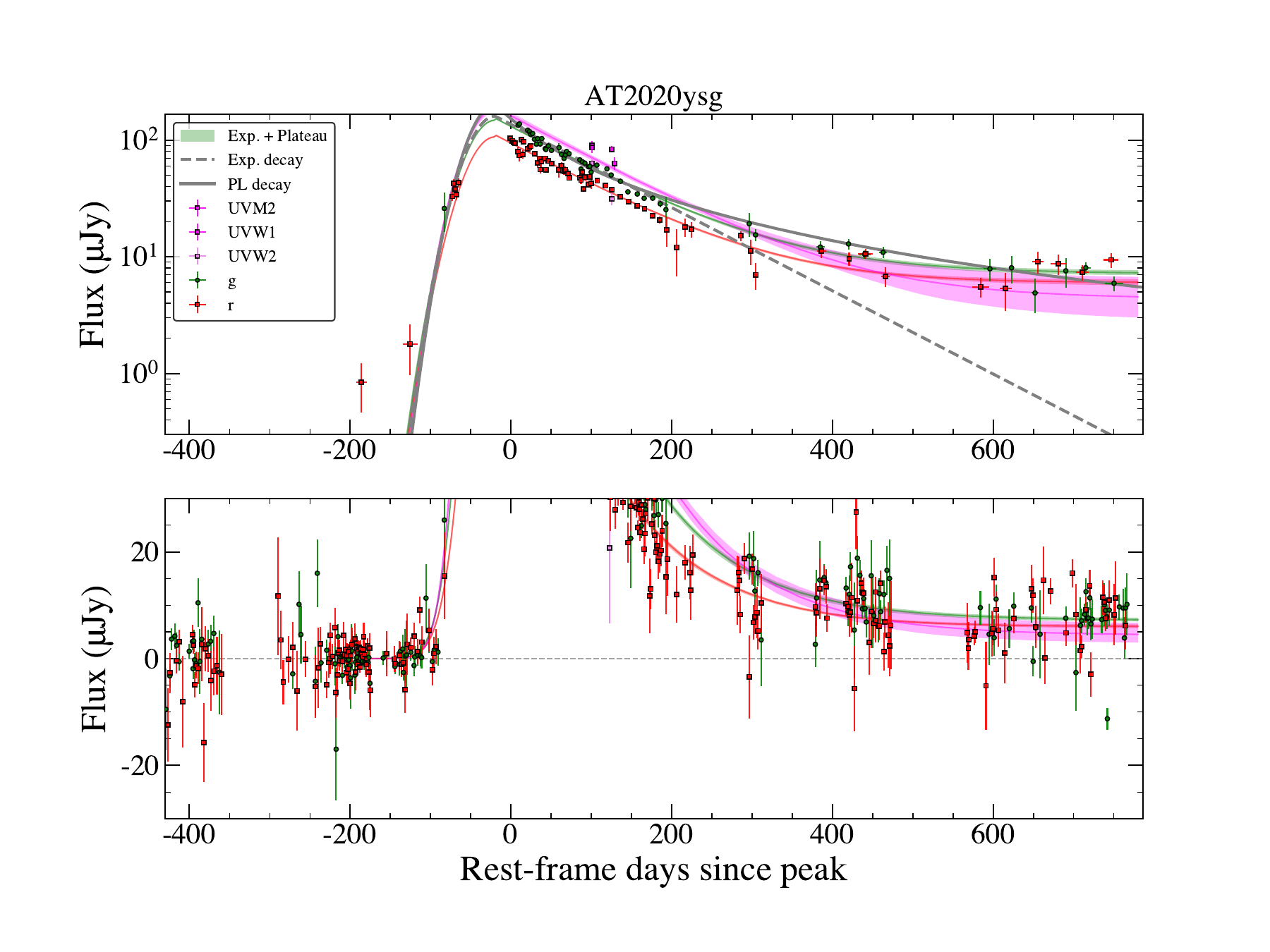}\quad
\includegraphics[width=.32\textwidth, clip=30 10 30 10, clip]{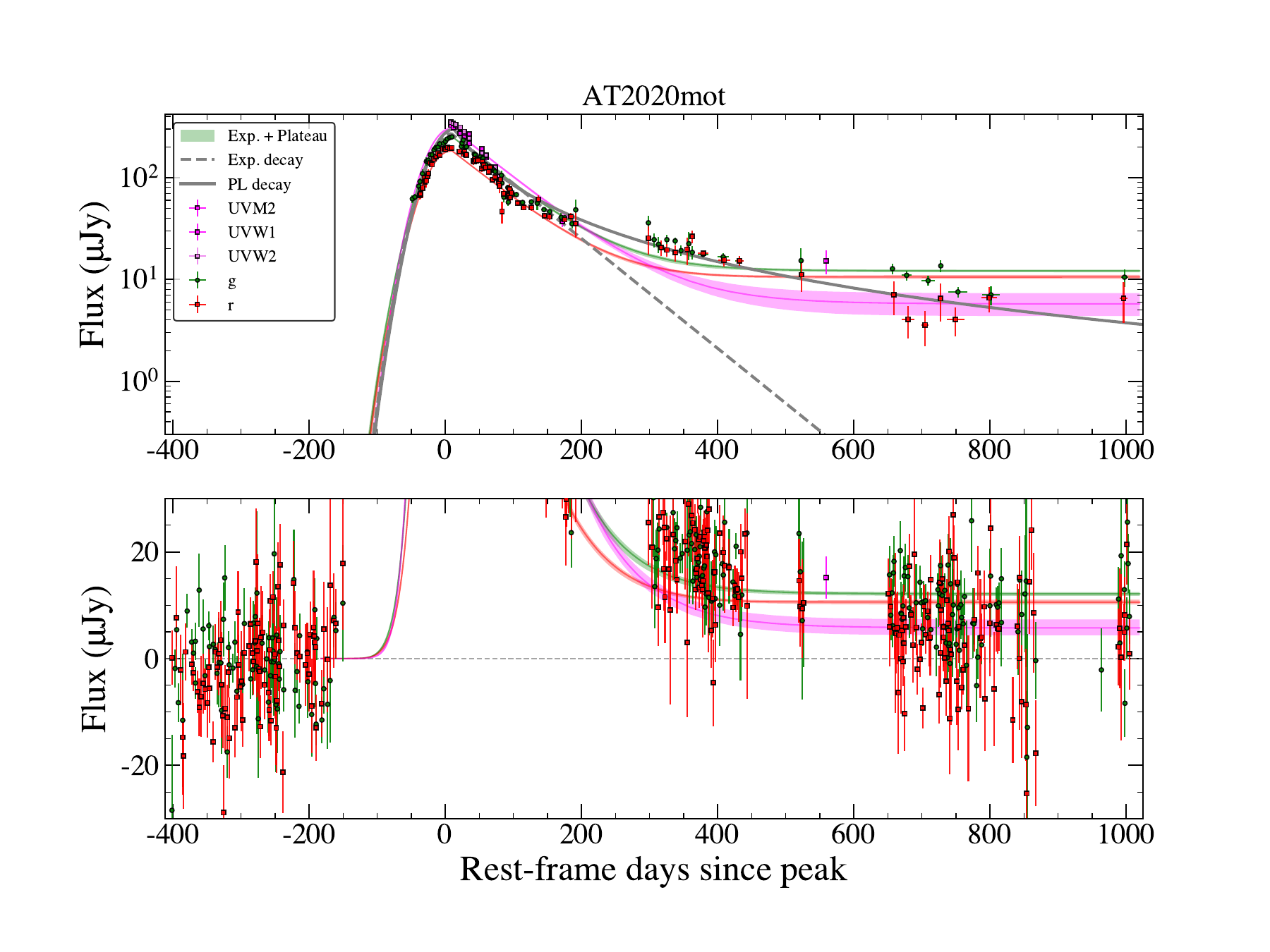}\\
\caption{
Observed and model lightcurves for TDEs with detected late-time plateaus. 
The colored lines show the 68\% credible level
for a model lightcurve curve described by an exponential decay 
plus a constant plateau.  We also show the best-fit exponential decay seperately, 
as obtained using only the first 100 days days of post-peak observations (grey dashed line). 
This can be compared to the best-fit power-law decay (with index fixed to $-5/3$) 
obtained from the first year of post-peak observations (grey solid line). 
{
For each TDE, the data in the upper panel are binned and show 
only positive flux detections. Relative to the time of peak, the following binwidths are used: 
30 days for $\Delta t \in \{\min(\Delta t), -100\}$; 
1 day for $\Delta t \in \{-100, +100\}$; 
10 days for $\Delta t \in \{+100, +365\}$;
30 days for $\Delta t \in \{+365, \max(\Delta t)\}$.
The lower panel shows the unbinned data on a linear scale, which was used to find the posterior distribution of the model lightcurves.}
}
\label{fig:lcfits}
\end{figure*}

In Fig.~\ref{fig:lcfits} we show the lightcurve data and model posterior curves for the TDEs with a detected plateau. 

\label{lastpage}

    \renewcommand{\arraystretch}{1.25}
    \onecolumn 

    \begin{longtable}{ |p{2.0cm}|p{1.6cm}|p{1.2cm}|p{1.6cm}|p{1.cm}|p{1.5cm}|p{1.cm}|p{.7cm}|p{1.cm}|p{1.cm}|p{1.cm}|  }
    \hline
    TDE Name & $L_{\rm plat}$ & $\M$ & $L_{\rm peak}$ & $\M$ & $E_{\rm rad}$ & $\M$ & $ \sigma $ & $\M$ & $M_{\rm gal}$ & $\M$ \\
    \hline
    & $ {\rm erg\, s^{-1}}$ & $ M_\odot$ & $  {\rm erg\, s^{-1}}$ & $  M_\odot$ & $  {\rm erg}$ & $  M_\odot$ & km/s & $  M_\odot$  & $  M_\odot$ & $  M_\odot$ \\
    \hline
     SDSS-TDE1 & $42.01^{+0.14}_{-0.13}$ & $6.85^{+ 0.50}_{- 0.40}$  & $42.67^{+0.04}_{-0.04}$ & $6.19$  & $49.37^{+0.11}_{-0.10}$ & $6.17$  & $126$ & $7.42$  & $10.16$ & $6.92$ \\ \hline 
 SDSS-TDE2 & $42.55^{+0.27}_{-0.24}$ & $7.48^{+ 0.33}_{- 0.36}$  & $43.40^{+0.02}_{-0.05}$ & $6.92$  & $50.55^{+0.11}_{-0.11}$ & $7.32$  & -- & --  & $10.48$ & $7.44$ \\ \hline 
 PS1-10jh & $41.85^{+0.08}_{-0.08}$ & $6.63^{+ 0.53}_{- 0.40}$  & $43.16^{+0.02}_{-0.02}$ & $6.68$  & $49.86^{+0.04}_{-0.04}$ & $6.64$  & $65$ & $6.15$  & $9.60$ & $6.02$ \\ \hline 
 PTF-09ge & $41.64^{+0.17}_{-0.15}$ & $6.36^{+ 0.55}_{- 0.42}$  & $43.30^{+0.01}_{-0.00}$ & $6.82$  & $49.99^{+0.02}_{-0.02}$ & $6.77$  & $82$ & $6.60$  & $10.33$ & $7.20$ \\ \hline 
 PTF-09djl & $42.11^{+0.11}_{-0.10}$ & $6.98^{+ 0.47}_{- 0.41}$  & $43.55^{+0.05}_{-0.04}$ & $7.06$  & $50.13^{+0.12}_{-0.11}$ & $6.91$  & $64$ & $6.13$  & $9.93$ & $6.56$ \\ \hline 
 ASASSN-14ae & $41.48^{+0.07}_{-0.07}$ & $6.13^{+ 0.55}_{- 0.42}$  & $43.15^{+0.04}_{-0.04}$ & $6.67$  & $49.63^{+0.05}_{-0.05}$ & $6.42$  & $53$ & $5.77$  & $9.90$ & $6.50$ \\ \hline 
 ASASSN-14li & $41.49^{+0.01}_{-0.01}$ & $6.14^{+ 0.55}_{- 0.42}$  & $42.41^{+0.02}_{-0.02}$ & $5.94$  & $49.14^{+0.03}_{-0.03}$ & $5.94$  & $81$ & $6.57$  & $9.73$ & $6.23$ \\ \hline 
 ASASSN-15oi & $41.29^{+0.04}_{-0.03}$ & $5.86^{+ 0.55}_{- 0.44}$  & $43.09^{+0.03}_{-0.03}$ & $6.61$  & $49.45^{+0.04}_{-0.04}$ & $6.25$  & $61$ & $6.03$  & $10.07$ & $6.77$ \\ \hline 
 ASASSN-15lh & $43.25^{+0.08}_{-0.11}$ & $8.02^{+ 0.32}_{- 0.35}$  & $44.68^{+0.04}_{-0.04}$ & $8.17$  & $51.66^{+0.06}_{-0.06}$ & $8.41$  & $225$ & $8.52$  & $10.85$ & $8.02$ \\ \hline 
 AT2018dyb & $41.96^{+0.02}_{-0.02}$ & $6.78^{+ 0.51}_{- 0.41}$  & $43.26^{+0.05}_{-0.04}$ & $6.78$  & $49.94^{+0.05}_{-0.04}$ & $6.72$  & -- & --  & $10.10$ & $6.83$ \\ \hline 
 AT2018fyk & $42.61^{+0.02}_{-0.02}$ & $7.53^{+ 0.32}_{- 0.36}$  & $43.02^{+0.03}_{-0.03}$ & $6.54$  & $50.11^{+0.05}_{-0.05}$ & $6.89$  & $158$ & $7.85$  & $10.61$ & $7.65$ \\ \hline 
 AT2019ahk & $42.03^{+0.03}_{-0.04}$ & $6.88^{+ 0.49}_{- 0.41}$  & $43.60^{+0.02}_{-0.01}$ & $7.11$  & $50.37^{+0.02}_{-0.02}$ & $7.15$  & -- & --  & $10.21$ & $7.00$ \\ \hline 
 iPTF-15af & $41.46^{+0.13}_{-0.18}$ & $6.10^{+ 0.55}_{- 0.43}$  & $42.32^{+0.02}_{-0.01}$ & $5.85$  & $49.17^{+0.04}_{-0.03}$ & $5.97$  & $106$ & $7.09$  & $10.35$ & $7.23$ \\ \hline 
 iPTF-16axa & $41.88^{+0.16}_{-0.21}$ & $6.68^{+ 0.53}_{- 0.41}$  & $42.91^{+0.02}_{-0.02}$ & $6.44$  & $49.51^{+0.03}_{-0.04}$ & $6.30$  & $82$ & $6.60$  & $10.24$ & $7.05$ \\ \hline 
 iPTF-16fnl & $41.32^{+0.04}_{-0.04}$ & $5.91^{+ 0.56}_{- 0.44}$  & $42.39^{+0.29}_{-0.07}$ & $5.92$  & $48.86^{+0.29}_{-0.08}$ & $5.66$  & $55$ & $5.84$  & $9.50$ & $5.86$ \\ \hline 
 OGLE16aaa & $42.68^{+0.14}_{-0.15}$ & $7.60^{+ 0.32}_{- 0.35}$  & $43.81^{+0.06}_{-0.06}$ & $7.32$  & $50.70^{+0.07}_{-0.08}$ & $7.46$  & -- & --  & $10.39$ & $7.29$ \\ \hline 
 AT2018zr & $42.13^{+0.04}_{-0.04}$ & $7.01^{+ 0.47}_{- 0.40}$  & $43.28^{+0.02}_{-0.02}$ & $6.79$  & $50.34^{+0.04}_{-0.04}$ & $7.12$  & $50$ & $5.65$  & $10.03$ & $6.71$ \\ \hline 
 AT2018hco & $42.19^{+0.04}_{-0.04}$ & $7.08^{+ 0.44}_{- 0.40}$  & $43.23^{+0.02}_{-0.02}$ & $6.75$  & $50.19^{+0.05}_{-0.05}$ & $6.97$  & -- & --  & $9.95$ & $6.58$ \\ \hline 
 AT2018hyz & $42.05^{+0.05}_{-0.05}$ & $6.91^{+ 0.49}_{- 0.41}$  & $43.52^{+0.03}_{-0.15}$ & $7.03$  & $50.25^{+0.04}_{-0.16}$ & $7.02$  & $67$ & $6.20$  & $9.53$ & $5.91$ \\ \hline 
 AT2018lni & $42.33^{+0.04}_{-0.05}$ & $7.26^{+ 0.37}_{- 0.39}$  & $43.43^{+0.02}_{-0.02}$ & $6.95$  & $50.15^{+0.04}_{-0.04}$ & $6.92$  & $56$ & $5.88$  & $9.84$ & $6.41$ \\ \hline 
 AT2018lna & $41.83^{+0.04}_{-0.04}$ & $6.62^{+ 0.53}_{- 0.40}$  & $43.16^{+0.01}_{-0.01}$ & $6.67$  & $49.77^{+0.04}_{-0.04}$ & $6.56$  & $36$ & $5.05$  & $9.41$ & $5.70$ \\ \hline 
 AT2019cho & $41.81^{+0.10}_{-0.09}$ & $6.58^{+ 0.54}_{- 0.41}$  & $43.47^{+0.01}_{-0.01}$ & $6.98$  & $50.28^{+0.02}_{-0.02}$ & $7.05$  & -- & --  & $10.04$ & $6.72$ \\ \hline 
 AT2019bhf & $41.81^{+0.05}_{-0.04}$ & $6.59^{+ 0.54}_{- 0.41}$  & $43.39^{+0.04}_{-0.04}$ & $6.90$  & $50.09^{+0.06}_{-0.06}$ & $6.87$  & -- & --  & $10.38$ & $7.28$ \\ \hline 
 AT2019azh & $41.70^{+0.01}_{-0.01}$ & $6.43^{+ 0.55}_{- 0.41}$  & $43.37^{+0.02}_{-0.01}$ & $6.88$  & $50.16^{+0.03}_{-0.02}$ & $6.94$  & $68$ & $6.24$  & $9.79$ & $6.32$ \\ \hline 
 AT2019dsg & $42.03^{+0.02}_{-0.02}$ & $6.88^{+ 0.49}_{- 0.41}$  & $42.99^{+0.02}_{-0.02}$ & $6.51$  & $49.79^{+0.03}_{-0.03}$ & $6.57$  & $87$ & $6.71$  & $10.54$ & $7.54$ \\ \hline 
 AT2019ehz & $41.36^{+0.05}_{-0.04}$ & $5.94^{+ 0.56}_{- 0.43}$  & $43.12^{+0.01}_{-0.01}$ & $6.64$  & $49.70^{+0.02}_{-0.02}$ & $6.49$  & $47$ & $5.52$  & $9.69$ & $6.17$ \\ \hline 
 AT2019meg & $41.40^{+0.15}_{-0.24}$ & $6.02^{+ 0.56}_{- 0.43}$  & $43.38^{+0.01}_{-0.01}$ & $6.90$  & $49.98^{+0.02}_{-0.02}$ & $6.76$  & -- & --  & $9.59$ & $6.00$ \\ \hline 
 AT2019qiz & $40.98^{+0.02}_{-0.01}$ & $5.42^{+ 0.58}_{- 0.45}$  & $42.84^{+0.02}_{-0.02}$ & $6.36$  & $49.34^{+0.03}_{-0.03}$ & $6.13$  & $72$ & $6.35$  & $10.02$ & $6.70$ \\ \hline 
 AT2020mot & $42.07^{+0.05}_{-0.05}$ & $6.93^{+ 0.48}_{- 0.41}$  & $43.18^{+0.01}_{-0.01}$ & $6.69$  & $50.02^{+0.02}_{-0.02}$ & $6.80$  & $77$ & $6.47$  & $10.28$ & $7.10$ \\ \hline 
 AT2020opy & $42.32^{+0.06}_{-0.05}$ & $7.24^{+ 0.38}_{- 0.39}$  & $43.54^{+0.01}_{-0.01}$ & $7.05$  & $50.41^{+0.02}_{-0.02}$ & $7.18$  & -- & --  & $10.00$ & $6.66$ \\ \hline 
 AT2020zso & $41.14^{+0.13}_{-0.17}$ & $5.64^{+ 0.57}_{- 0.45}$  & $43.15^{+0.04}_{-0.04}$ & $6.67$  & $49.71^{+0.06}_{-0.06}$ & $6.50$  & $62$ & $6.06$  & $10.16$ & $6.91$ \\ \hline 
 AT2020qhs & $43.36^{+0.02}_{-0.02}$ & $8.08^{+ 0.34}_{- 0.35}$  & $44.10^{+0.01}_{-0.01}$ & $7.60$  & $51.00^{+0.02}_{-0.02}$ & $7.76$  & $215$ & $8.44$  & $11.33$ & $8.81$ \\ \hline 
 AT2020ysg & $43.14^{+0.05}_{-0.05}$ & $7.94^{+ 0.35}_{- 0.36}$  & $44.07^{+0.01}_{-0.01}$ & $7.58$  & $51.09^{+0.01}_{-0.01}$ & $7.85$  & -- & --  & $10.90$ & $8.12$ \\ \hline 
 AT2020wey & $40.73^{+0.07}_{-0.08}$ & $5.07^{+ 0.62}_{- 0.46}$  & $42.28^{+0.02}_{-0.02}$ & $5.82$  & $48.61^{+0.03}_{-0.03}$ & $5.42$  & $39$ & $5.20$  & $9.85$ & $6.41$ \\ \hline 
 AT2020riz & $42.94^{+0.05}_{-0.06}$ & $7.80^{+ 0.33}_{- 0.35}$  & $44.27^{+0.02}_{-0.02}$ & $7.77$  & $50.65^{+0.04}_{-0.04}$ & $7.41$  & -- & --  & $10.84$ & $8.02$ \\ \hline 
 AT2020vwl & $41.45^{+0.03}_{-0.03}$ & $6.09^{+ 0.55}_{- 0.43}$  & $42.88^{+0.01}_{-0.01}$ & $6.40$  & $49.65^{+0.02}_{-0.02}$ & $6.44$  & $49$ & $5.60$  & $9.91$ & $6.51$ \\ \hline 
 AT2020acka & $43.42^{+0.04}_{-0.04}$ & $8.13^{+ 0.32}_{- 0.40}$  & $44.27^{+0.02}_{-0.02}$ & $7.77$  & $51.18^{+0.03}_{-0.03}$ & $7.93$  & $174$ & $8.03$  & $11.00$ & $8.28$ \\ \hline 
 AT2021crk & $41.79^{+0.17}_{-0.56}$ & $6.56^{+ 0.54}_{- 0.41}$  & $43.37^{+0.02}_{-0.01}$ & $6.88$  & $49.89^{+0.04}_{-0.03}$ & $6.67$  & $58$ & $5.93$  & $10.18$ & $6.95$ \\ \hline 
 AT2021axu & $42.62^{+0.04}_{-0.05}$ & $7.54^{+ 0.32}_{- 0.36}$  & $43.66^{+0.01}_{-0.01}$ & $7.17$  & $50.40^{+0.01}_{-0.01}$ & $7.17$  & $74$ & $6.39$  & $10.03$ & $6.71$ \\ \hline 
 AT2021ehb & $41.94^{+0.01}_{-0.01}$ & $6.76^{+ 0.52}_{- 0.41}$  & $42.35^{+0.01}_{-0.01}$ & $5.88$  & $49.34^{+0.02}_{-0.02}$ & $6.13$  & $100$ & $6.97$  & $10.11$ & $6.85$ \\ \hline 
 AT2021nwa & $41.99^{+0.02}_{-0.02}$ & $6.83^{+ 0.51}_{- 0.40}$  & $42.64^{+0.01}_{-0.01}$ & $6.17$  & $49.59^{+0.02}_{-0.02}$ & $6.38$  & $102$ & $7.02$  & $10.13$ & $6.86$ \\ \hline 
 AT2021mhg & $41.76^{+0.05}_{-0.06}$ & $6.52^{+ 0.55}_{- 0.41}$  & $42.89^{+0.01}_{-0.01}$ & $6.41$  & $49.66^{+0.02}_{-0.02}$ & $6.44$  & $58$ & $5.93$  & $9.71$ & $6.19$ \\ \hline 
 AT2021sdu & $42.29^{+0.03}_{-0.03}$ & $7.21^{+ 0.39}_{- 0.39}$  & $42.81^{+0.02}_{-0.02}$ & $6.33$  & $49.41^{+0.04}_{-0.03}$ & $6.21$  & -- & --  & $10.35$ & $7.23$ \\ \hline 
 AT2021uqv & $42.67^{+0.02}_{-0.03}$ & $7.59^{+ 0.32}_{- 0.35}$  & $43.09^{+0.01}_{-0.01}$ & $6.61$  & $49.91^{+0.02}_{-0.02}$ & $6.69$  & $62$ & $6.07$  & $9.90$ & $6.51$ \\ \hline 
 AT2021yte & $41.33^{+0.14}_{-0.20}$ & $5.92^{+ 0.56}_{- 0.44}$  & $42.74^{+0.03}_{-0.03}$ & $6.26$  & $49.41^{+0.07}_{-0.07}$ & $6.21$  & $34$ & $4.93$  & $9.61$ & $6.03$ \\ \hline 
 AT2020yue & $42.69^{+0.05}_{-0.04}$ & $7.61^{+ 0.32}_{- 0.36}$  & $43.92^{+0.01}_{-0.01}$ & $7.43$  & $50.84^{+0.02}_{-0.02}$ & $7.60$  & -- & --  & $10.44$ & $7.37$ \\ \hline 
 AT2018jbv & $43.35^{+0.02}_{-0.03}$ & $8.08^{+ 0.34}_{- 0.35}$  & $44.13^{+0.01}_{-0.02}$ & $7.63$  & $51.12^{+0.03}_{-0.03}$ & $7.88$  & -- & --  & $10.38$ & $7.28$ \\ \hline 
 AT2019cmw & $42.97^{+0.05}_{-0.05}$ & $7.80^{+ 0.35}_{- 0.35}$  & $44.61^{+0.01}_{-0.01}$ & $8.11$  & $51.33^{+0.01}_{-0.02}$ & $8.09$  & -- & --  & $11.09$ & $8.41$ \\ \hline 
 AT2020neh & $41.01^{+0.07}_{-0.07}$ & $5.47^{+ 0.57}_{- 0.46}$  & $43.06^{+0.02}_{-0.02}$ & $6.58$  & $49.69^{+0.05}_{-0.05}$ & $6.48$  & $40$ & $5.23$  & $9.43$ & $5.74$ \\ \hline 
 PTF-09axc & -- & --  & $43.24^{+0.04}_{-0.05}$ & $6.75$  & $50.31^{+0.15}_{-0.13}$ & $7.08$  & $60$ & $6.00$  & $10.07$ & $6.78$ \\ \hline 
 AT2018iih & -- & --  & $43.77^{+0.01}_{-0.01}$ & $7.27$  & $51.09^{+0.02}_{-0.02}$ & $7.85$  & $149$ & $7.73$  & $10.84$ & $8.02$ \\ \hline 
 AT2019mha & -- & --  & $43.29^{+0.01}_{-0.01}$ & $6.80$  & $49.44^{+0.03}_{-0.03}$ & $6.23$  & -- & --  & $10.03$ & $6.71$ \\ \hline 
 AT2019lwu & -- & --  & $43.35^{+0.01}_{-0.01}$ & $6.86$  & $49.71^{+0.03}_{-0.03}$ & $6.49$  & -- & --  & $9.83$ & $6.38$ \\ \hline 
 AT2019teq & -- & --  & $42.91^{+0.02}_{-0.02}$ & $6.43$  & $49.96^{+0.05}_{-0.04}$ & $6.74$  & -- & --  & $9.91$ & $6.52$ \\ \hline 
 AT2020pj & -- & --  & $42.83^{+0.02}_{-0.02}$ & $6.35$  & $49.42^{+0.04}_{-0.03}$ & $6.21$  & -- & --  & $10.07$ & $6.78$ \\ \hline 
 AT2019vcb & -- & --  & $43.25^{+0.16}_{-0.06}$ & $6.76$  & $49.63^{+0.18}_{-0.08}$ & $6.41$  & -- & --  & $9.51$ & $5.87$ \\ \hline 
 AT2020ddv & -- & --  & $43.33^{+0.01}_{-0.01}$ & $6.84$  & $50.03^{+0.03}_{-0.03}$ & $6.81$  & $58$ & $5.93$  & $10.22$ & $7.01$ \\ \hline 
 AT2020ocn & -- & --  & $42.44^{+0.01}_{-0.01}$ & $5.97$  & $49.22^{+0.02}_{-0.02}$ & $6.02$  & $81$ & $6.57$  & $10.32$ & $7.17$ \\ \hline 
 AT2020mbq & -- & --  & $43.15^{+0.02}_{-0.02}$ & $6.67$  & $49.73^{+0.04}_{-0.04}$ & $6.52$  & -- & --  & $9.64$ & $6.08$ \\ \hline 
 AT2021jjm & -- & --  & $43.59^{+0.02}_{-0.02}$ & $7.10$  & $50.13^{+0.06}_{-0.06}$ & $6.90$  & -- & --  & $9.50$ & $5.86$ \\ \hline 
 AT2021yzv & -- & --  & $44.02^{+0.01}_{-0.01}$ & $7.52$  & $51.00^{+0.02}_{-0.02}$ & $7.75$  & -- & --  & $10.86$ & $8.05$ \\ \hline 
 AT2020abri & -- & --  & $43.58^{+0.02}_{-0.02}$ & $7.09$  & $50.25^{+0.03}_{-0.03}$ & $7.02$  & -- & --  & $9.67$ & $6.13$ \\ \hline 
 AT2021utq & -- & --  & $43.50^{+0.01}_{-0.01}$ & $7.01$  & $50.11^{+0.04}_{-0.04}$ & $6.89$  & -- & --  & $9.80$ & $6.33$ \\ \hline 

    \caption{The black hole masses of the 63 TDEs in our sample, computed from various different scaling relationships. 
    The quoted error ranges correspond to $1\sigma$ uncertainties. All quantities other than the velocity dispersion are presented as logarithms $\log_{10}$. 
    The column directly to the right of each observed quantity corresponds to the black hole mass computed from that quantity. The plateau luminosity here is measured at $\nu = 10^{15}$ Hz. 
    We do not quote uncertainties on the black hole masses computed from the peak g-band luminosity, g-band radiated energy, 
    velocity dispersion and galactic mass scaling relationships. 
    The intrinsic scatter in each relationship is $\sim 0.4$ dex, $\sim 0.3$ dex, $\sim 0.3$ dex and $\sim 0.8$ dex respectively.   }
    \label{mass_plat_table}
    \end{longtable}

    \twocolumn


\renewcommand{\arraystretch}{1.}
\onecolumn 
\footnotesize

\begin{longtable}{ |p{2.0cm}|p{1.6cm}|p{1.6cm}|p{1.6cm}| }
\hline
TDE Name & $L_{\rm plat}$ & $\M$ & $a_\bullet$  \\
\hline
& $ {\rm erg\, s^{-1}}$ & $ M_\odot$ &  \\
\hline
 AT2020acka & $43.42^{+0.04}_{-0.04}$ & $8.13^{+ 0.32}_{- 0.39}$ & $0.68^{+0.26}_{-0.69}$  \\  
 AT2020qhs & $43.36^{+0.02}_{-0.02}$ & $8.09^{+ 0.33}_{- 0.35}$ & $0.68^{+0.26}_{-0.70}$  \\  
 AT2018jbv & $43.35^{+0.02}_{-0.03}$ & $8.08^{+ 0.33}_{- 0.34}$ & $0.70^{+0.25}_{-0.74}$  \\  
 ASASSN-15lh & $43.25^{+0.08}_{-0.11}$ & $8.00^{+ 0.34}_{- 0.35}$ & $0.65^{+0.28}_{-0.72}$  \\  
 AT2020ysg & $43.14^{+0.05}_{-0.05}$ & $7.94^{+ 0.34}_{- 0.36}$ & $0.64^{+0.29}_{-0.75}$  \\  
 AT2019cmw & $42.97^{+0.05}_{-0.05}$ & $7.80^{+ 0.35}_{- 0.34}$ & $0.55^{+0.35}_{-0.76}$  \\  
 AT2020riz & $42.94^{+0.05}_{-0.06}$ & $7.80^{+ 0.33}_{- 0.35}$ & $0.54^{+0.35}_{-0.75}$  \\  
 AT2020yue & $42.69^{+0.05}_{-0.04}$ & $7.60^{+ 0.32}_{- 0.35}$ & $0.38^{+0.45}_{-0.74}$  \\  
 OGLE16aaa & $42.68^{+0.14}_{-0.15}$ & $7.54^{+ 0.33}_{- 0.37}$ & $0.31^{+0.49}_{-0.74}$  \\  
 AT2021uqv & $42.67^{+0.02}_{-0.03}$ & $7.59^{+ 0.32}_{- 0.35}$ & $0.37^{+0.45}_{-0.74}$  \\  
 AT2021axu & $42.62^{+0.04}_{-0.05}$ & $7.55^{+ 0.32}_{- 0.36}$ & $0.31^{+0.49}_{-0.71}$  \\  
 AT2018fyk & $42.61^{+0.02}_{-0.02}$ & $7.54^{+ 0.32}_{- 0.36}$ & $0.31^{+0.49}_{-0.71}$  \\  
 SDSS-TDE2 & $42.55^{+0.27}_{-0.24}$ & $7.15^{+ 0.49}_{- 0.57}$ & $0.10^{+0.61}_{-0.70}$  \\  
 AT2018lni & $42.33^{+0.04}_{-0.05}$ & $7.26^{+ 0.37}_{- 0.39}$ & $0.09^{+0.60}_{-0.68}$  \\  
 AT2020opy & $42.32^{+0.06}_{-0.05}$ & $7.24^{+ 0.37}_{- 0.39}$ & $0.08^{+0.61}_{-0.68}$  \\  
 AT2021sdu & $42.29^{+0.03}_{-0.03}$ & $7.21^{+ 0.38}_{- 0.39}$ & $0.06^{+0.61}_{-0.68}$  \\  
 AT2018hco & $42.19^{+0.04}_{-0.04}$ & $7.08^{+ 0.44}_{- 0.40}$ & $0.00^{+0.63}_{-0.65}$  \\  
 AT2018zr & $42.13^{+0.04}_{-0.04}$ & $7.01^{+ 0.46}_{- 0.40}$ & $0.00^{+0.65}_{-0.66}$  \\  
 PTF-09djl & $42.11^{+0.11}_{-0.10}$ & $6.96^{+ 0.48}_{- 0.42}$ & $0.00^{+0.64}_{-0.66}$  \\  
 AT2020mot & $42.07^{+0.05}_{-0.05}$ & $6.93^{+ 0.48}_{- 0.41}$ & $0.00^{+0.65}_{-0.66}$  \\  
 AT2018hyz & $42.05^{+0.05}_{-0.05}$ & $6.91^{+ 0.48}_{- 0.40}$ & $0.00^{+0.65}_{-0.66}$  \\  
 AT2019ahk & $42.03^{+0.03}_{-0.04}$ & $6.88^{+ 0.49}_{- 0.40}$ & $0.00^{+0.65}_{-0.66}$  \\  
 AT2019dsg & $42.03^{+0.02}_{-0.02}$ & $6.88^{+ 0.48}_{- 0.40}$ & $0.00^{+0.65}_{-0.66}$  \\  
 SDSS-TDE1 & $42.01^{+0.14}_{-0.13}$ & $6.81^{+ 0.51}_{- 0.43}$ & $0.00^{+0.65}_{-0.66}$  \\  
 AT2021nwa & $41.99^{+0.02}_{-0.02}$ & $6.83^{+ 0.50}_{- 0.39}$ & $0.00^{+0.65}_{-0.66}$  \\  
 AT2018dyb & $41.96^{+0.02}_{-0.02}$ & $6.78^{+ 0.51}_{- 0.40}$ & $0.00^{+0.65}_{-0.66}$  \\  
 AT2021ehb & $41.94^{+0.01}_{-0.01}$ & $6.77^{+ 0.52}_{- 0.41}$ & $0.00^{+0.64}_{-0.66}$  \\  
 iPTF-16axa & $41.88^{+0.16}_{-0.21}$ & $6.63^{+ 0.54}_{- 0.45}$ & $0.00^{+0.66}_{-0.65}$  \\  
 PS1-10jh & $41.85^{+0.08}_{-0.08}$ & $6.62^{+ 0.53}_{- 0.41}$ & $0.00^{+0.66}_{-0.65}$  \\  
 AT2018lna & $41.83^{+0.04}_{-0.04}$ & $6.62^{+ 0.53}_{- 0.40}$ & $0.00^{+0.65}_{-0.66}$  \\  
 AT2019bhf & $41.81^{+0.05}_{-0.04}$ & $6.59^{+ 0.53}_{- 0.40}$ & $0.00^{+0.65}_{-0.66}$  \\  
 AT2019cho & $41.81^{+0.10}_{-0.09}$ & $6.57^{+ 0.54}_{- 0.41}$ & $0.00^{+0.66}_{-0.65}$  \\  
 AT2021crk & $41.79^{+0.17}_{-0.56}$ & $6.50^{+ 0.55}_{- 0.46}$ & $0.00^{+0.66}_{-0.65}$  \\  
 AT2021mhg & $41.76^{+0.05}_{-0.06}$ & $6.51^{+ 0.54}_{- 0.41}$ & $0.00^{+0.66}_{-0.65}$  \\  
 AT2019azh & $41.70^{+0.01}_{-0.01}$ & $6.43^{+ 0.53}_{- 0.39}$ & $0.00^{+0.66}_{-0.64}$  \\  
 PTF-09ge & $41.64^{+0.17}_{-0.15}$ & $6.29^{+ 0.56}_{- 0.46}$ & $0.00^{+0.66}_{-0.65}$  \\  
 ASASSN-14li & $41.49^{+0.01}_{-0.01}$ & $6.14^{+ 0.54}_{- 0.42}$ & $0.00^{+0.65}_{-0.67}$  \\  
 ASASSN-14ae & $41.48^{+0.07}_{-0.07}$ & $6.12^{+ 0.55}_{- 0.42}$ & $0.00^{+0.66}_{-0.65}$  \\  
 iPTF-15af & $41.46^{+0.13}_{-0.18}$ & $6.07^{+ 0.55}_{- 0.45}$ & $0.00^{+0.66}_{-0.66}$  \\  
 AT2020vwl & $41.45^{+0.03}_{-0.03}$ & $6.08^{+ 0.54}_{- 0.42}$ & $0.00^{+0.67}_{-0.65}$  \\  
 AT2019meg & $41.40^{+0.15}_{-0.24}$ & $5.97^{+ 0.56}_{- 0.47}$ & $0.00^{+0.66}_{-0.66}$  \\  
 AT2019ehz & $41.36^{+0.05}_{-0.04}$ & $5.95^{+ 0.56}_{- 0.43}$ & $0.00^{+0.65}_{-0.66}$  \\  
 AT2021yte & $41.33^{+0.14}_{-0.20}$ & $5.88^{+ 0.56}_{- 0.46}$ & $0.00^{+0.66}_{-0.66}$  \\  
 iPTF-16fnl & $41.32^{+0.04}_{-0.04}$ & $5.91^{+ 0.55}_{- 0.43}$ & $0.00^{+0.66}_{-0.66}$  \\  
 ASASSN-15oi & $41.29^{+0.04}_{-0.03}$ & $5.86^{+ 0.54}_{- 0.43}$ & $0.00^{+0.67}_{-0.65}$  \\  
 AT2020zso & $41.14^{+0.13}_{-0.17}$ & $5.61^{+ 0.57}_{- 0.47}$ & $0.00^{+0.66}_{-0.66}$  \\  
 AT2020neh & $41.01^{+0.07}_{-0.07}$ & $5.46^{+ 0.57}_{- 0.45}$ & $0.00^{+0.66}_{-0.66}$  \\  
 AT2019qiz & $40.98^{+0.02}_{-0.01}$ & $5.42^{+ 0.57}_{- 0.45}$ & $0.00^{+0.67}_{-0.66}$  \\  
 AT2020wey & $40.73^{+0.07}_{-0.08}$ & $5.06^{+ 0.60}_{- 0.46}$ & $0.00^{+0.65}_{-0.66}$  \\  
\hline
\caption{The black hole mass and spin constraints of the 49 TDEs in our sample with plateau luminosity measurements. The plateau luminosity here is measured at $\nu = 10^{15}$ Hz. 
The quoted error ranges correspond to $1\sigma$ uncertainties, and the black hole mass and plateau luminosities are presented as logarithms $\log_{10}$. For TDEs with plateau luminosities below $\sim 10^{42.5}$ 
we are unable to constrain the black hole spin of the TDE, and simply recover the input (flat) distribution.   }
\label{spin_plat_table}
\end{longtable}

\twocolumn


\begin{figure*}
\includegraphics[width=.32\textwidth, clip=30 10 30 10, clip]{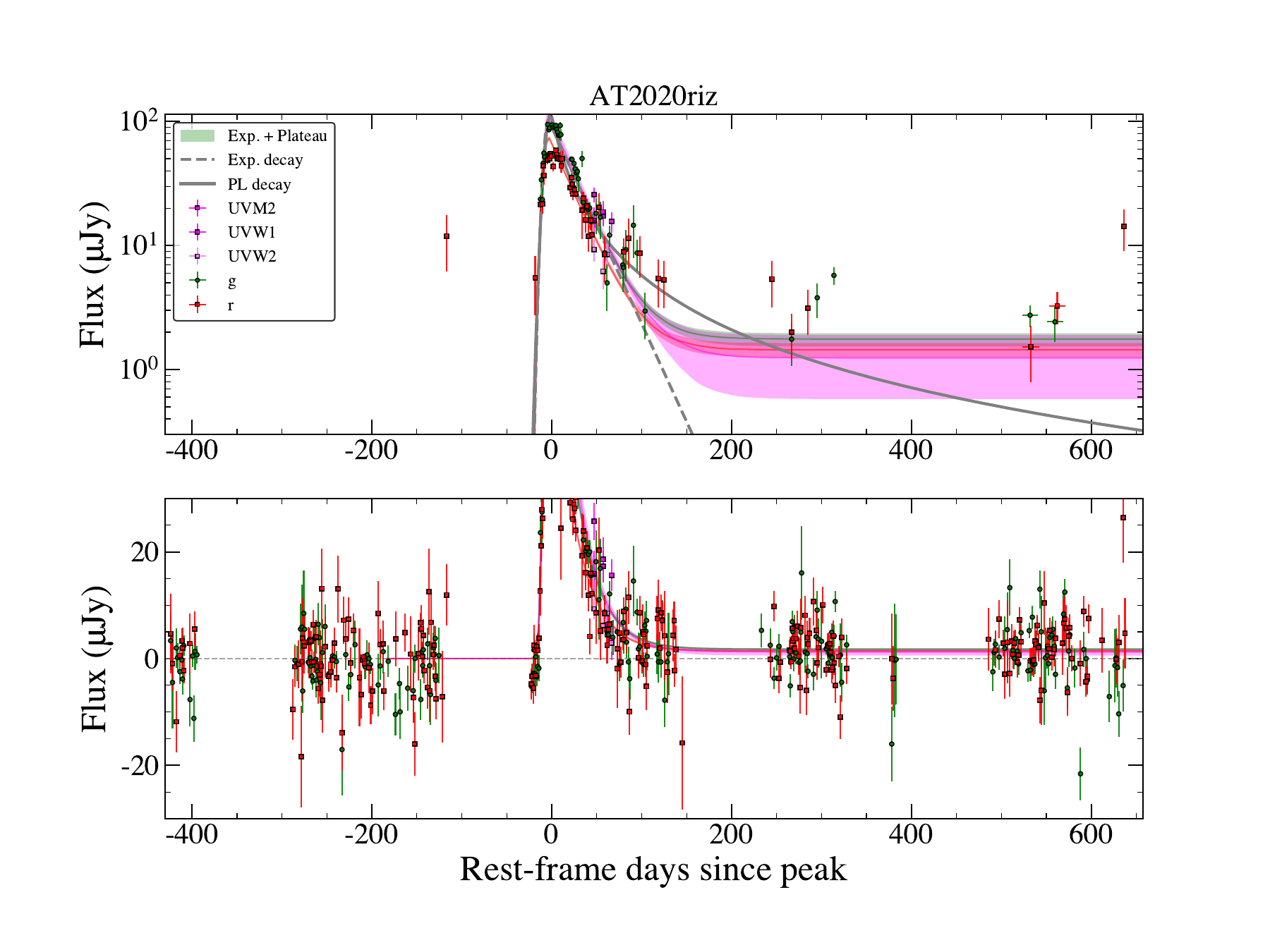}\quad
\includegraphics[width=.32\textwidth, clip=30 10 30 10, clip]{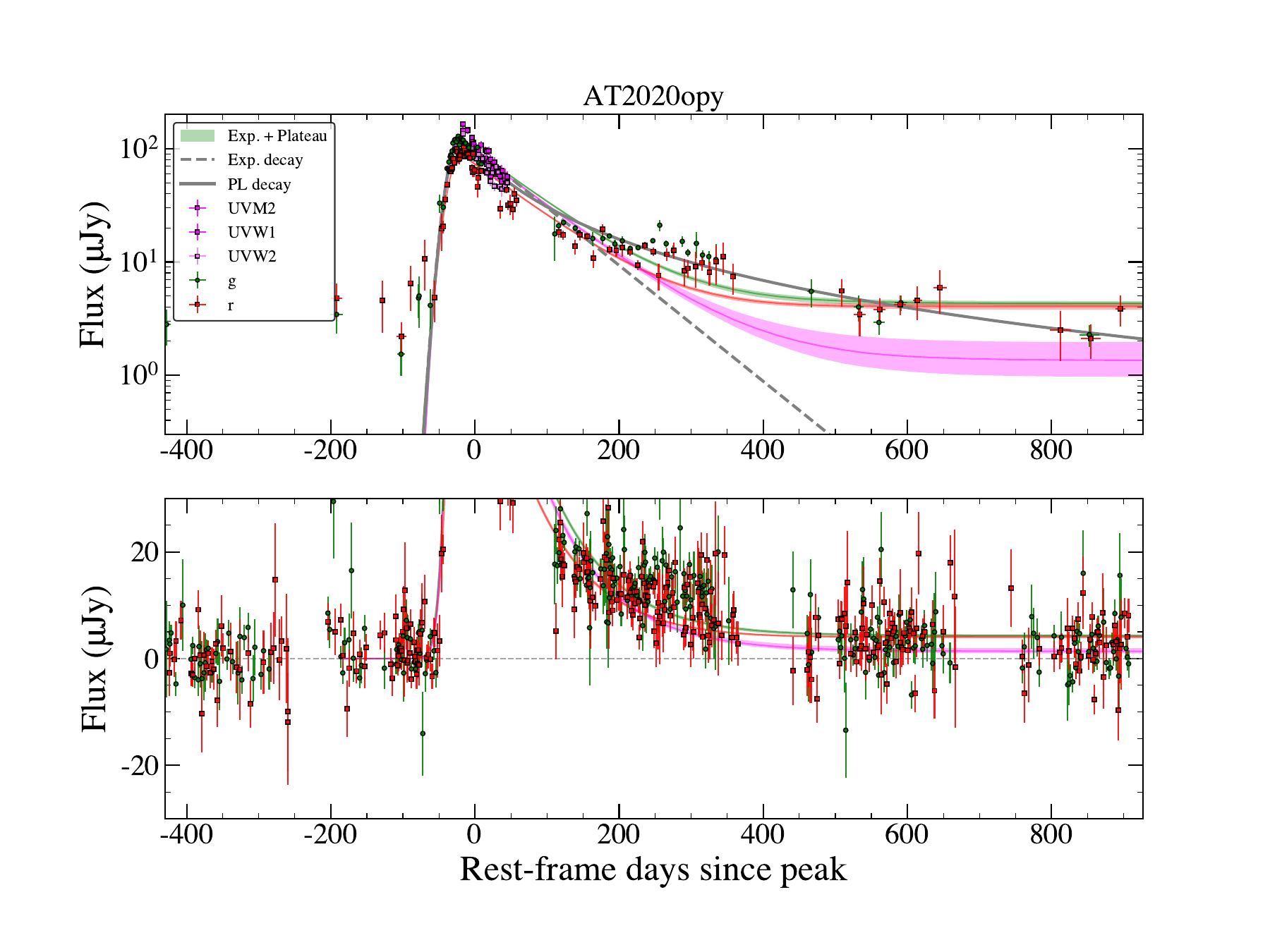}\quad
\includegraphics[width=.32\textwidth, clip=30 10 30 10, clip]{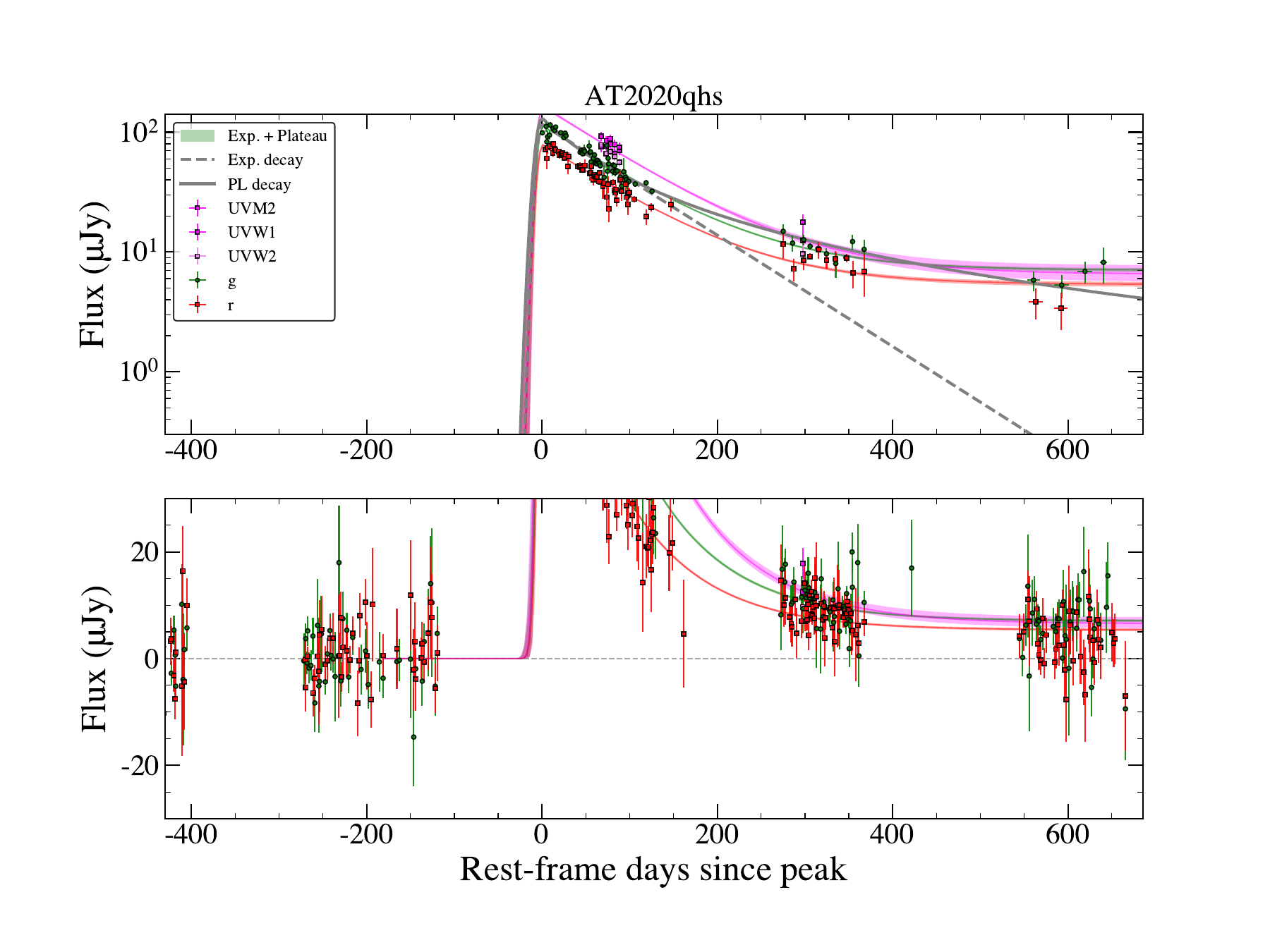}\\
\includegraphics[width=.32\textwidth, clip=30 10 30 10, clip]{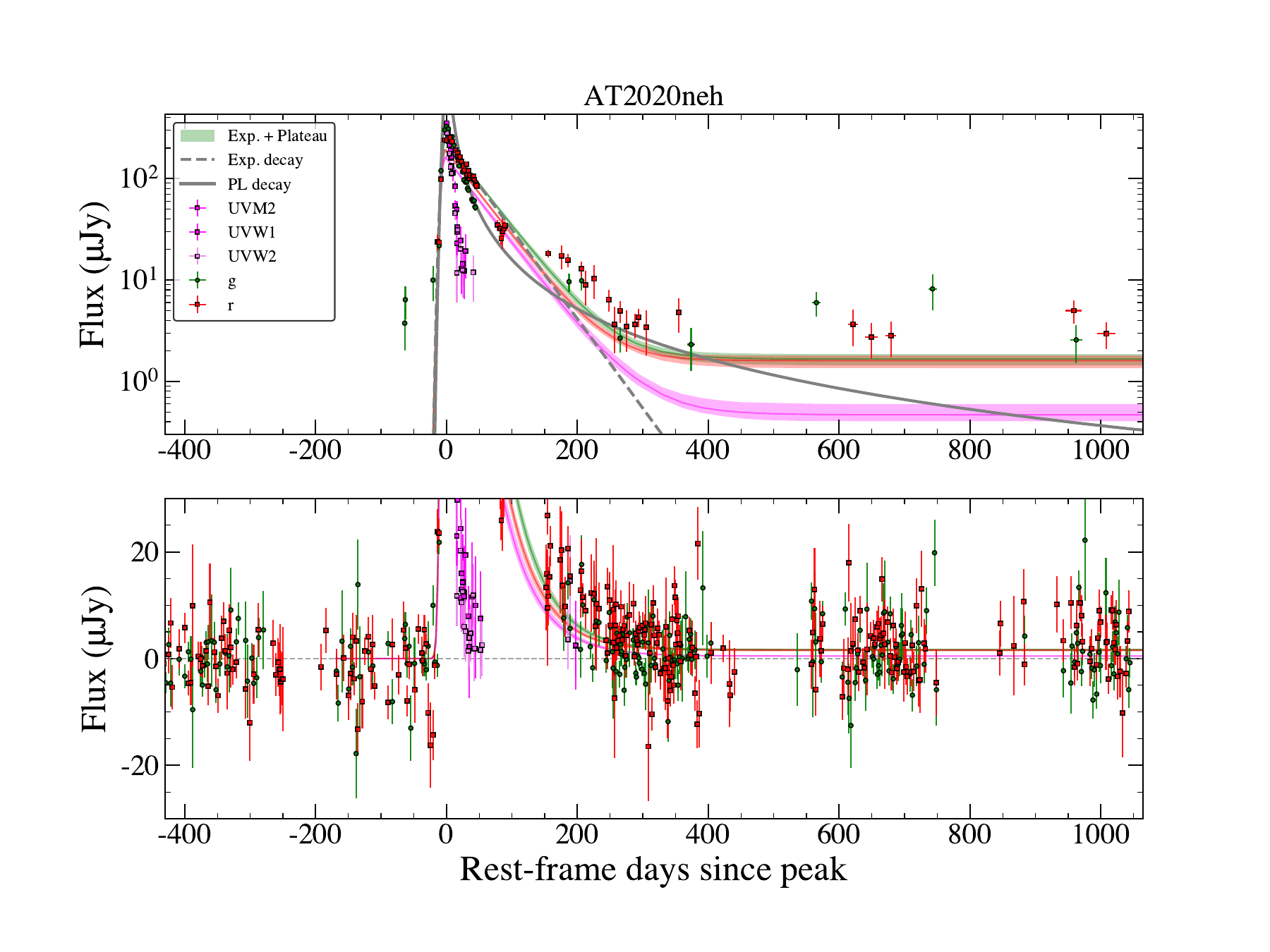}\quad
\includegraphics[width=.32\textwidth, clip=30 10 30 10, clip]{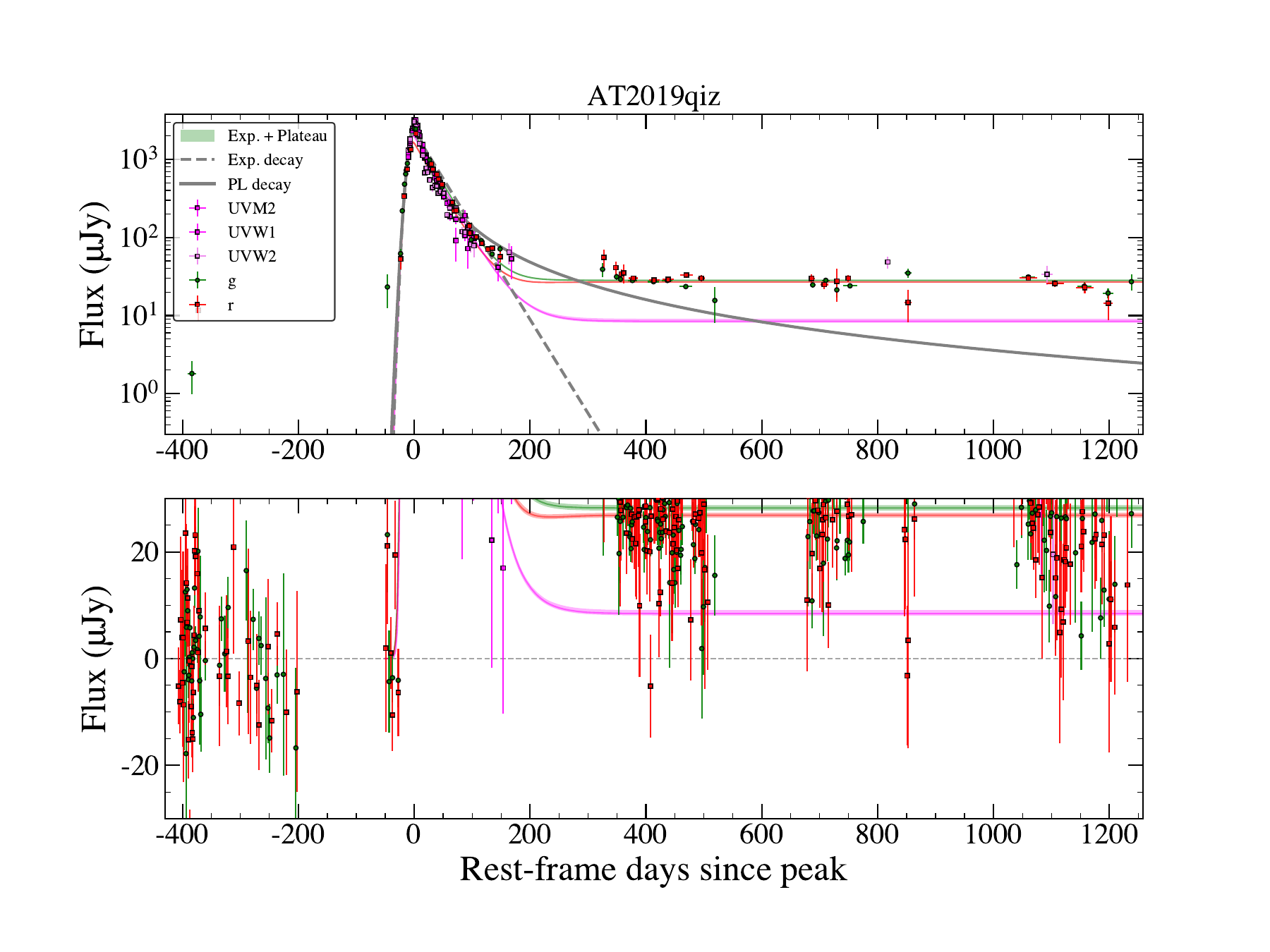}\quad
\includegraphics[width=.32\textwidth, clip=30 10 30 10, clip]{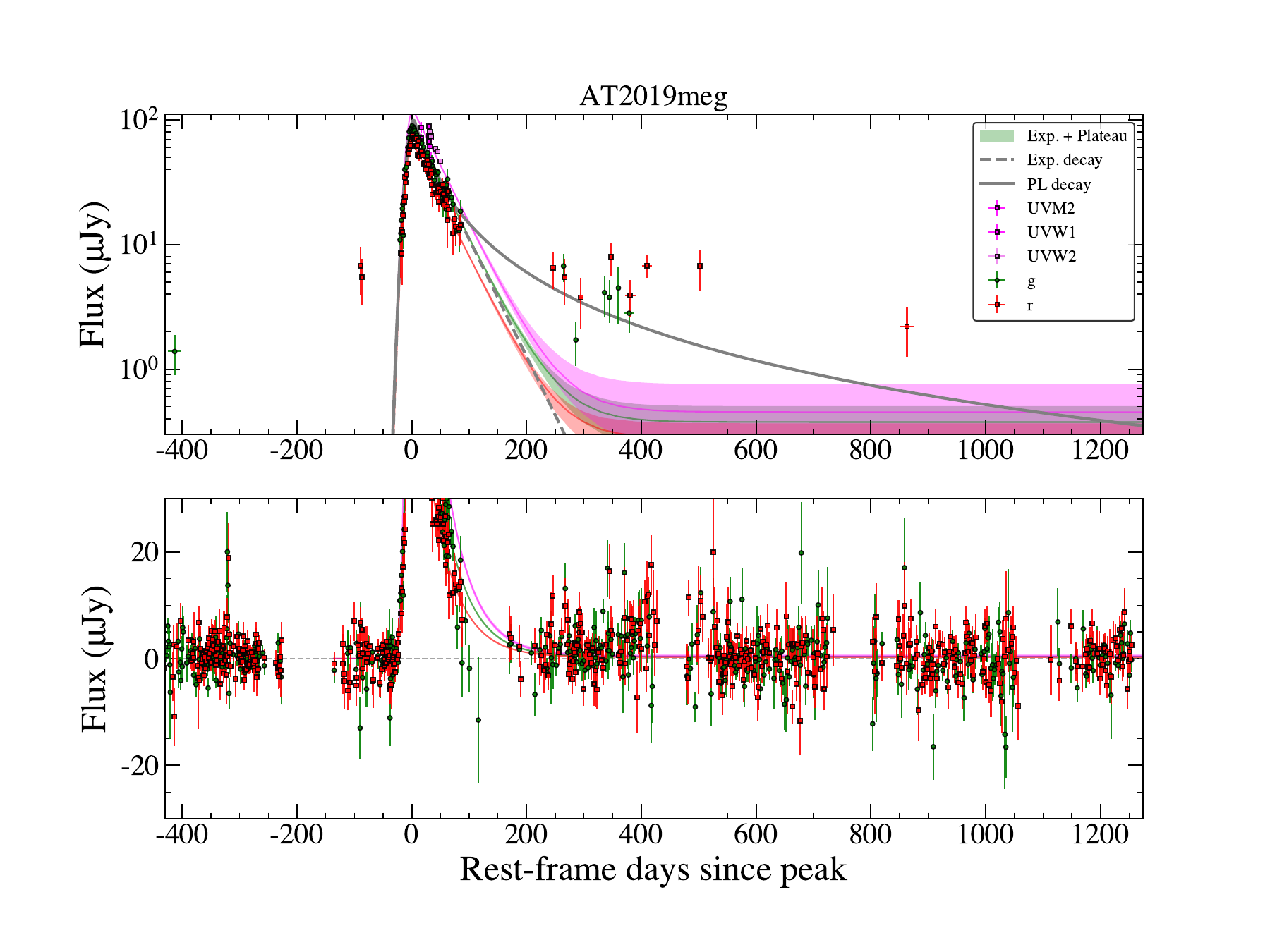}\\
\includegraphics[width=.32\textwidth, clip=30 10 30 10, clip]{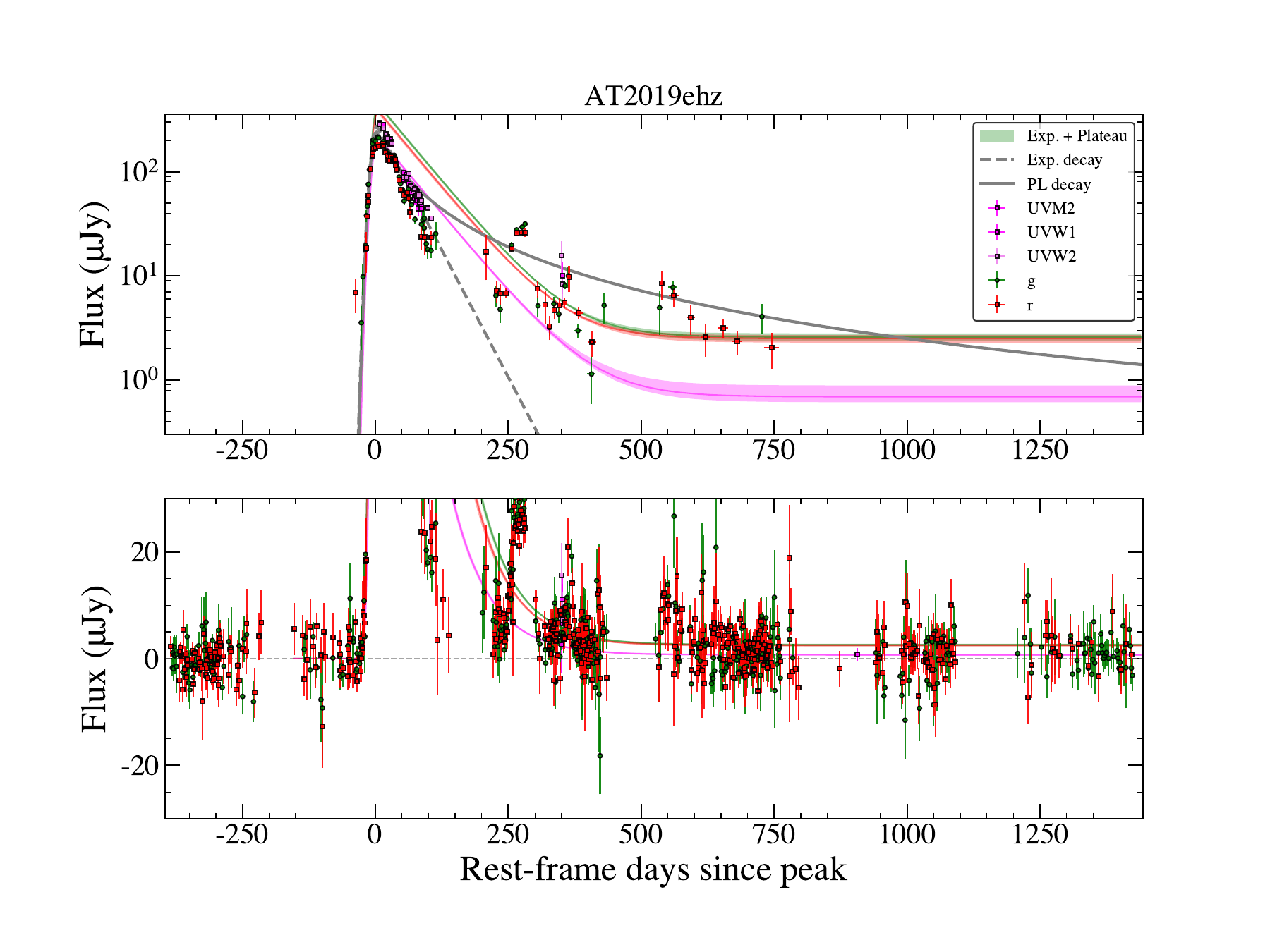}\quad
\includegraphics[width=.32\textwidth, clip=30 10 30 10, clip]{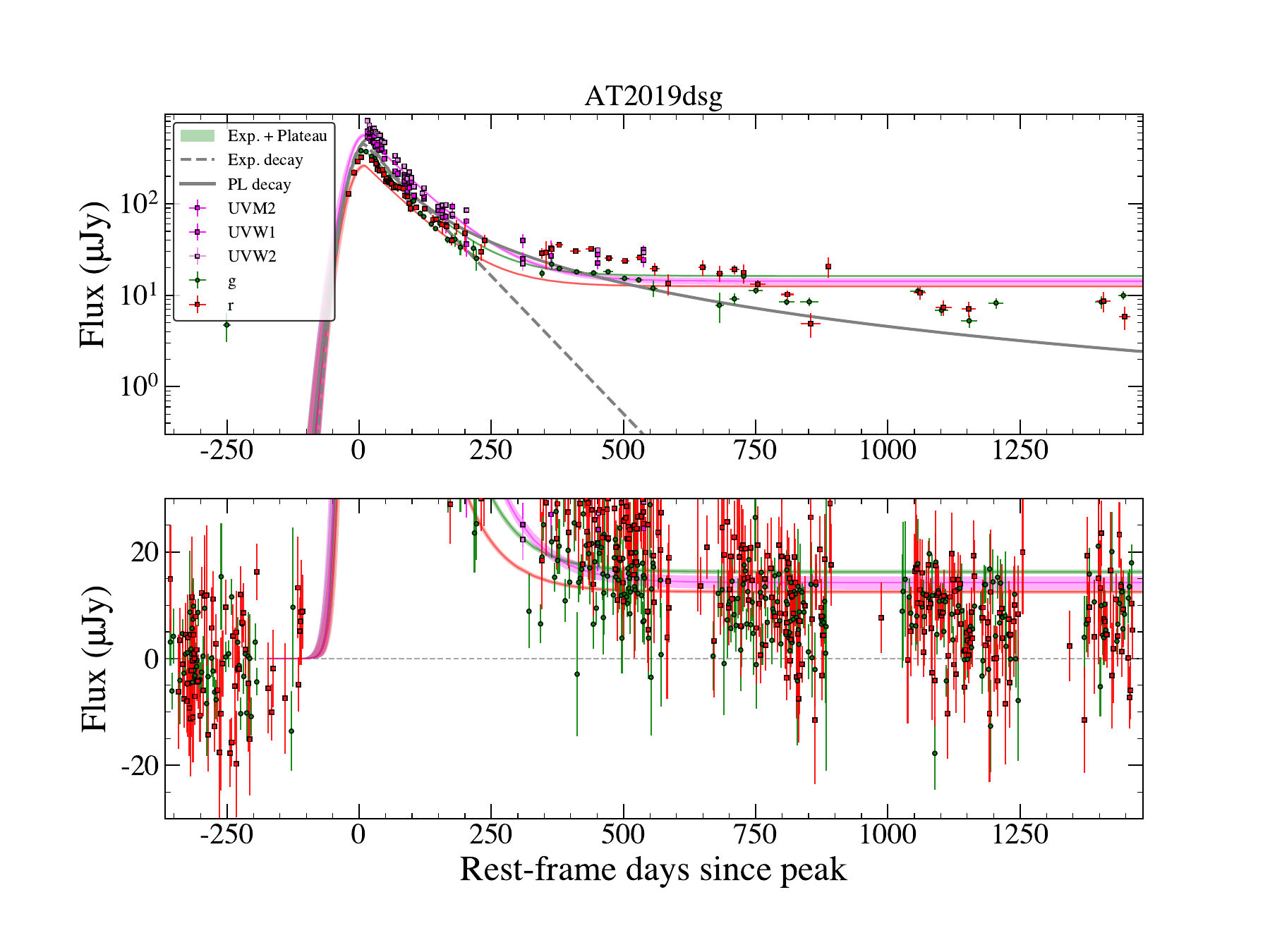}\quad
\includegraphics[width=.32\textwidth, clip=30 10 30 10, clip]{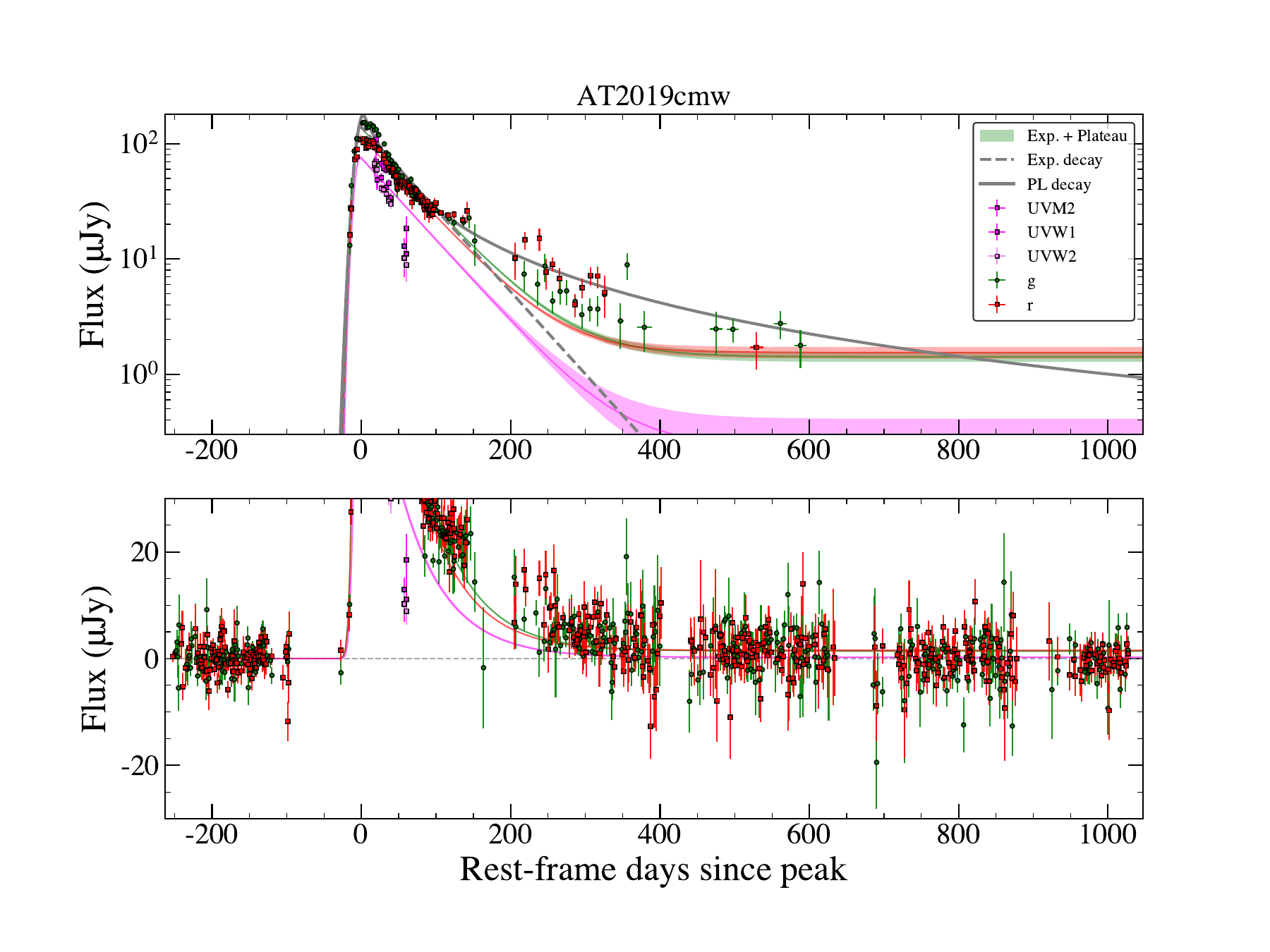}\\
\includegraphics[width=.32\textwidth, clip=30 10 30 10, clip]{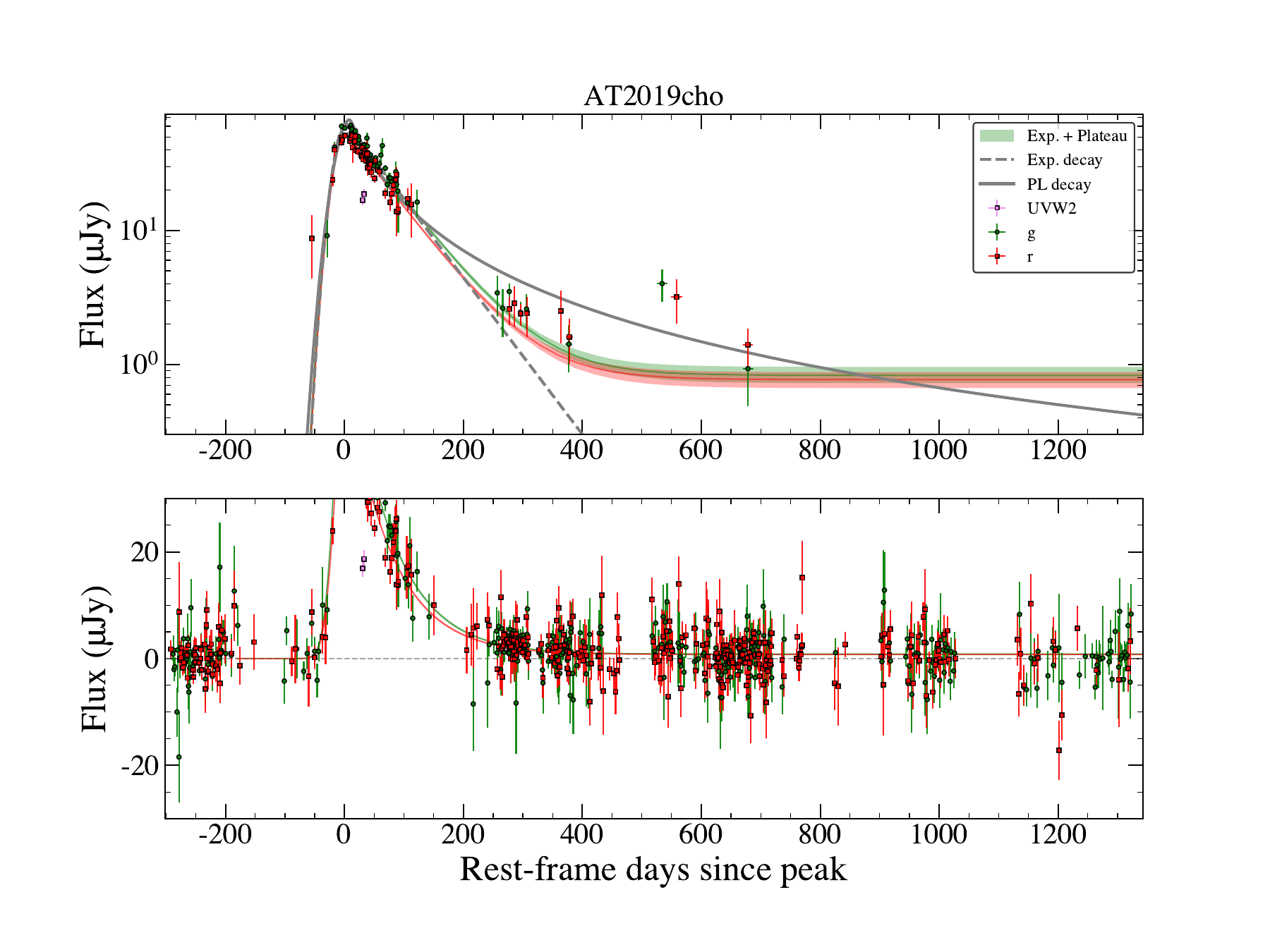}\quad
\includegraphics[width=.32\textwidth, clip=30 10 30 10, clip]{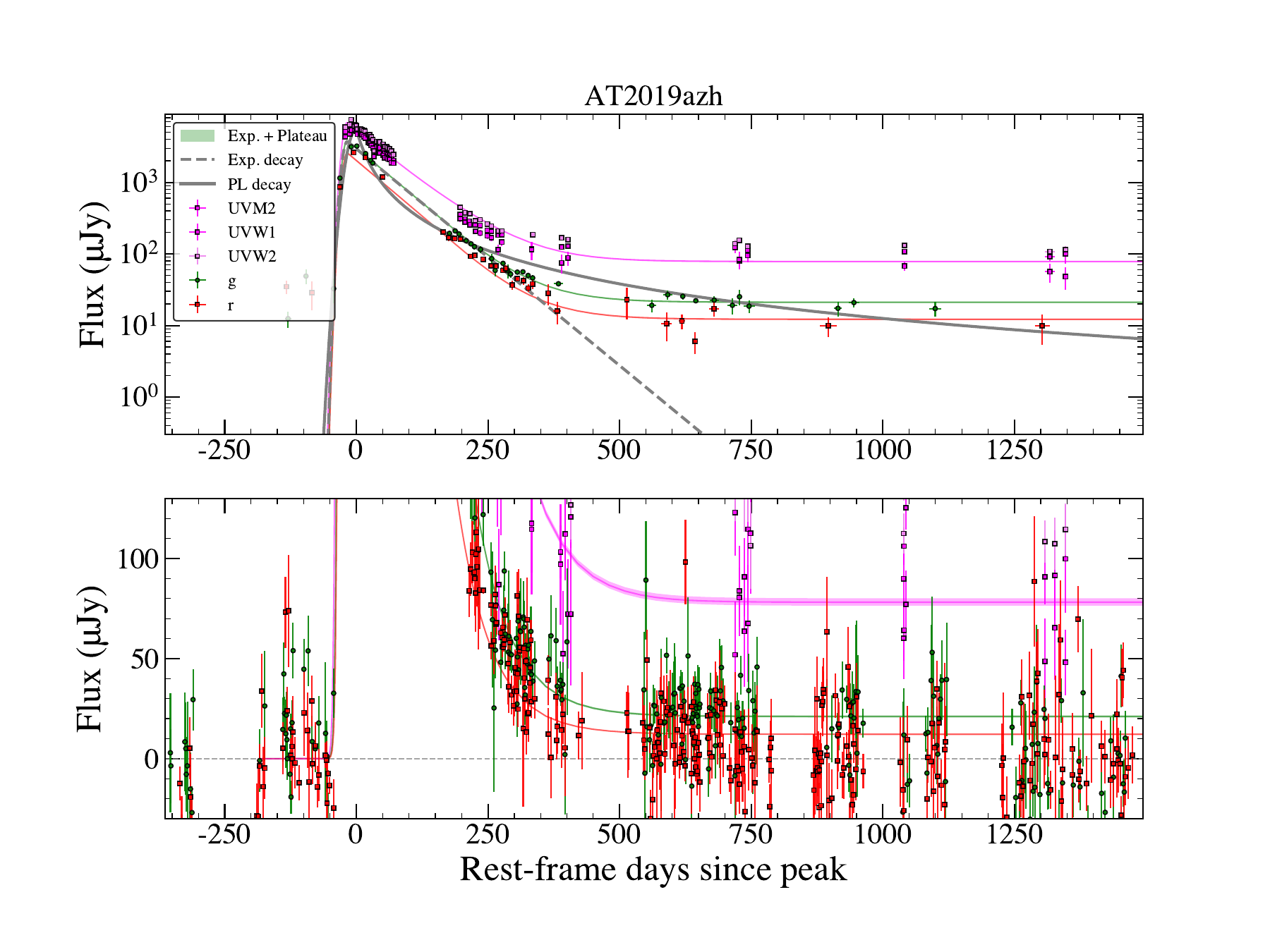}\quad
\includegraphics[width=.32\textwidth, clip=30 10 30 10, clip]{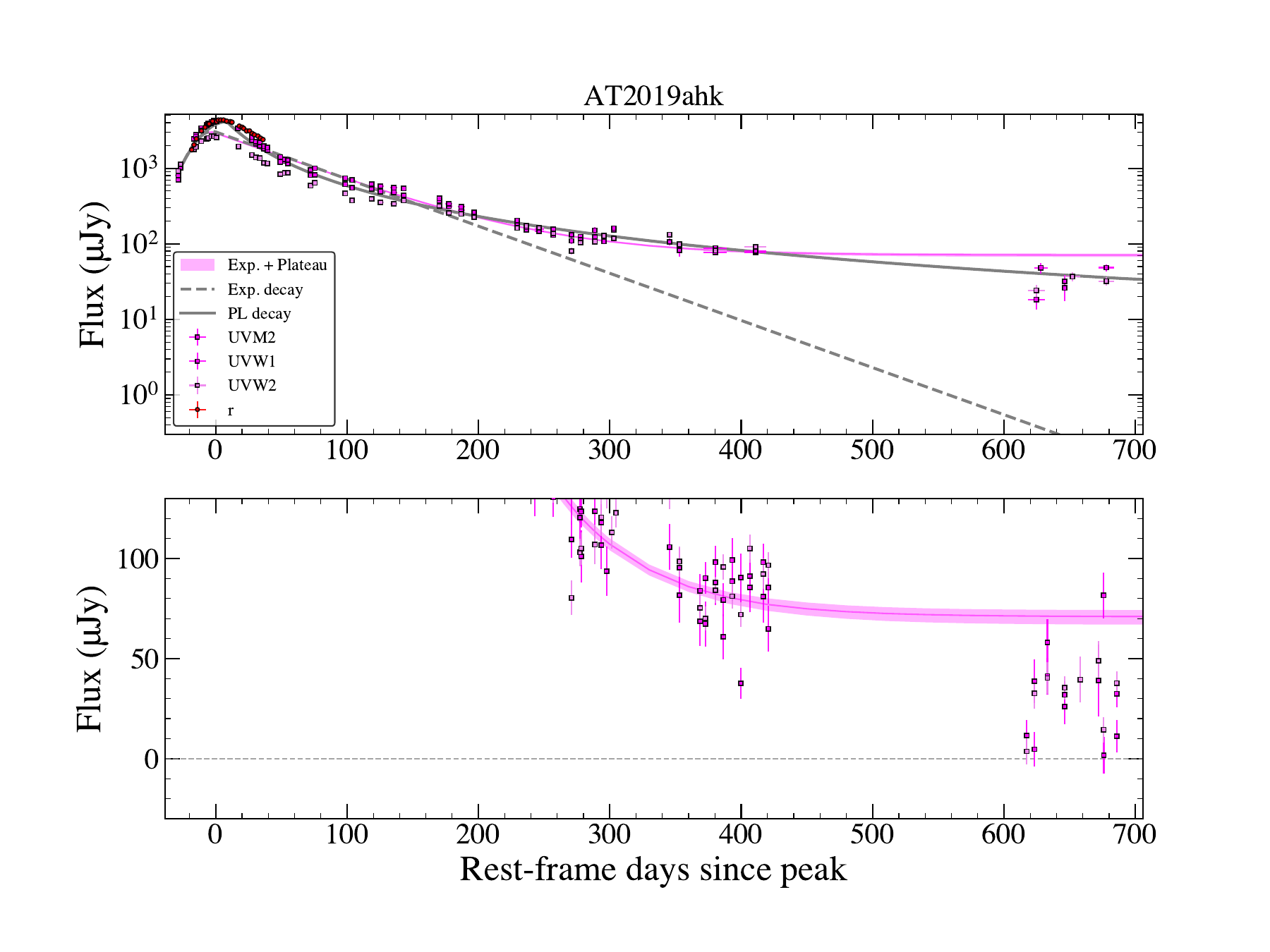}\\
\includegraphics[width=.32\textwidth, clip=30 10 30 10, clip]{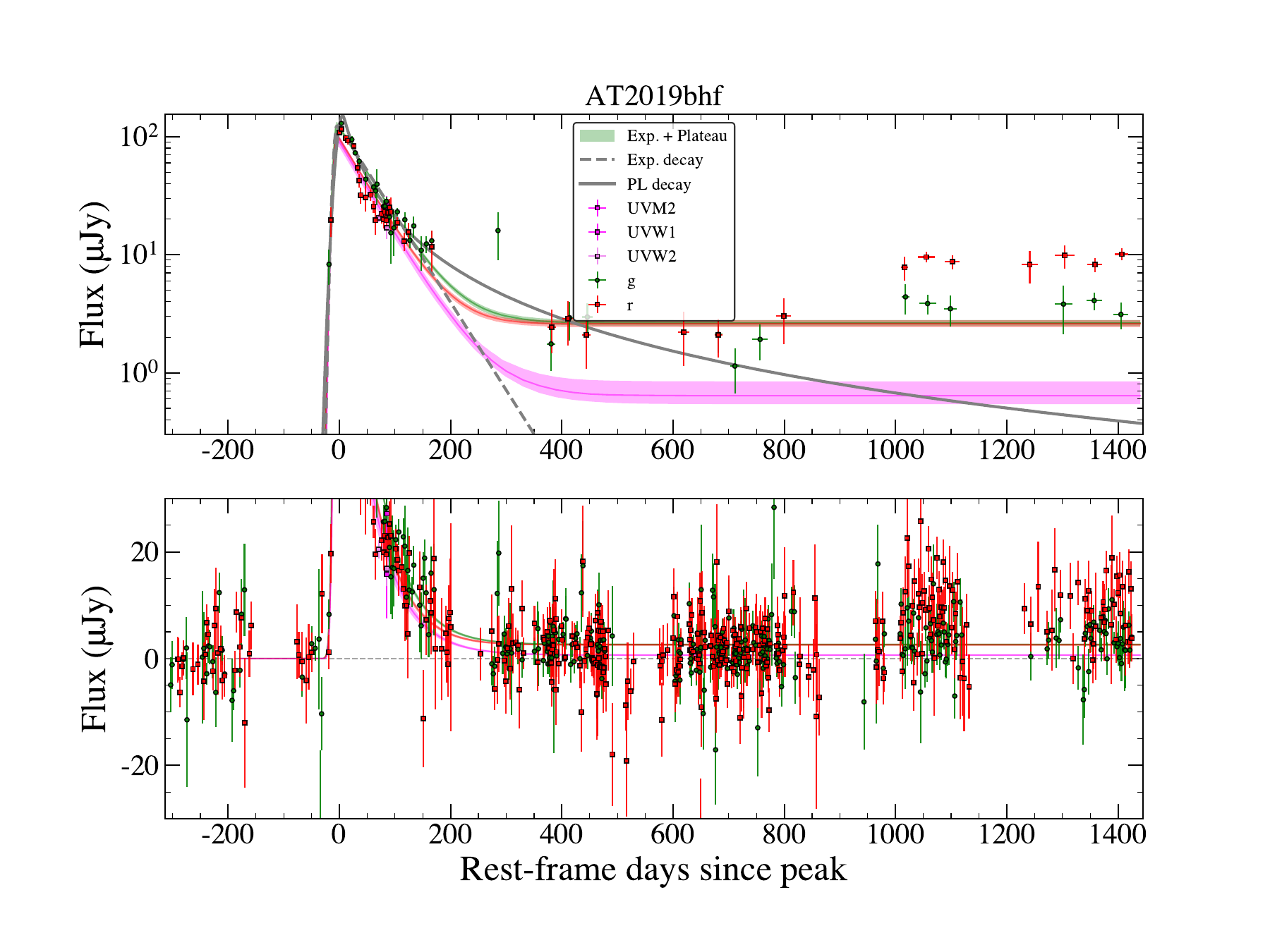}\quad
\includegraphics[width=.32\textwidth, clip=30 10 30 10, clip]{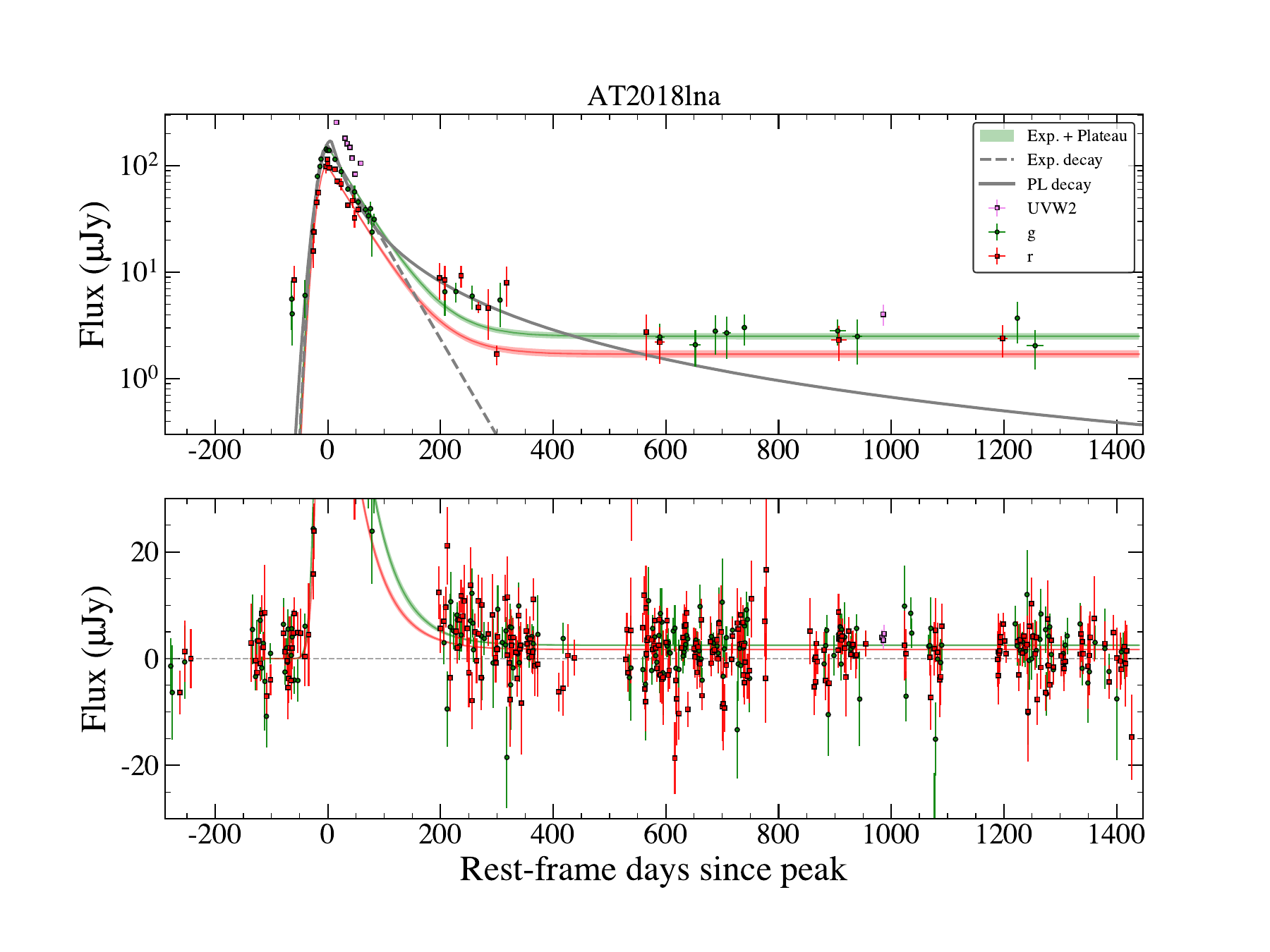}\quad
\includegraphics[width=.32\textwidth, clip=30 10 30 10, clip]{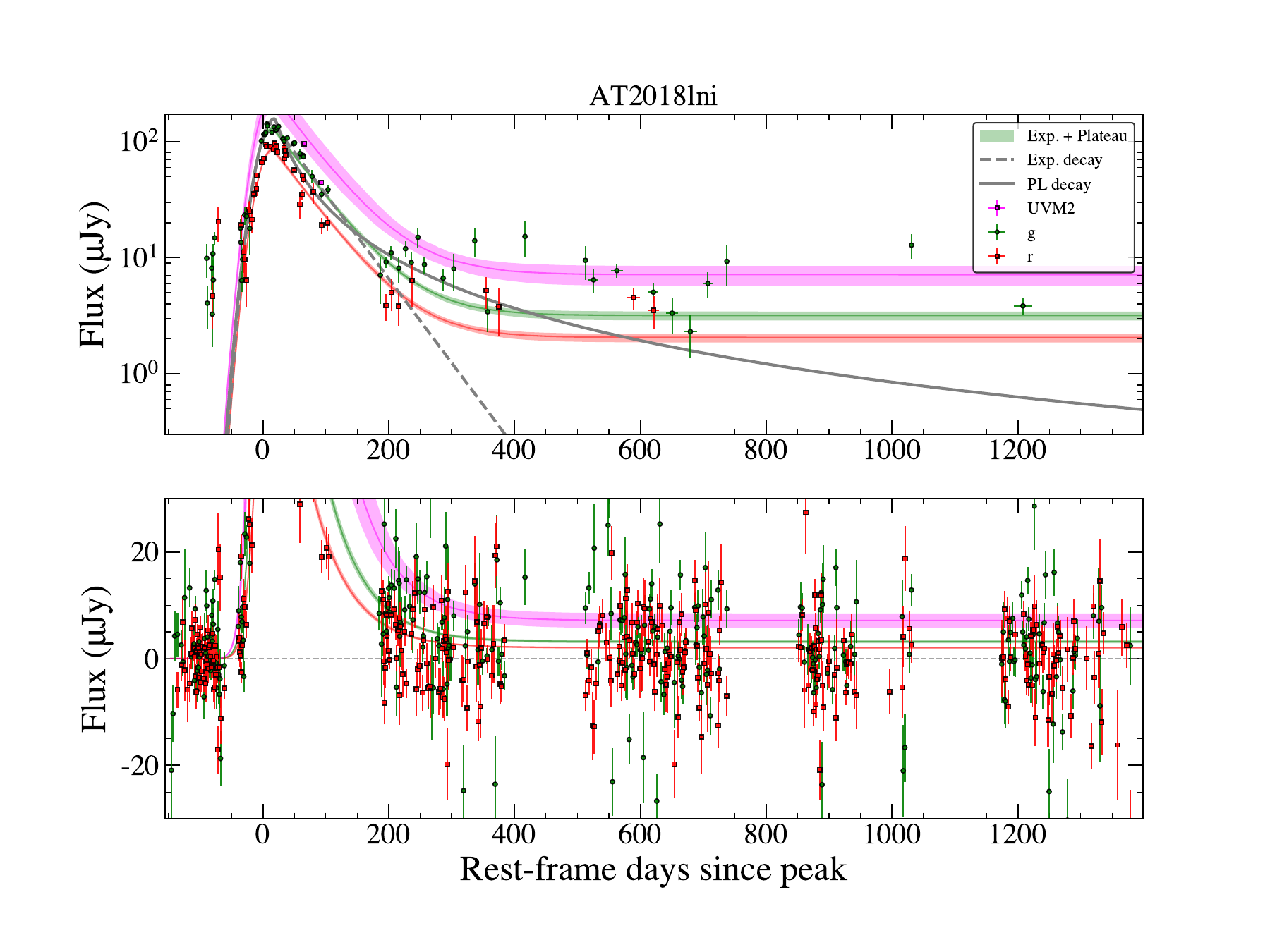}\\
\caption{Same as Fig.~\ref{fig:lcfits}.}
\end{figure*}

\begin{figure*}
\includegraphics[width=.32\textwidth, clip=30 10 30 10, clip]{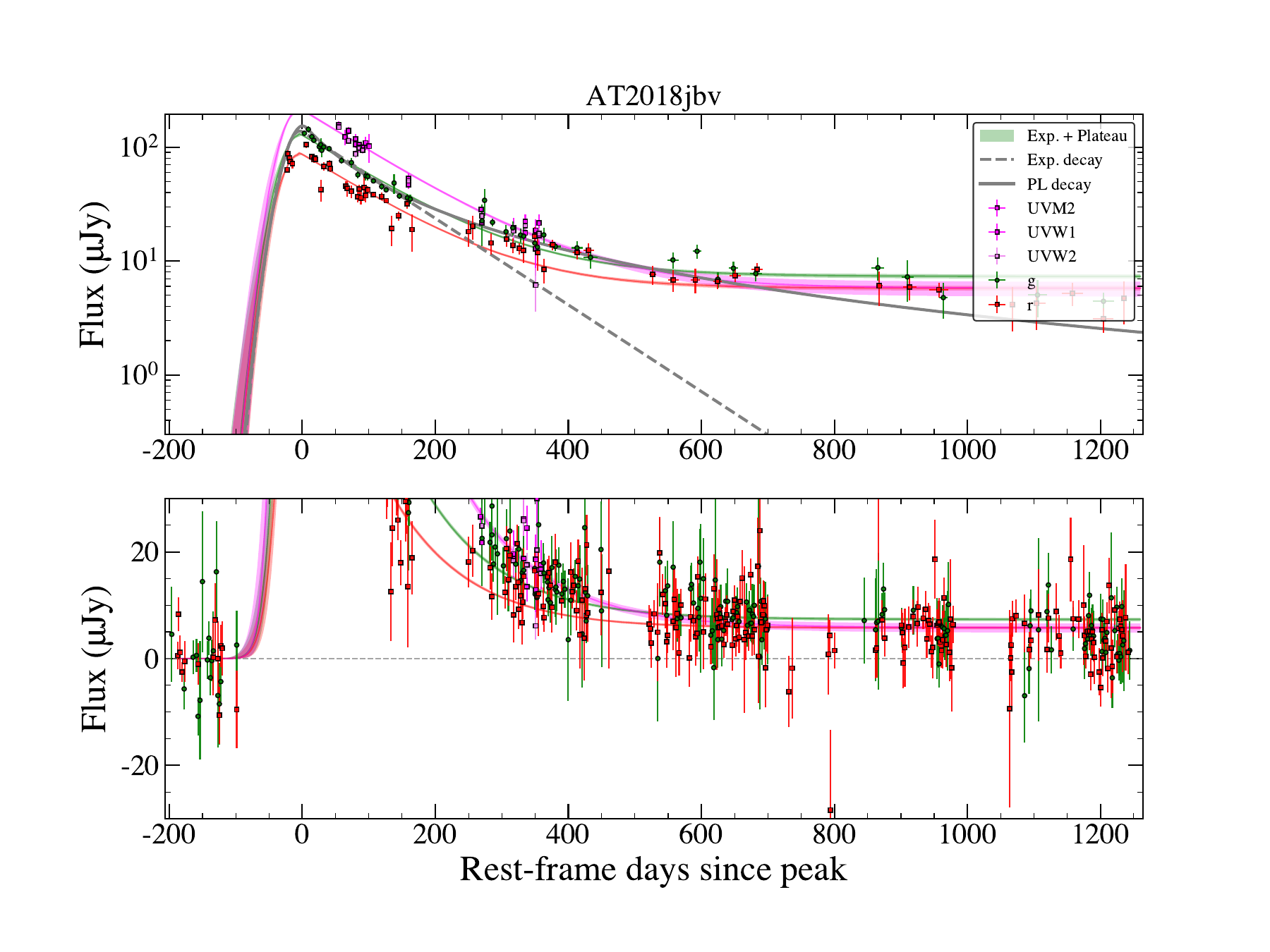}\quad
\includegraphics[width=.32\textwidth, clip=30 10 30 10, clip]{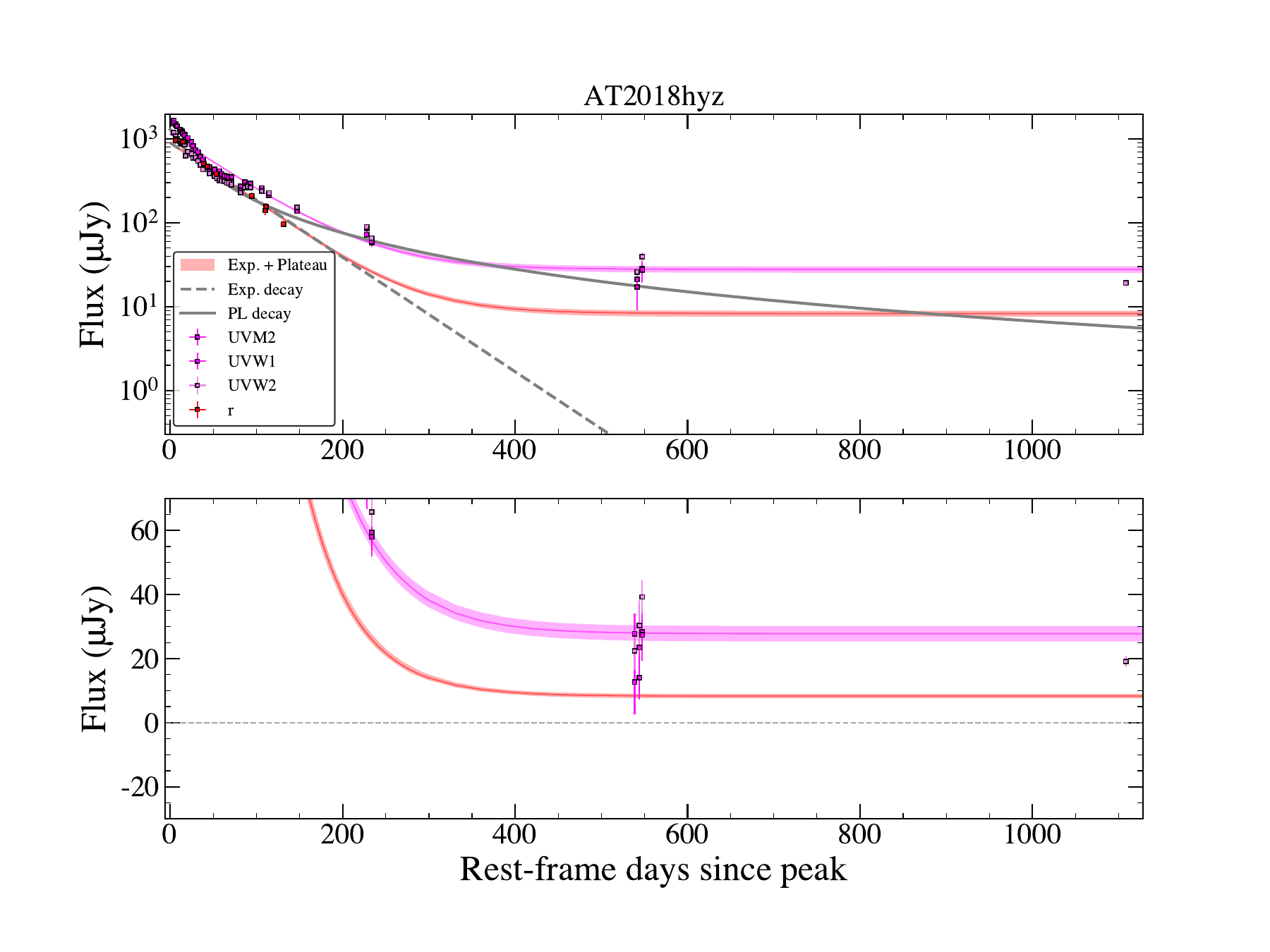}\quad
\includegraphics[width=.32\textwidth, clip=30 10 30 10, clip]{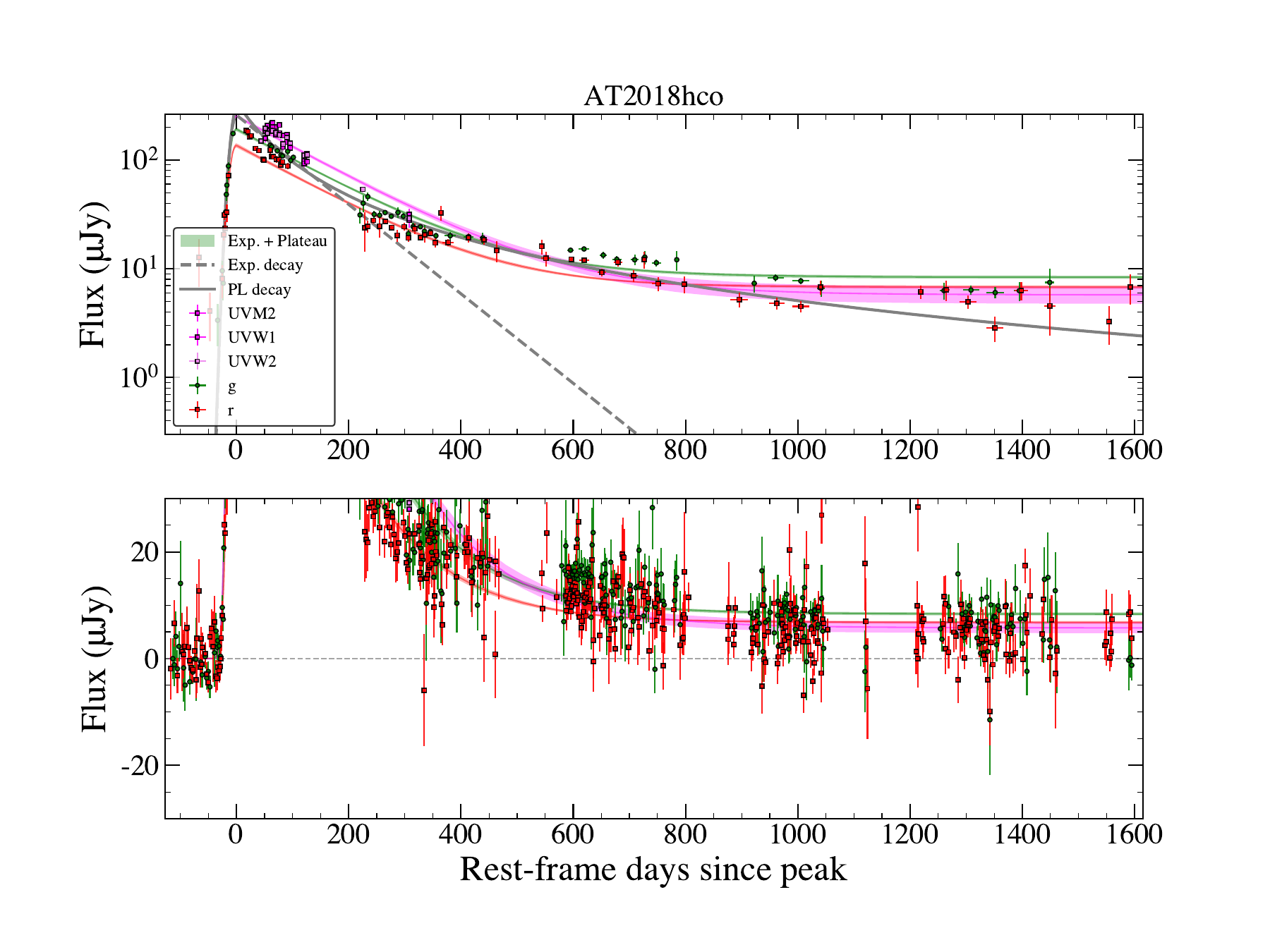}\\
\includegraphics[width=.32\textwidth, clip=30 10 30 10, clip]{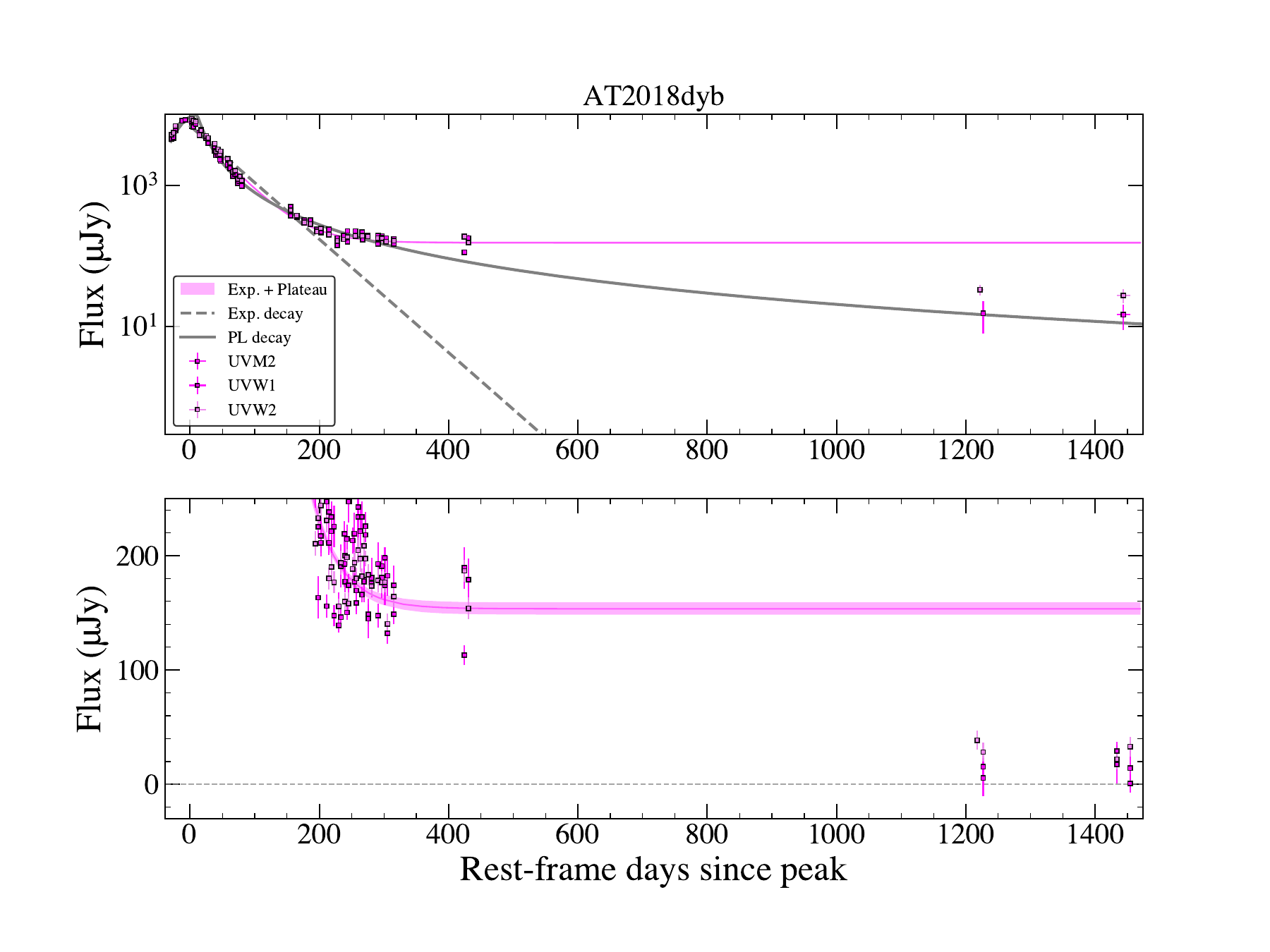}\quad
\includegraphics[width=.32\textwidth, clip=30 10 30 10, clip]{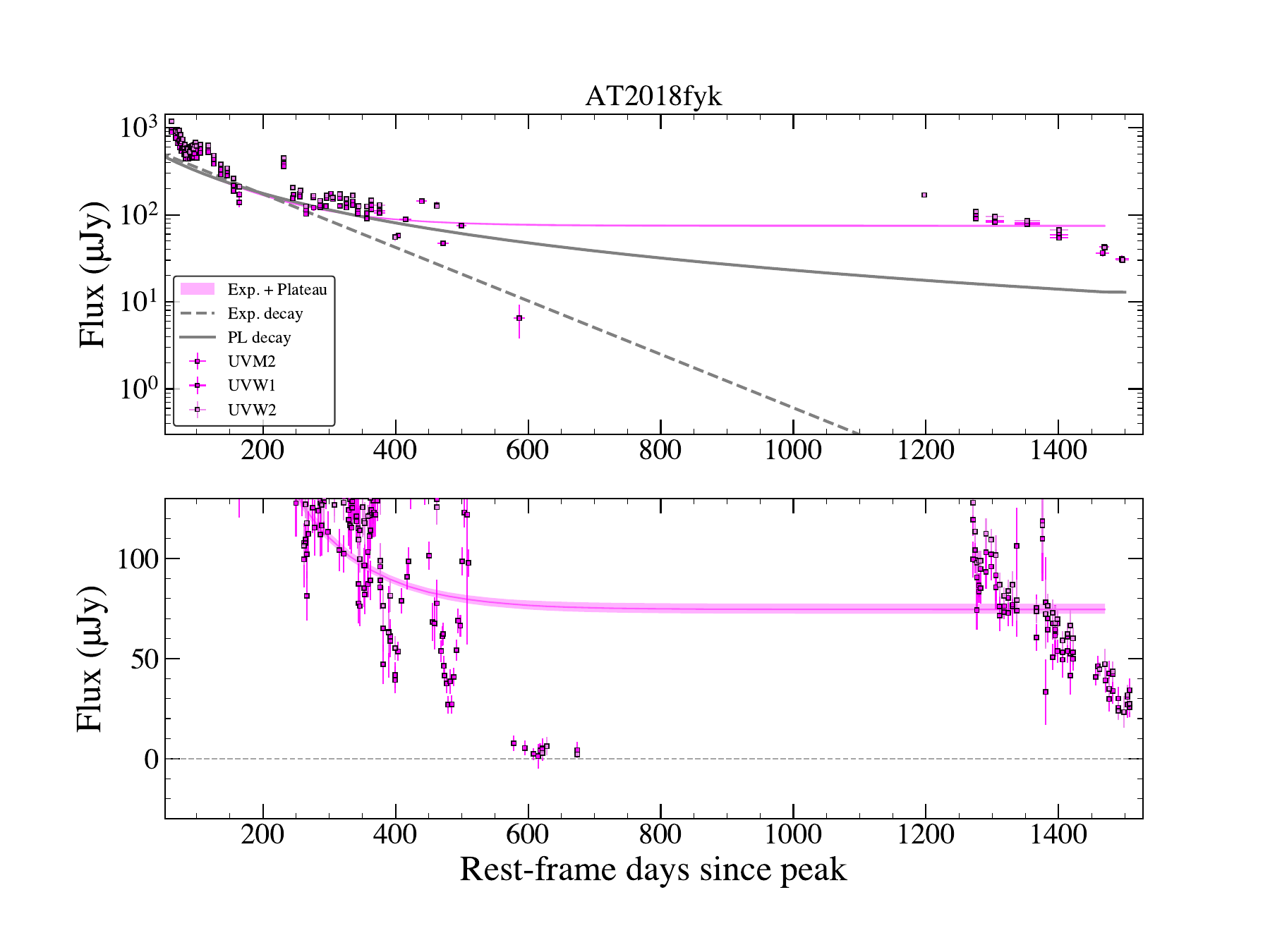}\quad
\includegraphics[width=.32\textwidth, clip=30 10 30 10, clip]{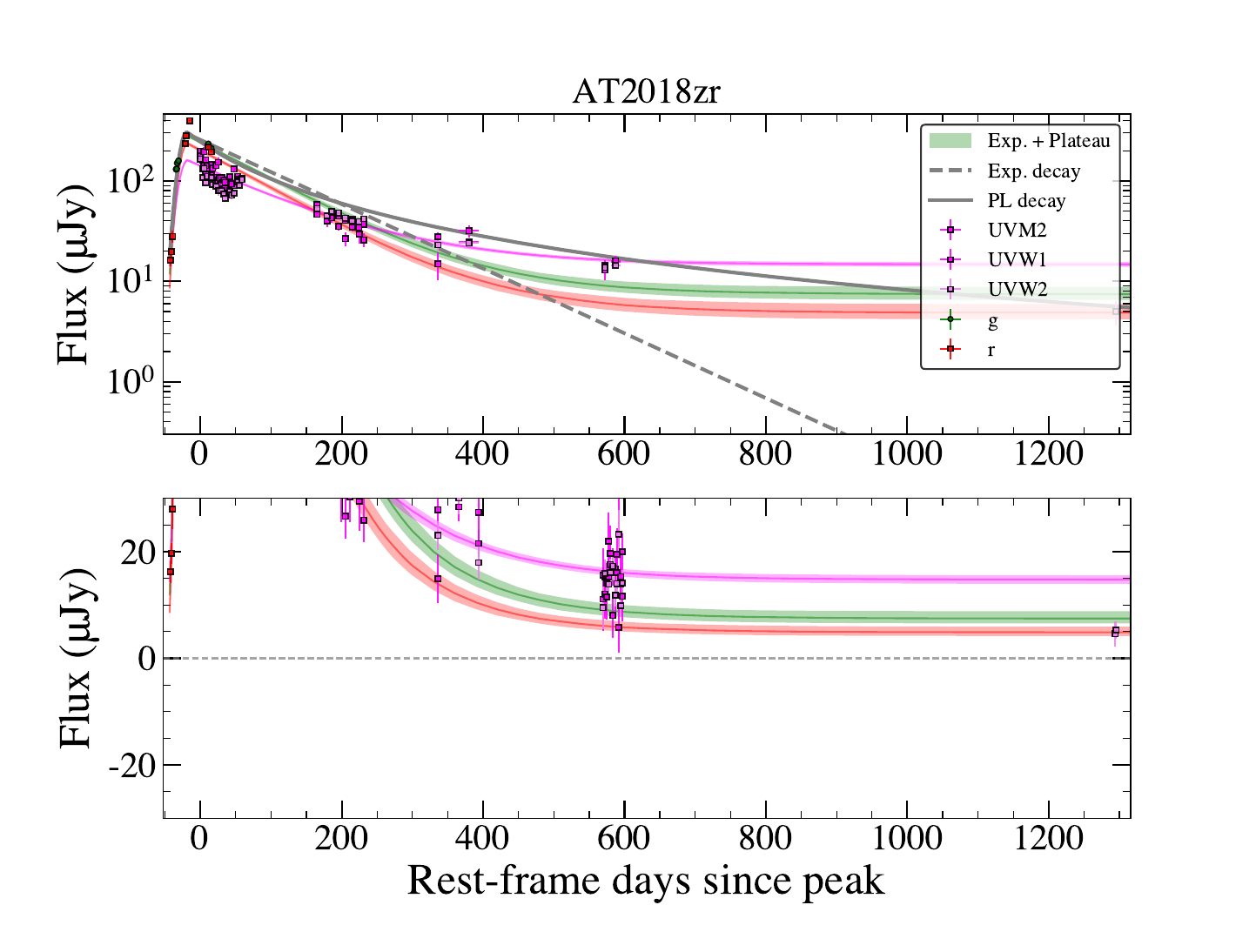}\\
\includegraphics[width=.32\textwidth, clip=30 10 30 10, clip]{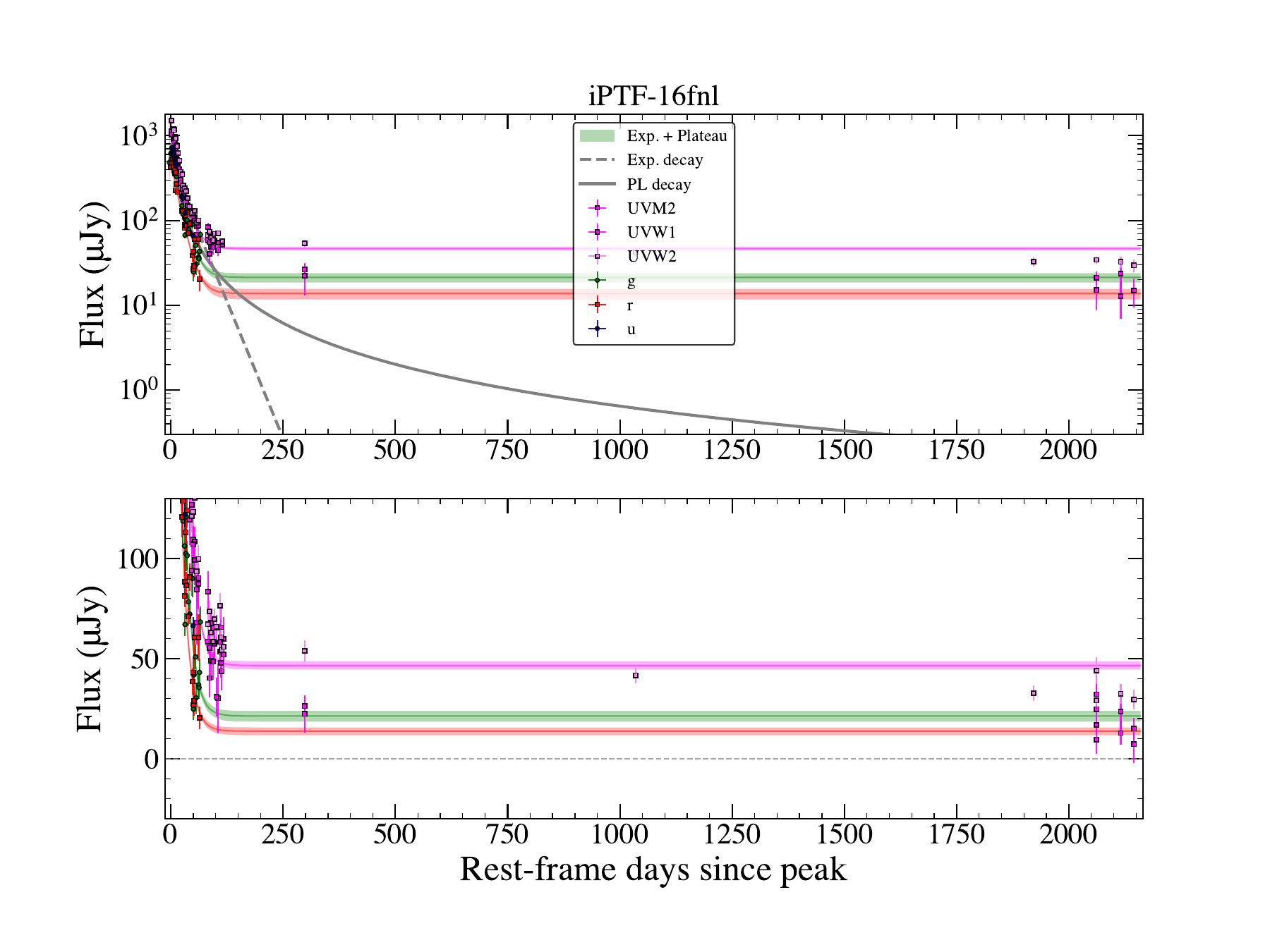}\quad
\includegraphics[width=.32\textwidth, clip=30 10 30 10, clip]{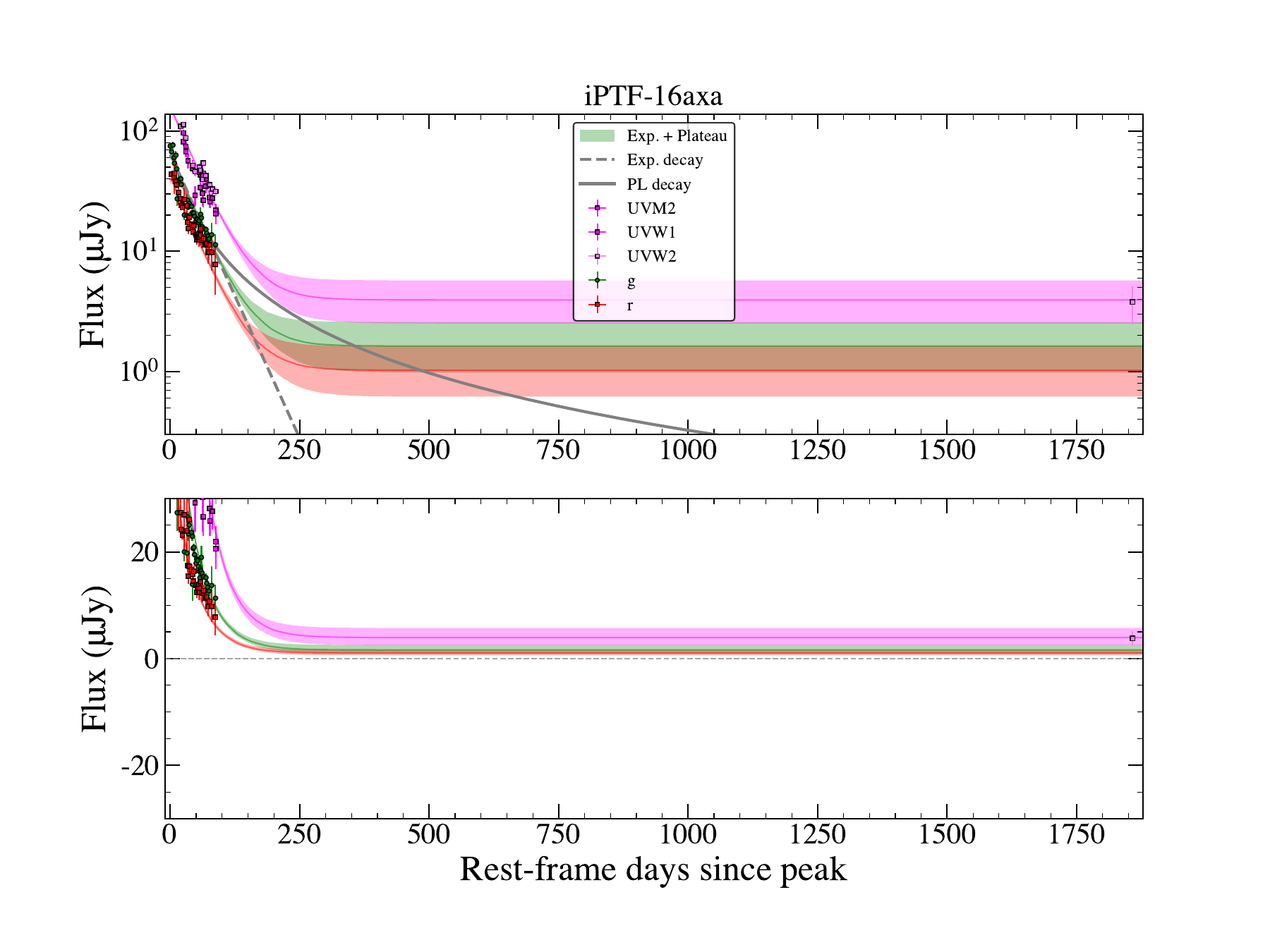}\quad
\includegraphics[width=.32\textwidth, clip=30 10 30 10, clip]{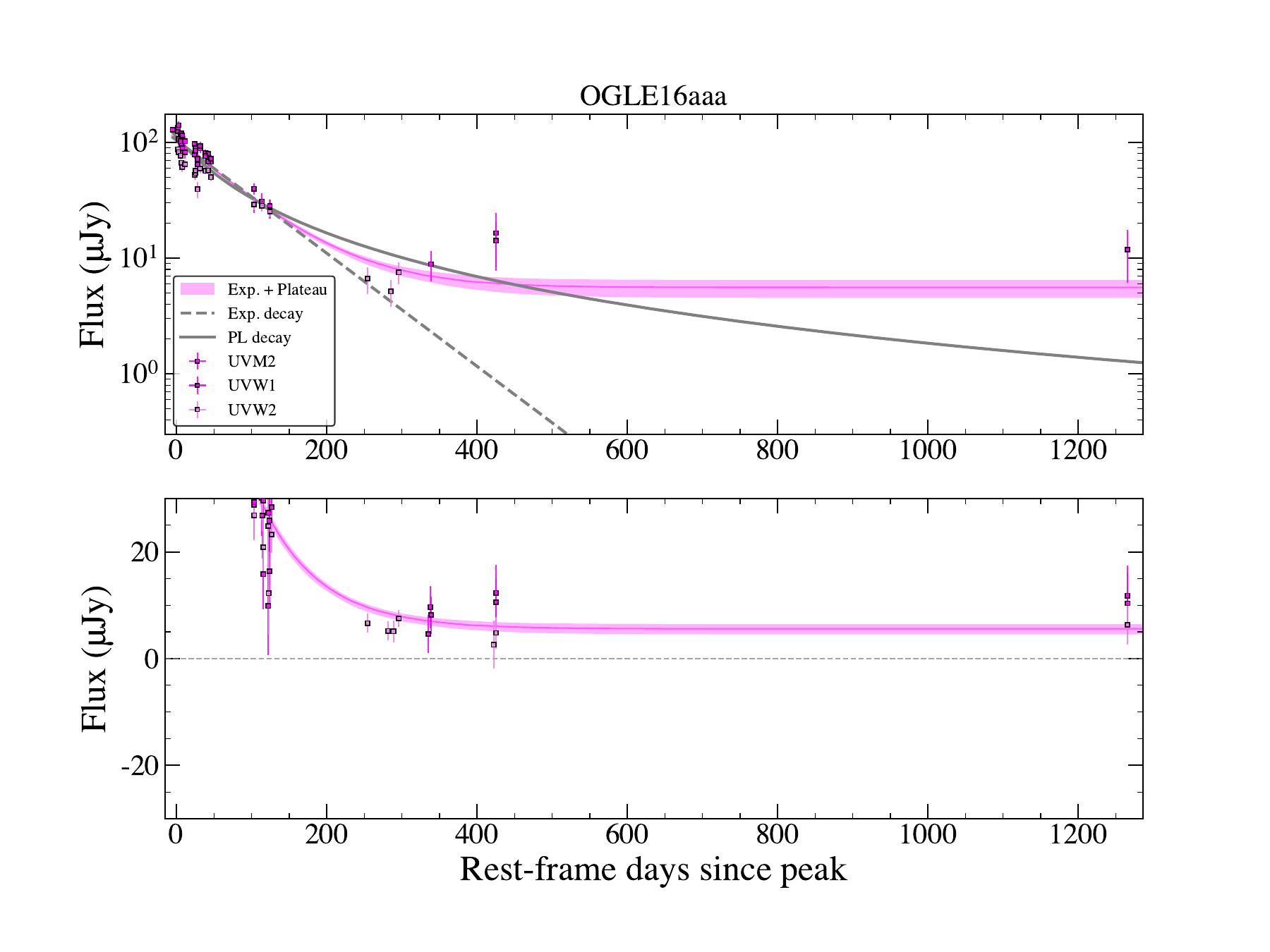}\\
\includegraphics[width=.32\textwidth, clip=30 10 30 10, clip]{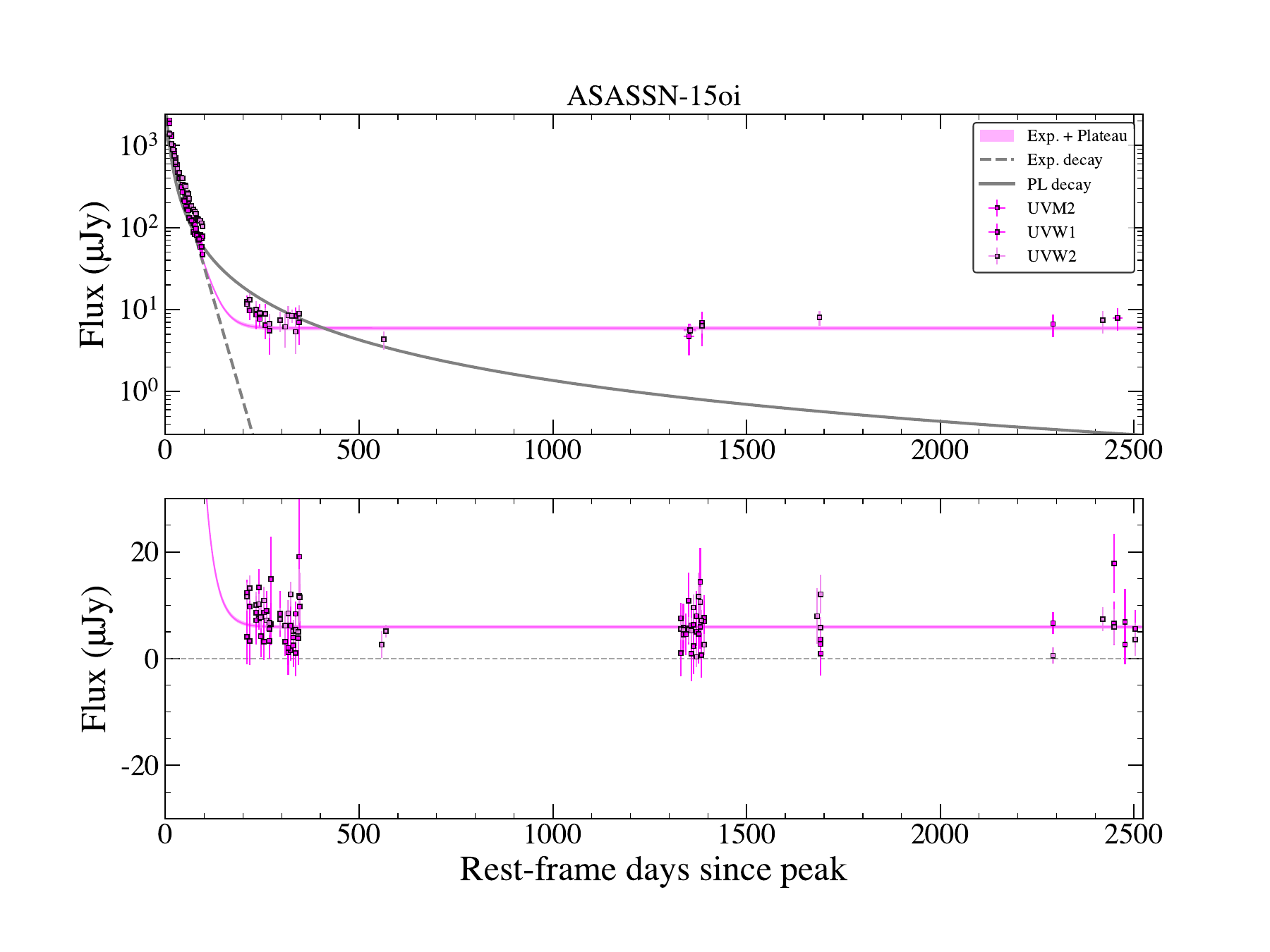}\quad
\includegraphics[width=.32\textwidth, clip=30 10 30 10, clip]{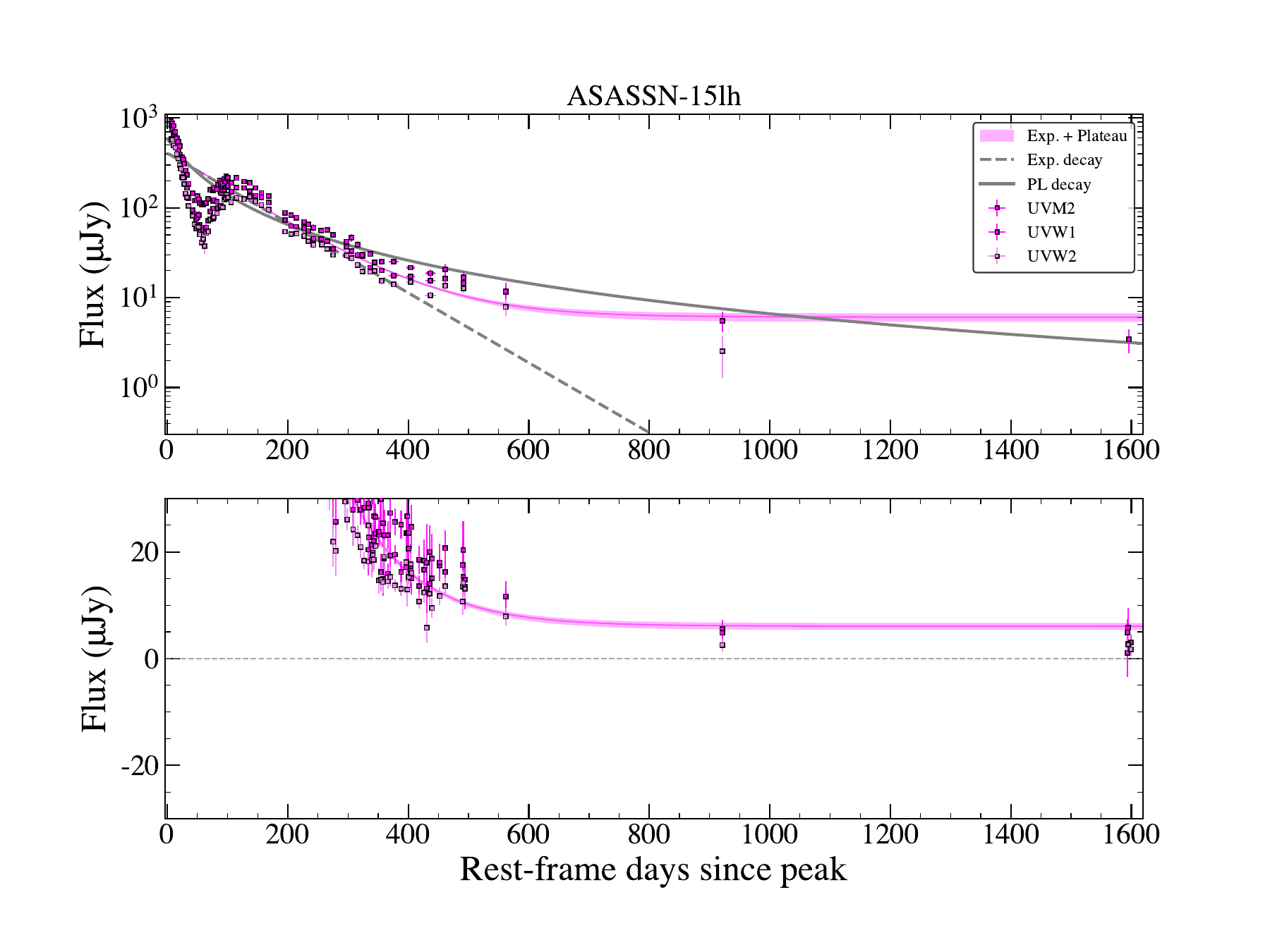}\quad
\includegraphics[width=.32\textwidth, clip=30 10 30 10, clip]{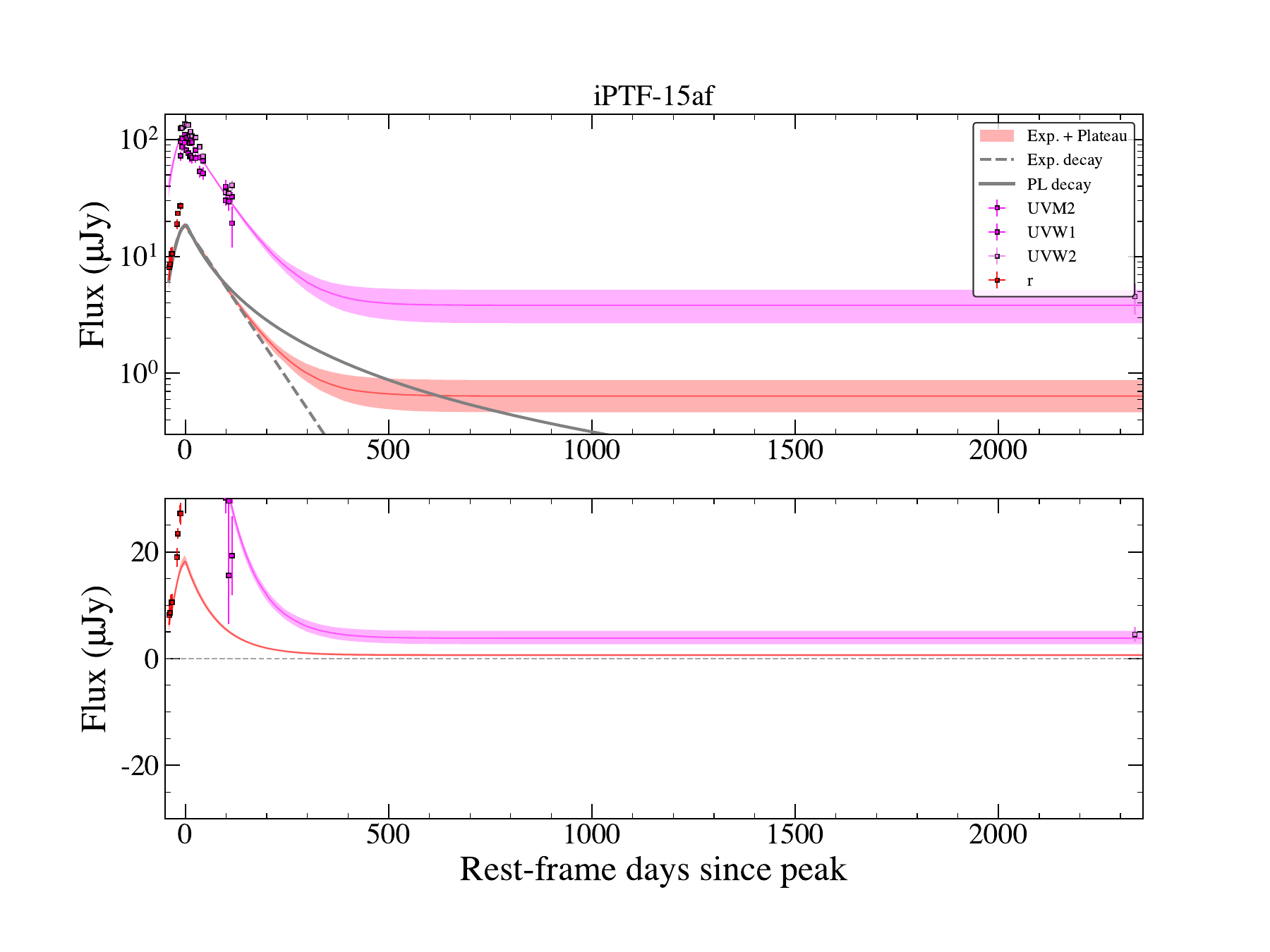}\\
\includegraphics[width=.32\textwidth, clip=30 10 30 10, clip]{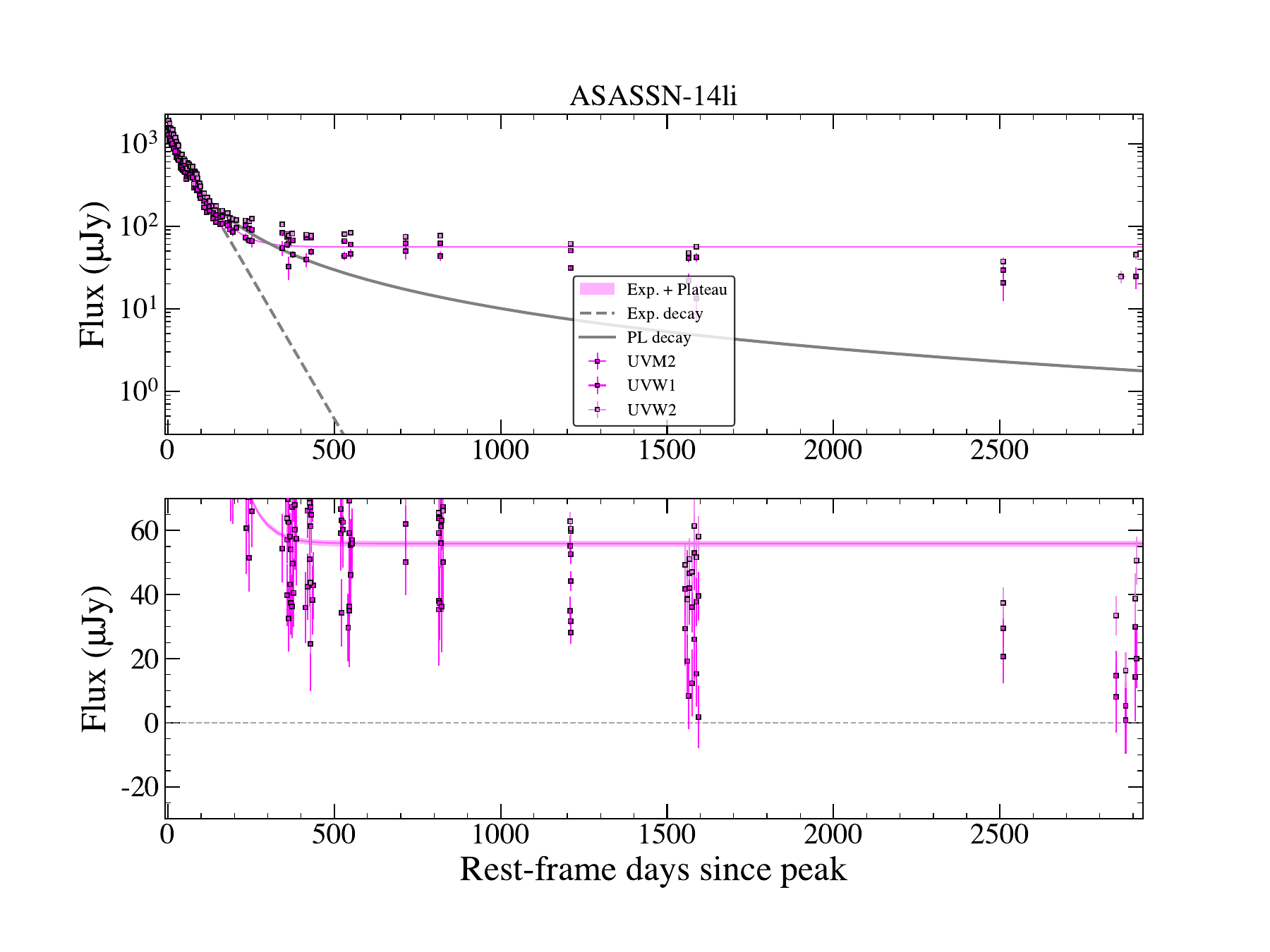}\quad
\includegraphics[width=.32\textwidth, clip=30 10 30 10, clip]{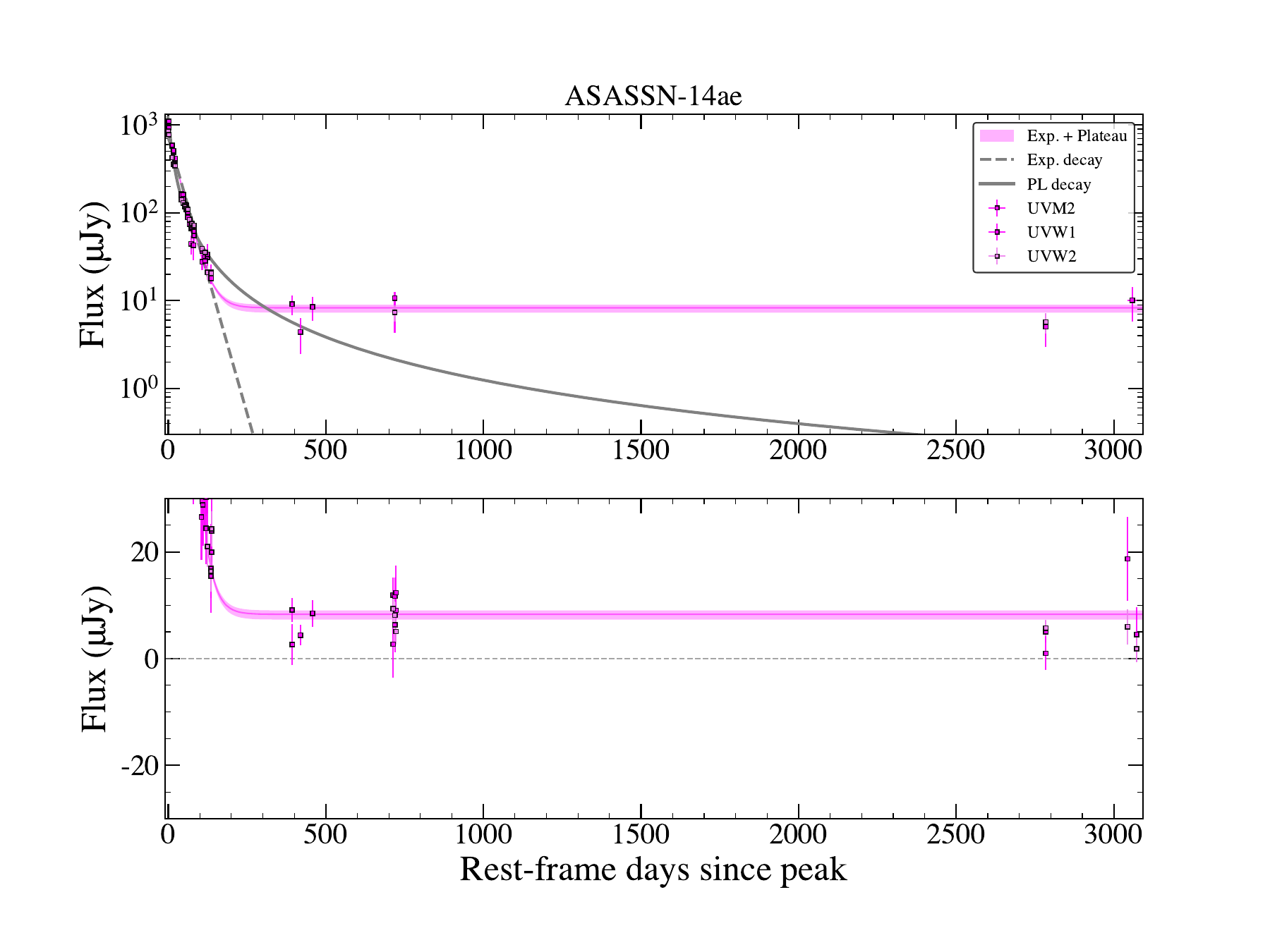}\quad
\includegraphics[width=.32\textwidth, clip=30 10 30 10, clip]{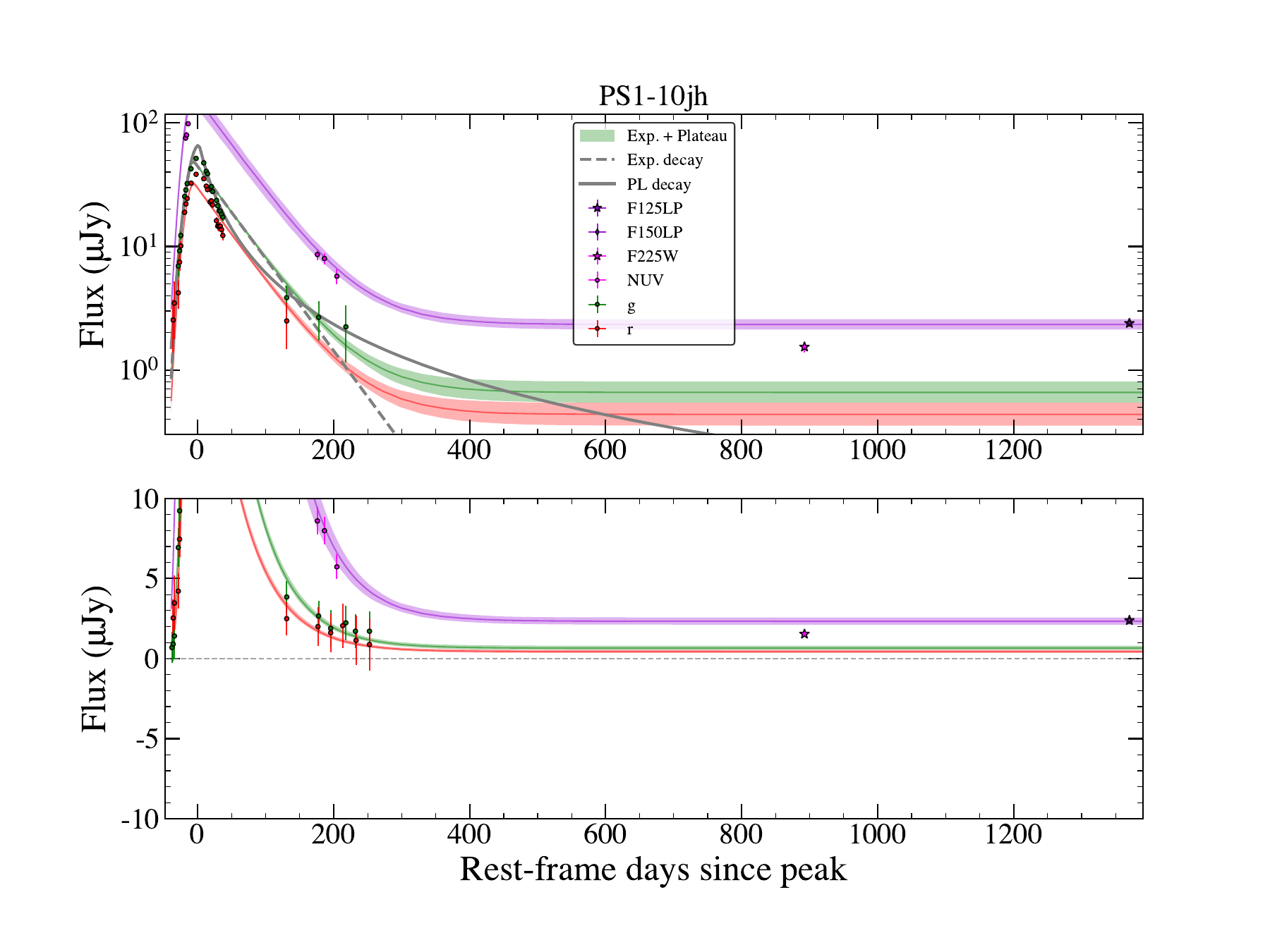}\\
\caption{Same as Fig.~\ref{fig:lcfits}.}
\end{figure*}

\begin{figure*}
\includegraphics[width=.32\textwidth, clip=30 10 30 10, clip]{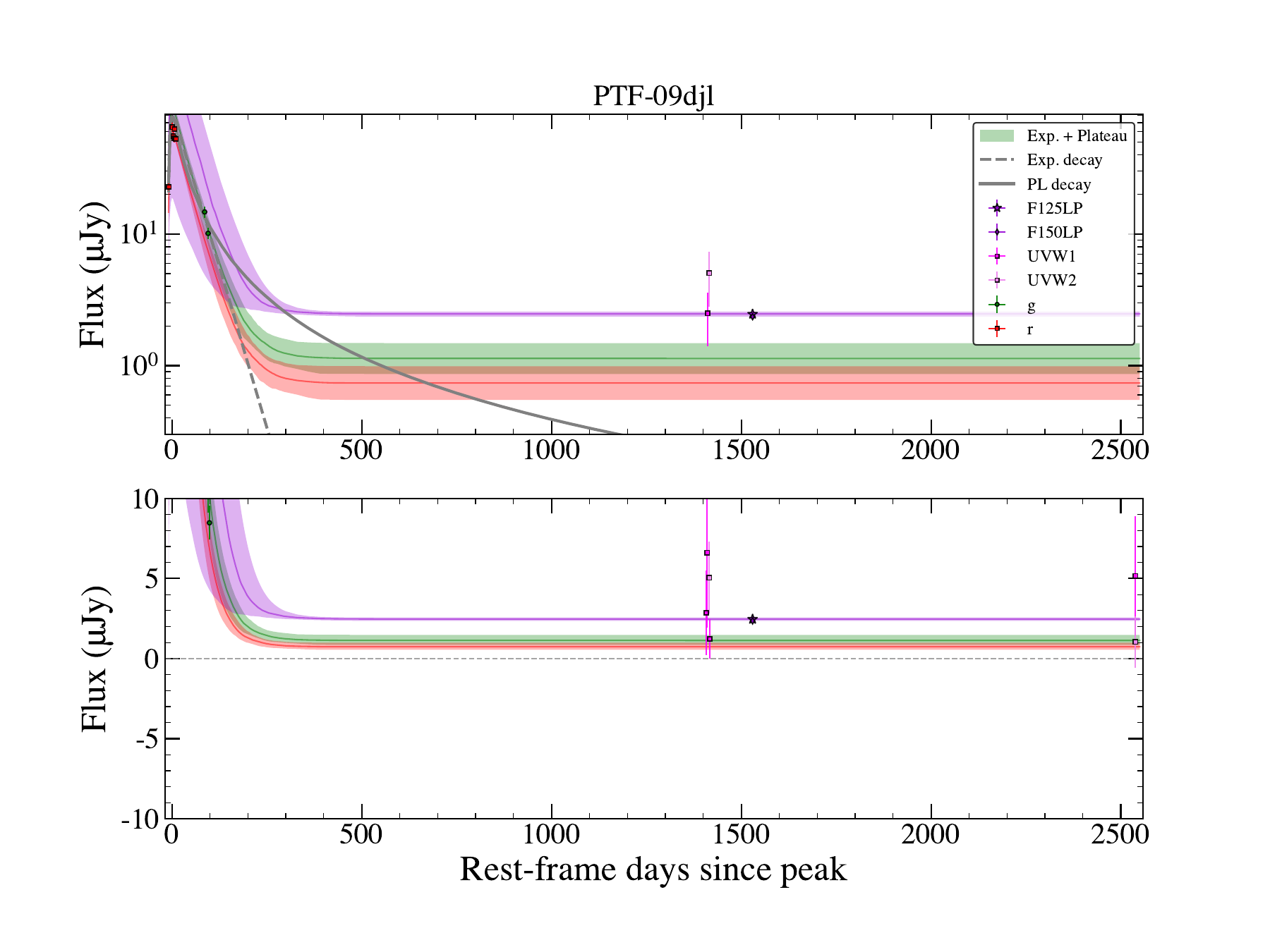}\quad
\includegraphics[width=.32\textwidth, clip=30 10 30 10, clip]{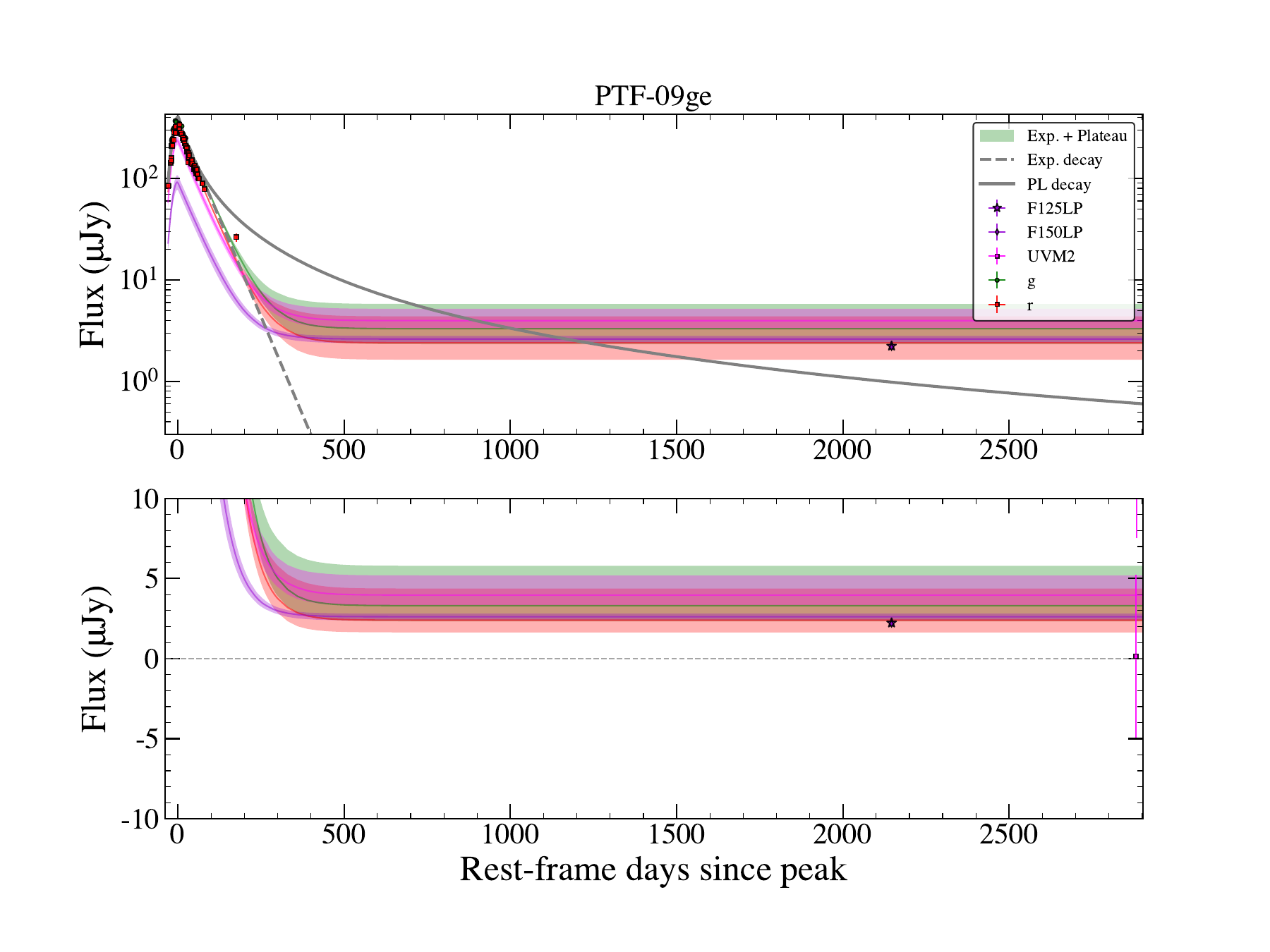}\quad
\includegraphics[width=.32\textwidth, clip=30 10 30 10, clip]{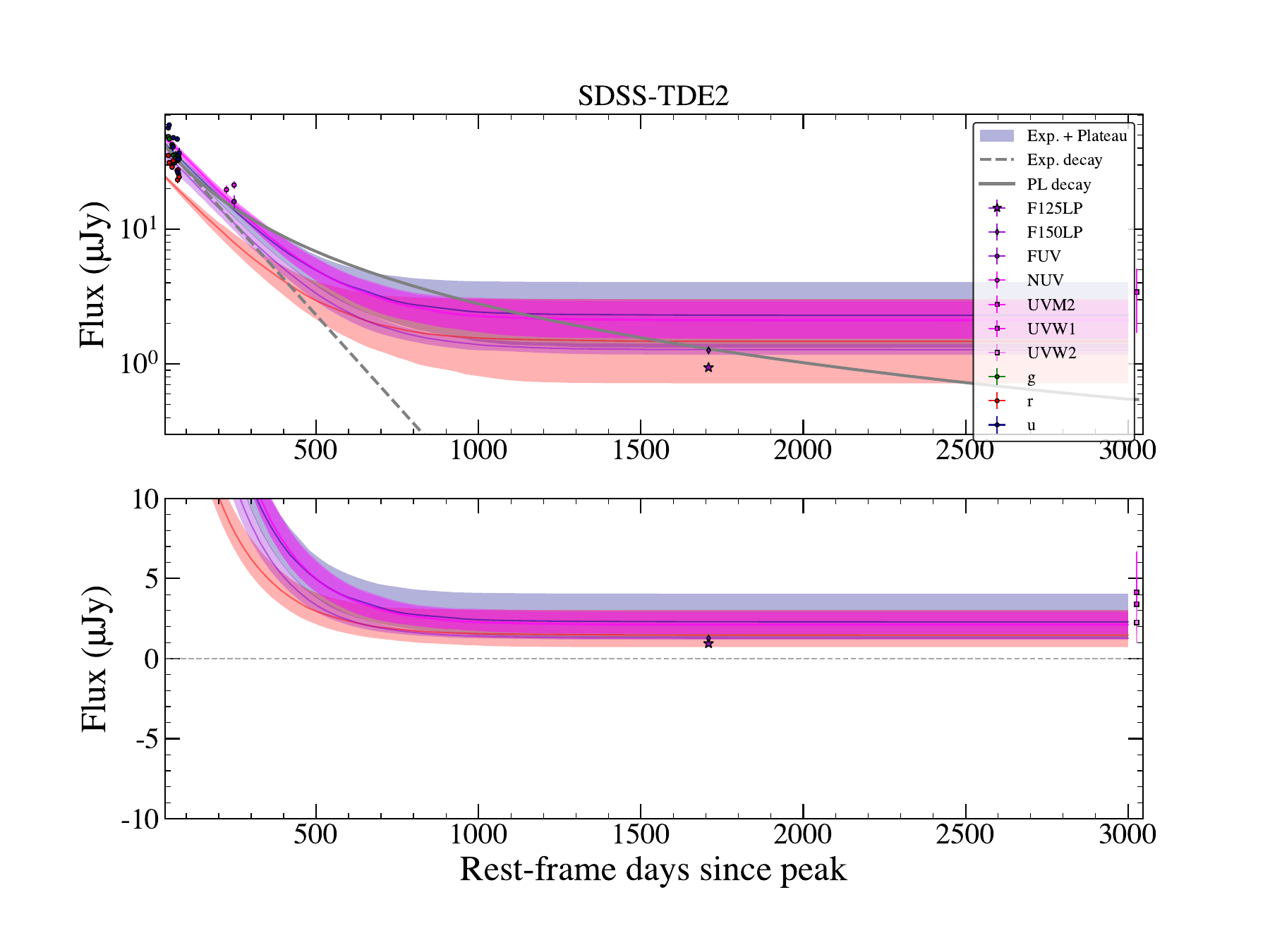}\\
\includegraphics[width=.32\textwidth, clip=30 10 30 10, clip]{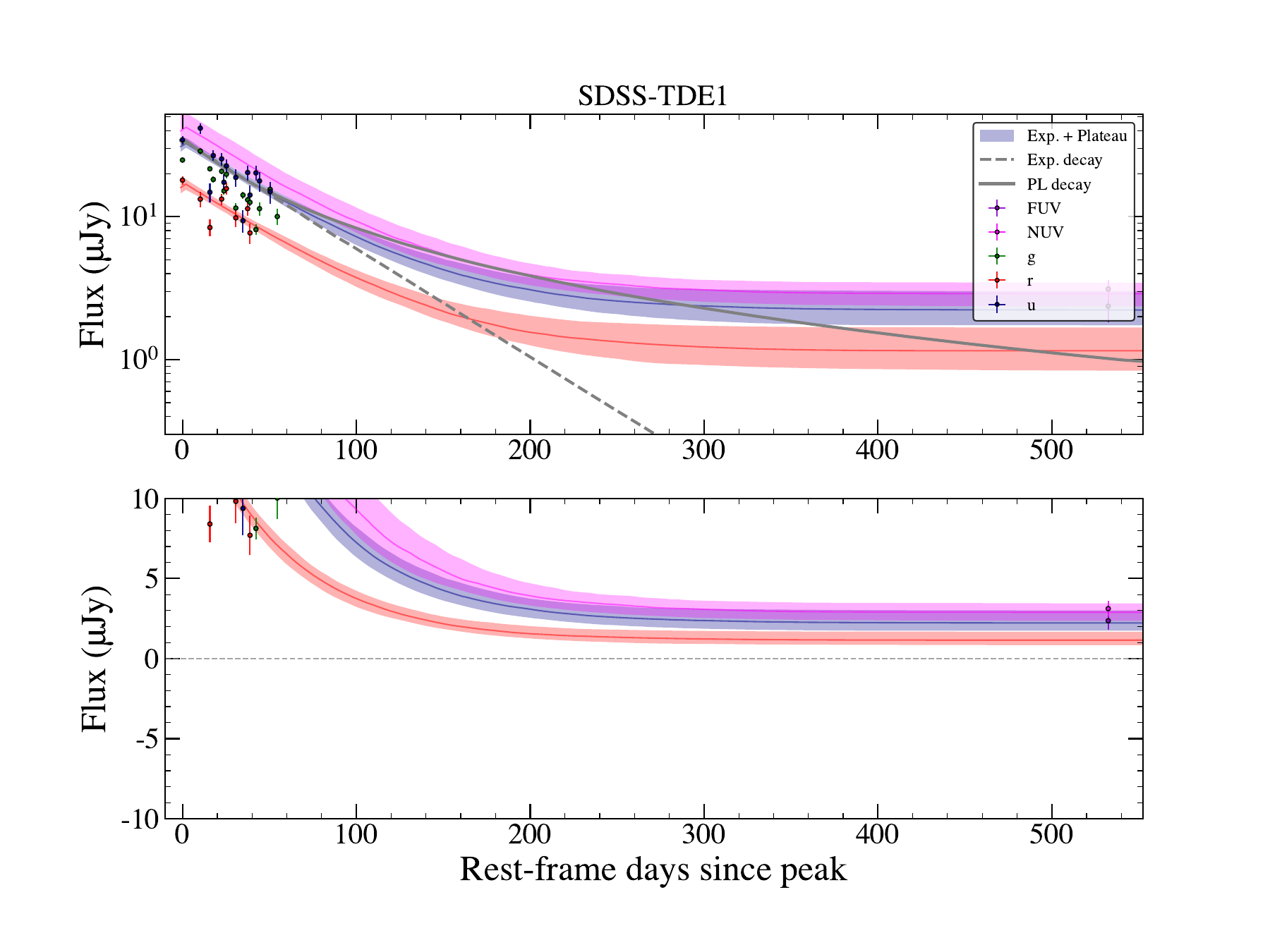}\quad
\caption{Same as Fig.~\ref{fig:lcfits}.}
\end{figure*}


\end{document}